\documentclass[12pt,preprint]{aastex}
\usepackage{graphicx}
\usepackage{epstopdf}
\usepackage{float}
\begin{document}

{ \centering Individual Elemental Abundances in Elliptical Galaxies\\
Abstract\\}

{\centering by Jedidiah Lee Serven, Ph.D.\\
 Washington State University\\
 August 2010\\}

 \vspace*{20 mm}

Using synthetic spectra to gauge the observational consequences of altering the abundance of individual elements, I determine the observability of new Lick IDS style indices designed to target individual elements. Then using these new indices and single stellar population models, I investigate a new method to determine Balmer series emission in a Sloan Digital Sky Surveys grand average of quiescent galaxies. I also investigate the effects of an old metal-poor stellar population on the near ultra violet spectrum through the use of these new indices and find that the presence of a small old metal-poor population accounts for discrepancies observed between index trends in the near UV and optical spectral regimes. Index trends for 74 indices and three data sets are presented and discussed. Finally, I determine the near nuclear line-strength gradients of 18 red sequence elliptical Virgo cluster galaxies for 74 indices.

\newpage
\pagestyle{plain}

{ \centering TABLE OF CONTENTS\\}
{\raggedleft{ Page\\}}
{\raggedright{

LIST OF TABLES \underline{\hspace{122mm} } 3\\
LIST OF FIGURES \underline{\hspace{119mm} } 4i\\

CHAPTER\\

{\hspace{5mm}}1. INTRODUCTION {\underline{\hspace{115mm} } 5\\

{\hspace{5mm}}2. THE OBSERVABILITY OF ABUNDANCE RATIO EFFECTS IN DYNAMICALLY 

{\hspace{10mm}}HOT STELLAR SYSTEMS {\underline{\hspace{98mm} } 9\\

{\hspace{5mm}}3. EMISSION CORRECTIONS FOR HYDROGEN FEATURES OF THE GRAVES ET

{\hspace{10mm}}AL. 2007 SLOAN DIGITAL SKY SURVEY AVERAGES OF EARLY TYPE, 

{\hspace{10mm}}NON-LINER GALAXIES {\underline{\hspace{100mm} } 13\\

{\hspace{5mm}}4. NH and Mg INDEX TRENDS IN ELLIPTICAL GALAXIES {\underline{\hspace{36mm} } 24\\

{\hspace{5mm}}5. THE EFFECTS OF VELOCITY DISPERSION ON INDEX TRENDS {\underline{\hspace{18mm} } 35\\

{\hspace{5mm}}6. LINE STRENGTH GRADIENTS IN VIRGO CLUSTER GALAXIES {\underline{\hspace{20mm} } 40\\

{\hspace{5mm}}7. RESULTS {\underline{\hspace{129mm}} 48\\

APPENDIX\\

{\hspace{5mm}}A. INDEX RESPONSE TABLE {\underline{\hspace{94mm} } 50\\

{\hspace{5mm}}B. INDEX VELOCITY DISPERSION TRENDS {\underline{\hspace{64mm} } 56\\

{\hspace{5mm}}C. VIRGO CLUSTER GALAXY INDEX GRADIENTS {\underline{\hspace{48mm} }  82\\}}

\newpage

{\centering LIST OF TABLES\\}

3.1 H$\alpha$ vs. Mg $b$ Line Fits {\underline{\hspace{110mm} } 16

3.2 Balmer Series Index Emission Corrections {\underline{\hspace{75mm} } 18

3.3 H$\alpha$ and H$\beta$ Z vs. Age Sensitivity Parameter (Zsp) {\underline{\hspace{60mm} } 21

4.1 NH and Mg Feature Index Definitions {\underline{\hspace{82mm} } 27

4.2 NH and Mg Indices Z vs. Age Sensitivity Parameter {\underline{\hspace{56mm} } 27

4.3 NH and Mg Index Response {\underline{\hspace{100mm} } 34

6.1 Virgo Cluster Galaxy Data {\underline{\hspace{102mm} } 42

6.2 Virgo Cluster Galaxy Data {\underline{\hspace{102mm} } 43

\newpage

{\centering LIST OF FIGURES\\}

2.1 Lick Style Index Illustration {\underline{\hspace{102mm} }10

3.1 Balmer Series Indices vs. Mg $b$ {\underline{\hspace{95mm} } 19

4.1 NH and Mg Index Element Sensitivities {\underline{\hspace{79mm} } 26

4.2 NH and Mg Index Age and Metallicity Sensitivities {\underline{\hspace{58mm} } 28

4.3 NH and Mg Index Observations {\underline{\hspace{93mm} } 30

6.1 Mg to Fe Ratio Plot {\underline{\hspace{115mm}} 45

6.2 H$\beta$ vs [MgFe] Age as a Function of Radius Plot {\underline{\hspace{67mm}} 46

\newpage

\pagestyle{plain}

{\centering CHAPTER ONE\\
INTRODUCTION\\}

The question of how galaxies form is a far from simple matter. Recent results seem to make it look simple, in the young universe (z $\sim$ 6) to the middle aged universe (z $\sim$ 1.5) the star formation rate for galaxies is approximately constant and then falls off (\cite{hart04} and references therein). However, a closer look at the properties of galaxies such as morphological type and stellar content reveals a complex evolutionary picture. Just some of the parameters that an evolutionary picture must fit are (1) spirals must have formed from an accretion disk of gas. Ellipticals, on the other hand, may have formed from any number of scenarios, including mergers of various numbers and types of progenitor galaxies. (2) There is a scatter in the mean ages of elliptical galaxies, which is less for larger galaxies \citep{wor07}. And, (3) there is a scatter in the elemental abundance ratios from galaxy to galaxy and a systematic enhancement in large ellipticals for light elements such as Mg, N and Na. The most likely cause of this element enhancement is a change in the ratio of Type {\sc I} to Type {\sc II} supernovae that are responsible for the chemical enrichment \citep{wor98}.

The chemical makeup of these stellar systems can be studied using an integrated light spectrum. These integrated-light spectra formed from the added contributions of all the stellar light captured by an observer in the absence of complications such as emission and dust screening are indicators of the ages and elemental abundances of the stars in that system. Stellar systems such as elliptical galaxies, SO galaxies, star clusters, and the bulges of spiral galaxies tend to be relatively free of gas and dust compared to disk galaxies (\cite{knapp99}; \cite{Oosterloo02}). These types of stellar systems are likely to contain small amounts of young stars, nebular gas, or dust making their spectra useful for comparison to stellar population models. For this thesis, a stellar population is defined as a single stellar population ( SSP ) of one age and composition, but all stellar masses. "Stellar populations" for a galaxy would be the weighted, collection of N-dimensional voxels, where the dimensions are age, average metallicity ($<$Z$>$), and the various individual metallicities $<$Z$_i$$>$.

Current models agree that age and abundance control the color and spectral feature strength of simple integrated-light systems (\cite{wor94}; \cite{bruzual03}; \cite{maraston05}). The "3/2 rule" of \cite{wor94b} states that if an age change is opposed by an abundance change such that $d$ log (age) = -(3/2)$d$ log (abundances), then very little change in colors, surface brightness fluctuation magnitudes, and absorption feature strengths will occur. \cite{wor94b} pointed out that this could be used to discover non-lockstep abundance ratio variations for individual elements, without having to know the absolute age or metallicity to very high precision if features could be found in the spectrum that were sensitive to such non-lockstep behavior.

\cite{wor92} presented models and observations that indicated that individual elemental abundances for individual galaxies could be found. This was shown by first plotting Mg and Fe absorption features for both models and observed elliptical galaxies. The models were for different ages and metallicity, and fell on a one-dimensional, highly degenerate trajectory where any reasonable change in age or metallicity simply moved an object back and forth along this line. The galaxies on the other hand showed considerable scatter about these model lines and showed that Mg features were stronger for larger galaxies. This indicated that there were nonsolar abundance ratios. The presence of this effect has been confirmed many times (\cite{wor98}; \cite{tan04}; \cite{thomas03}; \cite{proctor02}; \cite{maraston03}). 

\cite{trager00a} introduced the idea of using synthetic spectra to simulate the effects of elemental abundance ratio changes in integrated-light models. The technique involves generating many models of varying age, overall metallicity and individual abundance ratios then comparing those models with an observed galaxy to find a best fit model. This is commonly done by comparing measurements of  spectra indices and choosing the model that best fits the observed galaxy's index measurements. This best fit model is then used  to determine the age, metallicity and various elemental abundances. This technique is in common use today by those studying formation histories of elliptical galaxies.

\cite{ser05} employed this technique of using synthetic spectra to simulate the effects of elemental abundance ratio changes to expand the possible use of model index fitting to as many as 23 elements by using a crude model galaxy consisting of stellar models for a turnoff dwarf and a giant then combing them using equal contributions from the turnoff dwarf and giant at 5000 \AA\ to simulate a 5 Gyr stellar population. Then, by comparing this model with itself except with the abundance of a given element doubled, for example, the abundance of Cr would be increased by 0.3 dex, new indices were defined that were designed to be sensitive to this element .

The rest of this thesis explores the application of this synthetic spectra technique to determine elliptical galaxies characteristics such as ages, elemental abundances and elemental distributions.

Chapter 2 presents the \cite{ser05} expansion of index definitions to cover 23 elements and the results of that work.

Chapter 3 illustrates the use of index measurements of the hydrogen Balmer series and a new technique to determine emission corrections for the \cite{graves07} grand averages of elliptical galaxies form the Sloan Digital Sky Survey (SDSS).

Chapter 4 explores the use of these targeted indices to investigate discrepancies between index trends in indices in the near-ultraviolet and optical spectral regimes.

Chapter 5 presents index trends for 74 indices and three data sets.

Chapter 6 presents the near nuclear line-strength gradients for 74 indices and 18 Virgo Cluster galaxies.

\newpage
\pagestyle{plain}

{\centering CHAPTER TWO\\

The Observability of Abundance Ratio Effects in Dynamically Hot Stellar Systems\\}

\section{Introduction}
This chapter represents an update of \cite{ser05}. The purpose of that paper was to investigate the possibility of determining the abundances of C, N, O, Na, Mg, Al, Si, S, K, Ca, Sc, Ti, V, Cr, Mn, Fe, Co, Ni, Cu, Zn, Sr, Ba, and Eu in real galaxies by using synthetic spectra to define absorption feature indices sensitive to the individual abundances of these 23 elements. The methods used for that investigation are discussed in \S 2. Then, \S 3 discusses the updates used in this work.

\section{Analysis}
In \cite{ser05} to find places in the spectrum that show a noticeable response to the abundance change of an individual element, a "model" galaxy was constructed by combining two stellar synthetic spectra. One was the synthetic spectrum of a giant (T$_{eff}$ = 4250, log g = 1.90) and the other main sequence turnoff dwarf (T$_{eff}$ = 6200, log g = 4.10). These two spectra were then combined with equal contributions at 5000 \AA\/. The composite spectrum was then broadened to a simulated line-of-sight velocity field that is a Gaussian distribution with a width ($\sigma$) of 200 km s$^{-1}$ to simulate an elliptical galaxy a little more massive than the Milky Way. 

First this was done for scaled solar metallicity, then the same model was computed with the exception that an individual elemental abundance was increased by 0.3 dex. This was done for 22 of the 23 elements. The exception was C, which was increased by 0.15 dex due to the fact that for a 0.3 dex increase in C the C/O ratio approaches 1 and thus drastically changes the structure of the star toward that of a carbon star. Finally to locate places of significant response the ratio of an enhanced galaxy and the scaled solar galaxy was taken. This produced a spectrum where changes due to the element under study would show up as deviations from one in the ratio.

After finding these areas of spectral response, Lick IDS (Image Dissector Scanner) style indices were defined consisting of a central passband that spans the absorption feature itself and two continuum passbands, one on the blue side of the central band and one to the red. From these three passbands one can find a pseudo-equivalent width for any feature of interest. To do this one first constructs a pseudo-continuum by "drawing" a line between the wavelength midpoints and average flux values of the blue and red passbands. Then by integrating the difference between the pseudo-continuum and the galaxy flux over the central or index passband one gets an equivalent width characterizing the feature of interest (See Figure \ref{fig1}).

\renewcommand{\thefigure}{2.\arabic{figure}}
\begin{figure}[H]
\includegraphics[width=3in]{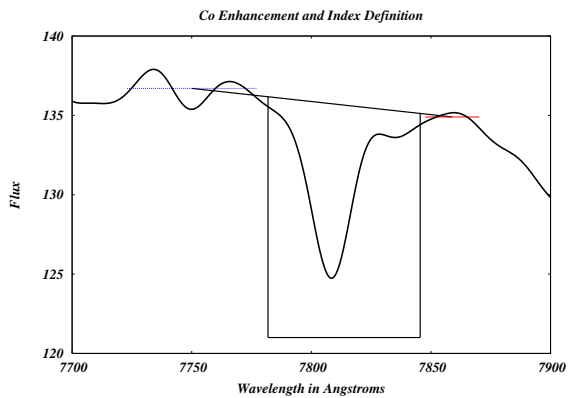}
\caption{An example of a Lick IDS style index is plotted. The blue continuum passband (blue), the red continuum passband (red), the central passband represented by the base of the trapezoid (black) and a model spectrum (black) are plotted. The pseudo-continuum line connecting the centers of the blue and red continuum passbands, which forms the top of the trapezoid (black) is also plotted. The area within the trapezoid and bound by the spectrum represents the difference between the pseudo-continuum and the galaxy flux. Integrating this difference gives the equivalent width characterizing the feature of interest.
\label{fig1} }
\end{figure}

Using this method \cite{ser05} defined 53 new indices for these 23 elements. The feasibility of measuring these elements using these indices was checked by determining index values and errors of these models and assuming a photon error such that S/N = 100 around 5000 \AA\ and proportional to F$_{\lambda}^{1/2}$. They determined that of the 23 elements 18 looked like they could be measured and 5 ( S, K, Cu, Zn, and Eu ) looked difficult to determine.

\section{Discussion}
\cite{ser05} then went on to determine the spectral response of these 53 new indices and 37 previously defined indices to these 23 elements shedding light on the influence any particular element may have on any given index. They summed up this work in two large tables giving the index response in angstroms of equivalent width. Since the rest of this paper will often be referring to these indices and their sensitivities those two tables have been recreated in the form of 6 tables located in Appendix A. In these tables the 1st column of the table gives the indices's names, the 2nd gives the index measurements ($I_o$) in angstroms of equivalent width, and the 3rd giving the error ($\sigma$) associated with S/N = 500 at 5000 \AA\/. The next 23 columns split between tables are the changes (enhanced minus unenhanced) in the index due to the elemental enhancement 0.3 dex (0.15 dex for C) of the element labeling the column in units of angstroms. The 24th column is the change brought about in the index by increasing all the alpha elements by a factor of 2.

One major difference between these two works is that some of the indices that are not used in this work have been left out. The index definitions that are used include the original 25 lick indices \citep{wor94,wor97}, 1 H$\alpha$ index defined in \cite{Cohen}, the 53 indices defined in \cite{ser05}, 2 indices defined in \cite{Davidage94} and 3 defined in chapter 3.

The other major difference is that the numbers tabulated here come from measuring these indices in full single stellar population models. These models are a version of the \cite{wor94} \& \cite{trag} models were that used a grid of synthetic spectra in the optical \citep{lee} in order to investigate the effects of changing the detailed elemental composition on an integrated spectrum. These models where then used to create synthetic spectra at a variety of ages and metallicities for single-burst stellar populations. The underlying isochrones for this paper were the \cite{wor94} isochrones, because they allow us "manual" HB (Horizontal Branch)morphology control. However, there are certain caveats to using these isochrones. Specifically the models are a bit crude by today's standards and the ages are about 2 Gyr too old, so that 17 Gyr should really be interpreted as 15 Gyr. Other isochrone sets were used to check the results.

For this thesis new index fitting functions were generated. The data sources include a variant of the original Lick collection of stellar spectra \citep{wor94b} in which the wavelength scale of each observation has been refined via cross-correlation, as well as the MILES (Medium-resolution Isaac Newton Telescope library of empirical spectra) spectral library \cite{san} with some zero point corrections, and the Coud{\'e} Feed Library (CFL) of \cite{valdes}. The CFL was used as the fiducial set, in the sense that any zero point shifts between libraries were corrected to agree with the CFL case. The MILES and CFL spectra were smoothed to a common Gaussian smoothing corresponding to 200 km s$^{-1}$. The rectified-Lick spectra were measured and then a linear transformation was applied to put it on the fiducial system.

Multivariate polynomial fitting was done in five overlapping temperature swaths as a function of $\Theta$$_{eff}$ = 5040/T$_{eff}$, log g, and [Fe/H]. The fits were combined into a lookup table for final use. As in \cite{wor94}, an index was looked up for each``star" in the isochrone and decomposed into ``index" and ``continuum" fluxes, which added, then re-formed into an index representing the final, integrated value after the summation. This gives us empirical synthetic spectra when variations in chemical composition are needed. The grid of synthetic spectra is complete enough to predict nearly arbitrary composition. For the remainder of this paper whenever the term model is used it is to these models that we are referring as they are the backbone of methods used in the following chapters.

\newpage
\pagestyle{plain}

\renewcommand{\thefigure}{3.\arabic{figure}}

{\centering CHAPTER THREE\\

Emission Corrections for Hydrogen Features of the Graves et al. 2007 Sloan Digital Sky Survey Averages of Early Type, Non-LINER Galaxies\\}

\section{Introduction}
One of the more important factors in determining the age of a stellar population such as an elliptical galaxy is being able to measure the hydrogen Balmer series \citep{wor94,oconnell76}. The first two Balmer lines H$\alpha$ and H$\beta$ are primarily used to determine the age of a stellar population because of their relative insensitivity to changes in metallicity (\cite{serv05,korn05}) making a measurement of these two lines a more reliable estimator of age then the higher order Balmer lines H$\gamma$ and H$\delta$, which are sensitive to changes in metallicity \citep{serv05,korn05} making a knowledge of the relative abundances of elements in a given spectra necessary in order to get a consistent agreement of ages from all the Balmer lines.

Another important factor in determining the age of a galaxy is hydrogen emission, which can give the impression of a much older galaxy by filling in and weakening the diagnostic Balmer absorption features. In the past, an emission correction is used that is based on the O{\sc II} or O{\sc III} emission lines \citep{schiavon06,gonzalez93} in which the difference in the measured equivalent width (EW) of a hydrogen line such as H$\alpha$ or H$\beta$ due to emission is some fraction of the EW of O{\sc II} or O{\sc III}. However, the correlation between hydrogen and oxygen emission strengths is weak, and indeed there is no astrophysical reason why the two should be tightly correlated. On the other hand, being able to use hydrogen recombination lines, which have ratios that are nearly fixed with respect to themselves \citep{oster} should give a more reliable result.

Toward that end, we compare 13 high-quality spectra of Virgo cluster early-type galaxies (Serven \& Worthey, in preparation) with a sample of SDSS spectra from \cite{graves07}.

The Virgo Cluster elliptical galaxies consist of 13 red galaxies with colors $0.75 < $B$-$V$< 0.97$. They also have a high S/N from S/N=150 to S/N=450 per pixel, are well fluxed using standard star flux calibrations, and have tight control over instrument resolution and velocity dispersion. Long slit spectra were obtained using the T2KB chip on the Cassegrain spectrograph on the Kitt Peak Mayall 4 meter telescope (2006 Jan 31 - Feb 5). A resolution of 1.4 \AA\ per pixel was selected to adequately sample the velocity dispersions of the program galaxies, which range from 50 to 350 km s$^{-1}$. The wavelength range covers from 3200 \AA\ to 7500 \AA\ in two wavelength swaths.

The \cite{graves07} sample was constructed from spectra of 22,501 SDSS galaxies that fall into the redshift range 0.06$< z <$0.08 and whose colors meet the criteria $^{(0.1)}$(g-r) $>$ -0.025($^{0.1}$M$_{r}$ - 5log h ) + 0.42, with $h = 0.70$ \citep{yan06} placing them firmly within the red sequence. These spectra are then further divided into those with H$\alpha$ and O{\sc II} emission lines and those without. Those with emission lines and high [O{\sc II}]/H$\alpha$ ratios were said to be "LINER-like" and those without "quiescent" where LINER stands for Low Ionization Nuclear Emission-line Region. The high [OII]/H$\alpha$ ratios are defined in \cite{yan06} as those with (EW([O{\sc II}])  $>$ 5EW(H$\alpha$) $-$ 7).

\cite{graves07} found that the [O{\sc II}] equivalent width distribution of their quiescent sample has a standard deviation of $\sigma$ = 1.56 \AA\/. In part to reduce the impact of outliers which may contain emission just below the detection threshold, they produced a random sub-sample of 2000 quiescent galaxies selected to conform to a Gaussian distribution centered on an [O{\sc II}] equivalent width of 0 and truncated at $\pm2\sigma$[O{\sc II}]. They then divided these 2000 galaxies into the following velocity bins: 70$< \sigma<$120 km s$^{-1}$, 120$< \sigma < $145 km s$^{-1}$, 145$< \sigma <$165 km s$^{-1}$, 165$< \sigma < $190 km s$^{-1}$, 190$< \sigma < $220 km s$^{-1}$, and 220$< \sigma < $300 km s$^{-1}$. For uniformity in comparison, each spectrum in every bin was smoothed to an equivalent velocity dispersion of $\sigma$ = 300 km s$^{-1}$, the highest velocity dispersion in the sample.  Finally, they coadded all of the individual spectra in each bin to produce six composite spectra corresponding to quiescent galaxies with different original dispersions. Along with these spectra the S/N at each resolution element was computed to produce an error spectrum for each of the six composite spectra. Factors that may contribute to the error spectrum estimate are age, individual abundances, and emission signal. The error spectrum is likely to be dominated almost entirely by measurement uncertainty given the poor S/N of the individual SDSS spectra.

We adopt the six composite spectra from \cite{graves07} for analysis and comparison with the Virgo spectra. The SDSS data points in the figures of this paper represent indices measured from those composite spectra.

The central aim of this chapter is to measure the strength of any residual hydrogen emission in these quiescent spectra. Noting the apparent residual H$\alpha$ emission of the SDSS spectra as seen in the first panel of Figure 3.1, we construct simple formulae for the correction of the relative intensities of the rest of the Balmer series for the Sloan Digital Sky Survey (SDSS) as a function of Mg $b$ (a Lick Style Index measurement for Mg) that includes continuum slope effects and the intrinsic decrement values for the Balmer lines. The corrections should find future use with integrated-light models to better predict stellar populations parameters, especially the mean metallicities and ages of galaxies.

Our analysis methods and results are set out in the following section with a discussion and summary in $\S$ 3.

\section{Analysis and Results}

For this work, the H$\alpha$ index is as defined by \cite{Cohen}. The first panel of Figure 3.1 shows that there is a discrepancy in the H$\alpha$ measurements between the SDSS and Virgo spectra. Plausibly, this is either due to hydrogen emission in the SDSS spectra or to nonspectrophotometric wavelength-dependent flux calibration errors that would propagate into Lick index measurements depending on how the various passbands lay across the various spectral ``lumps.'' We infer that the bulk of the discrepancy is most likely due to hydrogen emission in the SDSS spectra because the continuum-shape differences between the data sets are too small to account for much of the discrepancy, as we now show.

To determine if differences in the continuum of the two data sets could explain this discrepancy, the ratio of similar spectra (similar as regards velocity dispersion) from the two data sets was computed and a continuum fit was found for this ratio using the ``continuum'' routine in IRAF \footnote{IRAF is distributed by the National Optical Astronomy Observatories, which are operated by the Association of Universities for Research in Astronomy, Inc., under Cooperative agreement with the National Science Foundation.}. Lick style indices were measured from the {\em fit} as one would measure spectra. The difference in the index thus discovered was found to be small: about 0.05 \AA\/, insufficient to explain the SDSS-Virgo discrepancy. It should be noted that the effects of the continuum differences {\em could} be more severe if differences in the continuum shapes exist that are of the same wavelength span as the features of interest. Most would likely lay with the SDSS spectra, since our fluxing of the Virgo data was careful, but unfortunately we do not have the data to check the significance of these differences.

We characterized the SDSS-Virgo H$\alpha$ - Mg $b$ trend by best fit lines calculated using the routine fitexy.f \citep{numrec}. This is a fortran program for finding the best fit line for data with errors in both the x and y coordinate. This routine minimizes the distance of each point from the line while taking into account weighting by the precision of the individual measurements in both the x and y coordinates. The choice to characterize the trends in H$\alpha$ as functions of Mg $b$ was made due to the fact that it is easy to model galaxies of various Mg abundance and easy to measure the strong Mg $b$ feature. Also, since there exists a tight correlation between Mg $b$ and $\sigma$ (velocity dispersion) these conclusions can be implied for galaxies of various $\sigma$.  Those line fits and the root mean square (RMS) of the distances of the points from their fit lines are listed in Table 3.1. Also shown in Table 3.1 is the form of the correction term (${{ j_\alpha }\over{F_{c,\alpha}}}$) determined from the following derivation.

\renewcommand{\thetable}{3.\arabic{table}}
\begin{table}[H]
\begin{center}
\begin{tabular}{ l c c }
\multicolumn{3}{c}{Table 3.1} \\
\hline
\hline
Data set & Line Fit in \AA\ & RMS of Fit in \AA\ \\
\hline
Virgo      & $3.0132-0.3768 *$Mg $b$ & $0.066$ \\
SDSS       & $1.3476-0.0532 *$Mg $b$ & $0.020$ \\
Correction Term  &$1.6656-0.3236 *$Mg $b$ $= {{ j_\alpha }\over{F_{c,\alpha}}}$ & $0.069$ \\
\hline
\end{tabular}
\caption{Shown in this table are the best fit H$\alpha$ vs. Mg $b$ lines for the Virgo and Sloan data sets and their associated RMS values (lines 1 and 2). Line 3 is the difference between these two best fit lines, which represents the H$\alpha$ correction term along with its RMS value.}
\end{center}
\end{table}

Using the linear correction term for the hydrogen emission in H$\alpha$ a correction for the subsequent Balmer lines was constructed under the assumption that the entire shift is due to Balmer emission fill-in. To determine the form of the correction term we start with the definition of equivalent width (EW; see Eq. 1). In Eq. 1, $\mathbf{\lambda_{1}} $ and $\mathbf{\lambda_{2}} $ are defined as the blue and red wavelength bounds of the index passband, ${F_{i}(\lambda)}$ is the flux at each wavelength across the passband, and F$_{c}$ is the pseudocontinuum flux; see \cite{worthey94}.

\begin{equation}
  EW=  \int_{\lambda 1}^{\lambda 2}   ( 1 - {{F_{i}(\lambda)}\over{F_c}} ) d\lambda = \Delta \lambda ( 1 - {\overline{F_{i}}\over{F_c}} )   
  \end{equation}

With more generality, and including a term for the flux due to an emission feature,

\begin{equation}
  EW=  \int_{\lambda 1}^{\lambda 2}   ( 1 - {{F_{s}(\lambda) + F_{j}(\lambda)}\over{F_c}}   ) d\lambda = \Delta \lambda ( 1 - {\overline{F_{s}}\over{F_c}} - {\overline{F_j}\over{F_c}} )   
  \end{equation}

where $F_j$ is the flux of the emission feature and $F_{s}$ is the flux of the stellar light and $\Delta \lambda = \lambda_2 - \lambda_1$. If we call the emission line's power $j$ then the average emission line flux is defined by

\begin{equation}
 j = \int_{\lambda 1}^{\lambda 2} F_{j}(\lambda) d\lambda = \Delta \lambda  \overline{F_j}
\end{equation}

Thus, the average stellar flux inside the continuum band is $\overline{F_{s}}$ and the correction term for the equivalent width is $\Delta \lambda \overline{F_j}/F_c =j / F_c$. This gives us the equation for the equivalent width of an index with emission as

\begin{equation}
  EW=  \int_{\lambda 1}^{\lambda 2}   ( 1 - {{F_\lambda + F_j}\over{F_c}}   ) d\lambda = \Delta \lambda - \Delta \lambda{\overline{F_\lambda}\over{F_c}} - {j\over{F_c}}   
  \end{equation}

To extend to Balmer lines other than H$\alpha$, we exploit the fact that the decrements $j_\beta/j_\alpha$, $j_\gamma/j_\alpha$, and $j_\delta/j_\alpha$ are known, and relatively constant.

For example, if $j_\alpha$ is known, then extending from an H$\alpha$ index to an H$\beta$ index is accomplished by adding the correction term 

\begin{equation}
{{ j_\beta }\over{F_{c,\beta}}} = {{ j_\alpha }\over{F_{c,\alpha}}}  \times {{F_{c,\alpha}}\over{F_{c,\beta}}} {{j_\beta}\over{j_\alpha}}
\end{equation}

Having determined the H$\alpha$ correction from best line fits (see Table 3.1) all that is left to do is determine the conversion factors ${{F_{c,\alpha}}\over{F_{c,\beta}}}$ and ${{j_\beta}\over{j_\alpha}}$.

Models (chapter 2 section 3) were used to determine the continuum level conversion factors(${{F_{c,\alpha}}\over{F_{c,\beta}}}$). The first step was to take the ratio of the continuum levels near H$\alpha$ with the continuum levels near the other Balmer lines as measured from these models. The next step is to plot these continuum ratios against Mg $b$. The last step is to fit a least squares line to this data, giving a prescription for the change of the relative continuum levels as a function of Mg $b$. 

The last piece of the puzzle is that the native relative intensities of the Balmer lines need to be accounted for. This is taken into account by the relative Balmer line intensities as calculated by \cite{oster}. We adopt case B, 10000 K conditions, because they are near the middle of the range for star formation regions ($j_\alpha/j_\beta = 2.85$), and they are not too drastically different than LINER type spectra (j$_\alpha/j_\beta = 3.27$) from \cite{graves07}. Another reason to use these theoretical decrements is that observed decrements may suffer from asymmetric systematic errors due to local dust. The final correction formulae were applied to H$\beta$ defined in \cite{worthey94}, H$\gamma_A$, H$\gamma_F$, H$\delta_A$, and H$\delta_F$ as defined in \cite{worthey97} and are found in Table 3.2.

\renewcommand{\thetable}{3.\arabic{table}}
\begin{table}[H]
\begin{center}
\scriptsize
\begin{tabular}{l c c c c c}
\multicolumn{6}{c}{Table 3.2} \\
\hline
\hline
Balmer&Correction&H$\alpha$ Correction&Continuum Correction&Decrement&Correction term \\
Index& Term in \AA\ & in \AA\ (${{ j_\alpha }\over{F_{c,\alpha}}}$)& (${{F_{c,\alpha}}\over{F_{c,i}}}$)&(${{j_i}\over{j_\alpha}}$)& Uncertainty in \AA\ \\
\hline
H$\beta$&${{ j_\beta }\over{F_{c,\beta}}}=$   &$(1.666-0.324$(Mg $b$)) $\times$& $(0.838+0.076$(Mg $b$)) $\times$& $0.351$ & $0.076$ \\
H$\gamma_A$&${{ j_\gamma }\over{F_{c,\gamma}}}=$  &$(1.666-0.324$(Mg $b$)) $\times$& $(0.847+0.149$(Mg $b$)) $\times$& $0.165$ & $0.093$ \\
H$\gamma_F$&${{ j_\gamma }\over{F_{c,\gamma}}}=$ &$(1.666-0.324$(Mg $b$)) $\times$& $(0.847+0.149$(Mg $b$)) $\times$& $0.165$&$0.093$ \\
H$\delta_A$&${{ j_\delta }\over{F_{c,\delta}}}=$  &$(1.666-0.324$(Mg $b$)) $\times$& $(0.649+0.293$(Mg $b$)) $\times$& $0.091$ & $0.140$ \\
H$\delta_F$&${{ j_\delta }\over{F_{c,\delta}}}=$   &$(1.666-0.324$(Mg $b$)) $\times$& $(0.649+0.293$(Mg $b$)) $\times$& $0.091$ & $0.140$ \\
\hline
\end{tabular}
\caption{The most reliable results are obtained for 3.0 \AA\ $<$ Mg $b$ $<$ 4.3 \AA\/. This Mg $b$ range covers most elliptical galaxies. The correction term uncertainty is calculated by the propagation of errors for each of the three terms in the correction formulae. The first error term is that of the H$\alpha$ correction term found in Table 3.1. The second is the RMS of the continuum corrections term fit calculated in the same manner as that of the H$\alpha$ correction and the third is due to scatter in the possible theoretical values of the decrement.}
\end{center}
\end{table}

Figure 3.1 shows the measurements of Virgo and SDSS galaxy averages along with the corrected SDSS data (light blue line). In all the graphs the Virgo data is in red while the SDSS data is in green. Also in Figure 3.1 model grids are plotted in blue and pink. The blue corresponds to models of various ages from 1.5 to 17 Gyrs and the pink corresponds to models of various metallicities from -2.00 to 0.50. For the H$\beta$ correction the line fit is a little low for smaller galaxies with weaker Mg $b$, but not bad for the larger galaxies (See Figure \ref{fig1}).

\renewcommand{\thefigure}{3.1}
\begin{figure}[H]
\includegraphics[width=3.5in]{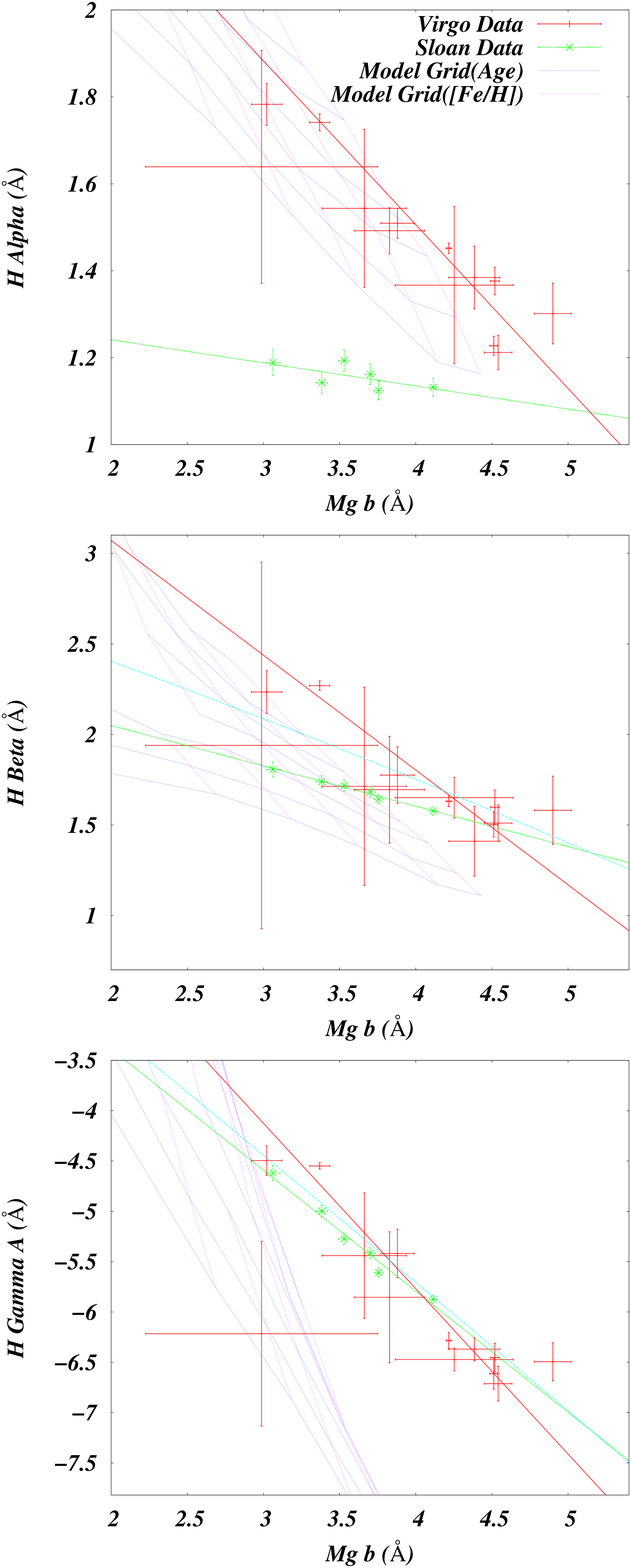}
\includegraphics[width=3.5in]{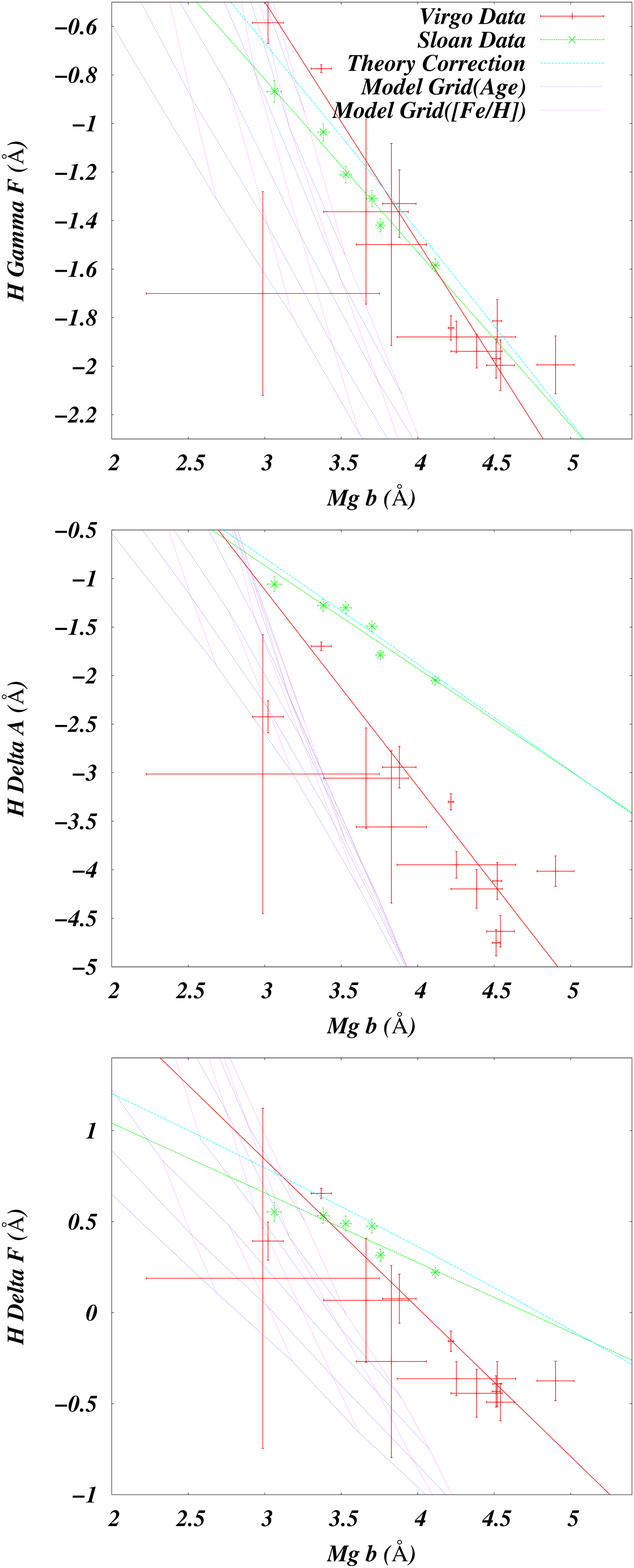}
\caption{In panel 1 H$\alpha$, H$\beta$, and H$\gamma_A$ against Mg $b$ indices for two data sets and models. In panel 2 H$\gamma_F$, H$\delta_A$, and H$\delta_F$ against Mg $b$ indices for two data sets and models. The Virgo spectra (red symbols with error bars), the SDSS spectra (green symbols with error bars), single stellar population models (blue grid lines) from bottom to top of ages 17, 12, 8, 5, 3, 2, and 1.5 Gyrs and models from right to left of metallicities $0.5, 0.25, 0, -0.25, -0.5, -1, -1.5$ and $-2$ (pink grid lines) are plotted. The fits to the index values for the SDSS galaxies after the correction for hydrogen emission are shown as a light blue line in each panel.
\label{fig1} }
\end{figure}

The fits for H$\gamma$ and H$\delta$ do not look as good when compared to the Virgo measurements (see Figure 3.1). The reason for this is most likely the sensitivity of the $H\gamma$ and $H\delta$ indices to changes in individual elemental abundances, with perhaps a non-negligible contribution from systematic errors between the two data sets. Elements N, C, and O have a profound and interdependent effect on this spectral region, and suggest themselves as candidates for further investigation.

\section{Discussion, Summary, and Conclusion}

Using the correction factor (${{ j_\beta }\over{F_{c,\beta}}}$) as derived above and associating it with the linear offset in H$\alpha$ this work has shown that for SDSS spectra cleaned like those in \cite{graves07}, a emission correction factor on the order of 0.5 \AA\ in H$\alpha$ and 0.2 \AA\ in H$\beta$ needs to be applied to the quiescent galaxies in order to better determine the mean age and metallicity of these galactic averages. These correction factors are much larger than those estimated in \cite{graves07}, where the estimated correction factors from measurements of O{\sc II} give H$\alpha$ = 0.082 \AA\ and H$\beta$ = 0.027 \AA\/. This work also shows that the higher-order Balmer lines H$\gamma$ and H$\delta$ have such modest emission corrections that other uncertainties dominate the error budget.  At least for our comparison sample of Virgo cluster elliptical galaxies, H$\gamma$ and H$\delta$ suffer from contamination from the varying of other elemental abundances, making age determination from these indices more complicated. This makes the need for correct H$\alpha$ and H$\beta$ measurements all the more important, since they are relatively insensitive to changes in metal abundances.

These results are in qualitative agreement with the results from \cite{eisen} who showed that in their SDSS spectra H$\beta$ suffers from interstellar emission lines and speculated that non-solar abundance ratios were to blame for the differences in age determinations made from H$\beta$, H$\gamma$ and H$\delta$. This is in agreement with these findings, where not only is there obvious hydrogen emission, but also it is clear that relative abundance ratios would certainly have to be taken into account in order to determine galactic ages from H$\gamma$ and H$\delta$.

For this work the H$\alpha$ correction was chosen as the basis for the corrections due to the facts that H$\alpha$ is $\sim$ 3 times more sensitive to hydrogen emission than H$\beta$ and that the line fit for H$\alpha$ as measured in the Virgo galaxies had a much tighter fit (RMS = 0.066 \AA\/) compared with H$\beta$ (RMS = 0.15 \AA\/) . This lead to a corrected SDSS line fit with much less uncertainty, especially for galaxies inside the range 3.0 $<$ Mg $b$ $<$ 4.3 \AA\ where the corrections for H$\beta$ yield the most reliable results. Another reason for choosing H$\alpha$ instead of directly using the correction that one could get from the H$\beta$ plot is that we wanted to preserve H$\beta$ for age determination.

The plot for H$\beta$ in figure 3.1 still shows a discrepancy between the corrected SDSS data (light blue line) and the Virgo data (red line). Although, statistically speaking, the discrepancy is of marginal significance, this residual difference could be due to a few variables such as varying abundance ratios within the Virgo galaxies, slightly different decrements values than the ones used here, or the possibility that the H$\alpha$ emission correction may be in part due to an difference in the mean ages of the two samples. \cite{thomas05} showed that cluster galaxies tend to be around 2 Gyr older than field galaxies.

A contributing factor might be that H$\alpha$ and H$\beta$ have different age sensitivities. In order to investigate the age sensitivities, the Z versus age sensitivity parameter (Zsp) was calculated for both H$\alpha$ and H$\beta$ as in \cite{worthey94}. The Zsp parameter is the modeled change in index strength due to a change in fractional metallicity (Z = $0.01689 \times 10^{\textrm{[Fe/H]}}$) divided by the change in index strength due to a change in fractional population age.

\begin{equation}
 Zsp = {[\delta \textrm{I}_m/(\delta \textrm{log(Z)})]\over[\delta \textrm{I}_a / (\delta \textrm{log(age)}]}
\end{equation}
 
Here $\delta$$\textrm{I}_m$/$\delta$log(Z) is the partial derivative of the index with respect to metallicity at age = 12 Gyrs. Similarly, $\delta$$\textrm{I}_a$/$\delta$log(age) is the partial derivative of the index with respect to age at solar metallicity. These sensitivities are shown in Table 3.3 along with the original \cite{worthey94} H$\beta$ sensitivity. Note that the models indicate that both H$\alpha$ and H$\beta$ are age indicators of the same sensitivity. This would mean that the difference between H$\alpha$ and H$\beta$ is more likely to be due to abundance ratios or the decrement values used than it is the age sensitivities of H$\alpha$ and H$\beta$.  However, since both H$\alpha$ and H$\beta$ are insensitive to changes in abundance ratio variation \citep{serv05} and decrement values can change from the effects of local dust, the difference would most likely stem from the choice of decrement. However, the measured extragalactic decrement values tend to be larger than those used here and that using these larger decrements would only increase the observed residual difference! We are therefore forced to postulate that at least one other systematic offset may be operating between the Virgo and SDSS data sets.

\renewcommand{\thetable}{3.\arabic{table}}
\begin{table}[H]
\begin{center}
\begin{tabular}{ l c }
\multicolumn{2}{c}{Table 3.3} \\
\hline
\hline
Index & Zsp \\
\hline
H$\alpha$      &  0.8 \\
H$\beta$      &  0.8 \\
H$\beta$ (Worthey et al. 1994)& 0.6\\
\hline
\end{tabular}
\end{center}
\end{table}

The applicability of this work for other grand SDSS averages may be limited due to the details of the sample selection. It is also unlikely to be applicable to individual red sequence galaxies due to the wide dispersion in index values and possible age effects. One possible avenue could be to scale other averages to match those of \cite{graves07}, but it would be a better idea to echo the work shown here by comparing those averages with the minimal-emission Virgo data set and determining a new correction.

It is conceptually possible to solve for both age and emission correction by considering both H$\beta$ and H$\alpha$ simultaneously, and increasing the emission correction until both indices give similar ages against a model grid. The obvious trouble with that scheme is that the solution then becomes model dependent. There is also a strong anticorrelation between derived age and emission correction. There is also the contribution of N emission near H$\alpha$, whose presence may cause spuriously large H$\alpha$ emission measurements in individual galaxies.

Speculation aside, it is safe to say that there is emission contamination in the SDSS spectra and that it is reasonably well accounted for by the linear fits presented in this work. In data sets that include H$\alpha$, observed OII and OIII do not need to be used as a proxy for Balmer emission.  Emission contamination is much less of a problem for H$\gamma$ and H$\delta$, but interpretation of these indices is complicated by the probable effects of individual elemental abundances.

\acknowledgements
We would like to thank Genevieve J. Graves for providing the Sloan Digital Sky Survey spectra, as well as her advice and input for this paper. Major funding for this work was provided by National Science Foundation grants 0307487 and 0346347.

\newpage
\pagestyle{plain}

\renewcommand{\thefigure}{4.\arabic{figure}}

{\centering CHAPTER FOUR\\

NH and Mg Index Trends in Elliptical Galaxies\\}

\section{Introduction}
One of the biggest problems with any attempt to determine the chemical make up of a stellar system is trying to disentangle the effects of C, N, and O in any given spectrum. This is due to the fact that in cool stars and elliptical galaxies, C, N, and O show themselves in any spectrum almost exclusively through molecular species such as NH, CN, C$_2$, CH, and CO rather than atomic species. The intertwining of these three elements starts with CO. The fact that CO has the highest dissociation energy of these molecules means that CO will form the most prolifically if given enough O and C. Since O is typically the most abundant of these three elements, C becomes incorporated into CO instead of the other molecular species. The rest of these molecules are connected through balancing of molecular equilibria. These interactions give a net effect of O acting like anti-CN since adding any O will decrease the amount of C available for the formation of other molecules \citep{serv05}.

To begin disentangling these three elements it should be possible to start with pseudo equivalent width indices that are sensitive to these elements such as C$_2$4668 \citep{wor94b}, which is sensitive to C abundance, and the indices CO5161 and CO4685 \citep{serv05} which are insensitive to N but react to C and O abundaces. After getting a grip on the C and O abundances, one can use the CN band to determine the N abundance. Unfortunately, disentangling C, N, and O this way turns out to be difficult \citep{bur03}. To low precision, most previous results agree that Mg, C, N, and Na appear to be enhanced in large elliptical galaxies and also correlated with velocity dispersion \citep{wor98a,sanchez03,Kelson06,Graves07}.

What is most often suggested as an alternative for determining the N abundance is the use of the NH feature at 3360 \AA\ since, as has been noted before, it is insensitive to C and O \citep{sne73,nor02} and it is also directly and sensitively measuring N abundances \citep{bes82,tom84}.

Below, I will show that this is not the whole story and that there are other contributors to the NH feature. Unfortunately, until recently there had been fewer than 15 early-type galaxies with published NH3360 values due in large part to relative insensitivity of detectors in the near-UV \citep{Toloba09}. So making use of the NH3360 feature was almost impossible until the introduction of NH3360 values of 35 early-type galaxies from \cite{Toloba09}. 

In \cite{Toloba09} this sample of 35 galaxies was measured using indices NH3360 \citep{Davidage94}, CNO3862, CNO4175, CO4685 \citep{serv05}, and Mg $b$ \citep{bur84}. Their findings included that there exists a flat relation between the NH3360 index and velocity dispersion. This seems to indicate that there does $not$ exist a velocity dispersion relation for nitrogen, contrary to work done previously. For example, since the CN relation is stronger and tighter than the C$_2$4668 relation, it seems that there should be a positive [N/Fe] trend with velocity dispersion \citep{trag}.

It is toward attempting to explain this disparity that the rest of this chapter is aimed. In the first section, the response of the NH3360 feature to various elements is calculated using models as done in \cite{serv05} but with better models. Then, after seeing Mg contamination in the index, two new indices are defined, one for the contaminate Mg, and the other for N. The responses for these new indices are then calculated. In \S 3 the NH3360 index along with NH3375, Mg3334, CN$_1$ and Mg $b$ are plotted and compared to models to determine if the effects of an old metal-poor stellar population can explain the observed index trends. Lastly, the results and conclusions are discussed in \S 4. We find that the presence of an old metal-poor population does account for the observed index trends.

\section{Analysis}
The method for determining the response of NH3360 as well as the two new indices was similar to that used in \cite{serv05}. In \cite{serv05} simple models of a galaxy spectrum were constructed, using a G dwarf and a K giant. These models varied in that there was one base model of solar metallicity and then 23 variations, each one with a particular elemental abundance doubled. Then, the ratios of these spectra were taken to find any spectral influence due to any particular element.

The difference here is that the numbers tabulated come from measuring these indices in full single stellar population models. These models are a version of the \cite{wor94} \& \cite{trag} models that use a grid of synthetic spectra in the optical \citep{lee}. Age, overall metallicity Z, and 23 individual elemental abundances can be varied independently in the models, and spectra for single-burst (simple) stellar populations produced. In this chapter plotted results are based on \cite{wor94} isochrones, but all results are confirmed with Padova \citep{bert94} evolution.
             
The grid of synthetic spectra is complete enough to predict nearly arbitrary composition. For the remainder of this paper whenever the term model is used it is with reference to these models, as they are the backbone of the methods used.

The responses of NH3360 and the two new indices NH3375 and Mg3334 can be found in Table 4.3. What can be seen is that the NH3360 index, while indeed insensitive to C and O, is sensitive to Mg and to a lesser degree Fe and Ni. The anti-correlation with Mg is due to a small Mg absorption feature located in the blue continuum passband. In order to remove this Mg dependence, a new index was defined; index NH3375, with an altered blue pseudocontinuum definition that avoids this feature. The new index NH3375 response, which is also found in Table 4.3, shows a smaller dependence on Mg with some small dependence on Ti and Ni. Unfortunately, the sensitivity of NH3375 to Ti and Ni comes from features of these two elements that overlap the NH feature itself. This limits the degree to which these sensitivities can be removed from any index definition.

For the sake of complementarity, a new index was defined for the small Mg feature as well: index Mg3334. The response of this index is surprisingly clean, showing only small sensitivity to O (see Table 4.3). The definitions of these new indices, as well as NH3360, can be found in Table \ref{Table1}. The index sensitivities to N and Mg are modeled in Figure \ref{Fig1}, where the first panel shows NH3360, NH3375, and Mg3334 as functions of N abundance and the second panel shows these indices as functions of the Mg abundance.

\renewcommand{\thefigure}{4.1}
\begin{figure}[H]
\includegraphics[width=3.5in]{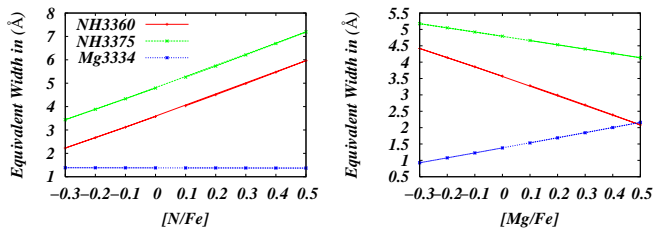}
\caption{In the first panel NH3360 (red), NH3375 (green), Mg3334 (blue) against [N/Fe]. In the second panel NH3360 (red), NH3375 (green), and Mg3334 (blue) against [Mg/Fe] for the second panel. This figure illustrates the sensitivity of these indices to changes in the N and Mg abundance at fixed [Z/H] = [Fe/H] = 0.0 and age = 8 Gyr with [Mg/Fe] = 0.0 for the first panel and [N/Fe] = 0.0 for the second.
\label{Fig1} }
\end{figure}

\renewcommand{\thetable}{4.1}
\begin{table}[H]
\begin{center}
\small
\begin{tabular}{ l c c c c}
\multicolumn{5}{c}{Table 4.1} \\
\hline
\hline
Index & Blue Passband & Index Passband & Red Passband & Reference\\
\hline
NH3360 & 3332-3350 & 3350-3400 & 3415-3435 & \cite{Davidage94}\\
NH3375 & 3342-3352 & 3350-3400 & 3415-3435 &  This work \\
Mg3334 & 3310-3320 & 3328-3340 & 3342-3355 &   This work \\
Mg $b$  & 5142.625-5161.375 & 5160.125-5192.625 & 5191.375-5206.375 & \cite{trag}\\
\hline
\end{tabular}
\caption{Index passband definitions in \AA\/.}
\label{Table1}
\end{center}
\end{table}

A Z versus age sensitivity parameter (Zsp) was calculated for NH3360, NH3375, Mg3334, and Mg $b$ as it was in \cite{wor94}. The Zsp is the ratio of the percentage change in age to the percentage change in Z ( Z $\approx 0.01689 \times 10^{\textrm{[Fe/H]}}$) of the index measured as shown below. 

\begin{equation}
 Zsp = {[\delta \textrm{I}_m/\delta \textrm{log(Z)}]\over[\delta \textrm{I}_a/\delta \textrm{log(age)}]}
\end{equation}
 
Here $\delta$$\textrm{I}_m$/$\delta$log(Z) is the partial derivative of the index with respect to metallicity at age = 12 Gyrs. Similarly, $\delta$$\textrm{I}_a$/$\delta$log(age) is the partial derivative of the index with respect to age at solar metallicity. These sensitivities are shown in Table \ref{table2}. Note that the models indicate that both NH3360 and NH3375 are far more sensitive to age than they are metallicity, especially NH3360 which has almost no metallicity sensitivity in the metal rich regime, while the Mg indices are more sensitive to metallicity.

\renewcommand{\thetable}{4.2}
\begin{table}[H]
\begin{center}
\begin{tabular}{ l c }
\multicolumn{2}{c}{Table 4.2} \\
\hline
\hline
Index & Zsp \\
\hline
NH3360  &  0.2 \\
NH3375  &  0.6 \\
Mg3334  &  1.1 \\
Mg $b$  &  1.7 \\
\hline
\end{tabular}
\end{center}
\caption{Table 4.2 shows the Z versus age sensitivity parameter (Zsp), which gauges how changes in metallicity and age effect various indices. A large Zsp indicates a larger dependence on the overall metallicity then on age with 1.0 indicating that age and metallicity effect the index equally.}
\label{table2}
\end{table}

A plot illustrating the sensitivity of these indices to metallicity and age can be found in Figure \ref{fig2}. Note that the NH3360 index, although age sensitive, goes nearly flat as a function of metallicity. The other indices exhibit behavior of increasing and plateauing out over time.

\renewcommand{\thefigure}{4.2}
\begin{figure}[H]
\includegraphics[width=3.5in]{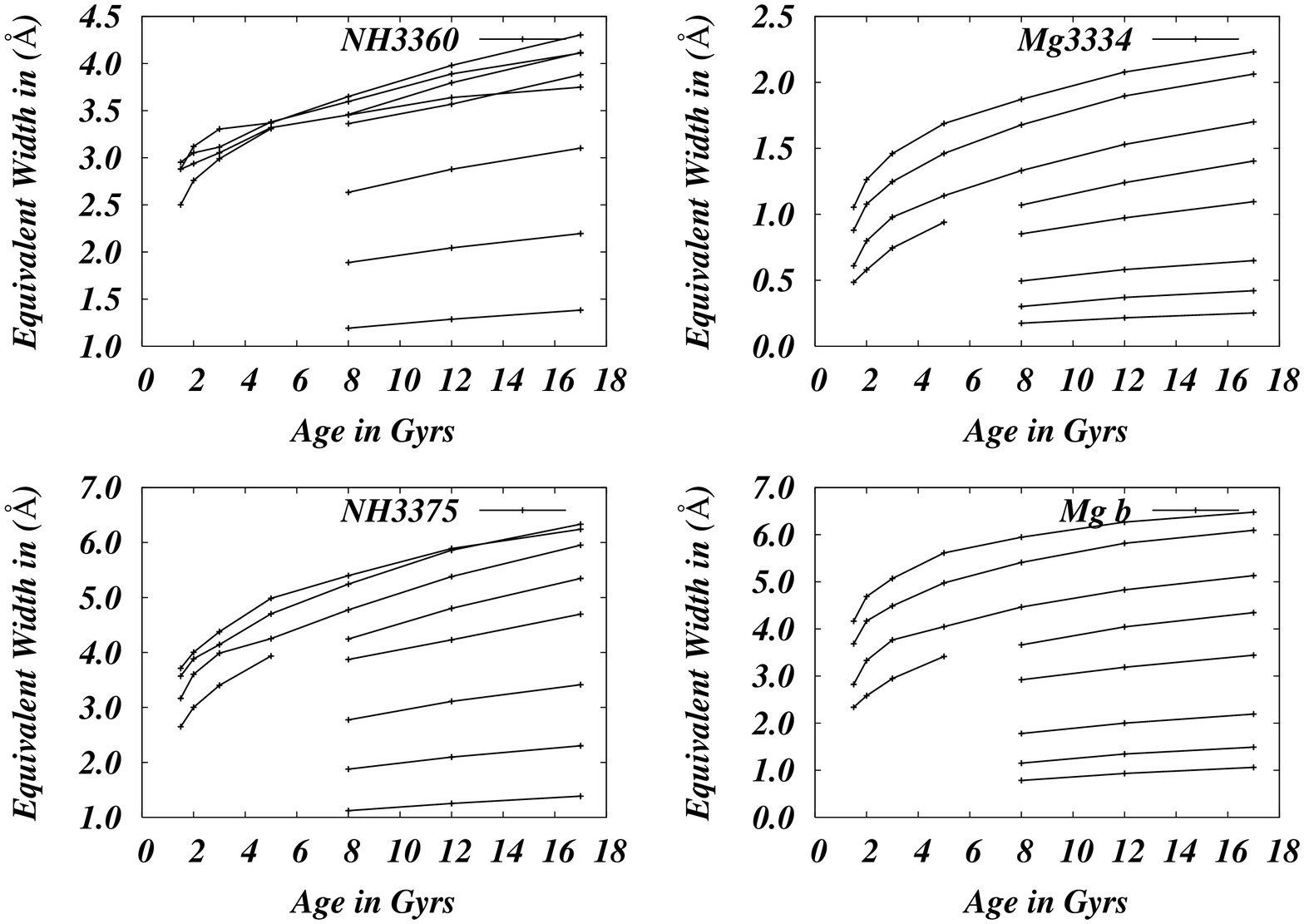}
\caption{NH3360, NH3375, Mg3334, and Mg $b$ against age for various metallicities. The NH3360 index (top left), the NH3375 index (bottom left), the Mg3334 index (top right), and the Mg $b$ index (bottom right) are plotted for metallicities from top to bottom $0.5, 0.25, 0, -0.25, -0.5, -1, -1.5, -2$ and ages 1.5, 2, 3, 5, 8, 12 and 17 Gyrs.
\label{fig2} }
\end{figure}

\section{Observations and Results}

The \cite{Toloba09} sample consists of long-slit spectra for 35 elliptical galaxies collected with the 4.2 m William Herschel Telescope at Roque de los Muchachos Observatory, using the ISIS spectrograph. The spectra were extracted within a central equivalent aperture of 4$^{\prime\prime}$ and broadened to a velocity dispersion of 300 km s$^{-1}$. Measurements for galaxies with velocity dispersions greater than 300 km s$^{-1}$ were corrected to a central velocity dispersion of 300 km s$^{-1}$ using their own models. The wavelength coverage is from 3140 to 4040 \AA\ with a resolution of 2.3 \AA\ (FWHM) and a typical signal to noise of $S/N=40$ per \AA\/. These spectra were chosen as a subset of those presented in \cite{sanchez06} so that this near-UV data could be supplemented with optical data in the wavelength range 3500 to 5250 \AA\/. The galaxies in this set were also chosen to include field, Virgo and Coma cluster ellipticals, which cover a range of velocity dispersions (130 $<  \sigma < $ 330 km s$^{-1}$).

From this sample the NH3360, NH3375, Mg3334, CN$_1$ and Mg $b$ Lick style indices were measured. The measurements for NH3360, NH3375 and Mg3334 were taken from the \cite{Toloba09} spectra, while the measurements for CN$_1$ and Mg $b$ were taken form the \cite{sanchez06} optical spectra. These measurements were then plotted against the Fe4383 index to compare the trends of these near-UV indices which are sensitive to N and Mg with the slightly redder CN$_1$ and Mg $b$ which are also sensitive to N and Mg (Fig. \ref{fig3}). For comparison, Mg3334 vs Mg $b$ is also plotted. The trend lines shown in Figure 4.3 are best-fit lines calculated using fitexy.f \citep{numrec}, a program for finding the best-fit line for data with errors in both the x and y coordinates. It minimizes the distance of each point form the line while taking into account weighting by the precision of the individual measurements in both the x and y coordinate.

The index plotted in the first panel of Fig. \ref{fig3} is NH3360, in the second NH3375 and in the third is Mg3334, all against the Fe4383 index. The index plotted in the fourth panel of Fig. \ref{fig3} is CN$_1$, in the fifth is Mg $b$ both against the Fe4383 index and in the sixth is a plot of Mg3334 vs Mg $b$. In each panel along with the index are plotted three lines. These lines represent the respective indices measured from a 12 Gyr galactic model of solar metallicity [Fe/H] = 0.0 to a metallicity of [Fe/H] = 0.25. The red line is a model of solar metallicity with no subpopulation. The green line is a model of solar metallicity, but with 5 percent of the galactic mass consisting of a 12 Gyr, metal-poor ([Fe/H] = $-$1.5) subpopulation. The blue line is the same model except that the subpopulation is now 10 percent of the total mass. The light blue points are galaxies suspected as having a spectral calibration issue, as they have a large CN$_1$ but a small Fe4383 or small CN$_1$ and large Fe4383, something not seen in normal elliptical galaxies.

What can be seen in Fig. \ref{fig3} is that, for all three near-UV indices the index trends look flat in contrast to the panels 4 \& 5 of Fig. \ref{fig3} which show a definite trend with increasing metallicity. What can also be seen is that, for the galactic models as the percent mass of the subpopulation is increased (red to green to blue) the model index trends flatten out; twisting to come into fairly good agreement with the observed indices within a metal-poor subpopulation fraction of 10 percent by mass. The redder indices of CN$_1$ and Mg $b$ show that the subpopulation has little to no effect on these index trends except lowering the average metallicity.

Further evidence for the existence of a old metal-poor population can be see in panel 6 of Fig. \ref{fig3}, what can be seen here is that Mg3334 and Mg $b$ which are fairly clean indicators of Mg (\cite{serv05} \& Table 4.3) do not appear to be measuring the same abundance trend. This is an indication that the near-UV indices do suffer from dilution due to an underlying bright and weak-lined near-UV population.

\renewcommand{\thefigure}{4.3}
\begin{figure}[H]
\includegraphics[width=6.5in]{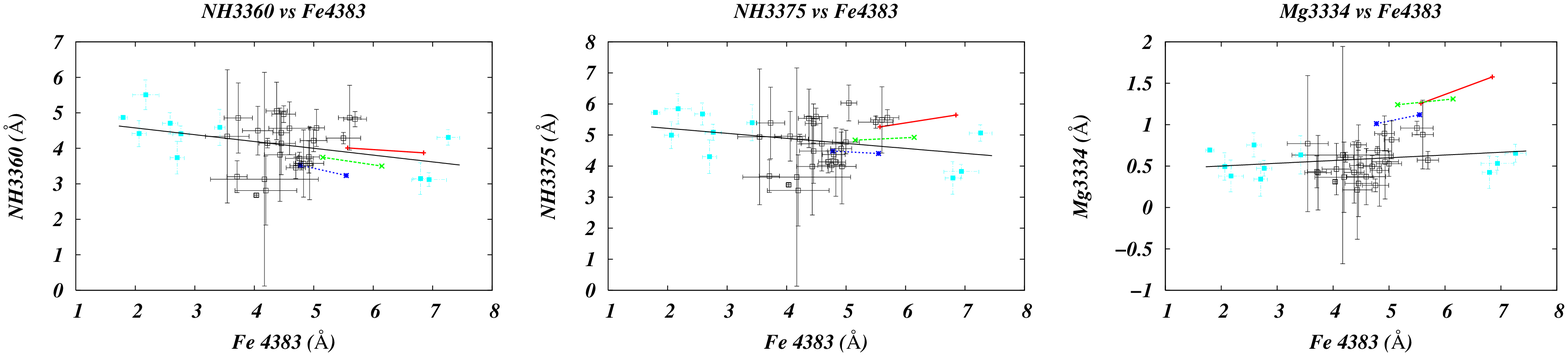}
\includegraphics[width=6.5in]{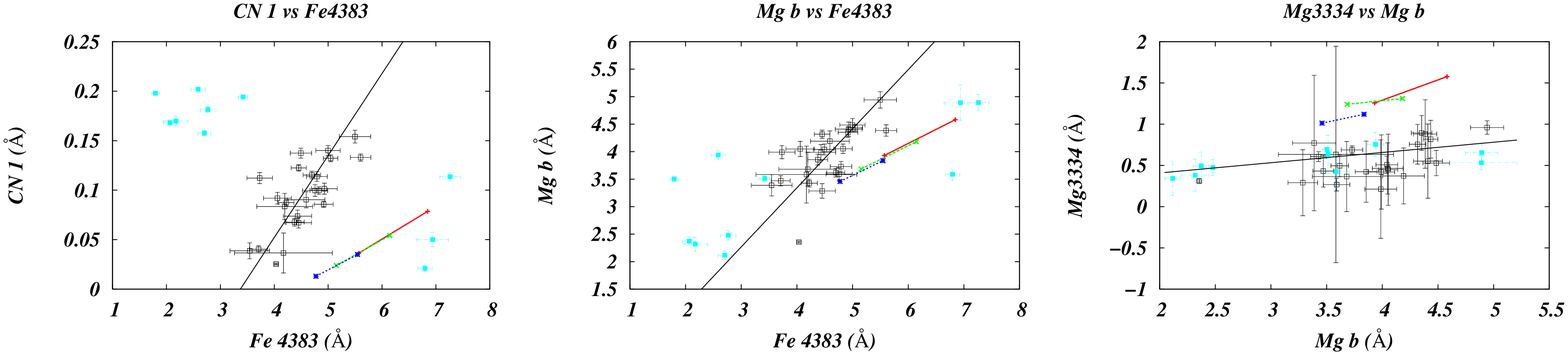}
\caption{In the first, second and third panels, NH3360, NH3375 and Mg3334 indices vs Fe4383 are plotted respectively. In the fourth and fifth panels, CN$_1$ and Mg $b$ indices vs Fe4383 are plotted respectively. In the sixth panel Mg3334 vs Mg $b$ is plotted. In each panel along with the index are plotted three lines. These lines represent the respective indices measured from a 12 Gyr galaxy model of metallicity [Fe/H] = 0.0 to a metallicity of [Fe/H] = 0.25. The red line is a model of solar metallicity with no subpopulation. The green line is a model of solar metallicity, but with 5 percent of the galaxy mass consisting of a 12 Gyr, metal-poor ([Fe/H] = $-$1.5) subpopulation. The blue line is the same model except that the subpopulation is now 10 percent of the total mass. The light blue points are galaxies suspected as having a spectral calibration issue, as they have a large CN$_1$ but a small Fe4383 or small CN$_1$ and large Fe4383; something not seen in normal elliptical galaxies.
\label{fig3} }
\end{figure}

\section{Discussion, Summary, and Conclusion}

With the use of simple stellar population models it has been demonstrated have that the NH3360 index, while being C and O insensitive, suffers from contamination from other elements, most notably Mg (see Table 4.3). This dependence of the NH3360 index on Mg can be effectively removed by a simple redefinition of the NH3360 index blue continuum passband to form a new index, index NH3375. This new index is now not only insensitive to C and O but also Mg (Table 4.3) and even though contaminations from other elements exist (e.g.,Ti), NH3375 is a cleaner measure of N than NH3360.

The predicted behavior of NH3375, as seen in Fig \ref{Fig1}, is an increased sensitivity to N abundance (Table 4.3). This prediction is borne out in the data (see Fig. \ref{fig3}) as an increase in the magnitude of the NH3375 measurements compared too that of NH3360. Also borne out in the data is that the index trends of NH3375 and NH3360 are nearly flat as is the Mg3334 index. This flat behavior is in contrast to the index trends seen in CN$_1$ and Mg $b$, the premise being that if these indices are measuring N and Mg, respectively, then these indices should show similar behaviors.

There appears to be a difference between the underlying stellar populations that produce the near-UV spectra and the visible spectra at least as regards these near-UV indices. This underlying near-UV spectrum is likely due to a fraction of metal-poor population on the order of 5-10 \% by mass \citep{wor96}. The addition of a metal-poor subpopulation to our models {\em does} account for the discrepancies observed in the index trends (Fig. \ref{fig3}) most notably that of Mg3334 and Mg $b$. The low metal-poor fractions tested here are of order those inferred by \cite{wor96} for the amount of mass locked into low metallicity stars in present day galaxies.

The Mg3334 index turns out to be a useful tool as it is a fairly clean index with very little contamination (Table 4.3), which allows for a direct comparison to Mg $b$ (Fig \ref{fig3}). The remarkable flat behavior of Mg3334 compared with Mg $b$ can therefore not be attributed to Mg abundance, but therefore must be due to some other effect, and the metal-poor subpopulations hypothesis explains the difference nicely. Indeed, the pair of the indices taken together can be used to characterize the amount and abundance spread in the underlying metal poor fraction of stellar mass.

With the inherently composite nature of the near-UV spectra \citep{Burstein88}, with metal-poor light mingling with metal-rich light, deriving a N abundance from NH3375 or NH3360 looks more difficult than heretofore suspected. It is outside the scope of this chapter, but by using of Mg3334 it may be possible to uncover N abundance by comparing derived Mg abundances from Mg3334 and Mg $b$ and then calibrating the underlying metal-poor population until the Mg3334 abundance agrees with that of Mg $b$. This metal-poor calibration could then be applied to NH3375, which in turn could be used to derive a N abundance for comparison with N abundances derived form C, N and O  indices. Hopefully these measurements will be in agreement with the N enhancements in large elliptical galaxies deduced by other authors \citep{Graves07,Kelson06,wor98a}. 

The existence of a small scatter Mg - $\sigma$ (velocity dispersion) relation among large elliptical galaxies implies increasingly effective Mg enrichment from Type II supernovae (i.e. massive star explosions) with increasing galaxy size. This might be because the initial mass function is stronger in large ellipticals (causing more Type II supernovae and light element enrichment), or a Type Ia delayed timescale is a cause, or because stellar binarism depends on environment \citep{wor98b}. The similarity of the N - $\sigma$ and Na - $\sigma$ relations to that of the Mg - $\sigma$ relation imply that the same mechanism (that of Type II supernovae) is responsible for the observed trends and that contributions from nucleosyathesis on the asymptotic giant branch do not contribute significantly. That is to say, N appears to be ``primary" rather than ``secondary" according to this line of reasoning \citep{hen99}. The C - $\sigma$ relation, on the other hand, seems to be intermediate between Mg, N, Na and Fe-peak elements (from Type Ia supernovae; \cite{trag}). This implies that the C abundance may come from contributions from both supernovae and the asymptotic giant branch \citep{hen99}.  

\acknowledgements
Major funding for this work was provided by National Science Foundation grants 0307487 and 0346347.

\newpage
\renewcommand{\thetable}{4.3}
\begin{table}[H]
\scriptsize
\begin{center}
\begin{tabular}{ l c c c c c c c c c c c c c}
\multicolumn{14}{c}{Table 4.3} \\
\hline
\hline
              &  $I_0$ & $\sigma$ & & & & & & & & & & & \\
Index   &   (\AA\/) &     (\AA\/) &     C &       N &      O &      Na &     Mg &    Al &     Si &      S &      K &      Ca &     Sc  \\
\hline
NH3360     &    3.221 &    0.285 &  -0.05 &   5.18 &  -0.52 &  -0.03 &  -3.43 &  -0.04 &  -0.30 &  -0.05 &   0.00 &  -0.20 &   0.13 \\
NH3375     &    4.688 &    0.322 &  -0.20 &   4.51 &  -0.87 &  -0.03 &  -0.71 &  -0.04 &  -0.04 &  -0.05 &   0.00 &  -0.50 &  -0.02 \\
Mg3334      &    1.542 &    0.105 &  -0.43 &  -0.22 &  -1.14 &  -0.04 &   5.91 &  -0.05 &   0.14 &  -0.04 &   0.00 &  -0.35 &  -0.24 \\
\hline
\end{tabular}
\end{center}
\caption{Table 4.3 shows the response of the NH3360, NH3375, and Mg3334 index definitions to various elements. The first column is the name of the index, the second column gives the index measurements in angstroms of equivalent width, and the third column gives the error associated with S/N = 100 at 5000 \AA\/. The remainder of the columns are the changes (enhanced minus unenhanced) in the index brought on by an element enhancement of 0.3 dex ( or 0.15 dex for C) in units of the error of the third column.}
\label{table3}
\end{table}

\begin{table}[H]
\scriptsize
\begin{center}
\begin{tabular}{ l c c c c c c c c c c c c c}
\multicolumn{14}{c}{Table 4.3 Continued} \\
\hline
\hline
              & & & & & & & & & & & & &  \\
Index   &      Ti &     V &     Cr  &     Mn &     Fe &     Co &     Ni &     Cu &     Zn &    Sr &     Ba &     Eu &    upX2 \\
\hline
NH3360           &   0.04 &   0.37 &  -0.18 &  -0.12 &  -1.70 &  -0.41 &   1.36 &  -0.04 &  -0.37 &   0.01 &   0.00 &   0.00 &  -3.82 \\
NH3375           &  -2.09 &   0.25 &  -0.08 &  -0.25 &  -0.78 &  -0.27 &   1.82 &   0.00 &  -0.57 &   0.00 &   0.00 &   0.01 &  -3.97 \\
Mg3334         &  -0.26 &  -0.52 &   0.18 &  -0.65 &  -0.29 &   0.03 &  -0.92 &   0.06 &  -0.26 &  -0.01 &   0.00 &   0.00 &   3.10 \\
\hline
\end{tabular}
\end{center}
\end{table}

\newpage
\pagestyle{plain}

{\centering
CHAPTER FIVE\\

The Effects of Velocity Dispersion on Index Trends\\}

\section{Introduction}

As stated in the Chapter 1, \cite{wor92} presented models and observations indicating that individual elemental abundances for individual galaxies could be found by taking advantage of the fact that measured indices showed considerable scatter when compared to models. This scatter indicated that there were nonsolar abundance ratios within individual galaxies. 

\cite{wor94} pointed out that this scatter along with the "3/2 rule" (Chap. 1 Sec. 1) for age and metallicity could be used to determine abundance ratio variations without having to know the age or metallicity to high precision if features could be found in the spectrum that were sensitive to such non-lockstep behavior.

In \cite{serv05} the task of finding non-lockstep features was undertaken with the use of synthetic spectra. The question of being to able to measure the effects of abundance ratio changes of 23 elements was explored with the use of element targeted indices, of those 23, 18 looked measurable in spectra of integrated light spectra of S/N $>$ 100. 

With this in mind, high quality spectra for 18 Virgo Cluster galaxies along with a Sloan Digital Sky Survey sample (Chapter 2) and a sample from \cite{Toloba09} (Chapter 3) were measured using the 25 lick indices \citep{wor94,wor97}, 1 H$\alpha$ index defined in \cite{Cohen} and 48 of the indices defined in \cite{serv05}. These measurements were then analyzed to look for abundance trends with velocity and individual scatter in the abundance measurements as an indication of individual abundance variation.

The rest of this chapter is organized as follows. The observations are discussed in \S 2, then the analysis of the spectra are outlined in \S 3. In \S 4, the results are presented and discussed.

\section{Observations}

The sample used for the measurements in this paper consist of new spectroscopic data for 18 Virgo cluster galaxies, a set of averaged Sloan Digital Sky Survey spectra which have been coadded and binned by internal velocity dispersion into six stacks (Chapter 2) and a sample of the 35 Toloba spectra (Chapter 3). Both the Virgo and Sloan data were chosen to be quiet, red galaxies, and without concern for orientation of the galaxies axis with respect to the instrument. 

The Virgo cluster galaxy spectra are high quality long slit spectra obtained with the Kitt Peak National Observatory (KPNO) Mayall 4m telescope with the Cassegrain spectrograph. These spectra were chosen primarily to cover a velocity dispersion from 50 to 350 km/s. For each galaxy two overlapping spectrograph setups were used, one for the blue spectral range of 3200\AA\ to 5300\AA\ and one for the red spectral range of 5000\AA\ to 7500\AA .

For the blue spectra the KPC-007 grating/grism with a 2 arcsec slit was used along with the T2KB CCD. For each galaxy, multiple exposures were taken. In the blue the exposure time calculator gave S/N = 40 for a 600 second exposure time for a galaxy at the faint end of the velocity dispersion range, so 6 to 8 such exposures were taken in the blue in an attempt to reach S/N = 100 after stacking the individual exposures. The time and number of exposures varied from galaxy to galaxy, but not by much.

For the red spectra the Bl-420 grating/grism with a 2 arcsec slit was used along with the T2KB CCD. For each galaxy multiple exposures were taken. In the red the setup is more sensitive, and also the galaxies emit more red light so the 6 to 8 exposures were usually only 300 seconds in length, again with the idea to reach S/N =100 in the end.

These spectra where then reduced using standard long slit spectroscopy reduction techniques with the use of  IRAF \footnote{IRAF is distributed by the National Optical Astronomy Observatories, which are operated by the Association of Universities for Research in Astronomy, Inc., under Cooperative agreement with the National Science Foundation.}. This produced one dimensional spectra binned linearly in wavelength.

The SDSS spectra are high signal to noise spectra and are made up quiescent galaxy stacked spectra in 6 velocity dispersion bins. The bins are $70< \sigma <  $ 120 km s$^{-1}$, 120 $< \sigma < $ 145 km s$^{-1}$, 145 $< \sigma < $ 165 km s$^{-1}$, 165 $< \sigma < $ 190 km s$^{-1}$, 190 $< \sigma < $ 220 km s$^{-1}$, and 220 $< \sigma <$ 300 km s$^{-1}$. In each bin are several hundred stacked spectra of similar galaxies. To read more about selection criteria and image processing for the SDSS spectra see \cite{graves07}.

The \cite{Toloba09} spectra were chosen to include field, Virgo and Coma ellipticals, spanning a range of velocity dispersions (130 $<$ $\sigma$ $<$ 330 km s$^{-1}$). The index measurements come from two sources. For indices bluer than 3900 \AA\ the measurements were taken from spectra presented in \cite{Toloba09}. The remaining Toloba index measurements come from spectra presented in \cite{san06}. The same set of galaxies is represented in all index measurements.

\section{Analysis}

For the purpose of comparing trends in the SDSS, Toloba and Virgo spectra, the spectra where broadened to a velocity dispersion of 300 km s$^{-1}$. Measurements of the Lick indices and the Lick style indices introduced in \cite{serv05} where then taken. 

These measurements where then plotted first against the internal velocity dispersion of the individual galaxies. For the Virgo spectra, velocity dispersions and errors from \cite{davies1987} were used, while center values of the velocity dispersion bins and half of the bins interval was used for the velocity dispersion and errors of the SDSS data, respectively. The \cite{Toloba09} values were calculated from the spectra using templates and the MOVEL and OPTEMA fitting algorithms (\cite{san06} and references therein).

Next these measurements and their errors were plotted against the Mg $b$ and $<$Fe$>$ indices. The first set should reveal correlations with alpha element like behavior and also with velocity dispersion, given the positive correlation between Mg $b$ and velocity dispersion. The second set should reveal correlations of these indices with Fe. Note, however, that young age will push galaxies to weak metal line strength, regardless of element. 

It is also important to note that for errors in the large index measurements of the Virgo sample a error floor was imposed. This error floor was imposed due to the fact that for large indices the errors tend to average out giving unrealistically small errors. The error floor was constructed by fitting a line to the index measurements vs. index size, and then correcting those indices that fell below this line. This was done for CN$_1$, CN$_2$, Mg$_1$, Mg$_2$, TiO$_1$, TIO$_2$ (corrections in magnitudes $\approx$ 0.03) and CaHK (corrections in angstroms $\approx$ 0.3 \AA\/).

In these plots a trend line was then fit in order to characterize the correlations. This was done with fitexy.f \citep{numrec} which is a fortran program for finding the best fit line from data with errors in the x and y coordinate. Where the data for the two data sets is in agreement a single trend line is used, but there are indices for which two trend lines must be used due to differences in the data sets. Differences of note are the Balmer series, which is most likely due to hydrogen emission, TiO and CN data shifts most likely due to aperture differences or spectral shape errors. Aperture differences come from the fact that the Virgo spectra were collected using a narrow slit and thus contain light from the galactic nucleus. The SDSS spectra on the other hand contain more light from non-nuclear regions of the galaxies. The spectral shape errors may come from differences in reduction techniques. While a distinct worry, this only seems to effect a few of the 74 indices.

Plotted alongside everything else are 5 of the Virgo cluster galaxies with definite hydrogen emission. These can be seen as the gray crosses in the Appendix B figures. Also in the graphs of index vs Mg $b$ or $<$Fe$>$ a model grid is plotted. These models are from \cite{wor94} plus my own additions, and span a range of age from 1.5 to 17 Gyr and metallicity from [Fe/H]=$-2.00$ to 0.50. The 74 index plots can be found in Appendix B.

\section{Discussion, Summary, and conclusion}

What can be seen in every plot of Appendix B is that there is always a scatter in the data. This is a clear indication that there are individual elemental abundances for each galaxy and is to be expected as it indicates the non-lockstep behavior pointed out in \cite{wor94}.

Also evident from the plots is that there is by no means an overall consistency in the $\sigma$, Mg $b$, or Fe trends. Even indices designed to target the same element can have greatly different trends, such is the case for Mg (App. B Figures 16 - 20), Ca (App. B Figures 26 - 28) and Ni (App. B Figures 60 - 65). Since these index sets generally range from blue wavelengths below 4100 \AA\ to much redder wavelengths, a probable explanation is that of the old metal poor population discussed in Chapter 3, which would have the effect of flattening out the bluer indices. A mere flattening, however, is not enough to explain the complete reversal in trends seen for Mg3835 (App. B Fig. 16) and CaHK (App. B Fig. 26). 

It is important too remember that often these indices are not pure indicators of a given element so inferring to much about a given element's behavior in general should be avoided. However, there are indices that are very clean and do give a fair representation of an element's behavior, such as Mg $b$, CaHK, Fe5406 and Na D.

In the case of H$\alpha$ and H$\beta$ (App. B Figs. 1 and 2) the trends indicate that the relative intensity of the absorption feature decreases as a function of $\sigma$, Mg and Fe. This general indicates that larger galaxies (i.e. larger velocity dispersion, $\sigma$) tend to be older as hydrogen absorption decreases as stellar population grows older. This conclusion is supported by the Fe5406 index (Fig. 54), which indicates that the overall Fe content increases with increasing galaxy size.

From Mg $b$ it can be inferred that Mg tends to be enhanced with respect to Fe in larger older stellar populations. The Mg enhancement can be seen in Figure 20 as the index trend in the third panel pulls up and away from the model grid.

The CaHK index is of particular interest due to its apparent underabundance in larger galaxies. This effect has been seen before in \cite{thomas03ca}, who speculate that the increase of Ca underabundance with galaxy mass balances the higher total metallicities of more massive galaxies, so that calcium abundance is constant and does not increase with increasing galaxy mass. They argue that metallicity dependent supernova yields are the most promising explanation for the under abundance. 

Although there is a wealth of information that these indices hint at, full modeling of these galaxies is required to decouple the various elemental influences present in each index before anything definite can be said about the individual elemental abundances. The modeling of these galaxies through the use of these indices will be the subject of future work.

\newpage
\pagestyle{plain}

{\centering CHAPTER SIX\\
Line Strength Gradients in Virgo Cluster Galaxies\\}
\section{Introduction}

It has been well established that line-strength profiles in elliptical galaxies vary with radius. Starting with \cite{faber73} the number of line strength gradients has grown considerably \citep{dav92,davies93,fish95,gon93,kunt98,kunt06}. The results of these investigations and others show some general results.

One of these results is that Mg and Fe line-strength gradients decline with increasing radius although they are generally shallow. This is mostly interpreted as a decline in the overall metallicity and suggest that abundance ratios stay more or less constant within a given elliptical galaxy \citep{gon93,kunt98}. 

Results also indicate that the H$\beta$ index, an age indicator, is generally constant as a function of radius \citep{dav92,gon93} with large scatter. \cite{davies93} concluded that this implied no gradients in the age of the stellar population within individual ellipticals. That conclusion was contradicted by \cite{hen99}, who argued that the flatness of H$\beta$ together with the fact that the metallic absorption features increase in strength toward the centers of galaxies implied that galaxy centers are mildly younger. Differences in the strengths of the H$\beta$ measurements led \cite{gon93} to suggest that significant differences in ages of elliptical galaxies exist, even within a single galaxy.

However, these results have been limited by the number of element sensitive indices and the chemical flexibility of the available population models to accurately determine the age and metallicity gradients of these galaxies.

In this chapter the 18 Virgo cluster galaxies and 74 indices presented in Chapter 4 have been measured and analyzed to look for index trends within these galaxies as a function of radius.

The rest of this chapter is organized as follows. The basic galaxy sample data is presented in \S 2, then the analysis of the spectra are outlined in \S 3. In \S 4 the results are presented and discussed.

\section{Observations \& Galaxy Data}

The sample used for the measurements in this chapter consist of the spectroscopic data presented in Chapter 4 for the 18 Virgo cluster galaxies, with the exception of processing the data to look for radial trends. The spectra were collected with the Kitt Peak National Observatory (KPNO) Mayall 4m telescope with the Cassegrain spectrograph.

To look for radial trends the spectra were reduced using standard long slit spectroscopy reduction techniques with the use of  IRAF \footnote{IRAF is distributed by the National Optical Astronomy Observatories, which are operated by the Association of Universities for Research in Astronomy, Inc., under Cooperative agreement with the National Science Foundation.}. This produced two dimensional spectra binned linearly in the wavelength and spacial directions.

The basic galaxy parameters used can be found in Table 6.1 and Table 6.2. The data has been gathered from various sources.

\noindent
Table 6.1 has the following columns.

Column 1: Galaxy catalog number.

Column 2: Adopted effective radius (R$_e$) in arc seconds (calculated from De Vaucouleurs 

\hspace{20mm}effective diameter (A$_e$) \citep{faber89}).

Column 3: Adopted effective radius (R$_e$) error in arc seconds (calculated from error 

\hspace{20mm}estimates in \cite{faber89}).

Column 4: Velocity dispersion ($\sigma$) in km s$^{-1}$ \citep{faber89}.

Column 5: Velocity dispersion ($\sigma$) error in km s$^{-1}$ \citep{faber89}.

Column 6: Indicates if the galaxy shows hydrogen emission in H$\alpha$ (this work).

\noindent
Table 6.2 has the following columns.

Column 1: Galaxy catalog number.

Column 2: Position angle (PA) in degrees.

Column 3: Minor to major axis ratio (b/a).

Column 4: Position angle and minor to major axis ratio references

\renewcommand{\thetable}{6.1}
\begin{table}[H]
\begin{center}
\begin{tabular}{ l c c c c c}
\multicolumn{6}{c}{Table 6.1} \\
\hline
\hline
NGC    &  R$_e$           & R$_e$ Error& $\sigma$  &  $\sigma$ Error& emission\\
             &   arc seconds &  arc seconds &km s$^{-1}$&  km s$^{-1}$ & in H$\alpha$  \\
\hline
1400          &    37.77 &    $\leq$3.3      &  250       &    25.0   &emission\\
1407          &    71.96 &     $\leq$3.7      & 285         &   28.5  &\\
2768          &    53.35 &      $\leq$3.7    & 198         &     27.7 &emission\\
3156          &    49.79 &     $\leq$3.7    & 112         &     15.7 &emission\\
3610          &    12.80 &     $\geq$3.7      &159         &     22.3 &\\
4278          &    32.89 &       $\leq$3.3   & 266         &      37.2& emission\\
4308          &    1.11   &       $\geq$3.7    & 88           &      12.3&\\
4365          &    57.16 &      $\leq$3.3    & 248         &      24.8 &\\
4406          &    90.60 &      $\leq$3.3    & 250         &      25.0&\\
4434          &    18.50 &       $\leq$3.3   &117          &      11.7&\\
4458          &    26.74 &       $\leq$3.3    &106         &       10.6&\\
4472          &   104.02 &      $\leq$3.3   &287          &      28.7 &\\
4473          &    24.95 &        $\leq$3.3  &178           &     17.8&\\
4478          &    14.03 &        $\leq$3.7   &149         &      14.9 &\\
4486B       &    3.07    &       $\geq$3.7  &161           &      22.5&\\
4486          &   104.02 &      $\leq$3.3   &361          &     36.1 &emission\\
4489          &    32.15 &        $\leq$3.7  &49             &    4.9    &\\
4564          &    21.73 &        $\leq$3.3  &153           &    15.3    &\\

\hline
\end{tabular}
\caption{Virgo Cluster galaxy parameters.}
\end{center}
\end{table}

\renewcommand{\thetable}{6.2}
\begin{table}[H]
\begin{center}
\begin{tabular}{ l c c c }
\multicolumn{4}{c}{Table 6.2} \\
\hline
\hline
NGC       &    PA &b/a &reference\\
             &(degrees) &       &\\
\hline
1400      &  35.0 & 0.900    &\cite{skru06}  \\
1407      &60.0 &   0.950      &  \cite{jar03} \\
2768      &92.5 &   0.460      &  \cite{jar03} \\
3156      &49.5 &   0.571       &  \cite{adel08} \\
3610     &134.2&   0.776       &  \cite{adel08}\\
4278      &21.8 &   0.925      &   \cite{adel08} \\
4308      &40.6 &   0.926        &  \cite{adel08}\\
4365      &45.0 &   0.740      &  \cite{jar03}   \\
4406       &125.0& 0.670      &   \cite{jar03}   \\
4434       &44.9 &  0.958      &   \cite{adel08} \\
4458      &11.3 &    0.900         &   \cite{adel08} \\
4472      &172.1&   0.399         &   \cite{adel08} \\
4473       &95.0 &    0.540         &  \cite{jar03}  \\
4478      &160.5&    0.942       &   \cite{adel08} \\
4486B    &101.3&   0.926      &  \cite{adel08} \\
4486     &152.5&     0.860         &  \cite{jar03}  \\
4489     &4.4   &        0.917          &  \cite{adel08} \\
4564     &50.0 &        0.440           &  \cite{skru06}   \\

\hline
\end{tabular}
\end{center}
\caption{Virgo Cluster galaxy parameters.}
\end{table}

\section{Analysis}

To analyze the radial properties of these 18 Virgo galaxies, the two dimensional spectra from the observations were further reduced by extracting them through 61 single pixel wide apertures, so that the 31st aperture was centered on the peak of the spectral flux. This produced 61 one dimensional spectra that sampled different radii symmetrically about the center of the galaxies. 

The one dimensional spectra were then measured using the 25 Lick indices \citep{wor94,wor97}, 1 H$\alpha$ index defined in \cite{Cohen} and 48 of the indices defined in \cite{serv05}. These measurements where then analyzed to check that they where indeed symmetric and then to look for index trends within individual galaxies.

To look for symmetry about the center of the galaxies (aperture 31) index measurements from symmetric pairs (e.g., apertures 30 and 32) where compared and found to agree confirming that the spectra where in fact symmetric about the center. The index measurements of symmetric pairs were then averaged together, except for the center aperture and its nearest neighbors (e.g., apertures 30, 31 and 32) which where averaged together as a set. The end result is a set of 14 data points for each index per galaxy. The 14 data points come as a result of cutting off data points with large errors.

The radial increments come from the Cassegrain spectrograph's angular resolution. The angular resolution is .69 arcsec / pixel in the spatial direction. Applying this to the aperture extracted spectra means that aperture 31 is at radius of 0 arcsec, aperture 32 is at a radius of 1.38 arcsec, etc. These intervals are what is represented in the figures of Appendix C where they have been transformed into the log of units of the effective radius. The exception is in the first data point where the average of the distance between the center aperture and its nearest neighbors has been used.

In plots of Appendix C a trend line was then fit in order to characterize the correlations. This was done with fitexy.f \citep{numrec} which is a fortran program for finding the best fit line from data with errors in the x and y coordinate.

\section{Discussion, Summary, and conclusion}

What can be seen in the plots of Appendix C is that index trends tend to vary considerably from galaxy to galaxy and from index to index making any sort of general statement about them difficult to make without a long list of exceptions. However, some trends can be seen. For example, the Fe Profiles (App. C Pages 153- 162) do tend to be shallow implying that the metallicity stays relatively constant throughout the galaxies in agreement with what others have found \citep{gon93,kunt98}. The most notable exception is that of NGC 4486 on page 48, where the inner most data point takes an exceptional dive. This behavior is not confined to this index and its presence in many others suggests the presence of a decoupled galaxy core.

Also in agreement with what others have found is that the Mg to Fe ratio is more or less constant with radius. This is implied by the slope of the radial measurements in Figure 6.1, where Mg $b$ vs. $<$Fe$>$ is plotted for different radii of four representative galaxies. The central measurements are in the top right with decreasing radii toward the bottom left, and they parallel constant age model sequences, as would be expected for global Mg enrichment. This observation, also implies that a mechanism for the chemical enrichment in these galaxies, most likely Type {\sc II} supernovae, is the same throughout the galaxy. This agrees with the conclusions found in \cite{wor98}, who speculated that the environment may play a role in three key ways. One, the upper initial mass function is stronger in large ellipticals (more Type {\sc II} supernovae and thus more light elements). Two, the Type {\sc I}a timescale is delayed in large ellipticals (less Fe peak enrichment), or that stellar binarism is keyed to environment.

\renewcommand{\thefigure}{6.1}
\begin{figure}[H]
\includegraphics[width=3.5in]{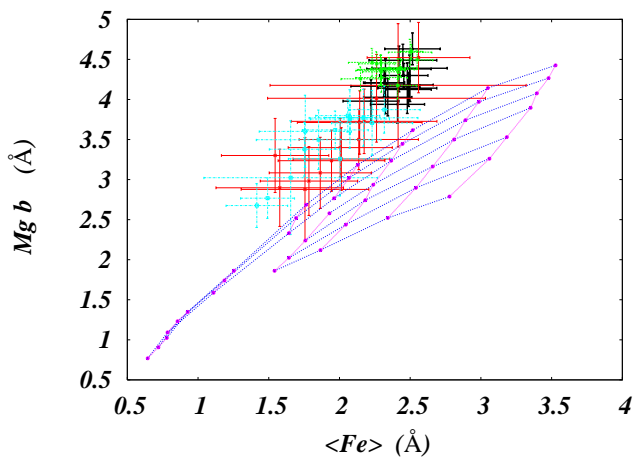}
\caption{Mg $b$ and $<$Fe$>$ measurements of four representative galaxies (black, red, green and light blue) for various radii are plotted. Single stellar population models (blue grid lines) from bottom to top of ages 17, 12, 8, 5, 3, 2, and 1.5 Gyrs and models from right to left of metallicities $0.5, 0.25, 0, -0.25, -0.5, -1, -1.5$ and $-2$ (pink grid lines) are also plotted. The central measurement for each galaxy is in the top right with decreasing radii toward the bottom left. The constant slope indicates that the [Mg/Fe] remains constant as a function of radius.
\label{Fig1} }
\end{figure}

Indices that are sensitive to C, N and O (App. C Pages 114 - 120) tend to decline with increasing radius suggesting that these elements are enhanced in the core of these galaxies.

H$\beta$ and H$\alpha$ are relatively flat showing little to no change with increasing radius. This coupled with the fact that most of the metallic indicators decline with increasing radius suggest that these galaxies are mildly younger toward the center as found by \cite{hen99}. This is illustrated in Figure 6.2 where the same 4 representative galaxies have been plotted as in Figure 6.1. In Figure 6.2, H$\beta$ vs. [MgFe] is plotted for different radii for the four representative galaxies. The central measurements are in the top right with decreasing radii toward the bottom left, and they cross constant age model sequences (horizontal blue lines), as would be expected for galaxy centers which are slightly younger than their outskirts.

\renewcommand{\thefigure}{6.2}
\begin{figure}[H]
\includegraphics[width=3.5in]{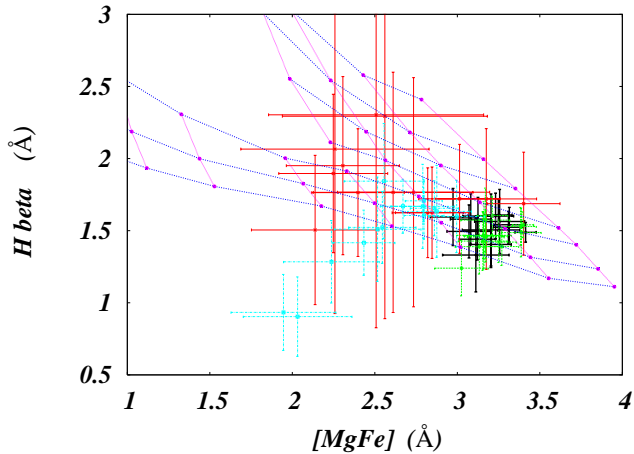}
\caption{H$\beta$ and [MgFe] measurements of four representative galaxies (black, red, green and light blue) for various radii are plotted. Single stellar population models (blue grid lines) from bottom to top of ages 17, 12, 8, 5, 3, 2, and 1.5 Gyrs and models from right to left of metallicities $0.5, 0.25, 0, -0.25, -0.5, -1, -1.5$ and $-2$ (pink grid lines) are also plotted. The central measurement for each galaxy is in the top right with decreasing radii toward the bottom left. The crossing of constant age lines indicates that the centers of these galaxies are slightly younger than the outskirts.
\label{Fig1} }
\end{figure}

The Mg indices (App. C Pages 122 - 126) are of interest as the index trend seen by other authors \citep{gon93,kunt98} is only clear in the Mg3835, Mg4780 and Mg $b$ indices. The other large Mg indices (Mg 1 and Mg 2) do show a decline with radius but only after an initial bump up from the center. The fact that this behavior is seen in almost all of the galaxies and not in Mg $b$ suggests that there is a wavelength dependent focal variation in the spectra that may effect this region of the spectrum for indices that have wavelength spans of 400 \AA\ or more. 

Although there is a wealth of information that these indices hint at, full modeling of these galaxies is required to decouple the various elemental influences present in each index before anything definite can be said about the individual elemental abundances. The modeling of these galaxies through the use of these indices will be the subject of future work.

\newpage
{\centering CHAPTER SEVEN\\
The Results\\}

Throughout the course of these investigations, several physical implications about the nature of elliptical galaxies and there evolutionary history have been discussed and implied. These implications range from age and metallically trends across galactic populations to considerations of individual galaxy structure.

In Chapter 2, the possibility of determining the existence of individual elemental abundances was investigated. It was found that individual elemental abundances for elliptical galaxies were in fact, observable. The presence of some of these individual abundance patterns was found in Chapter 5 and in Appendix B, where it is observable as scatter in any of the many figures found there. The implication is that, for each galaxy, an individual evolutionary history maybe needed to explain each galaxy's chemical profile. 

Also in Chapter 5 and Appendix B, what is apparent from the figures is that along with individual chemical profiles there are systematic chemical trends that cross galactic populations. These trends (i.e., Mg $b$ and CaHK) are found to correlate with galaxy size and suggests that environment plays a significant role in the evolution of elliptical galaxies. The evidence suggest that these trends are an artifact of supernovae enrichment that depends via mechanisms that are ambiguous on environment.

In Chapter 3, the important side topic of hydrogen emission in the elliptical spectra from the Sloan Digital Sky Survey was investigated. The presence of emission contamination was determined by direct comparison of index measurements between the survey spectra and spectra of 18 Virgo Cluster galaxies. It was found that the age determinations of the survey averages were 2 to 3 Gyr too young. This is an important constraint for the evolutionary picture due to the fact that the initial formation of galaxies must take place sooner on a universal time scale if galaxies are older, and puts tighter limits on late epoch star formation.

In Chapter 4, the presence of an old, metal-poor stellar subpopulation in elliptical galaxies was investigated and found to account for observed index trends, especially in the terrestrial ultraviolet. The presence of this subpopulation is important for two reasons. One it certainly needs to be accounted for in order to accurately model galactic properties such as age. Two, this blue population needs to be taken into account for an accurate determination of the N abundance from the NH feature.

In Chapter 6, individual Virgo Cluster galaxies were spatially dissected and with the use of index measurements several properties of elliptical galaxy structure was deduced. The physical implications are twofold. One is that the chemical make up of individual galaxies tend to be uniform form the center to the edge, with a few exceptions such as decoupled cores. Two is that the cores of these elliptical galaxies tend to be younger than the outskirts, meaning they had star formation later or for a more prolonged period.

The physical implications of this aid in the understanding of the history and evolution of elliptical galaxies.

\newpage
\pagestyle{plain}

{\centering APPENDIX A\\
INDEX RESPONSE TABLE\\}
These 6 tables give the sensitivity of each index to a given element. The 1st column of the table gives the indices names, the 2nd gives the index measurements in angstroms of equivalent width, and the 3rd giving the error associated with S/N = 500 at 5000 \AA\/. The next 23 columns split between tables are the changes (enhanced minus unenhanced in units of angstroms) in the index due to the elemental enhancement 0.3 dex (0.15 dex for C) of the element labeling the column. The 24th column is the change brought about in the index by increasing all the alpha elements by a factor of 2 (upX2).

\begin{table}[H]
\begin{center}
\scriptsize
\begin{tabular}{ l c c c c c c c c c c c c c}
\multicolumn{14}{c}{Table 1} \\
\hline
\hline
              &  $I_0$ & $\sigma$ & & & & & & & & & & & \\
Index   &   (\AA\/) &     (\AA\/) &     C &       N &      O &      Na &     Mg &    Al &     Si &      S &      K &      Ca &     Sc  \\
\hline
CN$_1$          &    0.036 &    0.005 &   3.69 &   6.52 &  -7.73 &  -0.23 &  -0.67 &  -0.27 &   2.40 &  -0.40 &  -0.02 &  -1.38 &   0.04 \\
CN$_2$          &    0.097 &    0.006 &   3.19 &   5.86 &  -7.58 &  -0.25 &  -0.81 &  -0.30 &   3.19 &  -0.40 &  -0.02 &  -1.44 &  -0.02 \\
Ca4227           &    2.394 &    0.084 &  -3.55 &  -2.41 &  -1.24 &  -0.05 &   1.61 &   0.09 &   0.19 &  -0.21 &   0.00 &   6.57 &  -0.43 \\
G4300              &    6.984 &    0.154 &   4.12 &  -0.05 &  -3.85 &  -0.43 &  -0.90 &  -0.14 &  -0.81 &  -0.16 &  -0.01 &   1.00 &   0.23 \\
Fe4383            &    7.526 &    0.222 &   0.23 &  -0.38 &  -3.09 &  -0.10 &  -1.18 &  -0.35 &  -0.80 &  -0.18 &  -0.02 &  -1.11 &   0.13 \\
Ca4455            &    1.826 &    0.127 &  -0.39 &  -0.19 &  -1.39 &  -0.04 &  -0.07 &  -0.03 &  -0.11 &  -0.10 &   0.00 &   0.23 &   0.00 \\
Fe4531             &    4.450 &    0.190 &  -0.40 &  -0.19 &  -1.75 &   0.06 &  -0.55 &  -0.06 &  -0.81 &  -0.11 &   0.00 &  -0.25 &  -0.01 \\
C$_2$4668      &    4.251 &    0.315 &   4.88 &  -0.45 &  -5.08 &  -0.23 &  -0.70 &  -0.02 &  -1.81 &  -0.24 &   0.00 &   0.02 &  -0.11 \\
H$\beta$           &    1.153 &    0.135 &  -0.04 &   0.13 &   1.55 &   0.24 &  -1.88 &   0.09 &  -0.68 &   0.10 &   0.01 &   0.03 &  -0.07 \\
Fe5015              &    5.375 &    0.298 &  -0.82 &  -0.17 &  -1.44 &   0.15 &  -1.72 &   0.07 &  -0.35 &  -0.15 &   0.00 &   0.13 &  -0.04 \\
Mg$_1$             &    0.139 &    0.003 &   7.73 &  -0.91 & -10.15 &  -0.67 &  13.39 &  -0.88 &   0.24 &  -0.48 &  -0.06 &  -1.18 &   0.03 \\
Mg$_2$             &    0.315 &    0.004 &   2.07 &  -1.05 &  -9.93 &  -0.51 &  19.32 &  -0.78 &   0.02 &  -0.56 &  -0.05 &  -0.68 &  -0.27 \\
Mg $b$              &    4.882 &    0.149 &  -1.10 &  -0.33 &  -2.03 &  -0.35 &  10.00 &  -0.21 &  -0.15 &  -0.14 &  -0.01 &   0.02 &  -0.01 \\
Fe5270              &    4.149 &    0.162 &  -0.06 &  -0.11 &  -2.98 &  -0.14 &  -0.68 &  -0.17 &  -0.08 &  -0.13 &  -0.01 &   0.80 &  -0.14 \\
Fe5335              &    3.700 &    0.184 &  -0.31 &  -0.27 &  -2.41 &  -0.13 &  -0.55 &  -0.19 &  -0.09 &  -0.14 &  -0.01 &  -0.14 &   0.08 \\
Fe5406              &    2.475 &    0.142 &  -0.21 &  -0.17 &  -2.07 &  -0.10 &  -0.32 &  -0.13 &  -0.07 &  -0.13 &  -0.01 &  -0.17 &   0.03 \\
Fe5709              &    1.176 &    0.119 &  -0.22 &  -0.19 &  -1.10 &  -0.45 &   0.08 &  -0.04 &  -0.04 &  -0.08 &   0.00 &  -0.07 &   0.15 \\
Fe5782              &    0.817 &    0.115 &  -0.18 &  -0.16 &  -1.25 &  -0.02 &   0.18 &  -0.03 &   0.03 &  -0.09 &   0.00 &  -0.06 &   0.00 \\
Na D                   &    4.602 &    0.143 &   0.22 &  -0.36 &  -5.04 &   7.99 &   0.49 &  -0.16 &   0.34 &  -0.26 &  -0.01 &  -0.36 &   0.01 \\
TiO$_1$             &    0.031 &    0.004 &  -1.67 &   0.17 &   0.42 &   0.14 &   0.31 &   0.14 &  -0.17 &  -0.14 &   0.00 &  -0.03 &   0.00 \\
TiO$_2$             &    0.068 &    0.003 &  -3.15 &  -0.03 &   0.15 &   0.21 &   0.52 &   0.36 &  -0.58 &  -0.36 &   0.00 &  -0.55 &   0.33 \\
H$\delta_A$      &   -5.490 &    0.187 &  -0.35 &  -0.23 &   5.79 &   0.22 &   0.60 &   0.36 &   1.00 &   0.39 &  -0.03 &   1.34 &  -0.26 \\
H$\gamma_A$ &   -9.781 &    0.207 &  -2.59 &   0.37 &   6.33 &   0.37 &   1.93 &   0.39 &   0.59 &   0.37 &   0.02 &  -0.41 &  -0.14 \\
H$\delta_F$      &   -0.600 &    0.120 &   0.47 &   0.23 &   2.01 &   0.05 &  -0.07 &   0.11 &   2.32 &   0.18 &   0.00 &   1.28 &  -0.05 \\
H$\gamma_F$ &   -2.258 &    0.120 &  -2.92 &   0.21 &   4.35 &   0.28 &   1.17 &   0.27 &   0.29 &   0.24 &   0.01 &  -0.41 &  -0.38 \\
H$_\alpha$       &    0.660 &    0.137 &   1.70 &   0.47 &  -0.22 &  -0.09 &  -0.71 &  -0.08 &  -0.06 &   0.15 &   0.00 &  -0.23 &  -0.04 \\
\hline
\end{tabular}
\end{center}
\end{table}

\begin{table}[H]
\begin{center}
\scriptsize
\begin{tabular}{ l c c c c c c c c c c c c c}
\multicolumn{14}{c}{Table 1 Continued} \\
\hline
\hline
              & & & & & & & & & & & & &  \\
Index   &      Ti &     V &     Cr  &     Mn &     Fe &     Co &     Ni &     Cu &     Zn &    Sr &     Ba &     Eu &    upX2 \\
\hline
CN$_1$   &   0.46 &  -1.42 &  -1.31 &  -0.69 &  -0.48 &  -0.42 &   0.46 &  -0.02 &   0.00 &  -0.04 &   0.06 &  -0.02 &  -7.81 \\
CN$_2$     &   0.58 &  -0.65 &  -0.96 &  -0.54 &  -0.61 &  -0.58 &   0.42 &   0.00 &   0.00 &   0.07 &   0.05 &   0.00 &  -7.32 \\
Ca4227      &   0.69 &   0.62 &  -0.24 &   0.19 &   1.60 &   0.10 &   0.11 &   0.00 &   0.00 &  -0.30 &   0.00 &   0.00 &   5.11 \\
G4300    &   2.34 &   0.05 &  -0.64 &  -0.03 &  -2.44 &   0.04 &   0.31 &  -0.02 &   0.00 &   0.00 &   0.00 &   0.00 &  -2.88 \\
Fe4383      &   0.40 &   0.76 &   0.17 &  -0.57 &   6.97 &  -0.01 &   0.15 &   0.00 &   0.00 &   0.00 &   0.02 &   0.00 &  -6.73 \\
Ca4455      &   0.42 &   0.89 &   1.18 &   0.77 &  -1.53 &   0.14 &   0.49 &   0.00 &   0.00 &   0.00 &   0.00 &   0.00 &  -1.28 \\
Fe4531        &   2.18 &   0.04 &   0.74 &  -0.18 &   0.48 &   0.23 &   0.20 &   0.02 &   0.00 &   0.01 &   0.76 &   0.02 &  -1.71 \\
C$_2$4668 &   1.29 &  -0.42 &  -0.33 &  -0.52 &   0.26 &  -0.33 &   0.20 &   0.04 &   0.04 &  -0.05 &   0.01 &  -0.02 &  -6.14 \\         
H$\beta$      &   0.33 &  -0.06 &  -0.45 &  -0.10 &  -0.62 &  -0.13 &   0.87 &   0.00 &   0.00 &  -0.02 &   0.00 &   0.00 &   0.40 \\
Fe5015        &   2.59 &   0.04 &  -0.21 &   0.08 &   1.02 &  -0.01 &   0.71 &   0.00 &   0.00 &  -0.03 &   0.00 &   0.00 &  -0.62 \\
Mg$_1$        &  -1.24 &  -0.06 &  -0.64 &   0.09 &  -5.09 &  -0.36 &   1.12 &   0.21 &   0.00 &   0.00 &  -0.15 &   0.00 &  -5.15 \\
Mg$_2$          &   0.78 &   0.05 &  -0.73 &   0.02 &  -3.95 &  -0.46 &   0.44 &   0.05 &   0.00 &   0.00 &  -0.07 &   0.00 &   2.80 \\
Mg $b$        &   0.32 &  -0.06 &  -2.20 &  -0.18 &  -1.55 &  -0.07 &  -0.02 &  -0.07 &   0.00 &   0.00 &   0.00 &   0.00 &   6.25 \\
Fe5270        &   0.27 &  -0.20 &   0.05 &   0.21 &   3.30 &   0.31 &   0.15 &  -0.01 &   0.00 &   0.02 &   0.00 &   0.00 &  -3.54 \\
Fe5335      &   0.19 &   0.01 &   0.79 &   0.11 &   3.27 &   0.38 &   0.03 &   0.00 &   0.00 &   0.00 &   0.00 &   0.00 &  -3.94 \\  
Fe5406      &  -0.07 &   0.07 &   0.61 &   0.26 &   3.02 &  -0.05 &   0.02 &   0.00 &   0.00 &   0.00 &  -0.01 &   0.00 &  -3.49 \\
Fe5709      &   0.26 &  -0.13 &   0.41 &  -0.03 &   0.91 &   0.00 &   0.21 &   0.13 &   0.00 &   0.00 &   0.00 &   0.00 &  -0.83 \\
Fe5782      &  -0.17 &   0.11 &   1.67 &   0.04 &  -0.28 &   0.00 &  -0.07 &   0.19 &   0.00 &   0.00 &   0.00 &   0.00 &  -1.46 \\
Na D       &  -0.63 &   0.02 &   0.00 &   0.00 &  -0.37 &  -0.01 &   0.08 &   0.00 &   0.00 &   0.00 &   0.00 &   0.00 &  -8.41 \\
TiO$_1$          &   3.44 &  -0.31 &  -0.14 &  -0.06 &  -0.22 &  -0.03 &  -0.25 &   0.00 &   0.00 &   0.00 &   0.03 &   0.00 &   3.92 \\
TiO$_2$          &   5.36 &   0.58 &  -0.09 &  -0.09 &   0.15 &  -0.12 &  -0.09 &   0.00 &   0.00 &   0.00 &  -0.03 &   0.00 &   4.45 \\
H$\delta_A$      &  -1.08 &   2.11 &   0.10 &  -1.39 &  -9.96 &   0.46 &  -0.34 &   0.00 &   0.00 &  -0.57 &  -0.06 &  -0.05 &   8.69 \\
H$\gamma_A$     &  -2.86 &  -0.50 &   0.85 &  -0.23 &  -4.99 &  -0.02 &  -0.41 &   0.00 &   0.00 &   0.00 &   0.00 &   0.00 &   7.25 \\
H$\delta_F$      &  -1.15 &   1.16 &  -0.12 &  -0.88 &  -4.96 &  -0.40 &  -0.17 &   0.00 &   0.00 &  -0.61 &  -0.09 &  -0.10 &   4.34 \\
H$\gamma_F$     &  -0.64 &  -0.08 &   0.98 &  -0.11 &  -2.98 &  -0.02 &  -0.27 &   0.00 &   0.00 &   0.00 &   0.02 &   0.00 &   5.66 \\
H$\alpha$        &  -1.67 &  -0.06 &  -0.03 &  -0.01 &  -0.65 &   0.04 &  -0.18 &   0.00 &   0.00 &   0.00 &  -0.02 &  -0.01 &  -2.47 \\
\hline
\end{tabular}
\end{center}
\end{table}

\begin{table}[H]
\begin{center}
\scriptsize
\begin{tabular}{ l c c c c c c c c c c c c c}
\multicolumn{14}{c}{Table 1} \\
\hline
\hline
              &  $I_0$ & $\sigma$ & & & & & & & & & & & \\
Index   &   (\AA\/) &     (\AA\/) &     C &       N &      O &      Na &     Mg &    Al &     Si &      S &      K &      Ca &     Sc  \\
\hline
CO4685             &    3.699 &    0.334 &   6.45 &  -1.14 &  -6.15 &   0.08 &  -0.28 &  -0.01 &  -2.34 &  -0.23 &   0.00 &  -0.77 &   0.15 \\
CO5161             &    0.352 &    0.086 &   4.24 &  -0.09 &  -3.42 &  -0.01 &   1.91 &  -0.13 &  -0.51 &  -0.03 &   0.00 &  -0.23 &   0.01 \\
CNO3862          &   10.767 &    0.196 &   9.39 &  15.11 & -18.54 &  -0.10 &   3.36 &  -0.13 &  -2.91 &  -0.87 &  -0.01 &  -6.44 &  -0.02 \\
CNO4175          &    3.109 &    0.286 &   3.92 &   7.09 &  -8.59 &  -0.20 &  -1.22 &  -0.33 &   2.47 &  -0.42 &  -0.01 &  -0.94 &  -0.14 \\
Na8190             &    1.834 &    0.190 &   0.11 &  -0.48 &  -1.77 &   1.94 &  -0.26 &  -0.14 &   0.27 &   0.08 &  -0.01 &  -0.11 &   0.02 \\
Mg3835             &    6.768 &    0.160 &  -3.40 &  -4.30 &   2.29 &   0.08 &   6.27 &   0.18 &   0.41 &   0.08 &   0.01 &  -0.09 &   0.11 \\
Mg4780             &    1.854 &    0.163 &  -0.66 &  -0.16 &  -0.76 &  -0.47 &   4.35 &  -0.25 &   0.49 &  -0.09 &  -0.02 &  -0.31 &  -0.26 \\
Al3953               &   -2.718 &    0.240 &  -0.85 &  -0.29 &  -1.52 &   0.00 &  -0.22 &   5.07 &  -0.44 &  -0.10 &   0.00 &   4.15 &  -0.06 \\
Si4101               &    0.273 &    0.096 &   0.38 &   0.11 &   1.83 &   0.02 &  -0.20 &   0.06 &   2.60 &   0.17 &   0.00 &   1.05 &   0.02 \\
Si4513               &   -1.400 &    0.166 &   0.35 &   0.36 &  -0.04 &   0.05 &  -0.50 &  -0.14 &   2.87 &   0.00 &  -0.01 &  -0.58 &  -0.04 \\
S4693                &    0.342 &    0.113 &   2.14 &  -0.19 &  -2.17 &  -0.10 &   0.73 &   0.00 &  -0.35 &   0.06 &   0.00 &   0.26 &  -0.17 \\
K4042                &    0.415 &    0.085 &  -0.14 &  -0.02 &  -0.14 &  -0.06 &  -0.32 &  -0.13 &   0.13 &  -0.01 &   0.18 &  -0.55 &  -0.35 \\
Cahk                  &   18.122 &    0.365 &  -1.67 &  -2.79 &   0.43 &  -0.24 &  -6.60 &   2.19 &  -2.29 &  -0.20 &  -0.12 &  25.97 &  -0.38 \\
Ca8542             &    3.196 &    0.174 &  -0.28 &  -0.24 &  -1.39 &  -0.16 &  -1.34 &  -0.19 &  -0.56 &  -0.11 &  -0.01 &   5.69 &   0.00 \\
Ca8662             &    2.339 &    0.149 &  -0.08 &  -0.29 &  -1.22 &  -0.09 &  -1.40 &  -0.19 &  -0.57 &   0.02 &  -0.01 &   5.18 &   0.00 \\
Sc3613             &    4.723 &    0.101 &  -3.20 &  -3.34 &   0.86 &  -0.01 &  -0.55 &   0.02 &   0.73 &  -0.04 &   0.00 &  -0.07 &   1.11 \\
Sc4312             &    0.553 &    0.116 &   2.03 &   0.11 &  -0.79 &  -0.04 &  -1.25 &  -0.15 &  -0.03 &   0.01 &  -0.01 &   0.77 &   0.33 \\
Sc6292             &    0.487 &    0.199 &  -1.51 &  -0.80 &   1.49 &   0.04 &   0.53 &   0.16 &   0.13 &   0.00 &   0.00 &   0.06 &   0.35 \\
Ti4296              &    5.974 &    0.111 &   4.49 &  -0.19 &  -5.64 &  -0.23 &  -0.88 &  -0.16 &  -1.02 &  -0.29 &  -0.01 &   1.36 &   0.32 \\
Ti4533              &    1.544 &    0.088 &  -0.23 &  -0.14 &  -1.80 &  -0.07 &   0.34 &  -0.03 &   0.26 &  -0.11 &   0.00 &   0.21 &  -0.03 \\
Ti5000              &    2.184 &    0.209 &  -1.40 &  -0.23 &  -0.87 &   0.31 &  -0.44 &   0.06 &  -0.21 &  -0.18 &   0.00 &  -0.06 &  -0.11 \\
V4112               &   -6.730 &    0.287 &  -1.11 &  -0.89 &   5.05 &   0.18 &   0.78 &   0.29 &   1.03 &   0.30 &   0.01 &   1.00 &  -0.18 \\
V4928               &    0.535 &    0.101 &   0.58 &   0.02 &  -0.71 &  -0.04 &  -0.34 &  -0.04 &  -0.08 &   0.01 &   0.00 &  -0.05 &   0.03 \\
V6604             &    0.199 &    0.117 &  -0.40 &  -0.55 &   0.36 &  -0.03 &  -0.02 &  -0.06 &  -0.04 &  -0.01 &   0.00 &  -0.04 &   0.16 \\
Cr3594           &    1.577 &    0.124 &  -1.77 &  -1.47 &   0.53 &   0.05 &   0.64 &   0.07 &   0.49 &  -0.04 &   0.00 &   0.08 &   0.07 \\
Cr4264           &    3.251 &    0.156 &   0.00 &  -3.21 &  -1.22 &  -0.13 &   0.44 &  -0.13 &   0.47 &  -0.11 &  -0.01 &  -0.50 &  -0.07 \\
Cr5206          &    1.688 &    0.085 &  -2.02 &  -0.38 &  -0.59 &  -0.15 &   2.03 &  -0.15 &   0.60 &  -0.13 &  -0.01 &  -0.22 &  -0.09 \\
Mn3794        &   -6.864 &    0.233 &   0.08 &  -0.13 &   3.40 &   0.06 &   1.39 &   0.07 &  -0.19 &   0.24 &   0.01 &  -0.19 &   0.00 \\
Mn4018         &  -14.657 &    0.305 &   0.18 &  -1.83 &   3.63 &   0.13 &   2.06 &  -3.82 &  -0.46 &   0.27 &   0.05 & -14.36 &   0.59 \\
Mn4461         &    0.494 &    0.152 &  -3.64 &  -0.29 &   0.62 &   0.07 &   1.74 &   0.32 &  -1.64 &  -0.04 &   0.02 &   1.61 &  -0.14 \\
Mn4757        &    0.345 &    0.141 &  -2.41 &   0.08 &   1.16 &   0.23 &   1.21 &  -0.06 &   0.47 &   0.06 &   0.00 &  -0.05 &   0.02 \\
Fe4058         &    6.046 &    0.165 &  -1.33 &  -1.25 &  -1.96 &  -0.17 &   0.03 &  -0.29 &  -1.26 &  -0.15 &   0.09 &  -1.67 &  -0.15 \\
Fe4930         &    3.160 &    0.168 &  -0.21 &  -0.20 &  -1.94 &  -0.06 &  -0.44 &  -0.08 &  -0.33 &  -0.12 &  -0.01 &  -0.11 &   0.08 \\
\hline
\end{tabular}
\end{center}
\end{table}

\begin{table}[H]
\begin{center}
\scriptsize
\begin{tabular}{ l c c c c c c c c c c c c c}
\multicolumn{14}{c}{Table 1 Continued} \\
\hline
\hline
              & & & & & & & & & & & & &  \\
Index   &      Ti &     V &     Cr  &     Mn &     Fe &     Co &     Ni &     Cu &     Zn &    Sr &     Ba &     Eu &    upX2 \\
\hline
CO4685           &   0.15 &  -0.74 &   0.59 &  -0.32 &   0.30 &   0.01 &   0.12 &   0.03 &  -0.01 &  -0.03 &  -0.13 &   0.02 &  -9.65 \\
CO5161           &  -1.85 &   0.01 &  -1.21 &  -0.15 &  -1.36 &   0.06 &   0.03 &   0.05 &   0.00 &   0.01 &   0.01 &   0.00 &  -4.45 \\
CNO3862          &  -0.83 &   0.59 &   0.57 &  -0.11 &  -2.62 &   0.40 &  -0.93 &   0.00 &   0.00 &   0.00 &   0.00 &  -0.08 & -25.83 \\
CNO4175          &  -0.22 &  -1.40 &  -1.75 &  -0.53 &   0.88 &  -0.59 &   0.43 &  -0.01 &   0.00 &   0.17 &   0.08 &   0.09 &  -9.67 \\
Na8190           &  -2.07 &   0.02 &   0.02 &   0.01 &   0.10 &   1.37 &   0.00 &  -0.03 &   0.00 &   0.00 &   0.00 &   0.00 &  -4.75 \\
Mg3835           &  -1.07 &   0.10 &  -0.73 &   0.56 &   0.06 &   0.46 &  -0.85 &   0.00 &   0.00 &   0.00 &   0.00 &   0.01 &   6.82 \\
Mg4780           &   0.84 &  -0.52 &   0.23 &   0.03 &   0.28 &   0.16 &  -0.32 &   0.00 &   0.00 &  -0.10 &   0.00 &   0.00 &   2.73 \\
Al3953           &  -0.24 &  -0.24 &  -0.35 &  -0.49 &  -2.78 &   0.01 &  -0.02 &   0.00 &   0.00 &   0.00 &   0.00 &  -0.04 &   0.15 \\
Si4101           &  -1.22 &   0.39 &  -0.29 &  -0.37 &  -4.20 &  -0.45 &  -0.22 &   0.00 &   0.00 &  -0.17 &  -0.06 &  -0.08 &   3.98 \\
Si4513           &   0.03 &  -0.81 &  -0.17 &  -0.70 &   0.64 &   0.10 &  -0.28 &   0.02 &   0.00 &   0.01 &  -0.50 &   0.02 &   1.13 \\
S4693            &   0.11 &  -0.23 &   0.37 &  -0.06 &  -1.49 &   0.03 &  -0.08 &  -0.02 &  -0.05 &  -0.02 &   0.02 &  -0.01 &  -1.55 \\
K4042            &  -1.09 &   0.29 &   0.12 &   1.71 &   2.80 &  -0.58 &  -0.56 &  -0.13 &   0.00 &  -0.01 &   0.00 &   0.00 &  -2.40 \\
Cahk             &   2.25 &  -0.30 &  -0.12 &  -7.49 &  -9.56 &   0.86 &  -0.39 &  -0.12 &   0.00 &   0.00 &   0.03 &   0.05 &  18.17 \\
Ca8542           &   0.08 &   0.00 &   0.03 &   0.00 &  -1.57 &  -0.15 &   0.00 &   0.00 &   0.00 &   0.00 &   0.01 &   0.00 &   2.56 \\
Ca8662           &  -0.55 &   0.00 &  -0.03 &  -0.03 &  -1.03 &  -0.01 &  -0.01 &   0.00 &   0.00 &   0.00 &   0.00 &   0.00 &   1.57 \\
Sc3613           &  -0.83 &  -0.75 &   0.72 &  -0.10 &   6.90 &  -0.45 &   1.53 &   0.00 &   0.00 &   0.00 &   0.00 &   0.00 &  -0.22 \\
Sc4312           &  -0.75 &  -0.27 &  -2.14 &   0.09 &   1.82 &  -0.03 &  -0.44 &   0.00 &   0.00 &   0.02 &  -0.01 &   0.00 &  -1.83 \\
Sc6292          &   0.81 &   0.13 &  -0.16 &  -0.01 &   0.21 &   0.06 &  -0.20 &   0.00 &  -0.03 &   0.00 &  -0.03 &   0.01 &   2.94 \\
Ti4296           &   3.15 &   0.05 &  -0.88 &  -0.18 &  -1.93 &   0.03 &   0.53 &  -0.01 &   0.00 &   0.01 &   0.00 &   0.00 &  -4.00 \\
Ti4533           &   1.95 &   0.14 &   0.68 &  -0.23 &   1.01 &   0.43 &  -0.50 &   0.04 &   0.00 &   0.01 &  -1.05 &   0.01 &  -0.12 \\
Ti5000           &   3.52 &   0.02 &  -0.36 &  -0.01 &   1.01 &  -0.01 &   0.07 &   0.00 &   0.00 &  -0.01 &  -0.16 &   0.00 &   1.68 \\
V4112            &  -0.95 &   1.91 &   0.42 &  -0.73 &  -8.03 &   0.17 &  -0.27 &   0.00 &   0.00 &  -0.25 &   0.07 &   0.09 &   7.62 \\
V4928            &  -0.55 &   0.13 &   0.16 &  -0.04 &  -0.18 &  -0.06 &   0.10 &   0.00 &   0.00 &  -0.04 &   0.39 &   0.00 &  -1.67 \\
V6604            &  -0.23 &   0.13 &  -0.10 &   0.00 &   0.79 &   0.03 &   0.18 &   0.00 &   0.00 &   0.00 &   0.00 &   0.00 &   0.26 \\
Cr3594           &   0.31 &   0.01 &   4.49 &  -0.15 &  -3.69 &  -0.41 &   1.83 &   0.00 &   0.00 &   0.00 &   0.00 &  -0.01 &   1.42 \\
Cr4264           &   0.35 &  -0.26 &   4.07 &   0.48 &   1.61 &   0.04 &  -0.30 &   0.02 &   0.00 &  -0.01 &   0.00 &   0.00 &  -1.75 \\
Cr5206           &   1.28 &  -0.05 &   5.26 &  -0.07 &  -1.19 &  -0.26 &  -0.30 &   0.03 &   0.00 &  -0.01 &   0.00 &   0.00 &   0.81 \\
Mn3794           &  -2.20 &   0.10 &  -0.07 &   0.47 &  -5.41 &  -0.44 &   0.93 &   0.00 &   0.00 &   0.00 &   0.00 &   0.00 &   3.16 \\
Mn4018           &   1.67 &  -1.19 &   0.40 &   4.72 &   6.00 &   0.16 &   0.45 &   0.11 &   0.00 &   0.00 &   0.03 &   0.00 &  -5.06 \\
Mn4461           &   0.82 &   0.80 &   0.50 &   2.11 &  -1.22 &  -0.11 &  -0.16 &  -0.01 &   0.00 &   0.00 &  -0.13 &  -0.02 &   3.13 \\
Mn4757           &  -0.43 &   0.09 &   0.44 &   1.36 &   0.02 &   0.21 &  -0.12 &   0.00 &  -0.25 &   0.00 &  -0.02 &   0.00 &   1.80 \\
Fe4058         &  -0.86 &  -0.75 &  -0.38 &   0.95 &  11.72 &  -0.40 &  -0.61 &  -0.15 &   0.00 &   0.77 &   0.00 &  -0.02 &  -6.37 \\
Fe4930          &   0.71 &  -0.01 &  -0.01 &   0.00 &   2.36 &  -0.14 &   0.17 &   0.00 &   0.00 &   0.00 &   0.23 &   0.00 &  -2.36 \\
\hline
\end{tabular}
\end{center}
\end{table}

\begin{table}[H]
\begin{center}
\scriptsize
\begin{tabular}{ l c c c c c c c c c c c c c}
\multicolumn{14}{c}{Table 1} \\
\hline
\hline
              &  $I_0$ & $\sigma$ & & & & & & & & & & & \\
Index   &   (\AA\/) &     (\AA\/) &     C &       N &      O &      Na &     Mg &    Al &     Si &      S &      K &      Ca &     Sc  \\
\hline
Co3701        &   -0.231 &    0.126 &  -0.11 &  -0.06 &  -0.28 &   0.00 &  -0.08 &   0.00 &  -0.24 &  -0.01 &   0.00 &   0.29 &  -0.12 \\
Co3840        &    3.279 &    0.143 &  -1.86 &  -2.69 &   1.68 &   0.04 &   4.55 &   0.09 &  -0.06 &   0.08 &   0.00 &  -0.05 &   0.11 \\
Co3876        &   -0.530 &    0.205 &   3.28 &   2.48 &  -2.23 &  -0.01 &   0.79 &   0.00 &  -0.29 &  -0.02 &   0.00 &  -3.78 &  -0.03 \\
Co7815       &    2.034 &    0.272 &  -0.65 &  -0.49 &  -0.37 &  -0.06 &  -0.44 &   0.14 &  -0.15 &  -0.06 &  -0.01 &  -0.15 &   0.00 \\
Co8185       &    1.753 &    0.199 &   0.15 &  -0.36 &  -1.82 &   1.83 &  -0.31 &  -0.12 &   0.19 &   0.05 &  -0.01 &  -0.07 &   0.01 \\
Ni3667        &   -5.338 &    0.205 &   1.09 &   0.16 &   2.50 &   0.07 &   1.04 &   0.12 &  -0.39 &   0.15 &   0.01 &  -0.07 &   0.11 \\
Ni3780        &   -7.051 &    0.255 &   2.13 &   1.03 &   3.09 &   0.01 &   0.88 &   0.04 &  -0.42 &   0.25 &   0.00 &  -0.16 &   0.00 \\
Ni4292        &    4.160 &    0.132 &   2.35 &  -0.16 &  -3.22 &  -0.24 &  -0.30 &  -0.05 &  -0.68 &  -0.18 &   0.00 &   0.72 &  -0.01 \\
Ni4910       &   -0.650 &    0.134 &   0.23 &   0.08 &   0.60 &   0.01 &  -0.01 &   0.01 &  -0.14 &   0.06 &   0.00 &  -0.07 &   0.10 \\
Ni4976       &    0.023 &    0.123 &  -0.06 &   0.04 &   0.19 &   0.46 &   1.02 &   0.07 &   0.26 &  -0.03 &   0.00 &   0.02 &  -0.08 \\
Ni5592       &    0.996 &    0.131 &   0.27 &  -0.27 &  -1.14 &  -0.07 &  -0.58 &  -0.07 &   0.22 &   0.01 &   0.00 &   2.51 &   0.01 \\
Cu5217      &   -2.304 &    0.108 &  -1.44 &   0.04 &   2.76 &   0.20 &  -6.18 &   0.28 &   0.12 &   0.07 &   0.01 &  -0.42 &  -0.02 \\
Cu5780      &    0.674 &    0.082 &  -0.12 &   0.00 &  -1.52 &  -0.02 &   0.08 &  -0.03 &   0.08 &  -0.11 &   0.01 &  -0.07 &  -0.04 \\
Zn4720       &   -0.635 &    0.118 &   1.97 &   0.02 &  -0.47 &  -0.07 &  -1.13 &   0.06 &  -0.39 &   0.01 &   0.00 &   0.08 &   0.20 \\
Ba4552      &    0.023 &    0.082 &  -0.04 &   0.11 &   0.18 &   0.08 &  -0.65 &   0.01 &  -0.86 &   0.00 &   0.00 &  -0.36 &   0.01 \\
Ba4933      &   -0.197 &    0.117 &   0.77 &   0.09 &  -0.16 &  -0.15 &  -0.54 &  -0.02 &  -0.03 &   0.06 &   0.00 &  -0.01 &   0.08 \\
Ba6142      &    0.143 &    0.117 &   0.72 &   0.01 &  -0.22 &   0.03 &  -0.40 &  -0.08 &   0.17 &   0.05 &   0.00 &  -0.96 &  -0.01 \\
Sr4076       &    1.235 &    0.085 &  -1.02 &  -0.52 &   0.50 &  -0.05 &  -0.39 &  -0.06 &  -0.90 &   0.02 &  -0.09 &  -0.33 &  -0.11 \\
Eu3970      &    6.259 &    0.080 &  -0.67 &   0.17 &  -0.68 &  -0.17 &  -2.07 &   0.85 &  -1.45 &  -0.16 &  -0.01 &   9.56 &  -0.06 \\
Eu4592      &    0.041 &    0.083 &  -1.05 &  -0.12 &   0.11 &  -0.11 &  -0.35 &  -0.01 &   0.36 &  -0.02 &   0.00 &   0.50 &   0.06 \\
OII               &    1.992 &    0.189 &  -0.54 &   0.57 &   0.19 &  -0.02 &   0.14 &  -0.02 &   0.71 &   0.01 &   0.00 &  -0.11 &  -0.01 \\
XX3580      &    6.517 &    0.282 &   2.68 &   4.89 &  -5.50 &   0.03 &  -0.75 &   0.06 &  -0.12 &  -0.29 &   0.00 &  -0.29 &   0.15 \\
NH3360     &    3.221 &    0.285 &  -0.05 &   5.18 &  -0.52 &  -0.03 &  -3.43 &  -0.04 &  -0.30 &  -0.05 &   0.00 &  -0.20 &   0.13 \\
NH3375     &    4.688 &    0.322 &  -0.20 &   4.51 &  -0.87 &  -0.03 &  -0.71 &  -0.04 &  -0.04 &  -0.05 &   0.00 &  -0.50 &  -0.02 \\
Mg3334      &    1.542 &    0.105 &  -0.43 &  -0.22 &  -1.14 &  -0.04 &   5.91 &  -0.05 &   0.14 &  -0.04 &   0.00 &  -0.35 &  -0.24 \\
\hline
\end{tabular}
\end{center}
\end{table}

\begin{table}[H]
\begin{center}
\scriptsize
\begin{tabular}{ l c c c c c c c c c c c c c}
\multicolumn{14}{c}{Table 1 Continued} \\
\hline
\hline
              & & & & & & & & & & & & &  \\
Index   &      Ti &     V &     Cr  &     Mn &     Fe &     Co &     Ni &     Cu &     Zn &    Sr &     Ba &     Eu &    upX2 \\
\hline
Co3701          &   0.07 &   0.39 &  -0.43 &   0.35 &  -0.24 &   0.52 &  -0.36 &   0.00 &   0.00 &   0.00 &   0.00 &  -0.05 &   0.02 \\
Co3840          &  -0.79 &   0.20 &  -0.71 &   0.30 &  -0.87 &   0.66 &  -0.88 &   0.00 &   0.00 &   0.00 &   0.00 &   0.00 &   4.74 \\
Co3876          &  -0.46 &   0.22 &  -0.28 &  -0.05 &  -0.82 &   0.44 &  -0.64 &   0.00 &   0.00 &   0.00 &   0.01 &  -0.01 &  -4.91 \\
Co7815         &  -0.20 &   0.00 &  -0.01 &  -0.02 &  -0.33 &   2.74 &   0.04 &   0.00 &   0.00 &   0.00 &   0.00 &   0.00 &  -1.03 \\
Co8185          &  -1.74 &   0.05 &   0.02 &   0.01 &  -0.15 &   1.38 &  -0.01 &   0.00 &   0.00 &   0.00 &   0.00 &   0.00 &  -4.53 \\
Ni3667           &  -1.16 &   0.10 &  -0.04 &  -0.14 &  -6.07 &  -0.64 &   0.75 &   0.00 &   0.00 &   0.00 &   0.00 &   0.01 &   3.17 \\
Ni3780         &  -2.63 &   0.16 &  -0.07 &  -0.61 &  -8.39 &  -0.40 &   1.76 &   0.00 &   0.00 &   0.00 &   0.00 &   0.00 &   2.31 \\
Ni4292           &   2.47 &   0.19 &   0.25 &   0.00 &  -2.62 &   0.07 &   0.78 &  -0.01 &   0.00 &  -0.01 &   0.01 &   0.00 &  -1.81 \\
Ni4910           &  -0.16 &  -0.24 &  -0.94 &  -0.06 &  -0.42 &  -0.03 &   0.51 &   0.00 &   0.00 &  -0.04 &   0.03 &   0.00 &   0.61 \\
Ni4976           &   1.07 &  -0.01 &  -0.34 &  -0.02 &  -3.22 &   0.00 &   0.62 &   0.00 &   0.00 &  -0.06 &   0.00 &   0.00 &   2.56 \\
Ni5592          &  -1.18 &   0.11 &   0.03 &  -0.05 &  -0.28 &   0.01 &   0.59 &   0.00 &   0.00 &  -0.01 &  -0.03 &   0.00 &  -0.70 \\
Cu5217          &   1.76 &   0.04 &   2.26 &  -0.14 &   0.06 &  -0.18 &  -0.05 &   0.46 &   0.00 &   0.02 &   0.00 &   0.00 &  -0.17 \\
Cu5780          &   0.19 &   0.02 &   1.52 &   0.01 &  -0.38 &  -0.02 &  -0.15 &   0.27 &   0.00 &   0.00 &   0.06 &   0.00 &  -1.38 \\
Zn4720          &  -0.72 &   0.36 &  -0.98 &  -0.62 &  -0.64 &  -0.25 &   0.43 &  -0.02 &   0.21 &   0.03 &   0.02 &  -0.02 &  -1.90 \\
Ba4552          &   0.71 &  -0.28 &  -0.11 &  -0.07 &  -0.97 &  -0.27 &   0.65 &  -0.02 &   0.00 &  -0.01 &   1.58 &  -0.01 &  -0.63 \\
Ba4933           &  -1.39 &   0.12 &   0.45 &   0.00 &  -0.38 &  -0.03 &  -0.13 &   0.00 &   0.00 &   0.00 &   0.40 &   0.00 &  -1.94 \\
Ba6142           &  -1.11 &  -0.06 &   0.11 &   0.02 &   1.18 &  -0.17 &  -0.44 &   0.00 &   0.00 &   0.00 &   0.32 &   0.00 &  -2.16 \\
Sr4076        &   0.18 &  -1.42 &  -0.53 &  -0.21 &   2.79 &   0.18 &   0.16 &   0.00 &   0.00 &   2.43 &   0.00 &  -0.01 &  -0.67 \\
Eu3970           &  -1.62 &  -0.34 &  -0.63 &  -0.14 &  -2.12 &  -0.07 &   0.26 &   0.00 &   0.00 &   0.00 &  -0.04 &   0.10 &   3.01 \\
Eu4592           &  -1.91 &   0.31 &   0.42 &  -0.04 &   1.82 &   0.30 &   0.07 &   0.00 &  -0.04 &   0.00 &  -0.11 &   0.00 &  -0.92 \\
OII              &  -1.30 &  -0.19 &  -0.12 &  -0.06 &   2.35 &  -0.39 &   0.23 &   0.00 &   0.00 &   0.00 &   0.00 &   0.14 &  -1.05 \\
XX3580           &  -0.99 &  -0.18 &   3.09 &   0.32 &   0.93 &  -1.66 &  -3.62 &   0.00 &   0.00 &   0.00 &  -0.03 &   0.00 &  -7.66 \\
NH3360           &   0.04 &   0.37 &  -0.18 &  -0.12 &  -1.70 &  -0.41 &   1.36 &  -0.04 &  -0.37 &   0.01 &   0.00 &   0.00 &  -3.82 \\
NH3375           &  -2.09 &   0.25 &  -0.08 &  -0.25 &  -0.78 &  -0.27 &   1.82 &   0.00 &  -0.57 &   0.00 &   0.00 &   0.01 &  -3.97 \\
Mg3334         &  -0.26 &  -0.52 &   0.18 &  -0.65 &  -0.29 &   0.03 &  -0.92 &   0.06 &  -0.26 &  -0.01 &   0.00 &   0.00 &   3.10 \\
\hline
\end{tabular}
\end{center}
\end{table}

\newpage
\pagestyle{plain}

{\centering APPENDIX B\\
INDEX VELOCITY DISPERSION TRENDS\\}
The Virgo spectra (red symbols with error bars), the Virgo emission spectra (black cross symbols with error bars), the SDSS spectra (green symbols with error bars) and the Toloba spectra (black box symbols with error bars) are plotted. Single stellar population models (blue grid lines) from bottom to top of ages 17, 12, 8, 5, 3, 2, and 1.5 Gyrs and models from right to left of metallicities $0.5, 0.25, 0, -0.25, -0.5, -1, -1.5$ and $-2$ (pink grid lines) are also plotted.

The following 74 figures consist of one index plotted against $\sigma$ in the first panel, Mg $b$ in the second and $<$Fe$>$ in the third. In general the index is fitted with a single black best fit line for all data sets. However if one data set is significantly offset from the others then it will have its own fit line colored the same as that data set and the other two will be fit by a black line. A red line is the fit for the Virgo data both emission and non. A green line is the fit for the SDSS data. There are plots for which no trend could be discerned for those plots there is no line fit.

The figures are arranged by order of atomic number starting with hydrogen; where there are more than one index for a given element or elements the indices are further arranged so that the bluest index is first.

The results of this arrangement puts hydrogen first followed by C, N and O then the alpha elements followed by the iron peak elements.

Note that in the H$\alpha$ plot (Fig. 1) the Virgo emission galaxies are not plotted because they fall well below the range of the other measurements. 

\renewcommand{\thefigure}{\arabic{figure}}
\setcounter{figure}{0}
\begin{figure}[H]
\includegraphics[width=7in,height=2in]{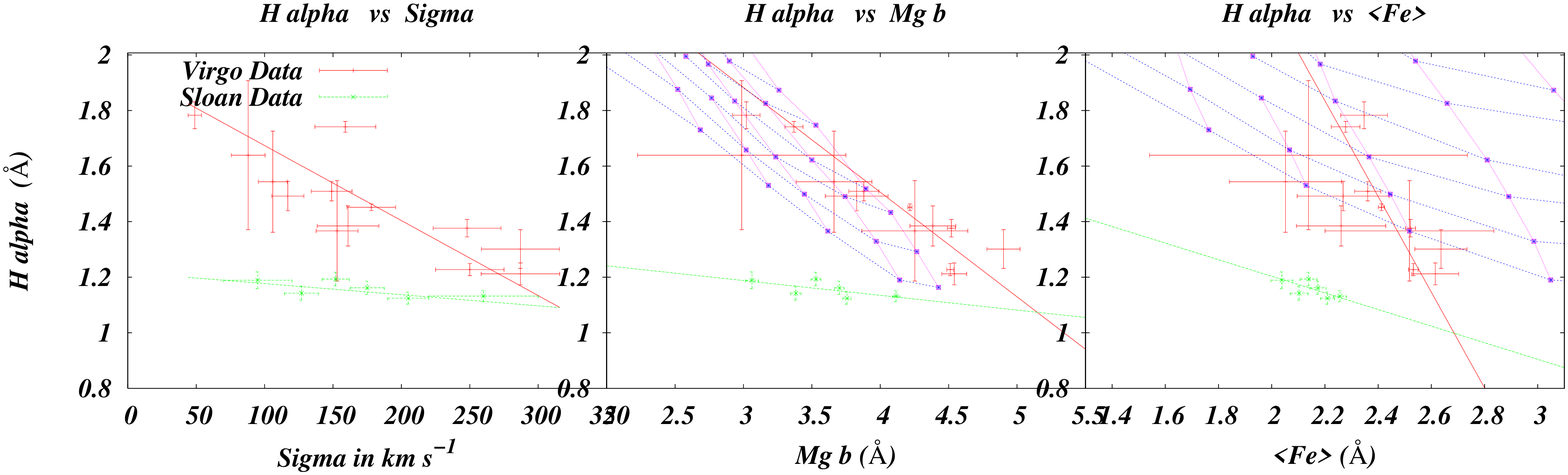}
\caption{
\label{Fig1}}
\end{figure}

\begin{figure}[H]
\includegraphics[width=7in,height=2in]{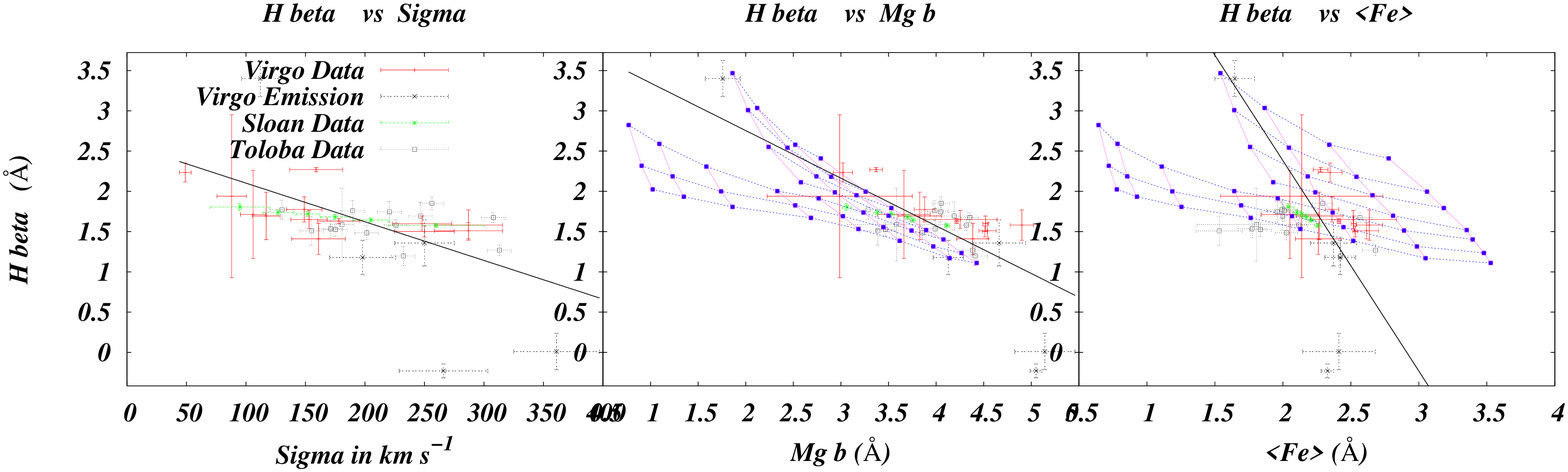}
\caption{}
\end{figure}

\begin{figure}[H]
\includegraphics[width=7in,height=2in]{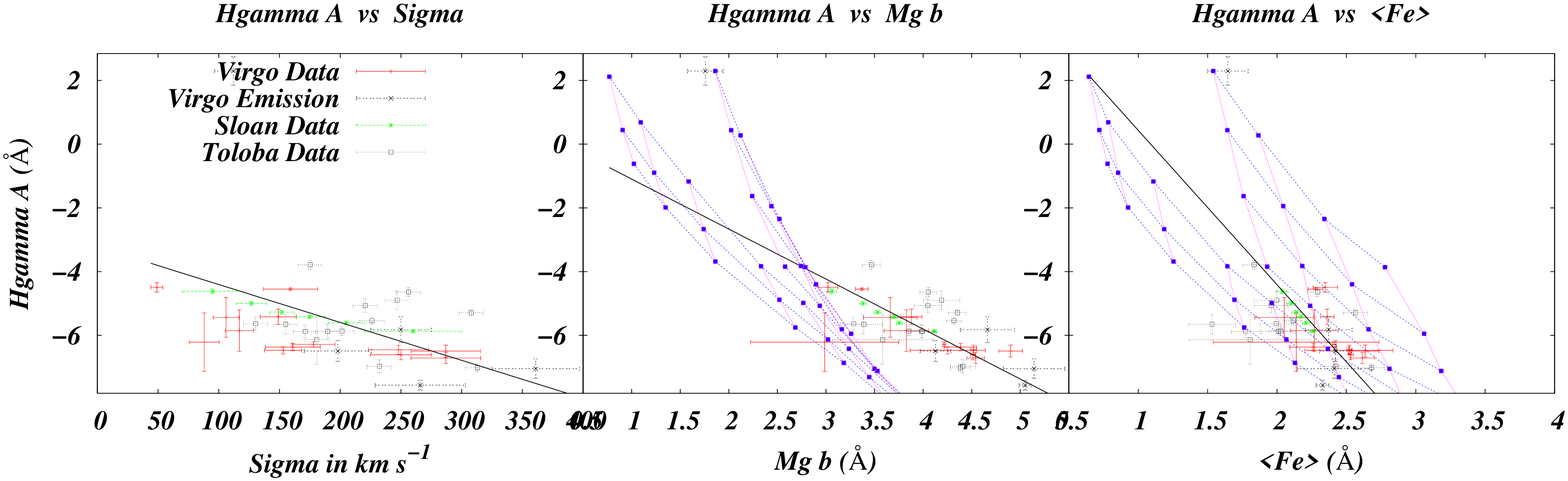}
\caption{
\label{Fig2} }
\end{figure}

\begin{figure}[H]
\includegraphics[width=7in,height=2in]{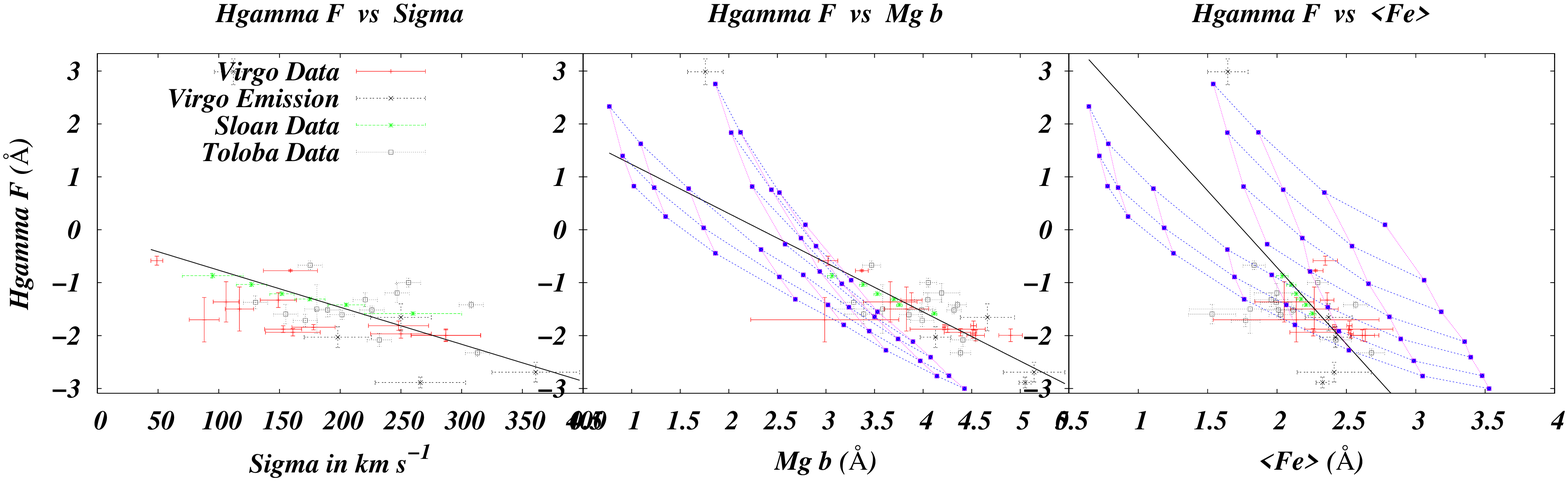}
\caption{
\label{Fig3} }
\end{figure}

\begin{figure}[H]
\includegraphics[width=7in,height=2in]{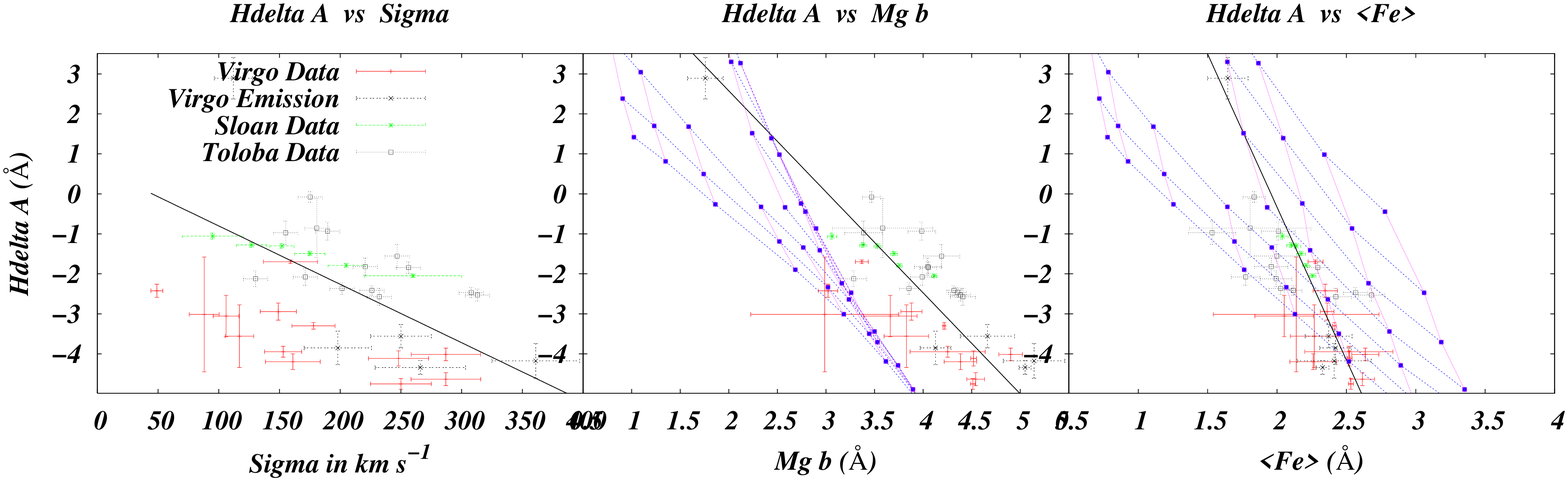}
\caption{
\label{Fig4} }
\end{figure}

\begin{figure}[H]
\includegraphics[width=7in,height=2in]{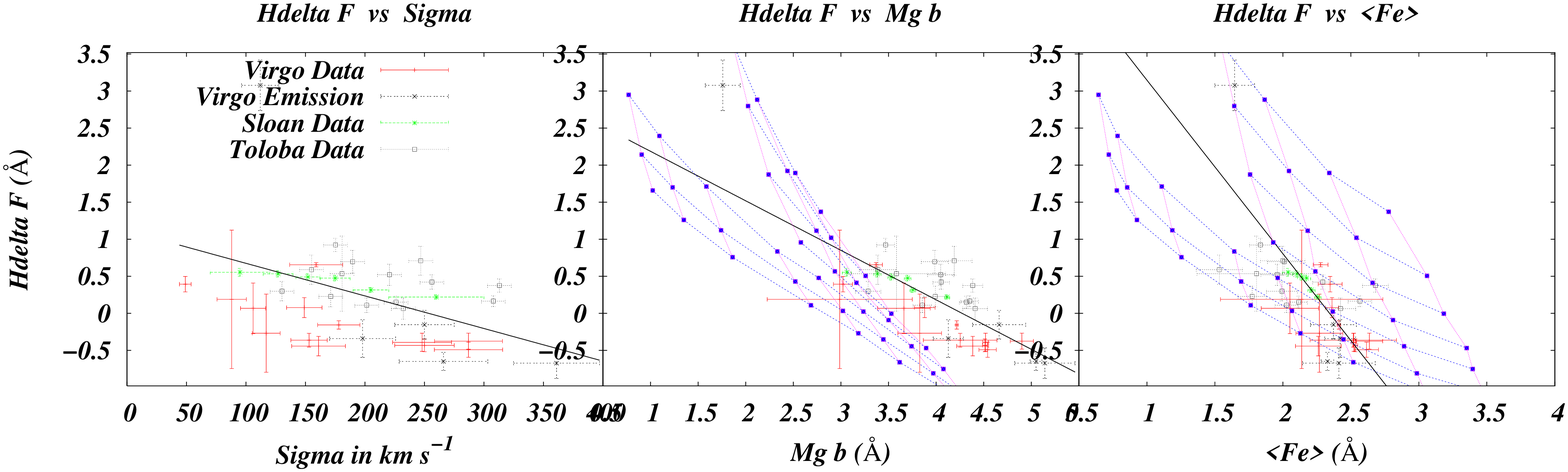}
\caption{
\label{Fig5} }
\end{figure}

\begin{figure}[H]
\includegraphics[width=7in,height=2in]{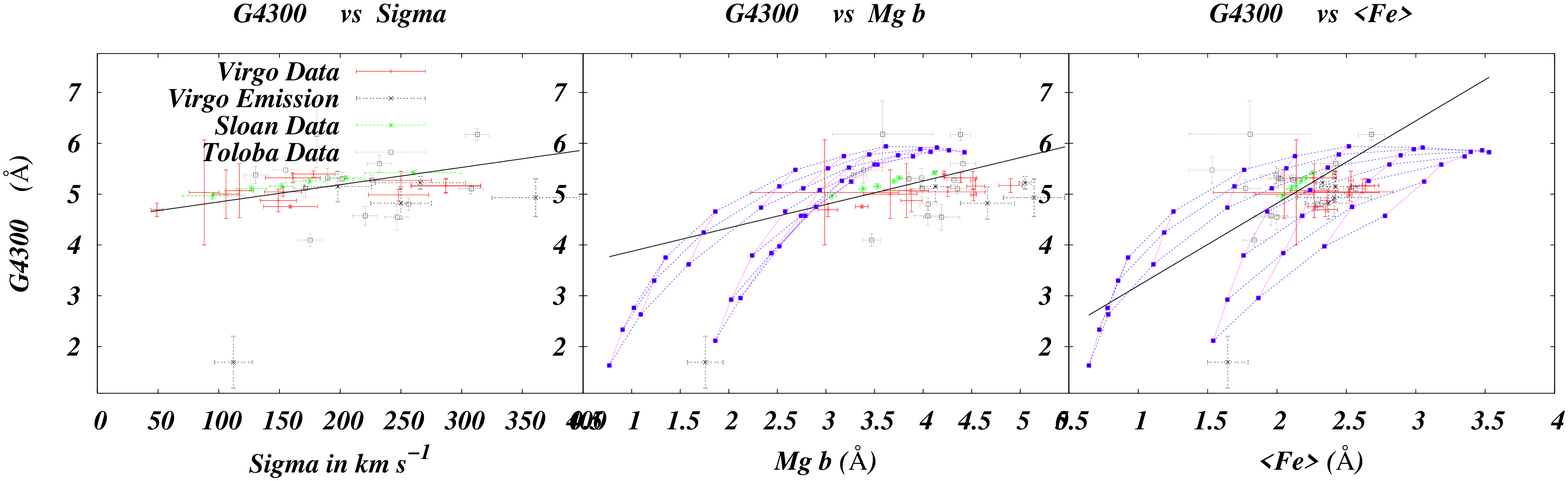}
\caption{}
\end{figure}

\begin{figure}[H]
\includegraphics[width=7in,height=2in]{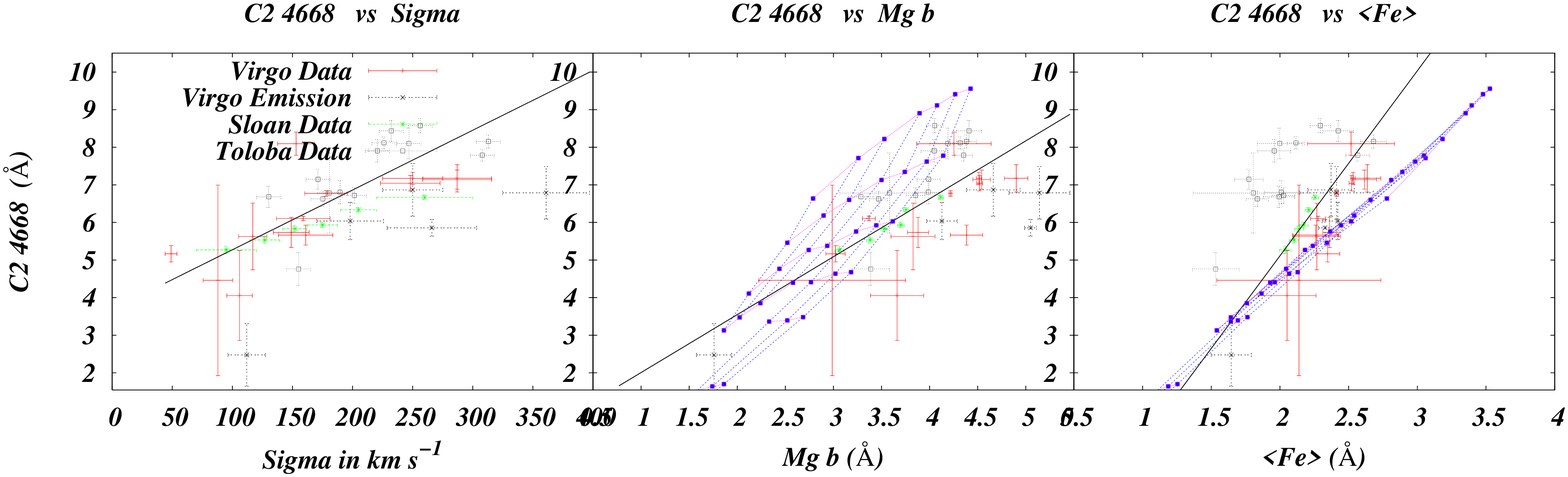}
\caption{}
\end{figure}

\begin{figure}[H]
\includegraphics[width=7in,height=2in]{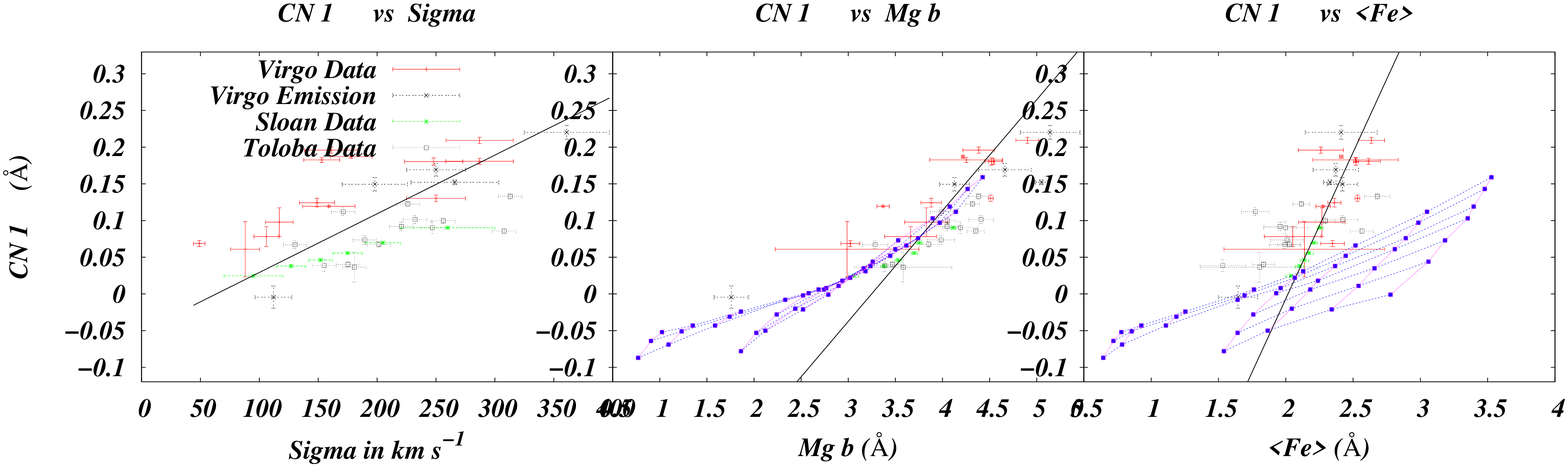}
\caption{}
\end{figure}

\begin{figure}[H]
\includegraphics[width=7in,height=2in]{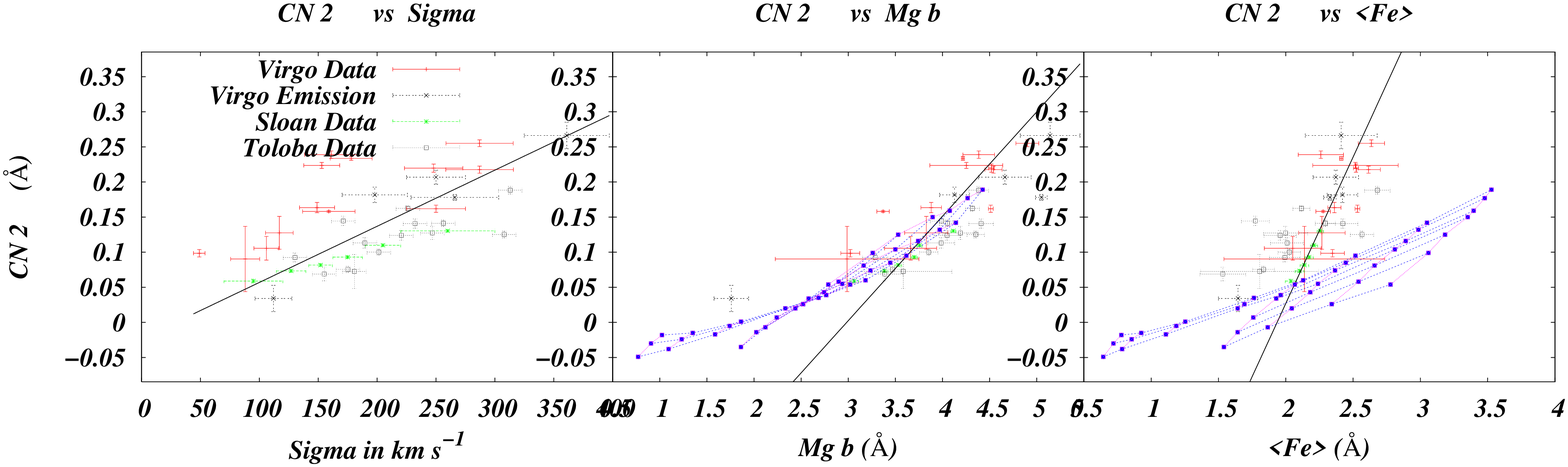}
\caption{}
\end{figure}

\begin{figure}[H]
\includegraphics[width=7in,height=2in]{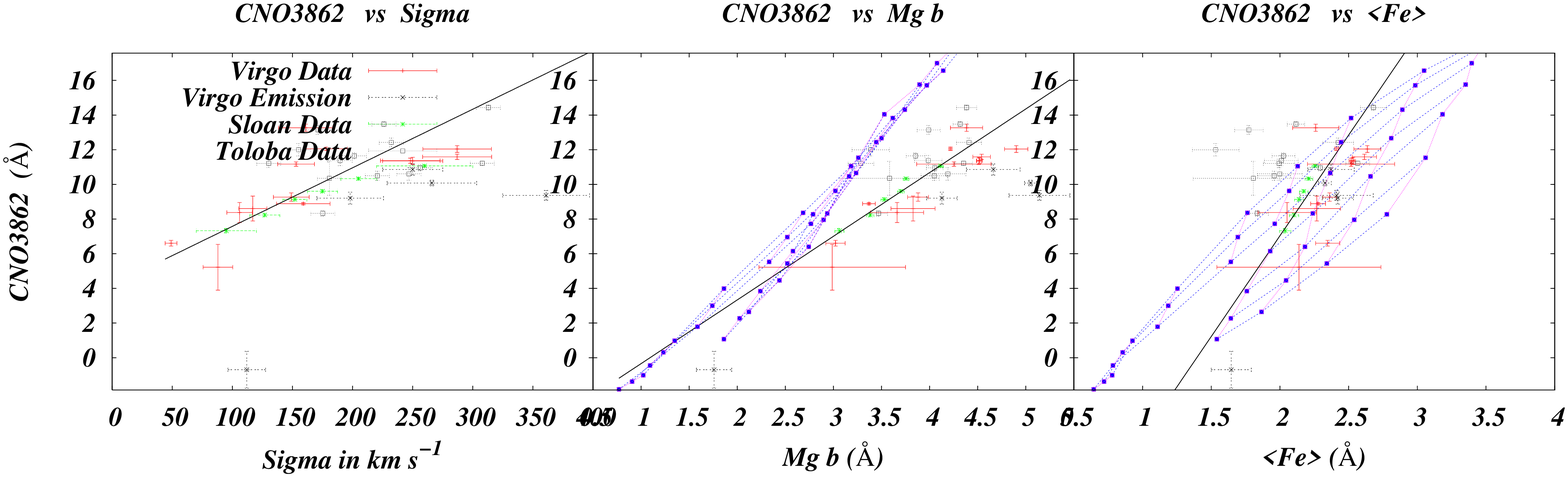}
\caption{}
\end{figure}

\begin{figure}[H]
\includegraphics[width=7in,height=2in]{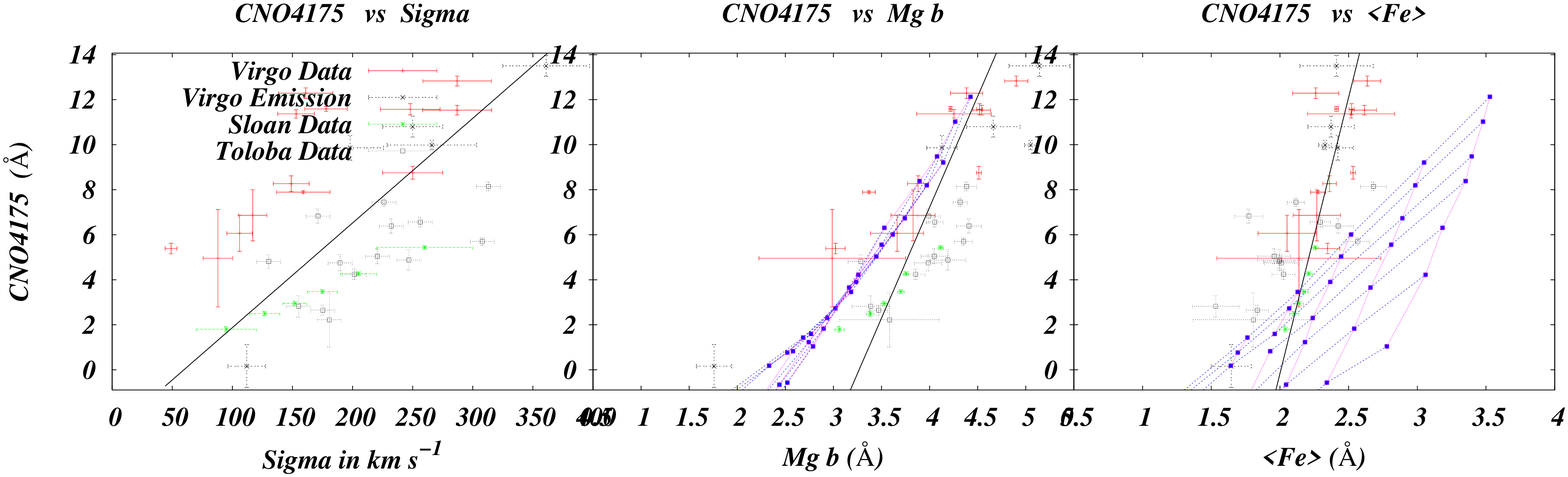}
\caption{}
\end{figure}

\begin{figure}[H]
\includegraphics[width=7in,height=2in]{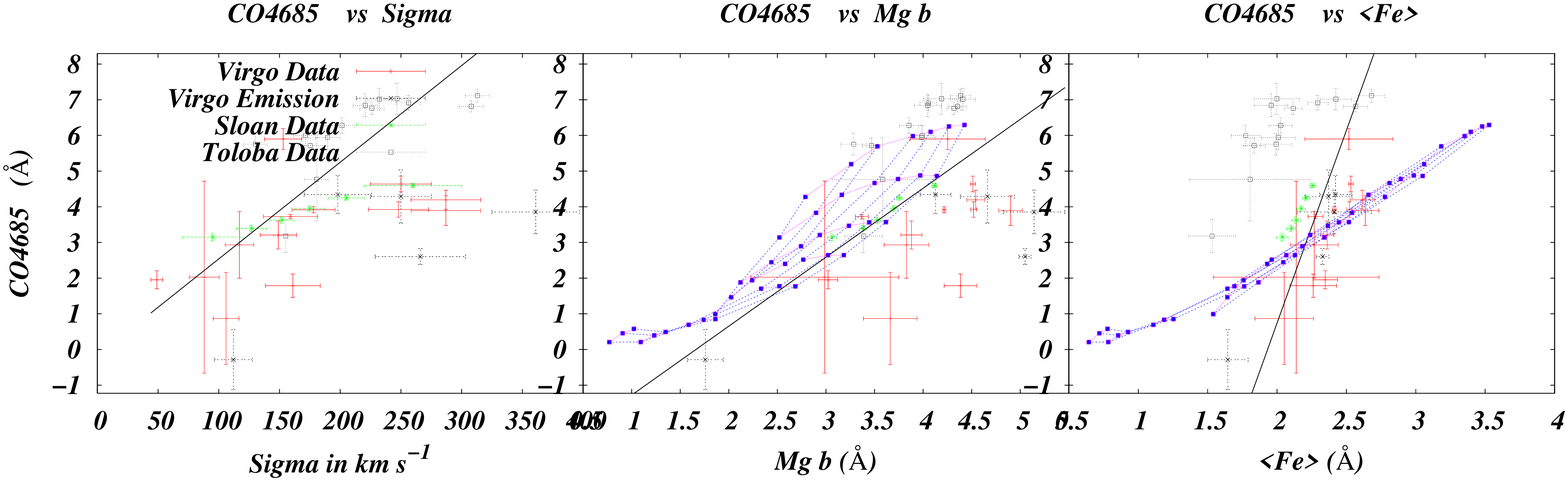}
\caption{}
\end{figure}

\begin{figure}[H]
\includegraphics[width=7in,height=2in]{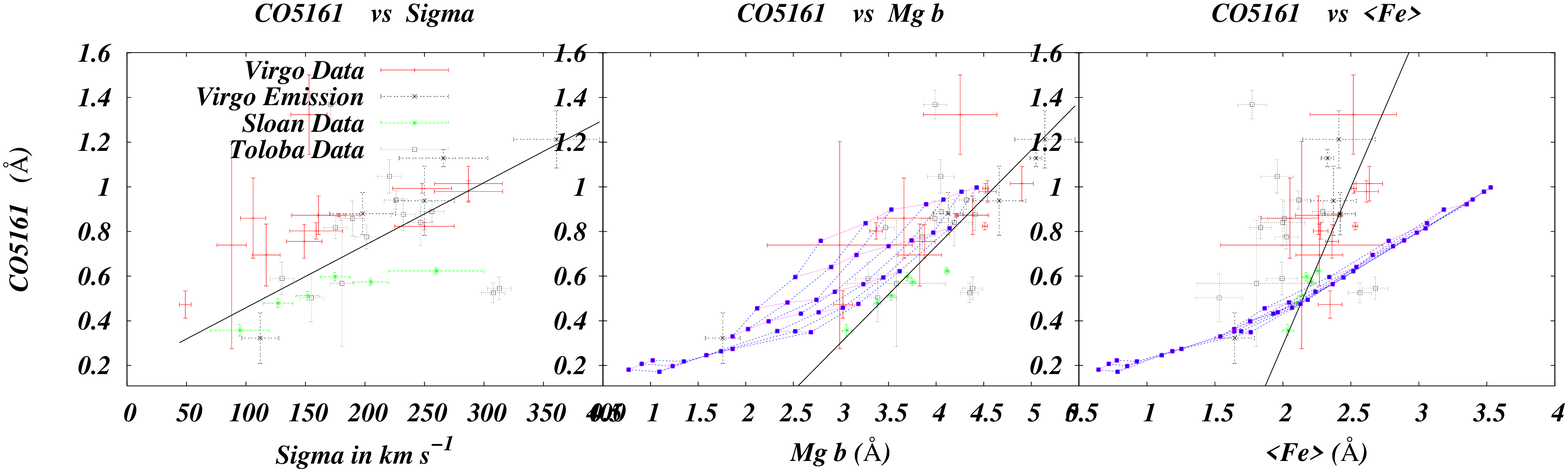}
\caption{}
\end{figure}

\begin{figure}[H]
\includegraphics[width=7in,height=2in]{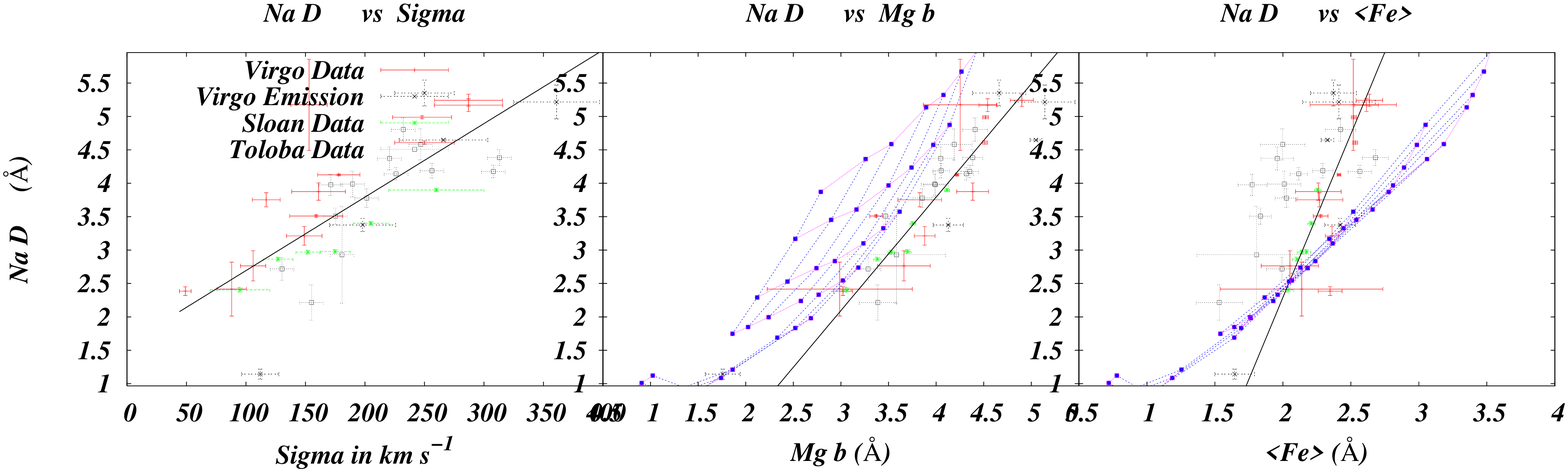}
\caption{}
\end{figure}

\begin{figure}[H]
\includegraphics[width=7in,height=2in]{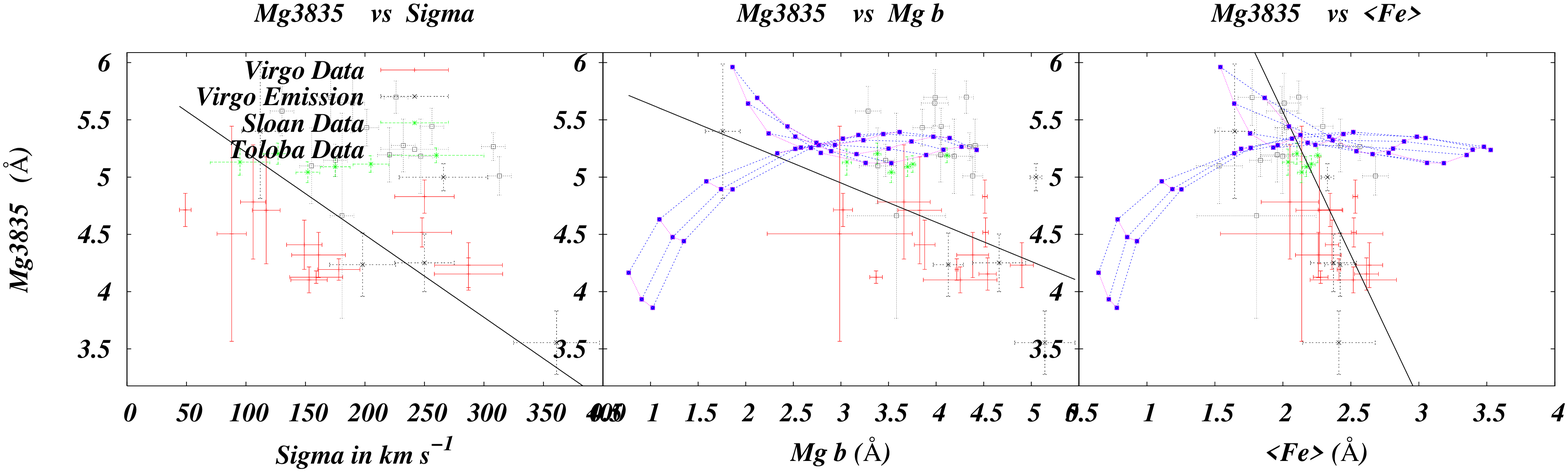}
\caption{}
\end{figure}

\begin{figure}[H]
\includegraphics[width=7in,height=2in]{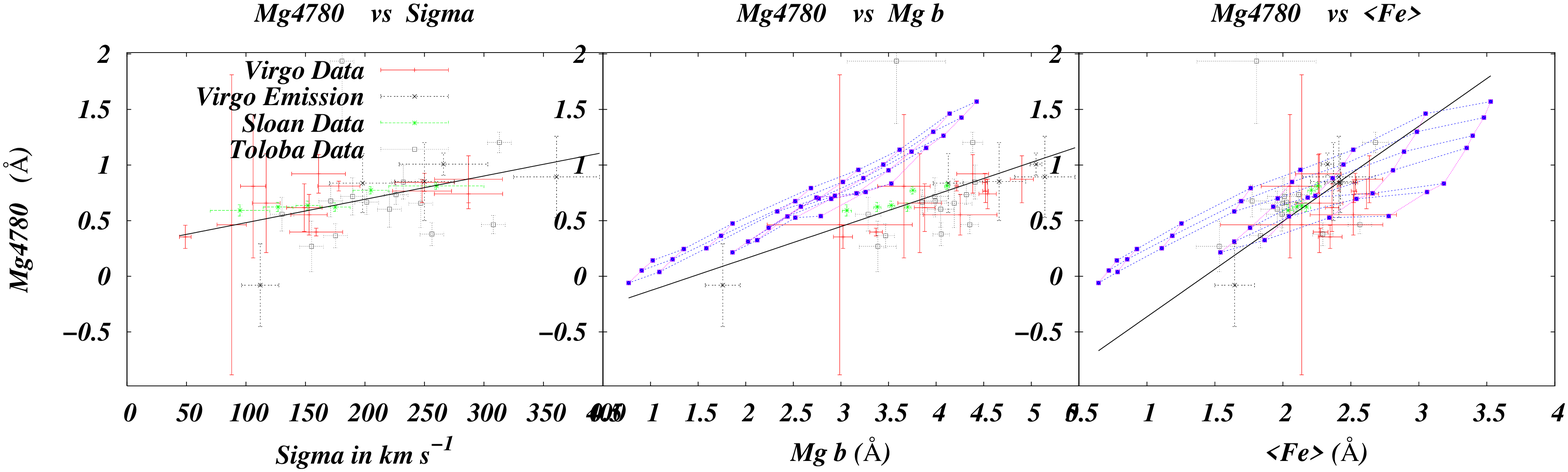}
\caption{}
\end{figure}

\begin{figure}[H]
\includegraphics[width=7in,height=2in]{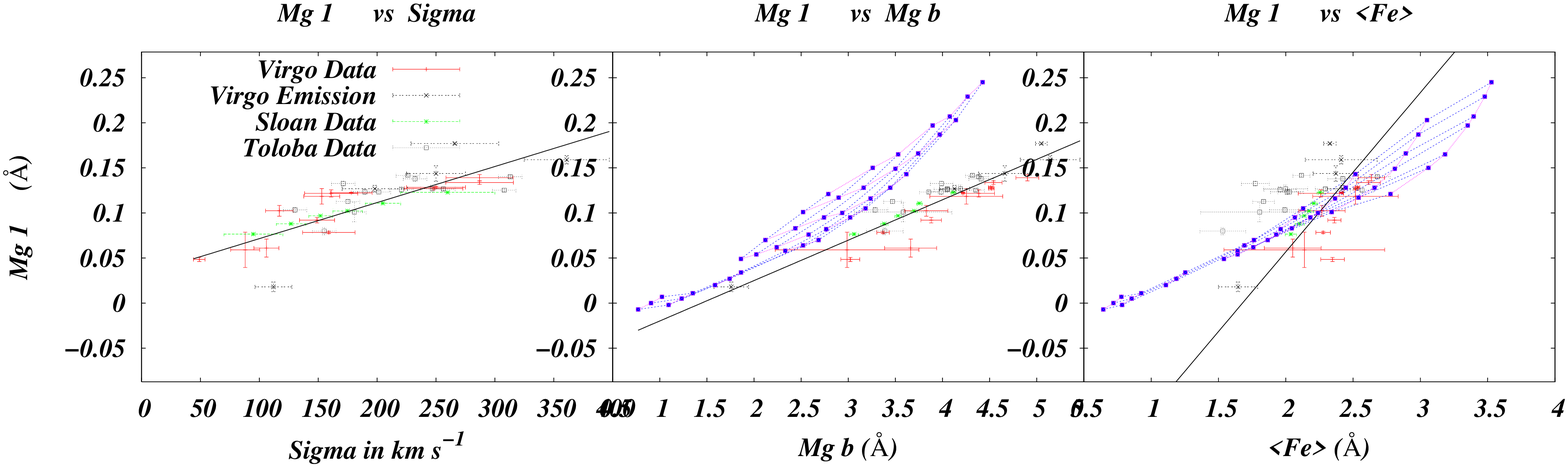}
\caption{}
\end{figure}

\begin{figure}[H]
\includegraphics[width=7in,height=2in]{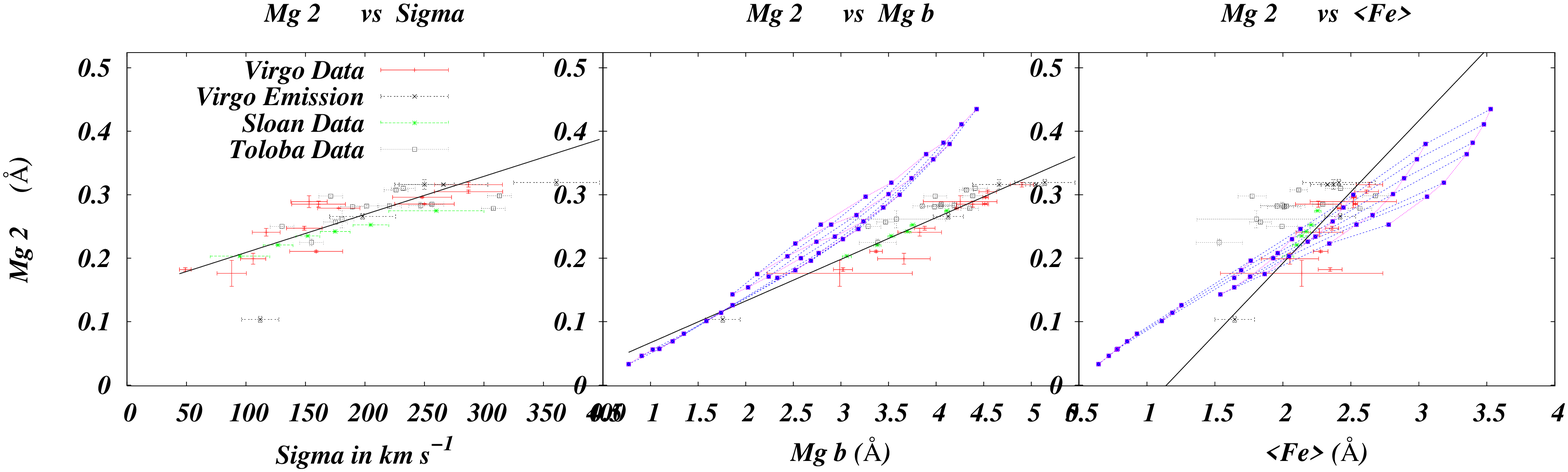}
\caption{}
\end{figure}

\begin{figure}[H]
\includegraphics[width=7in,height=2in]{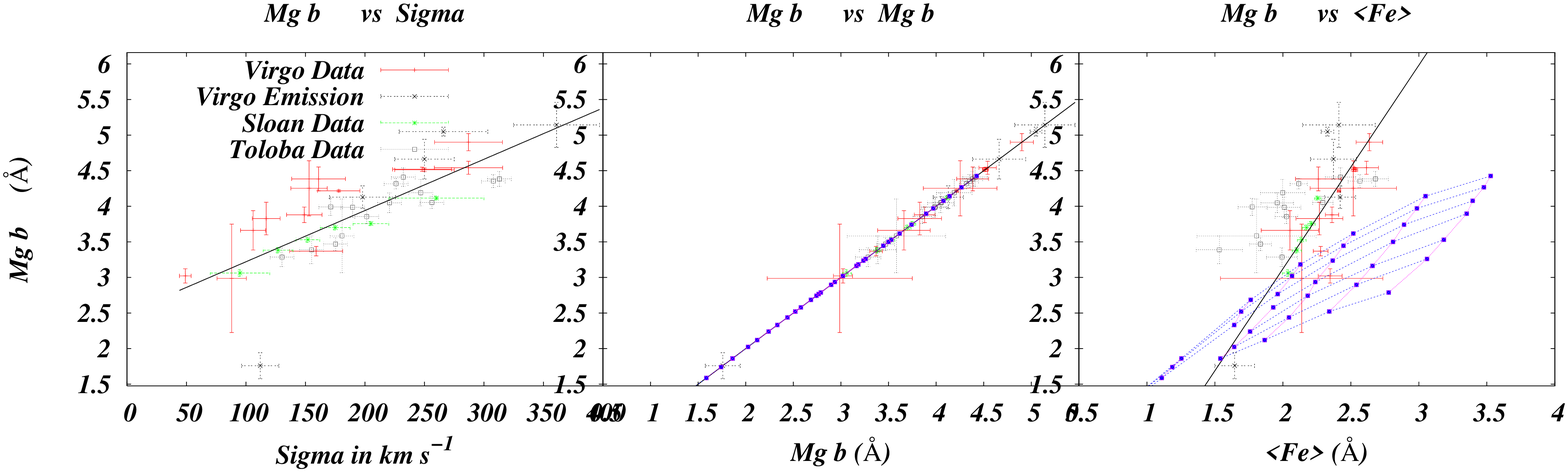}
\caption{}
\end{figure}

\begin{figure}[H]
\includegraphics[width=7in,height=2in]{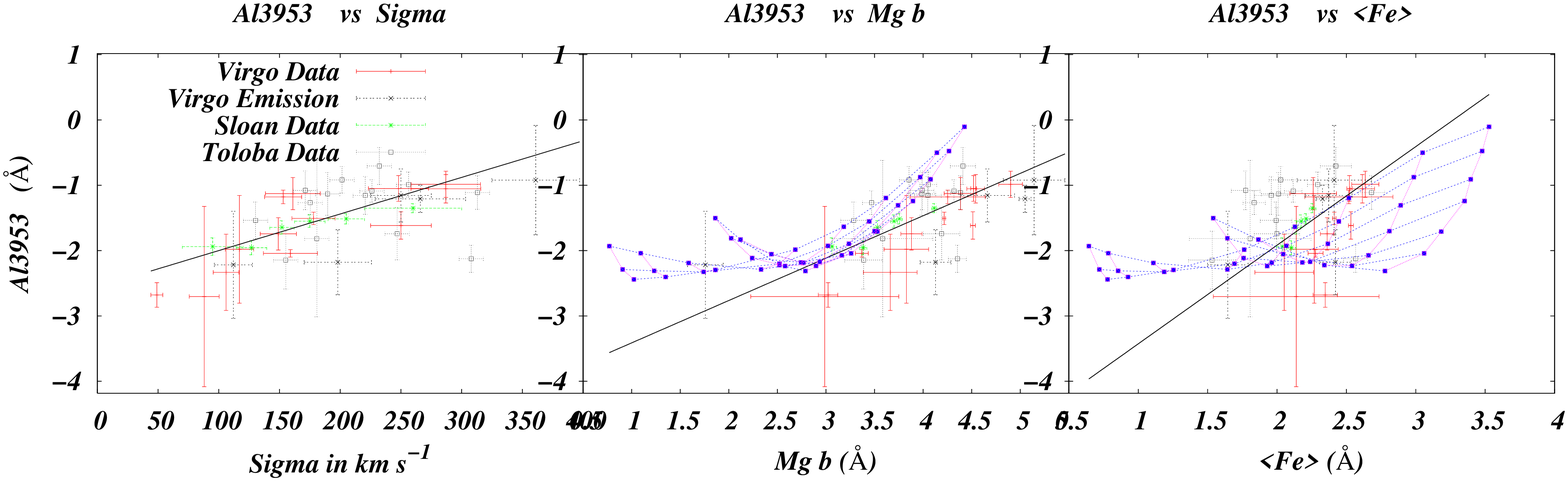}
\caption{}
\end{figure}

\begin{figure}[H]
\includegraphics[width=7in,height=2in]{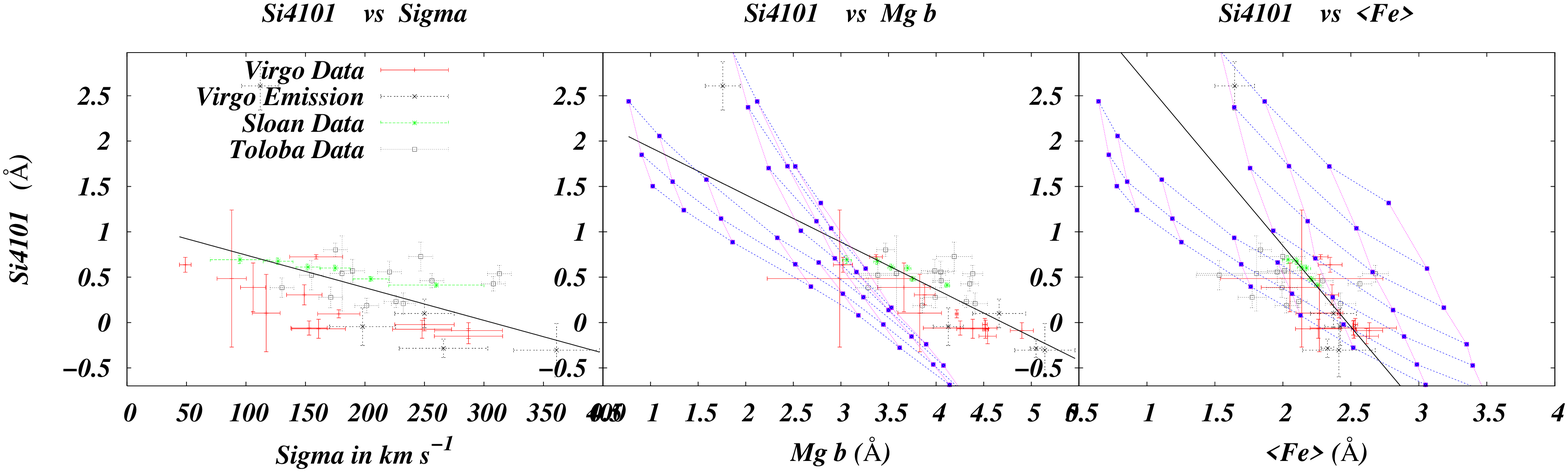}
\caption{}
\end{figure}

\begin{figure}[H]
\includegraphics[width=7in,height=2in]{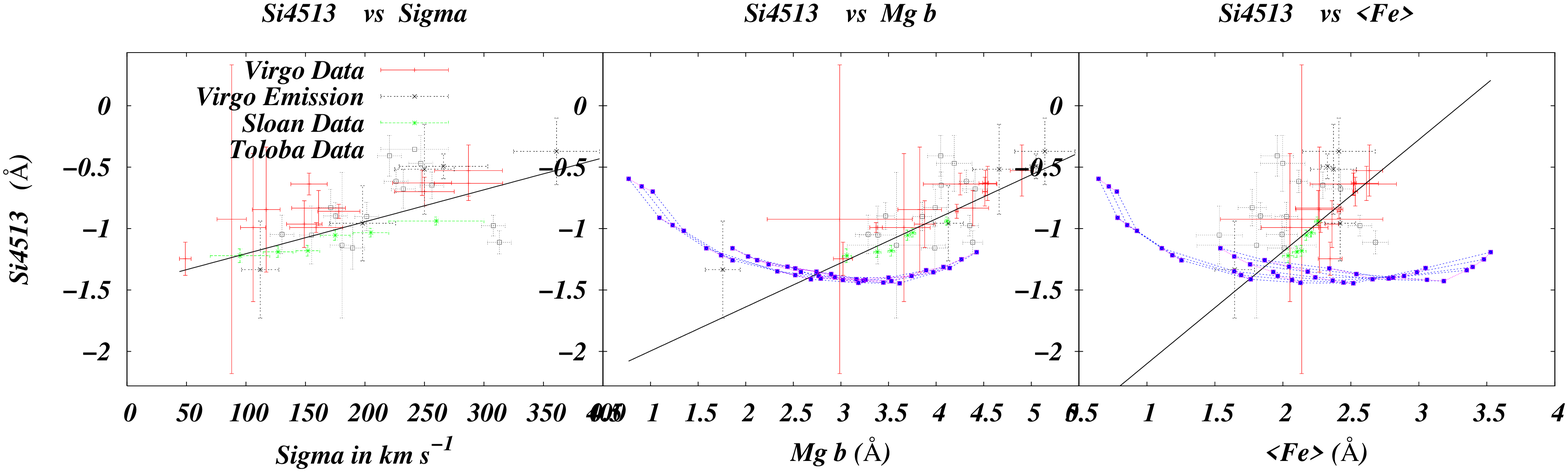}
\caption{}
\end{figure}

\begin{figure}[H]
\includegraphics[width=7in,height=2in]{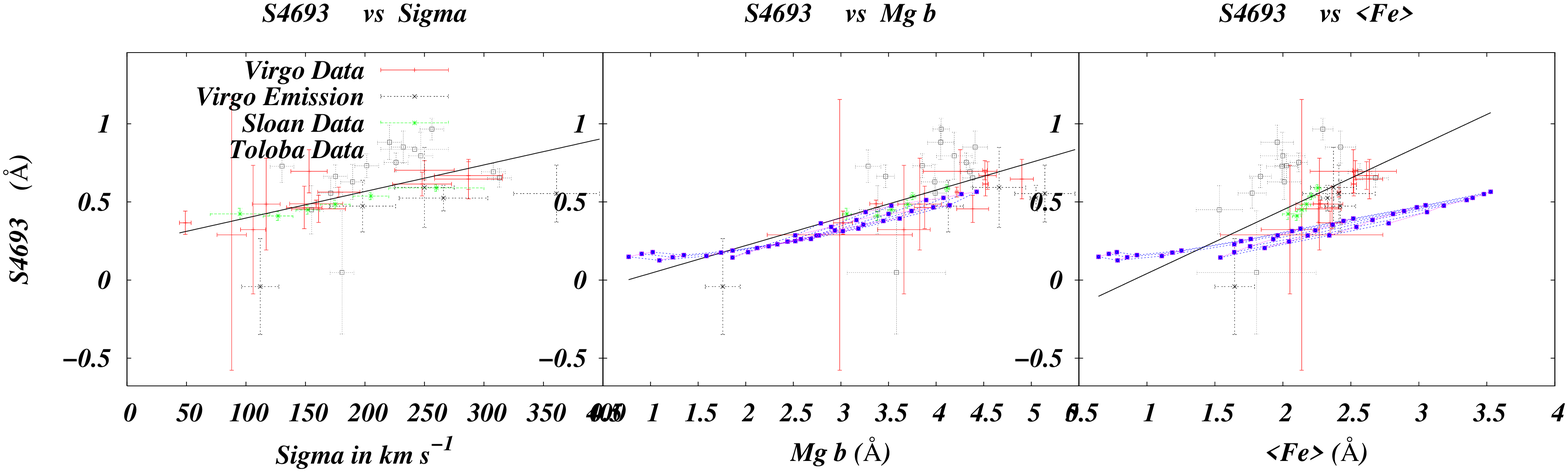}
\caption{}
\end{figure}

\begin{figure}[H]
\includegraphics[width=7in,height=2in]{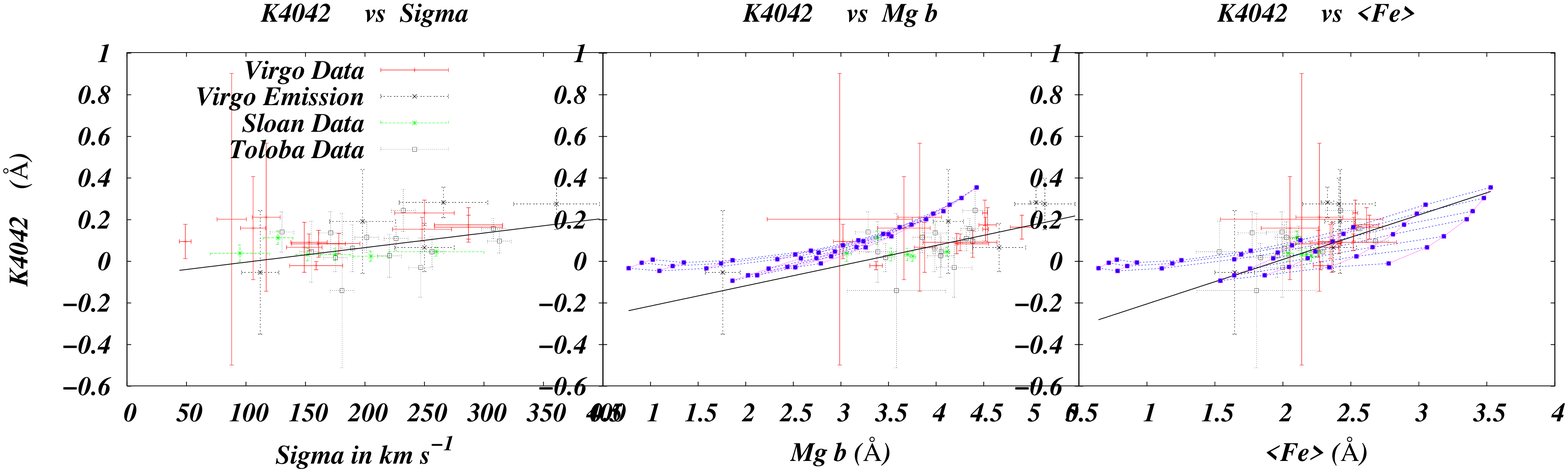}
\caption{}
\end{figure}

\begin{figure}[H]
\includegraphics[width=7in,height=2in]{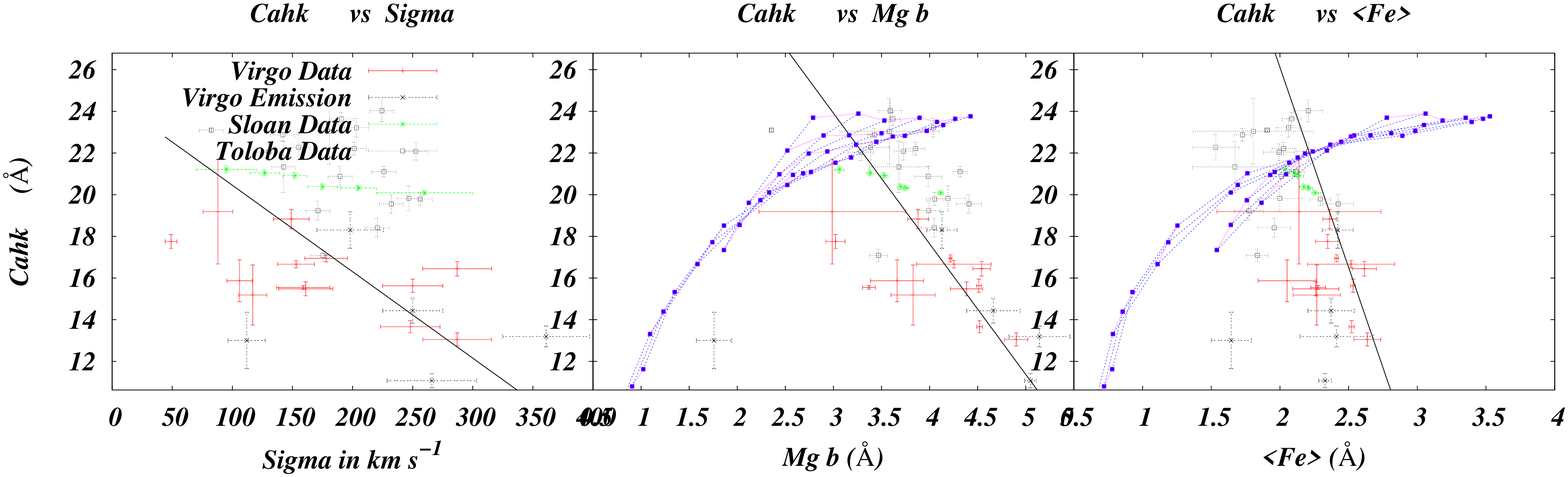}
\caption{}
\end{figure}

\begin{figure}[H]
\includegraphics[width=7in,height=2in]{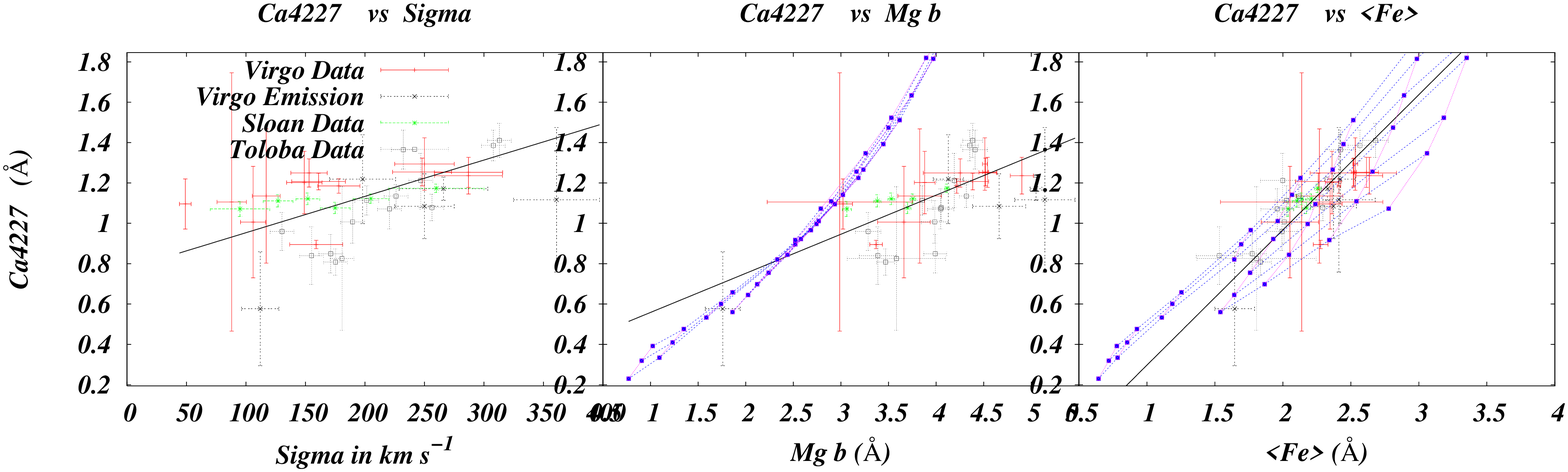}
\caption{}
\end{figure}

\begin{figure}[H]
\includegraphics[width=7in,height=2in]{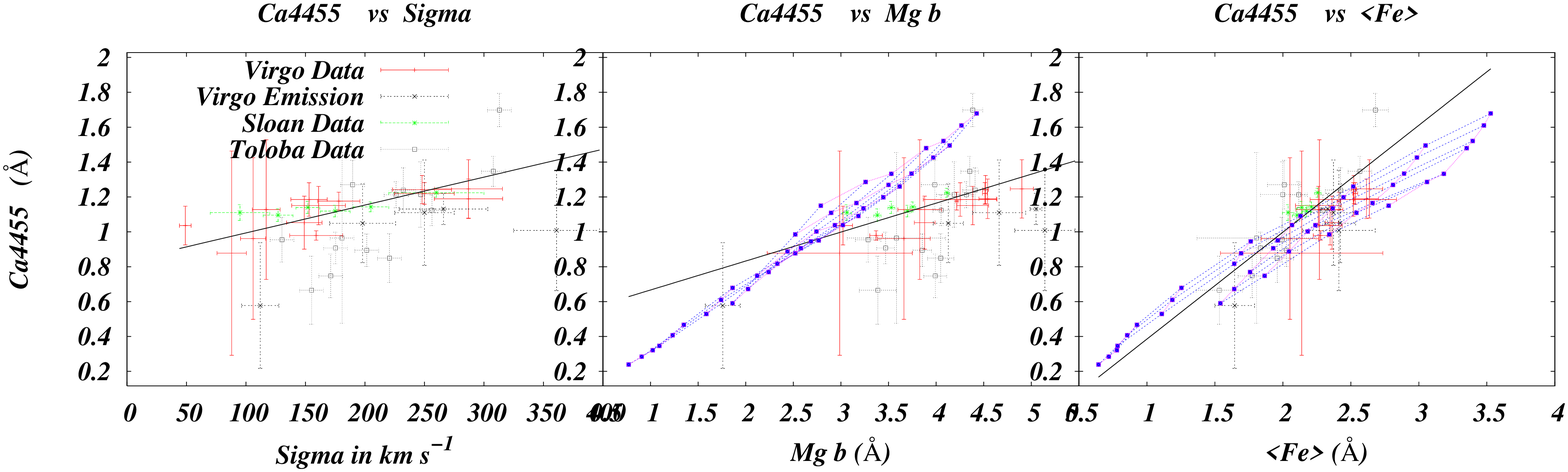}
\caption{}
\end{figure}

\begin{figure}[H]
\includegraphics[width=7in,height=2in]{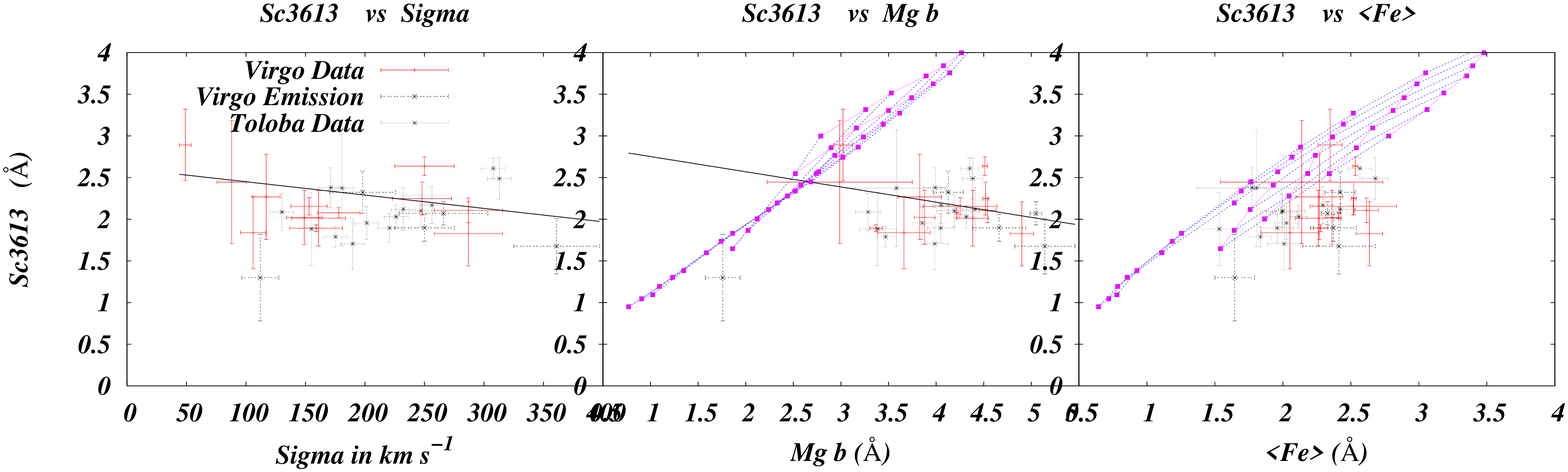}
\caption{}
\end{figure}

\begin{figure}[H]
\includegraphics[width=7in,height=2in]{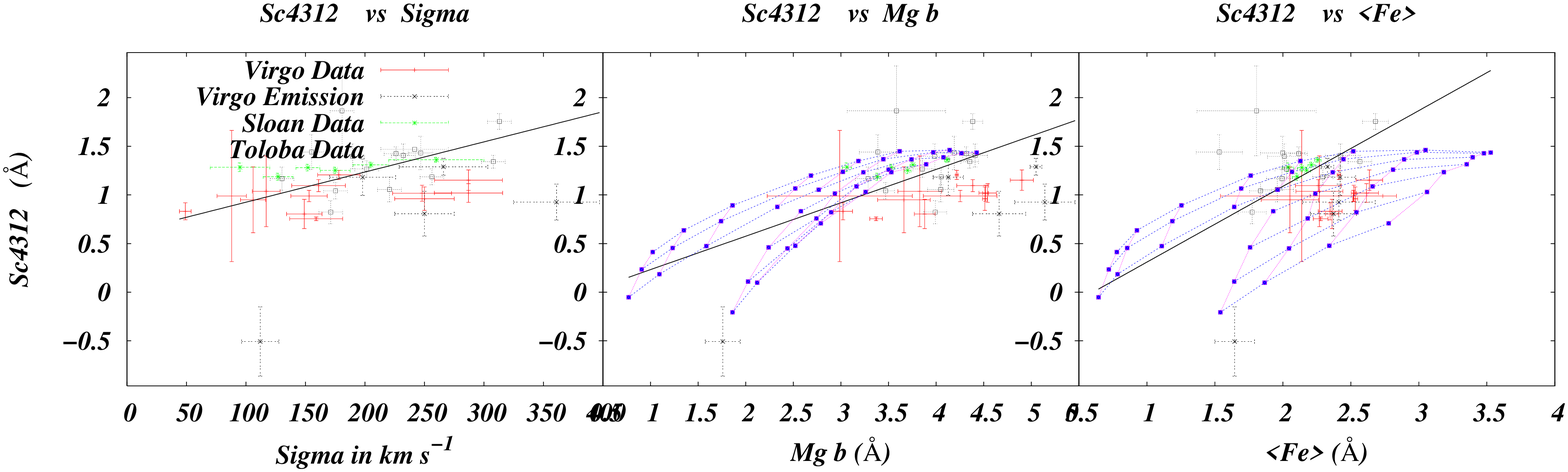}
\caption{}
\end{figure}

\begin{figure}[H]
\includegraphics[width=7in,height=2in]{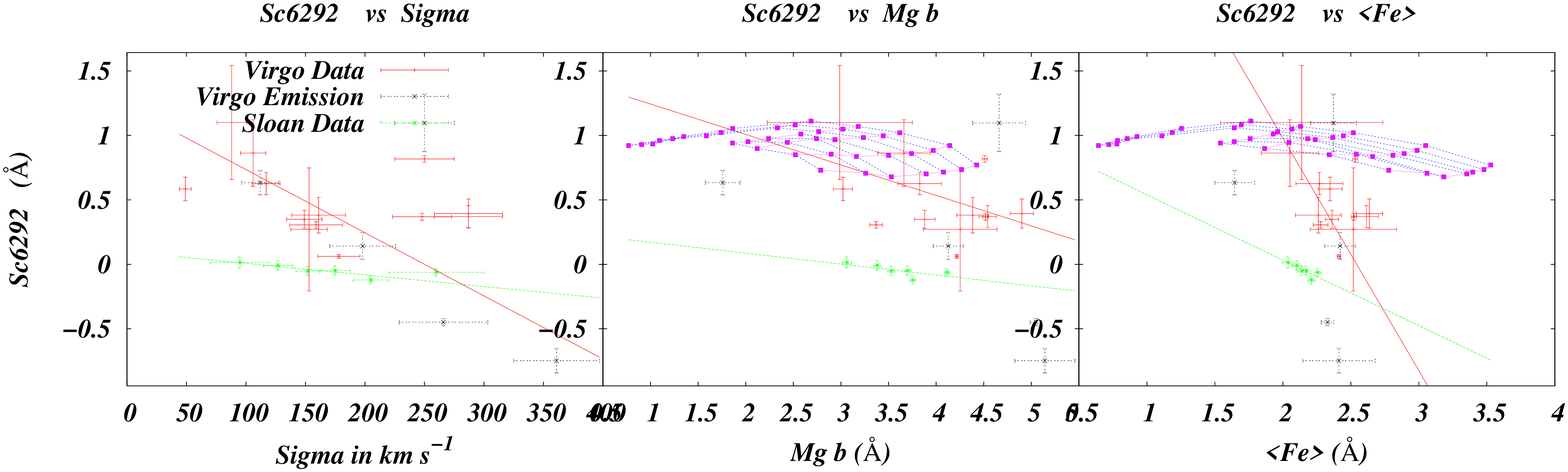}
\caption{}
\end{figure}

\begin{figure}[H]
\includegraphics[width=7in,height=2in]{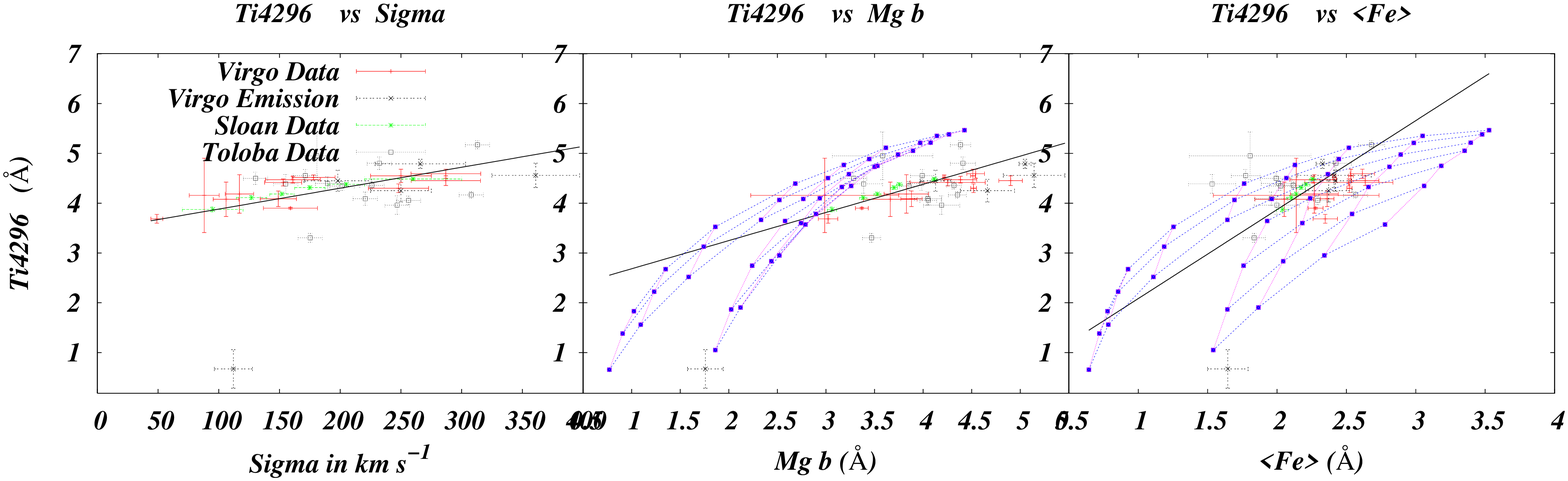}
\caption{}
\end{figure}

\begin{figure}[H]
\includegraphics[width=7in,height=2in]{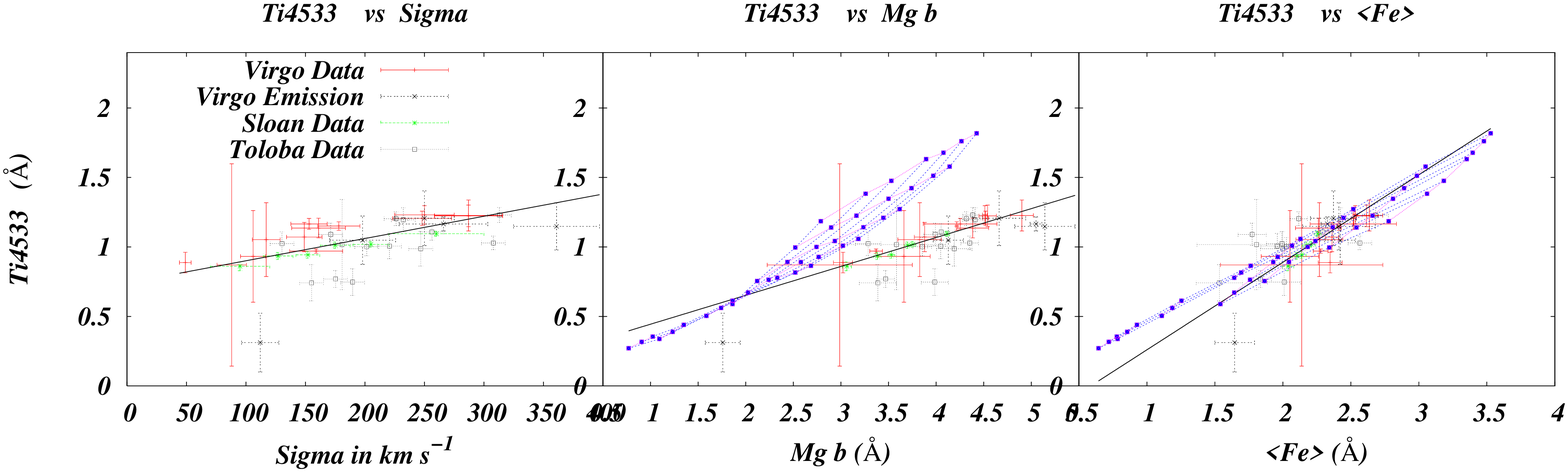}
\caption{}
\end{figure}

\begin{figure}[H]
\includegraphics[width=7in,height=2in]{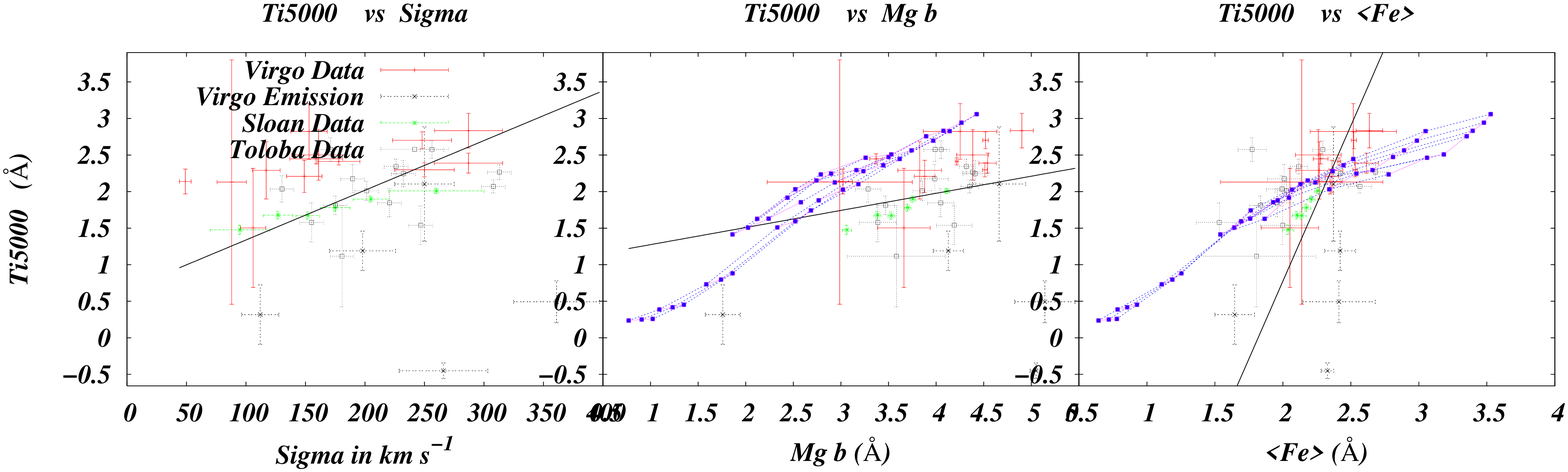}
\caption{}
\end{figure}

\begin{figure}[H]
\includegraphics[width=7in,height=2in]{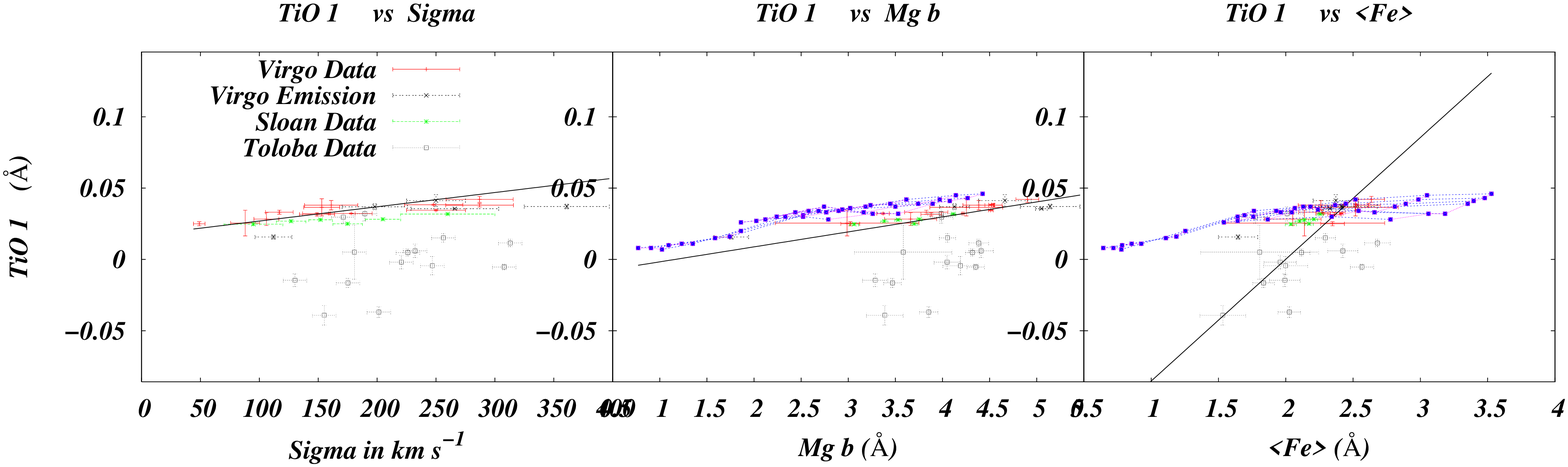}
\caption{}
\end{figure}

\begin{figure}[H]
\includegraphics[width=7in,height=2in]{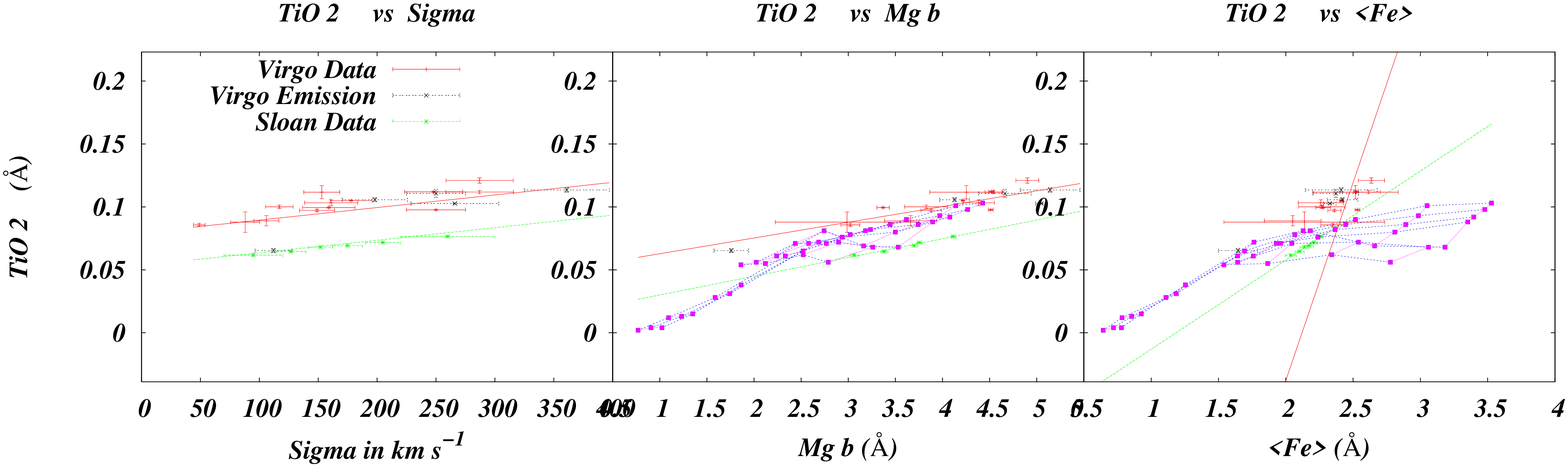}
\caption{}
\end{figure}

\begin{figure}[H]
\includegraphics[width=7in,height=2in]{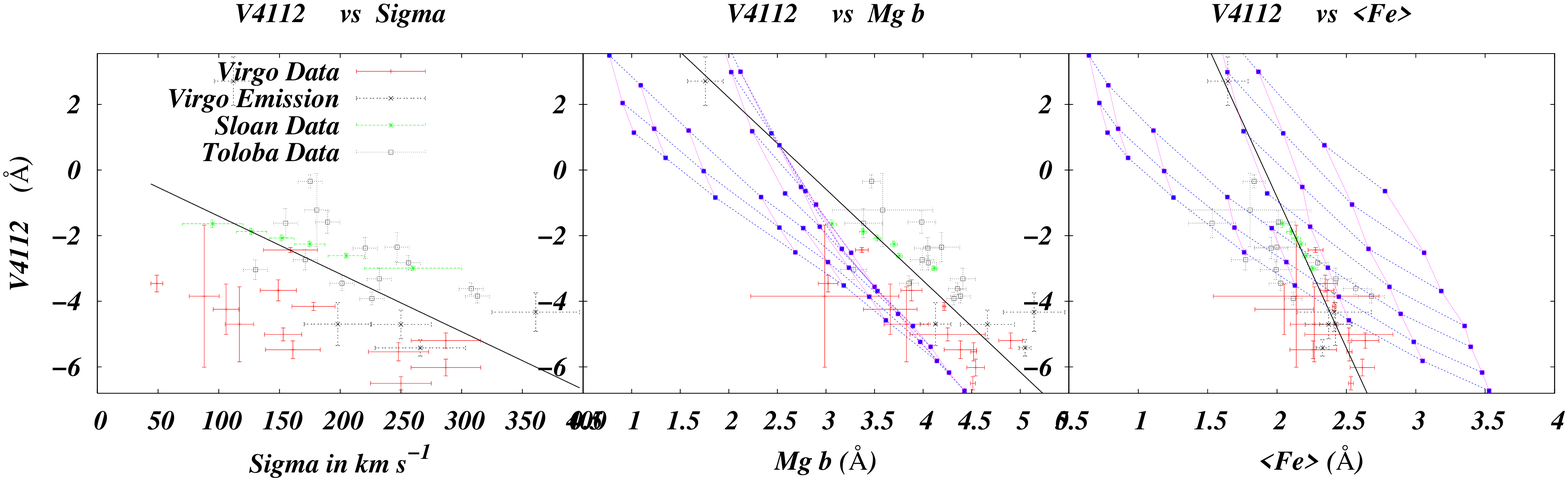}
\caption{}
\end{figure}

\begin{figure}[H]
\includegraphics[width=7in,height=2in]{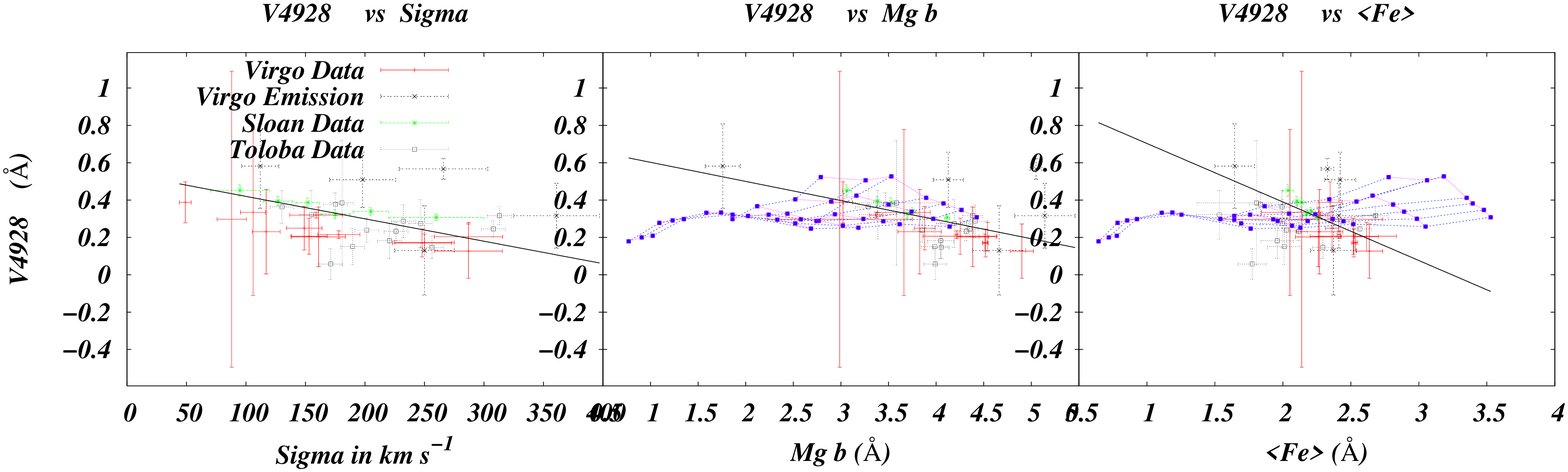}
\caption{}
\end{figure}

\begin{figure}[H]
\includegraphics[width=7in,height=2in]{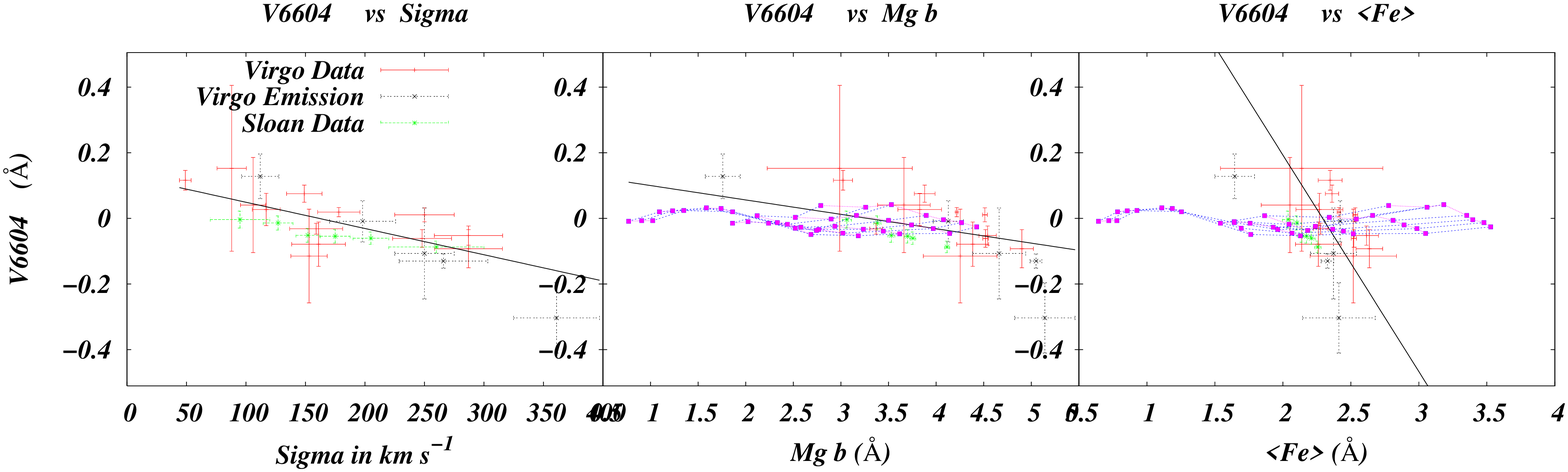}
\caption{}
\end{figure}

\begin{figure}[H]
\includegraphics[width=7in,height=2in]{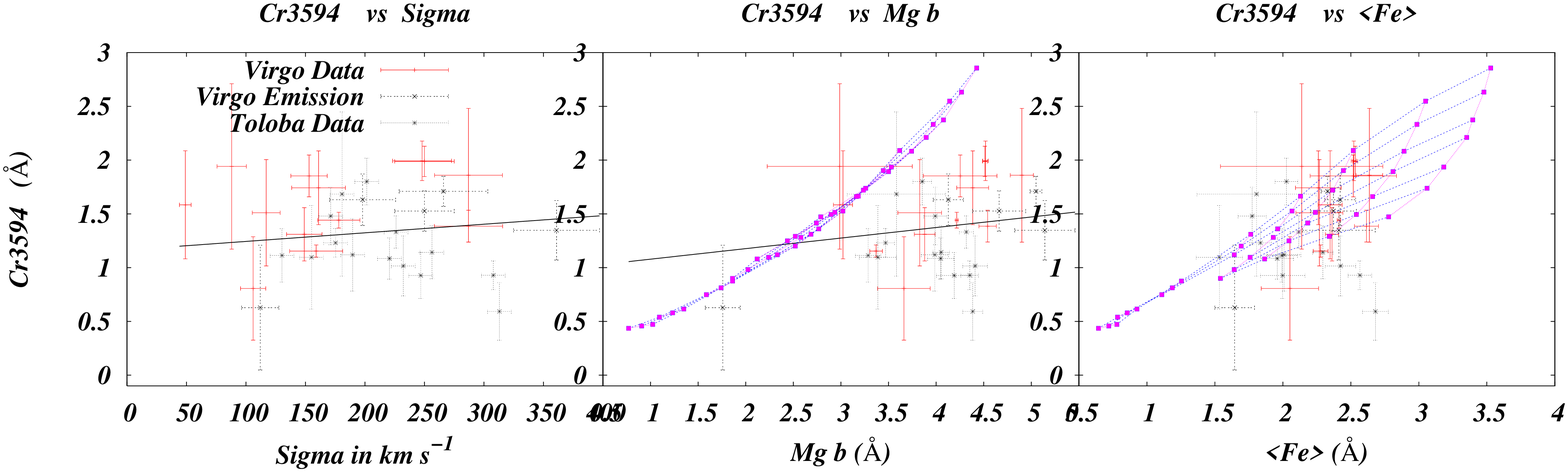}
\caption{}
\end{figure}

\begin{figure}[H]
\includegraphics[width=7in,height=2in]{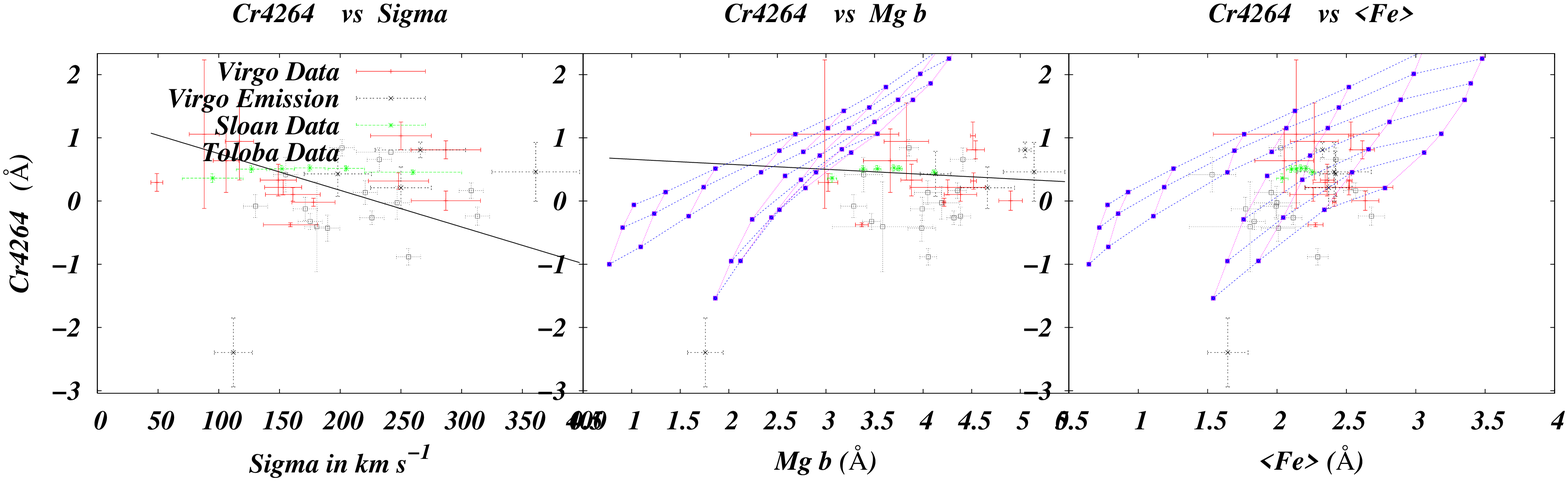}
\caption{}
\end{figure}

\begin{figure}[H]
\includegraphics[width=7in,height=2in]{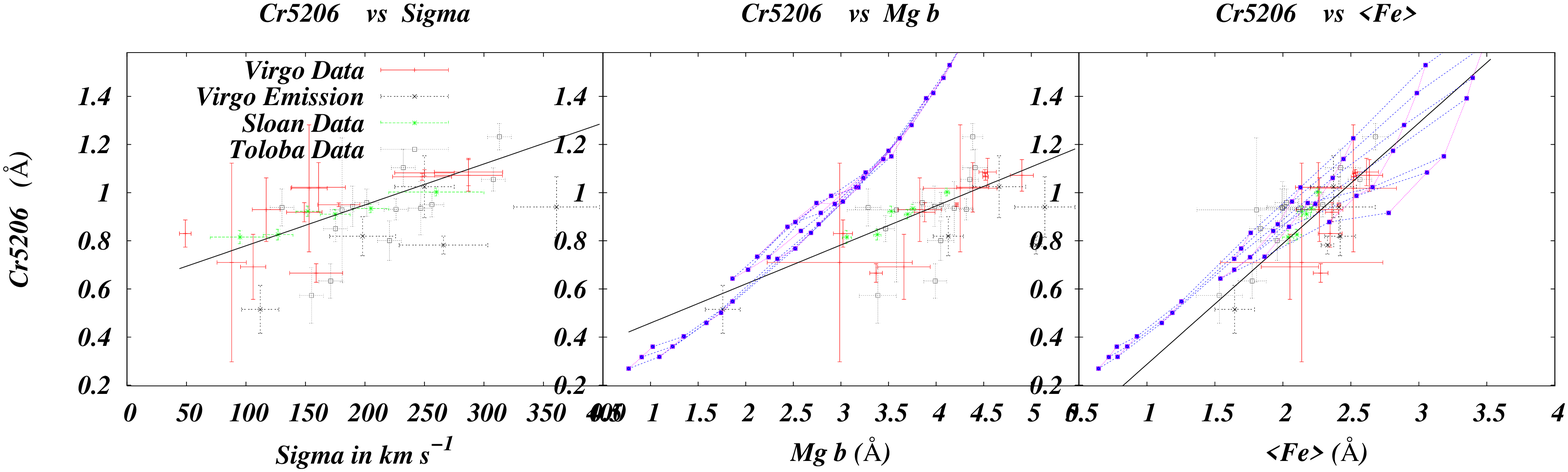}
\caption{}
\end{figure}

\begin{figure}[H]
\includegraphics[width=7in,height=2in]{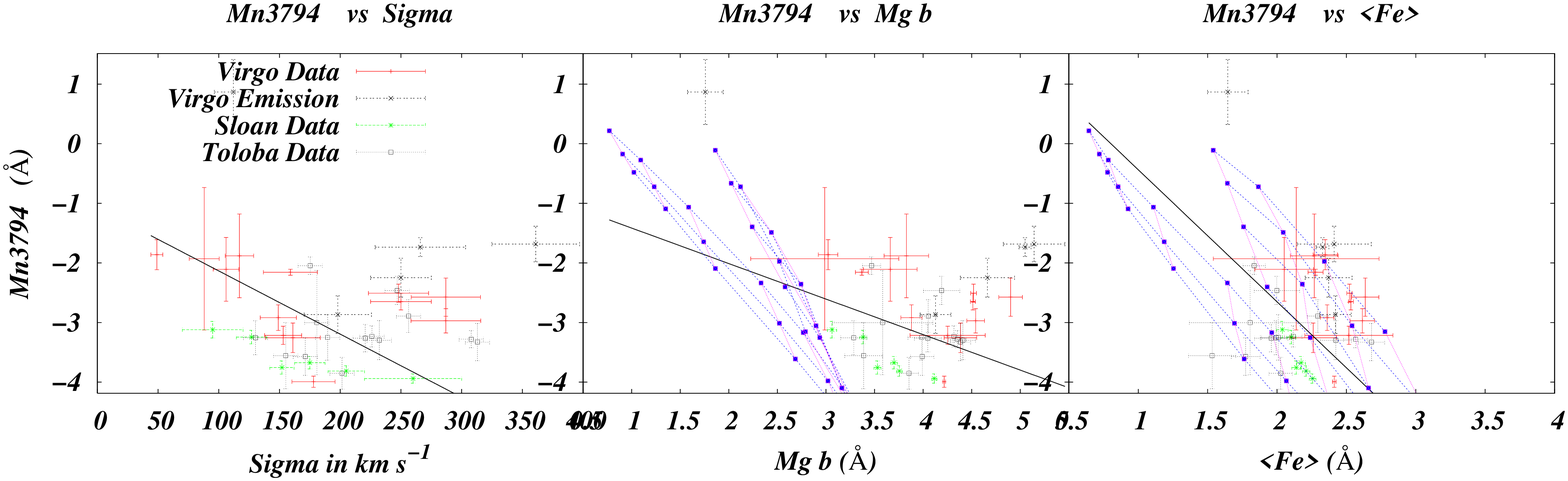}
\caption{}
\end{figure}

\begin{figure}[H]
\includegraphics[width=7in,height=2in]{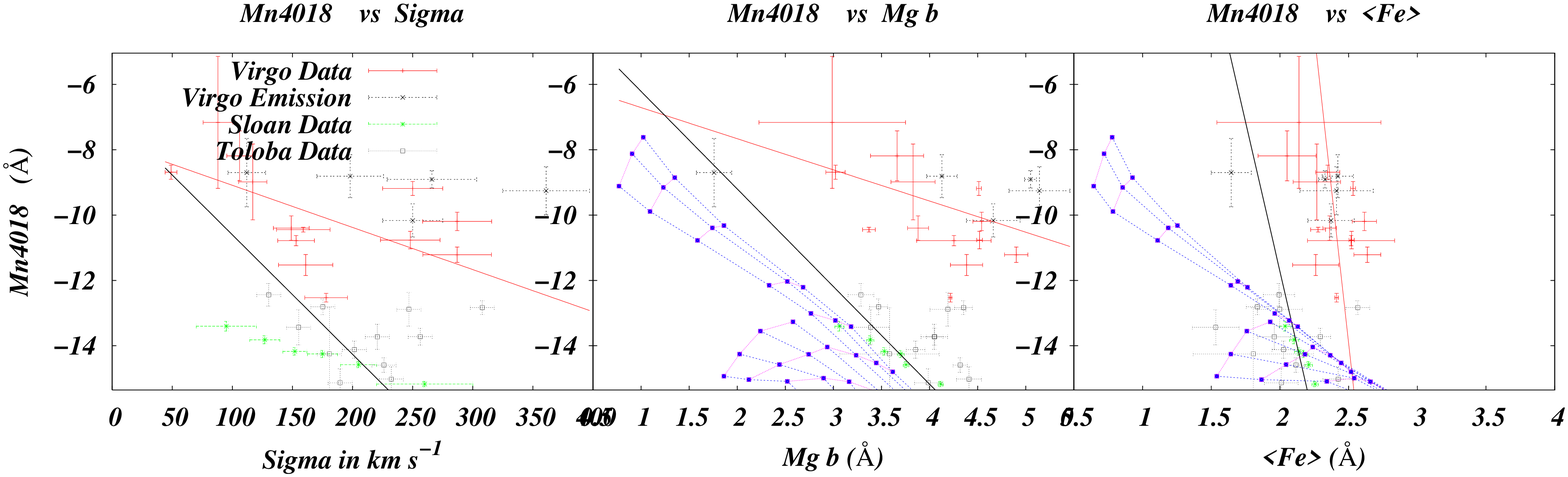}
\caption{}
\end{figure}

\begin{figure}[H]
\includegraphics[width=7in,height=2in]{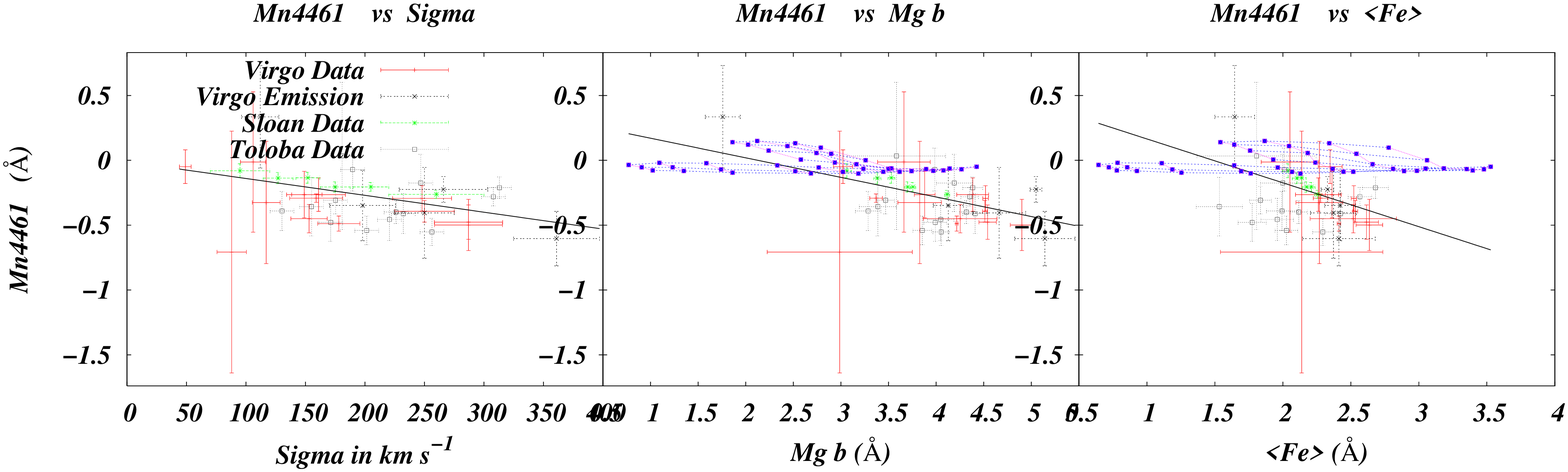}
\caption{}
\end{figure}

\begin{figure}[H]
\includegraphics[width=7in,height=2in]{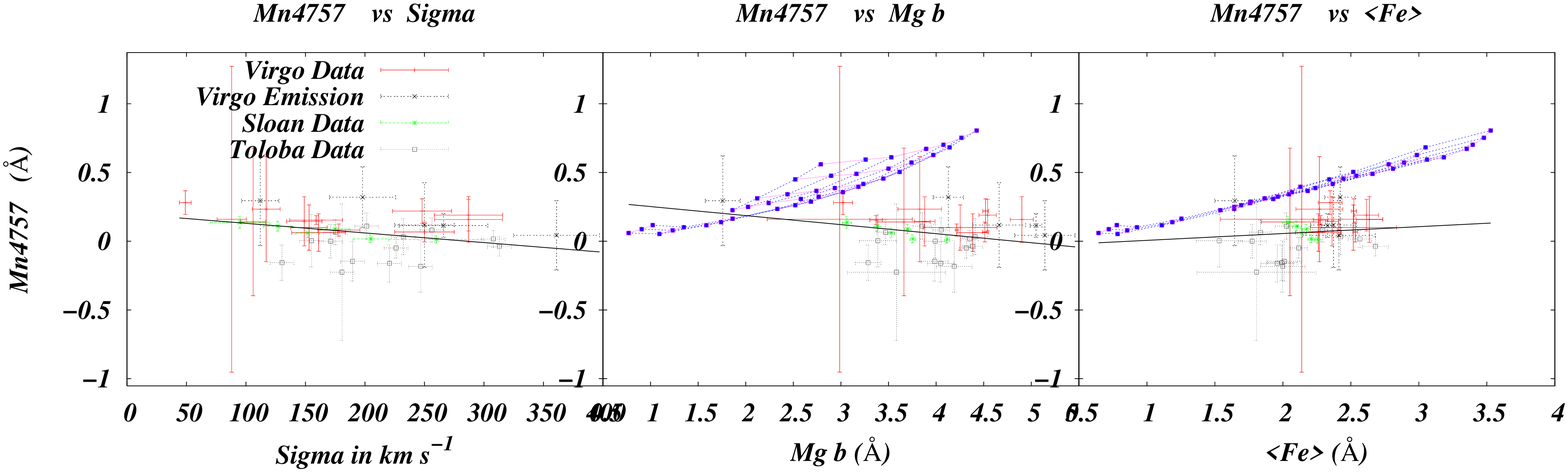}
\caption{}
\end{figure}

\begin{figure}[H]
\includegraphics[width=7in,height=2in]{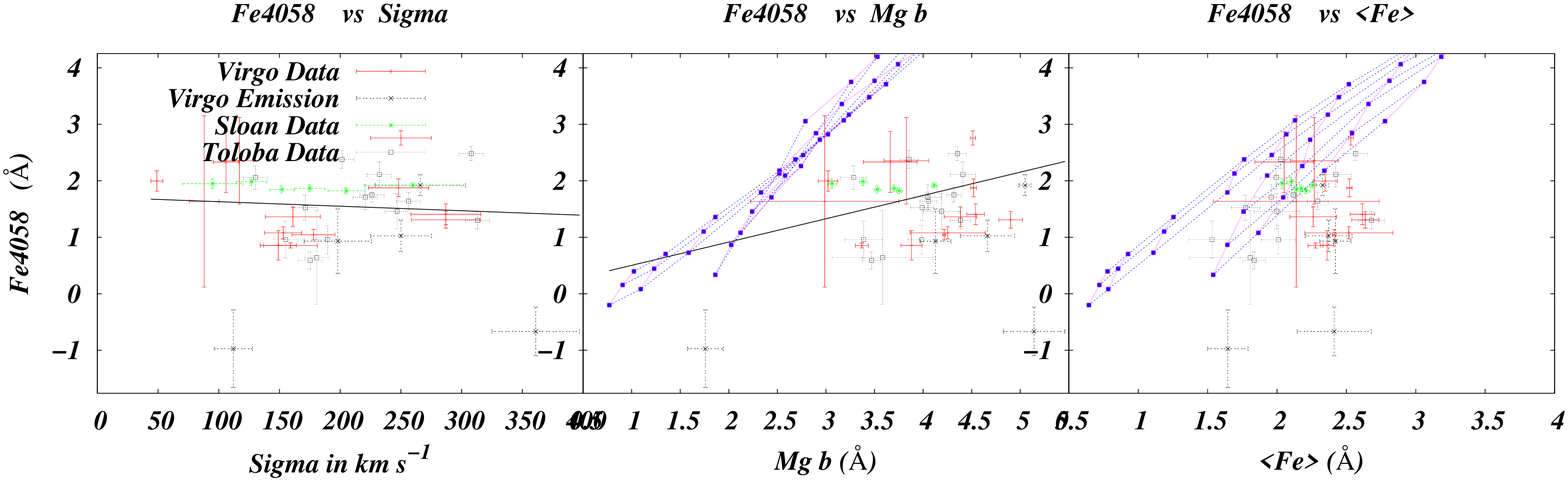}
\caption{}
\end{figure}

\begin{figure}[H]
\includegraphics[width=7in,height=2in]{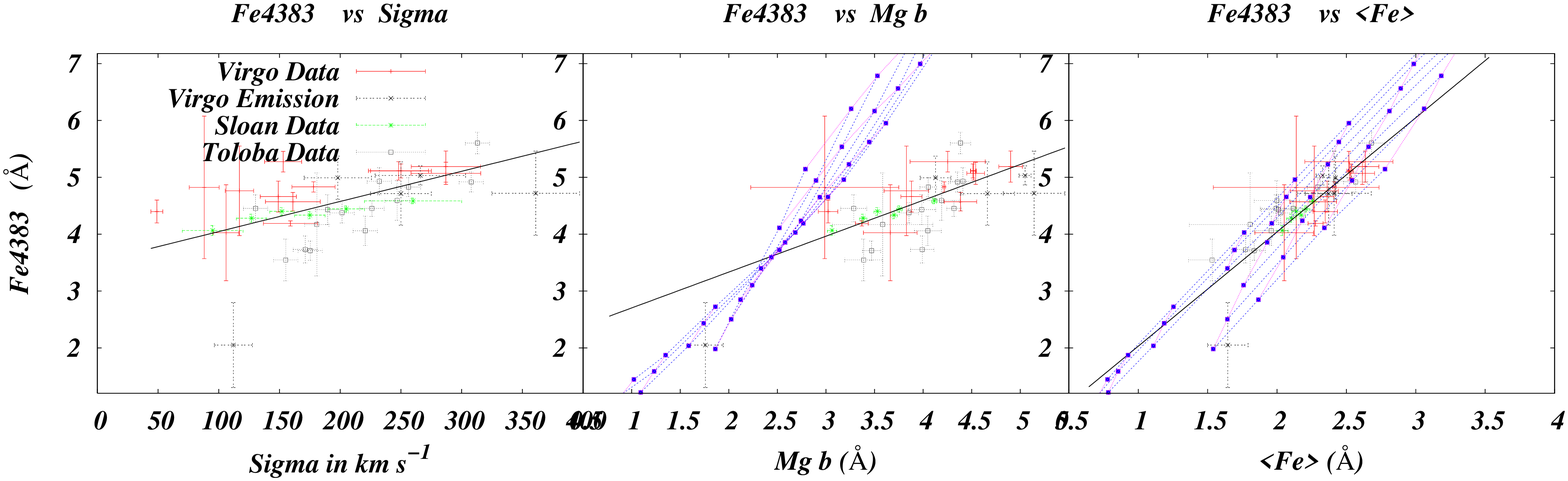}
\caption{}
\end{figure}

\begin{figure}[H]
\includegraphics[width=7in,height=2in]{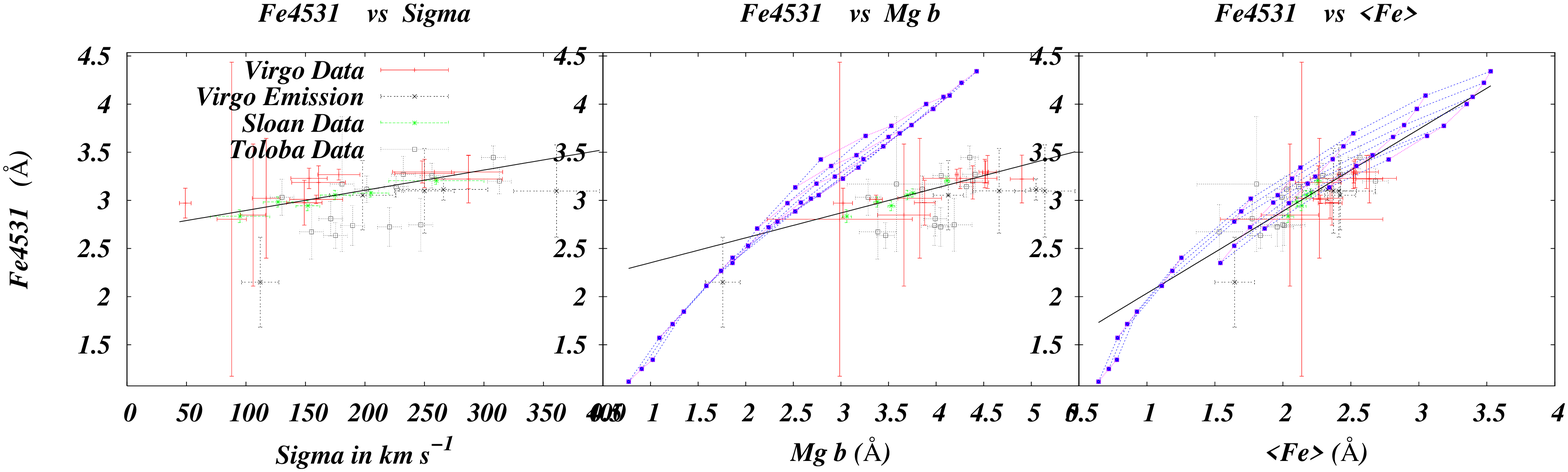}
\caption{}
\end{figure}

\begin{figure}[H]
\includegraphics[width=7in,height=2in]{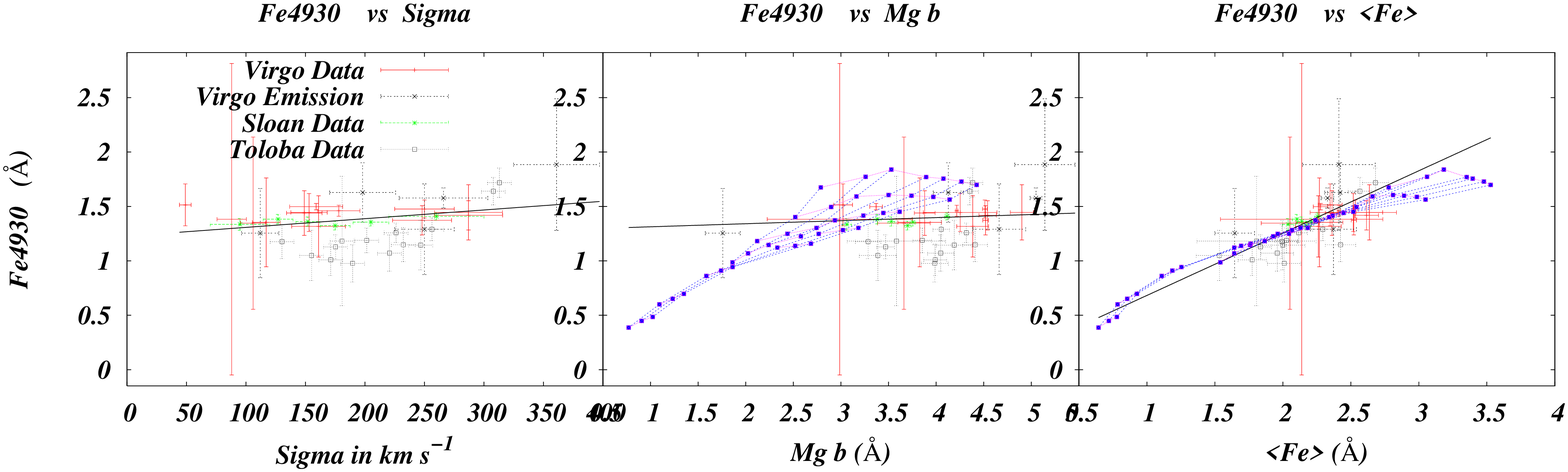}
\caption{}
\end{figure}

\begin{figure}[H]
\includegraphics[width=7in,height=2in]{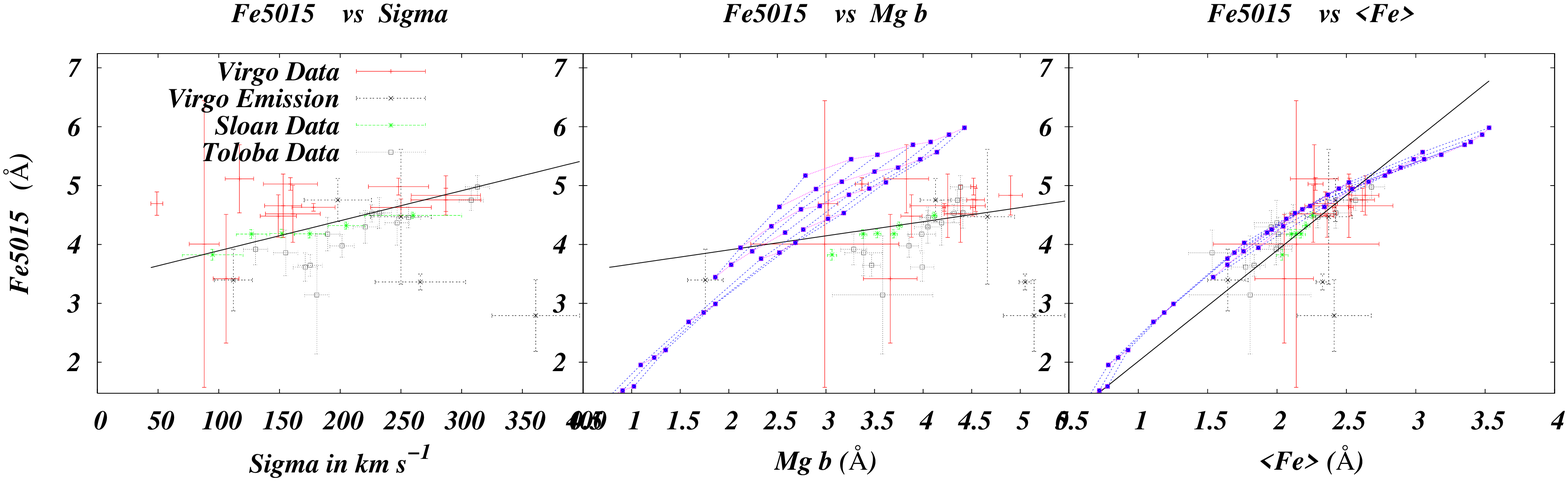}
\caption{}
\end{figure}

\begin{figure}[H]
\includegraphics[width=7in,height=2in]{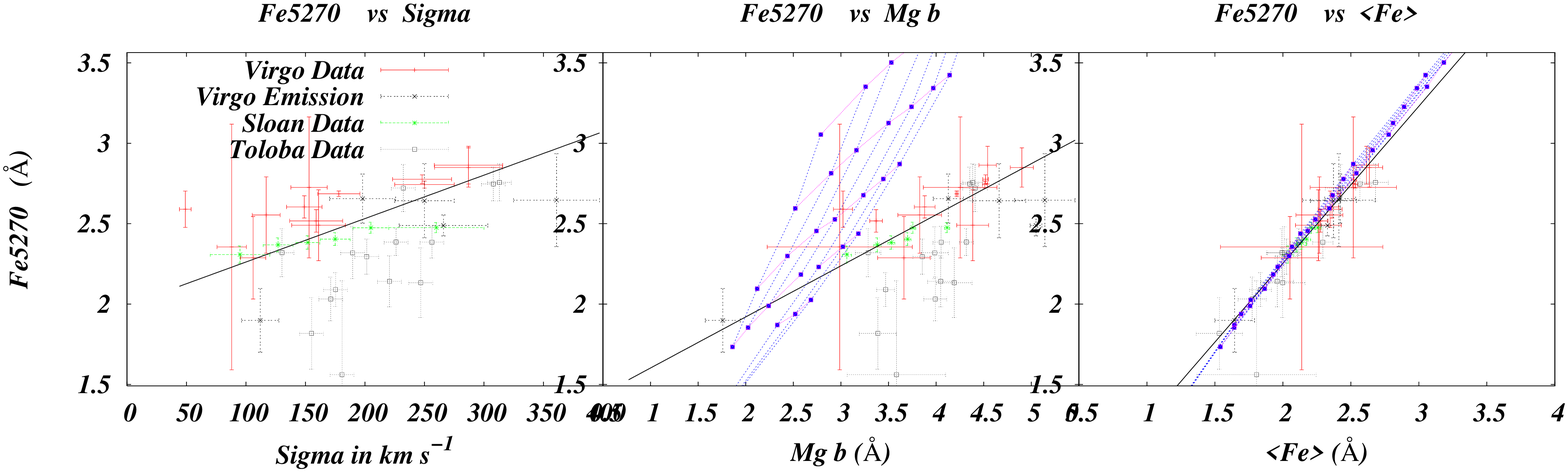}
\caption{}
\end{figure}

\begin{figure}[H]
\includegraphics[width=7in,height=2in]{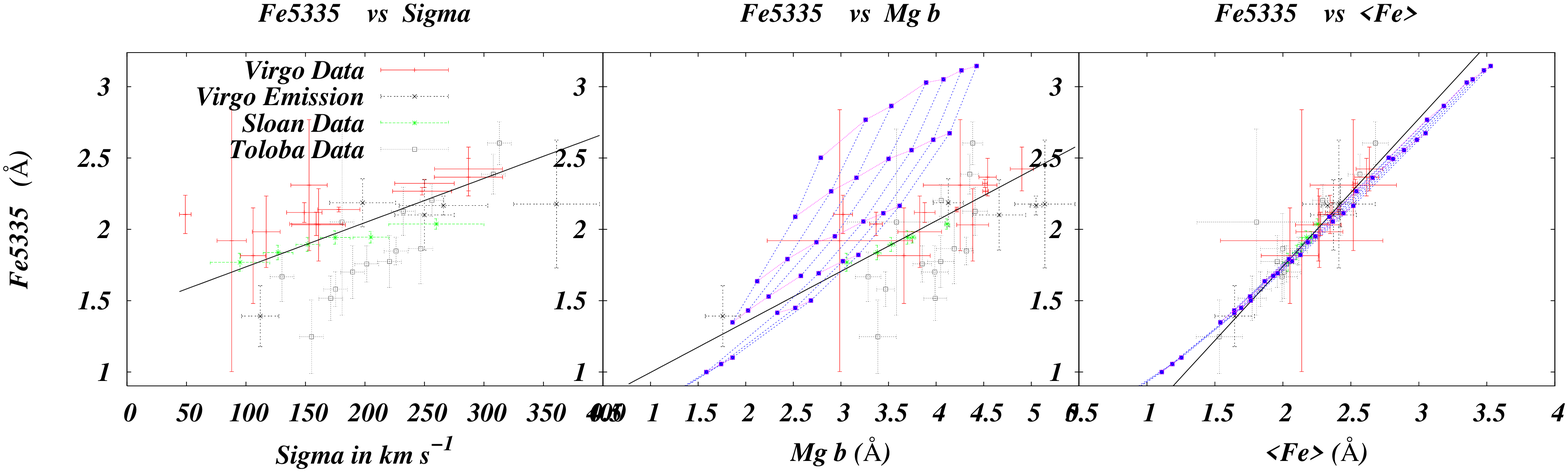}
\caption{}
\end{figure}

\begin{figure}[H]
\includegraphics[width=7in,height=2in]{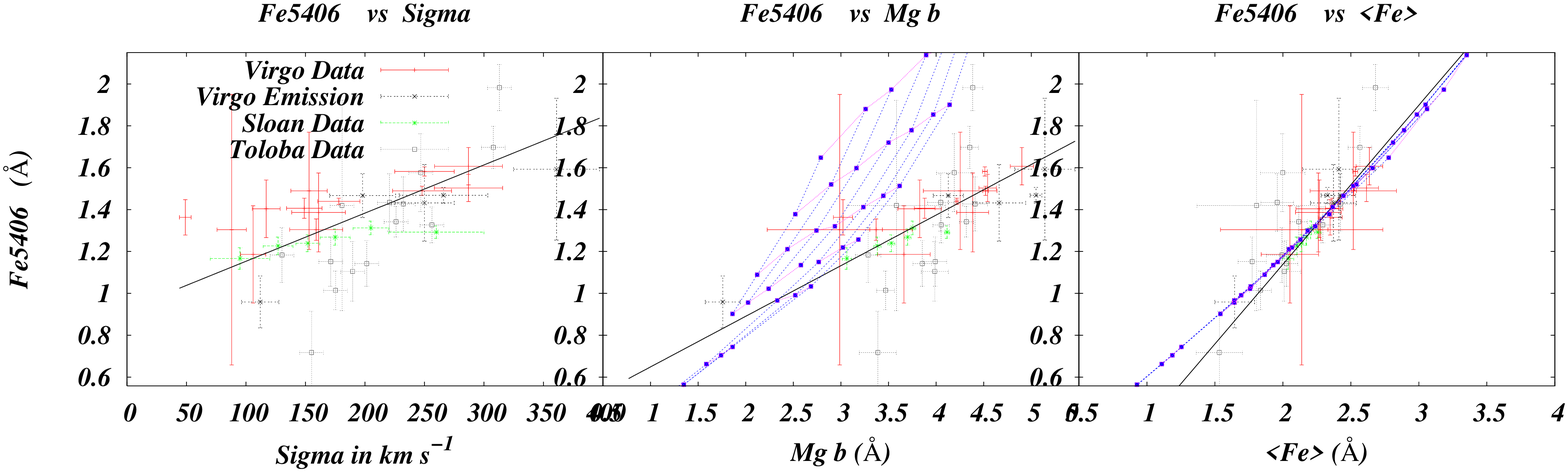}
\caption{}
\end{figure}

\begin{figure}[H]
\includegraphics[width=7in,height=2in]{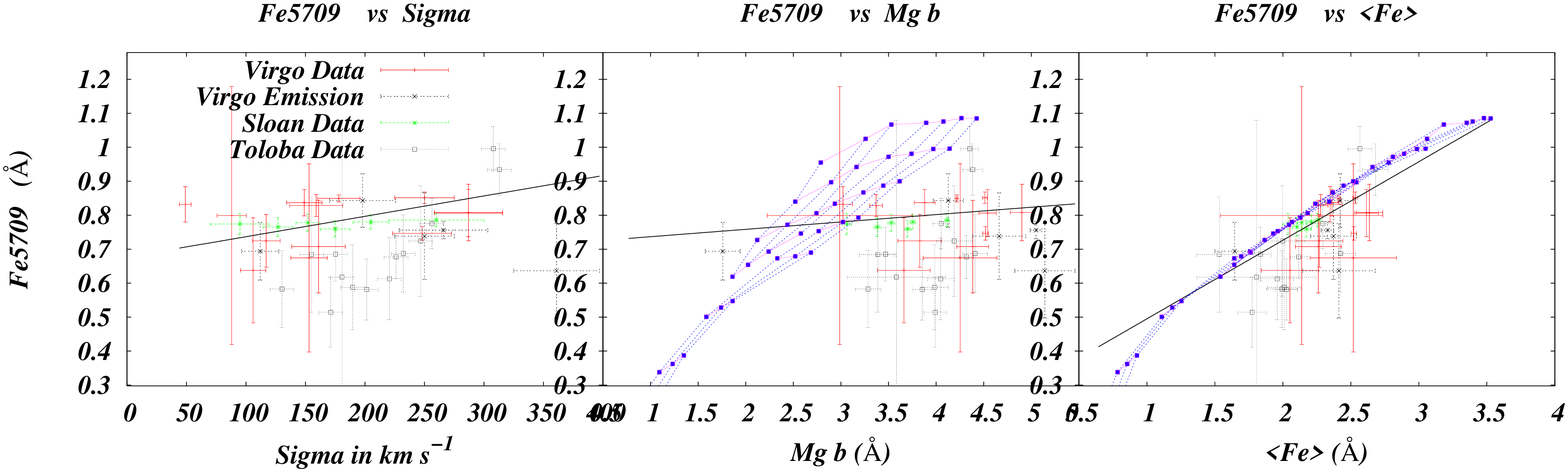}
\caption{}
\end{figure}

\begin{figure}[H]
\includegraphics[width=7in,height=2in]{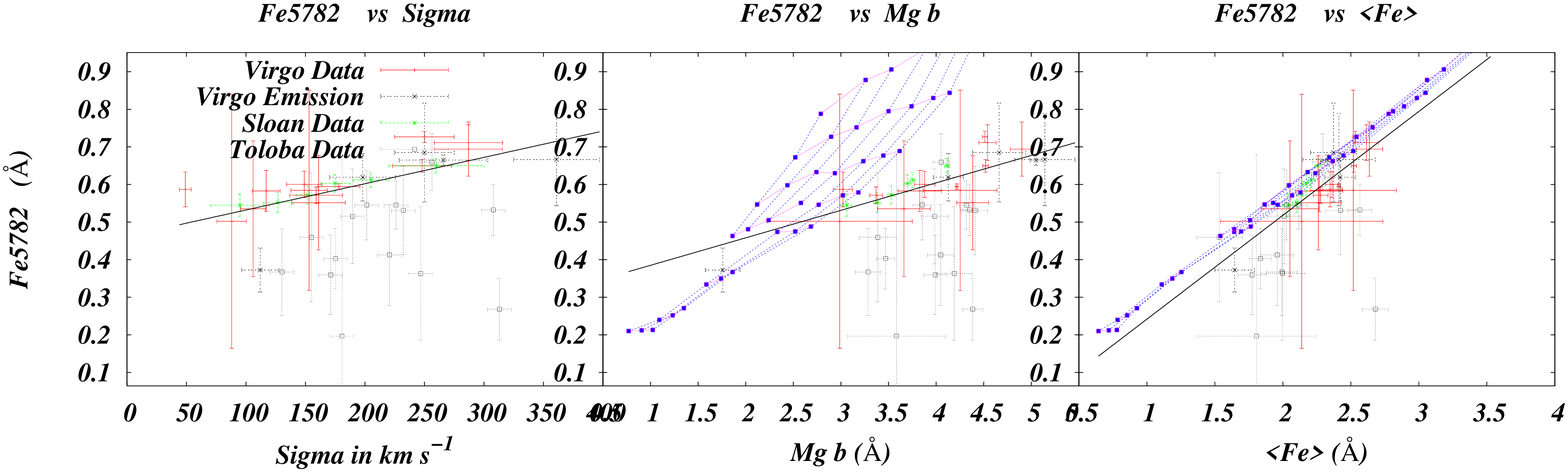}
\caption{}
\end{figure}

\begin{figure}[H]
\includegraphics[width=7in,height=2in]{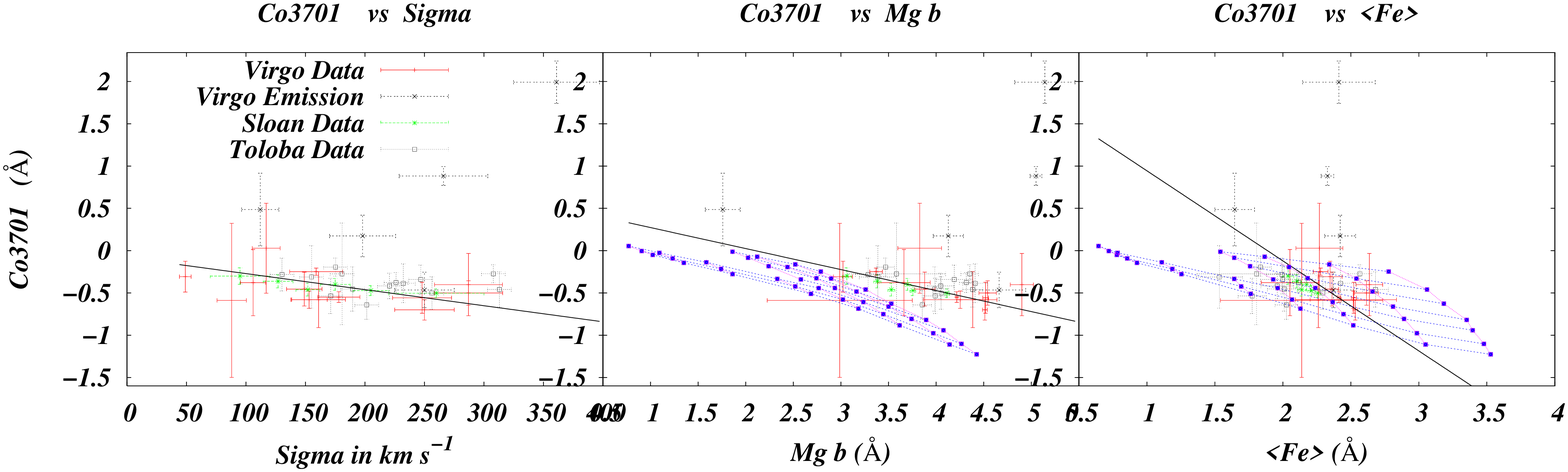}
\caption{}
\end{figure}

\begin{figure}[H]
\includegraphics[width=7in,height=2in]{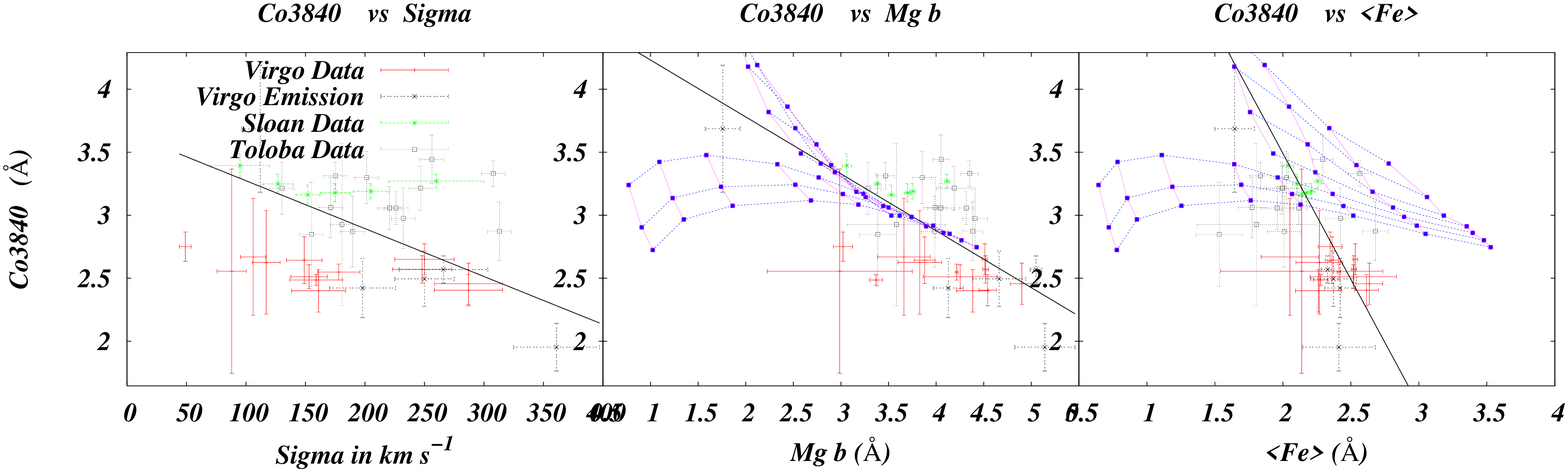}
\caption{}
\end{figure}

\begin{figure}[H]
\includegraphics[width=7in,height=2in]{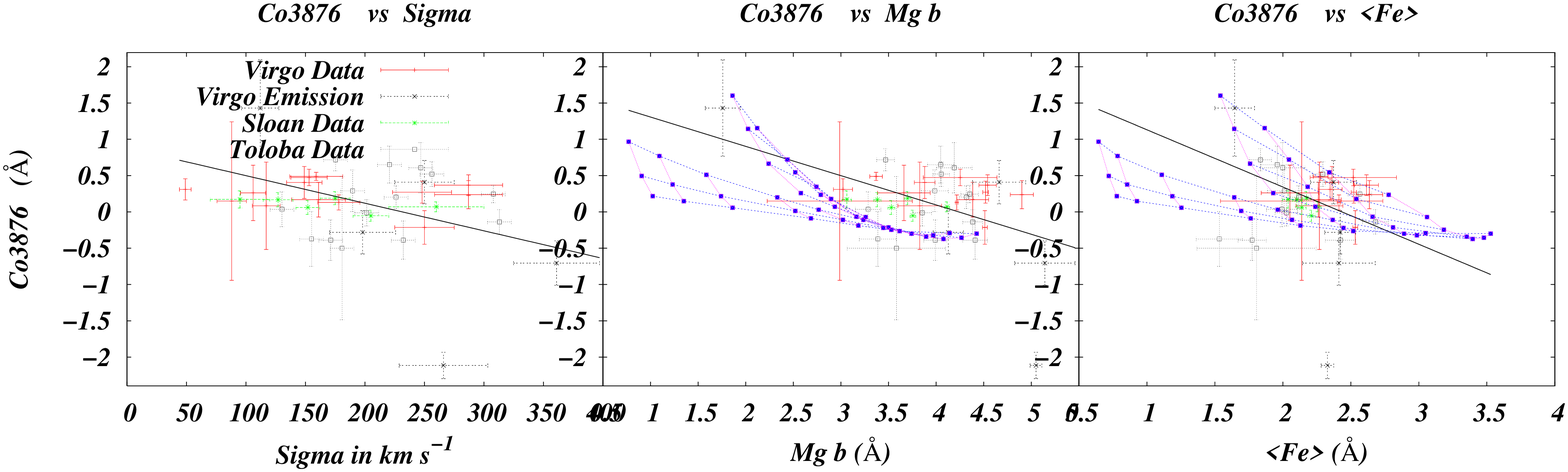}
\caption{}
\end{figure}

\begin{figure}[H]
\includegraphics[width=7in,height=2in]{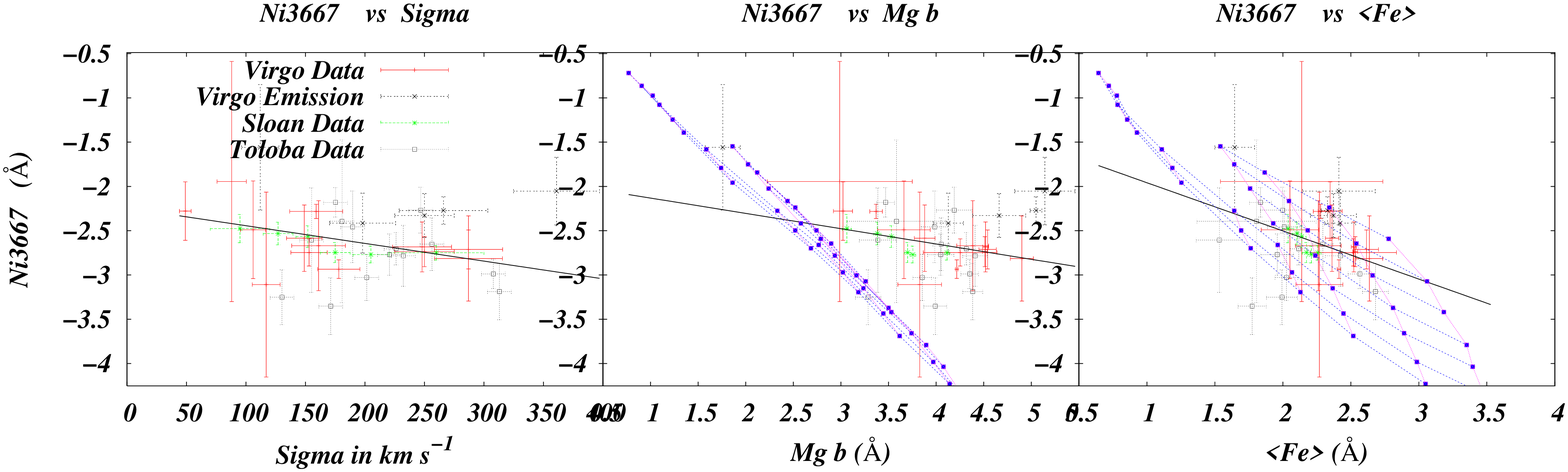}
\caption{}
\end{figure}

\begin{figure}[H]
\includegraphics[width=7in,height=2in]{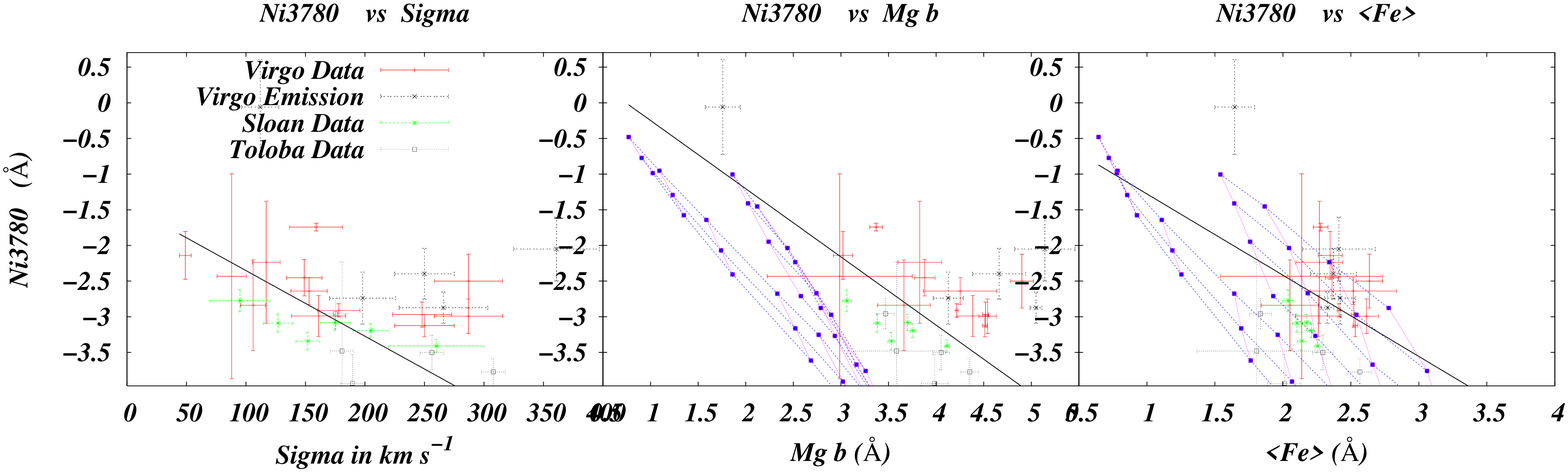}
\caption{}
\end{figure}

\begin{figure}[H]
\includegraphics[width=7in,height=2in]{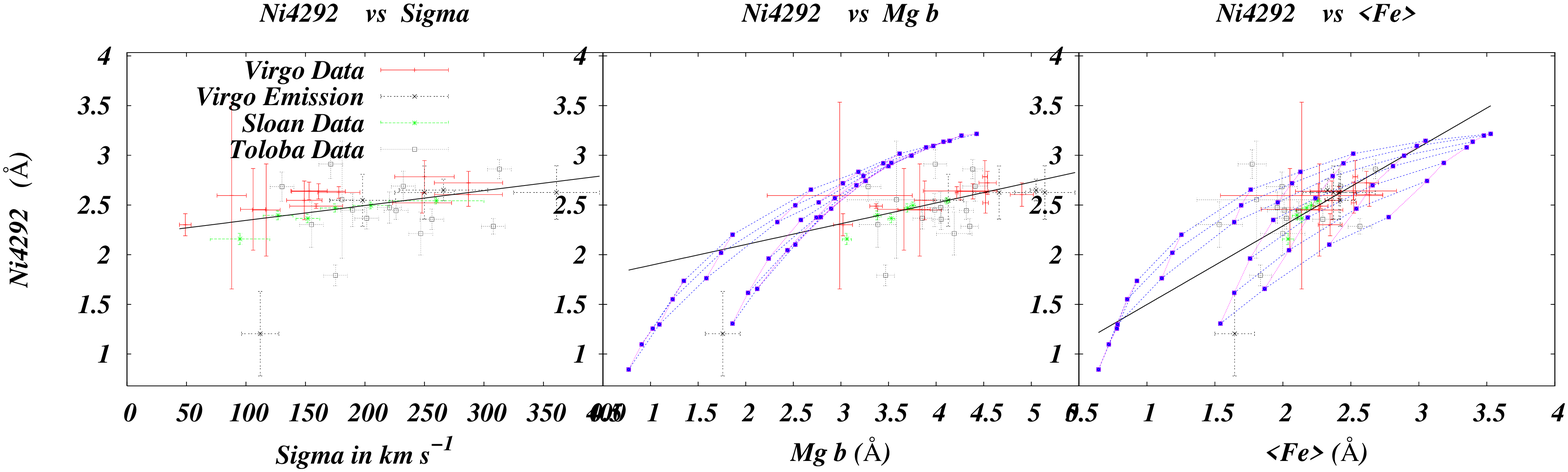}
\caption{}
\end{figure}

\begin{figure}[H]
\includegraphics[width=7in,height=2in]{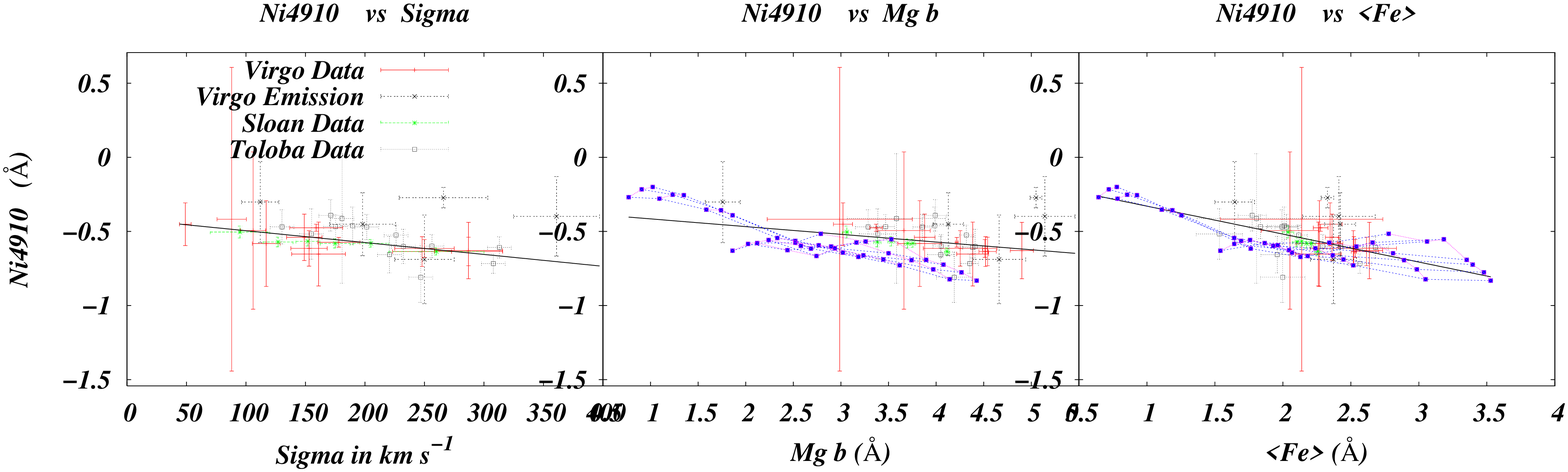}
\caption{}
\end{figure}

\begin{figure}[H]
\includegraphics[width=7in,height=2in]{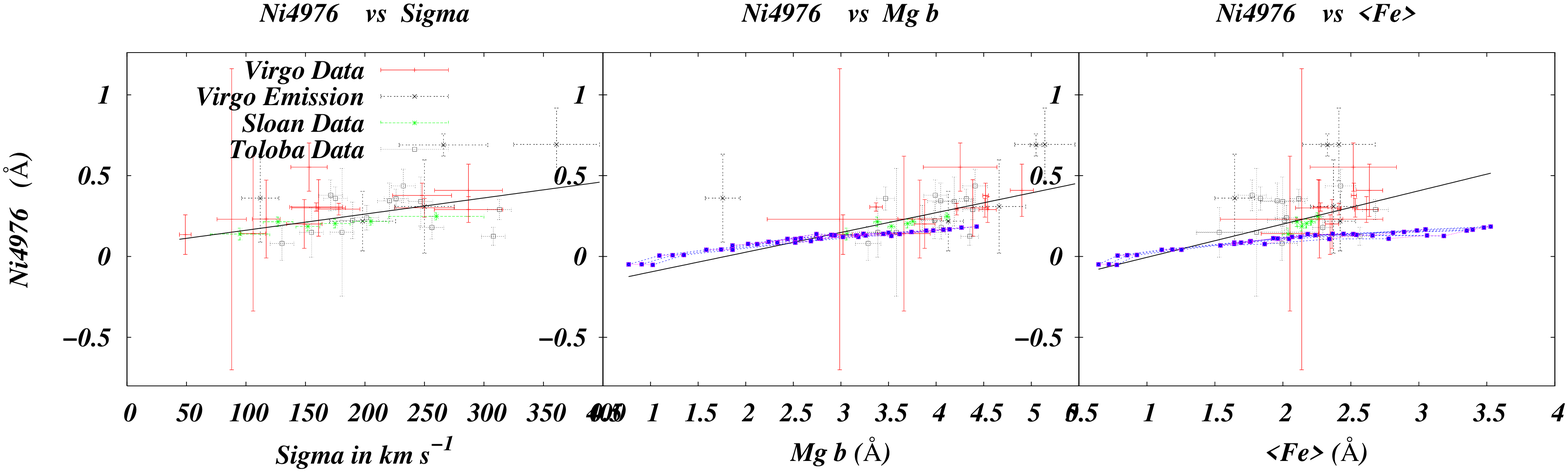}
\caption{}
\end{figure}

\begin{figure}[H]
\includegraphics[width=7in,height=2in]{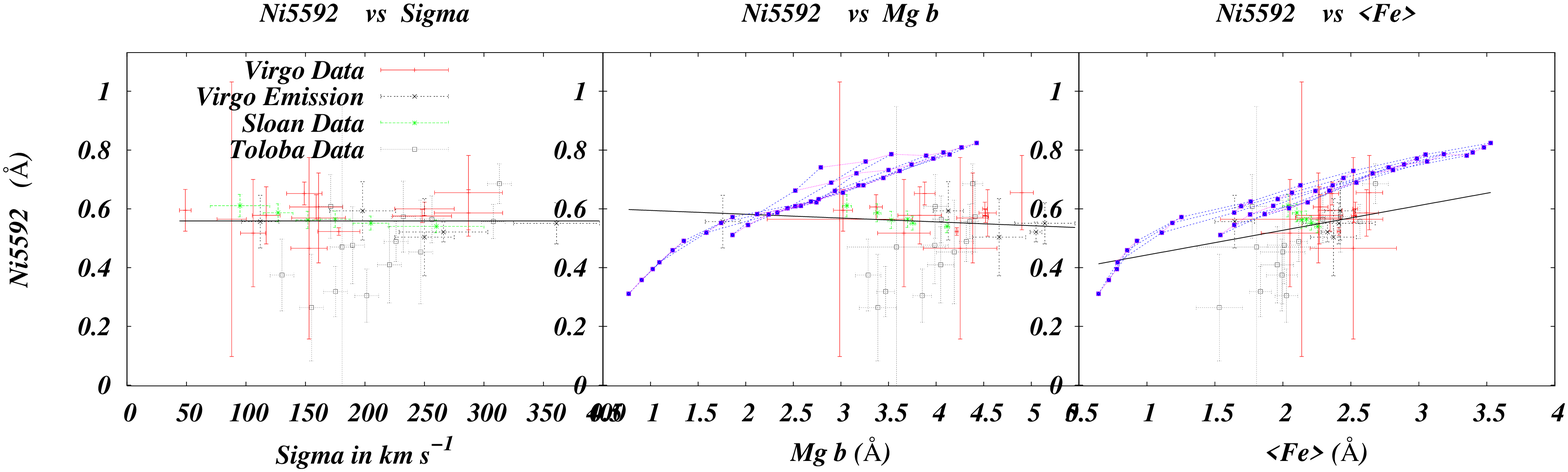}
\caption{}
\end{figure}

\begin{figure}[H]
\includegraphics[width=7in,height=2in]{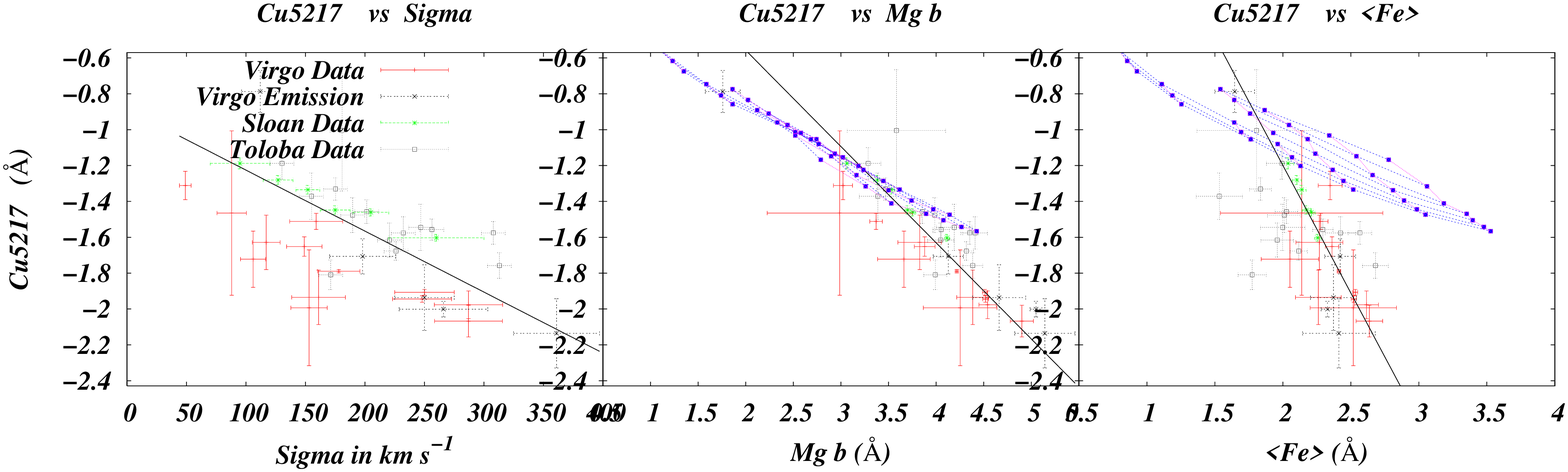}
\caption{}
\end{figure}

\begin{figure}[H]
\includegraphics[width=7in,height=2in]{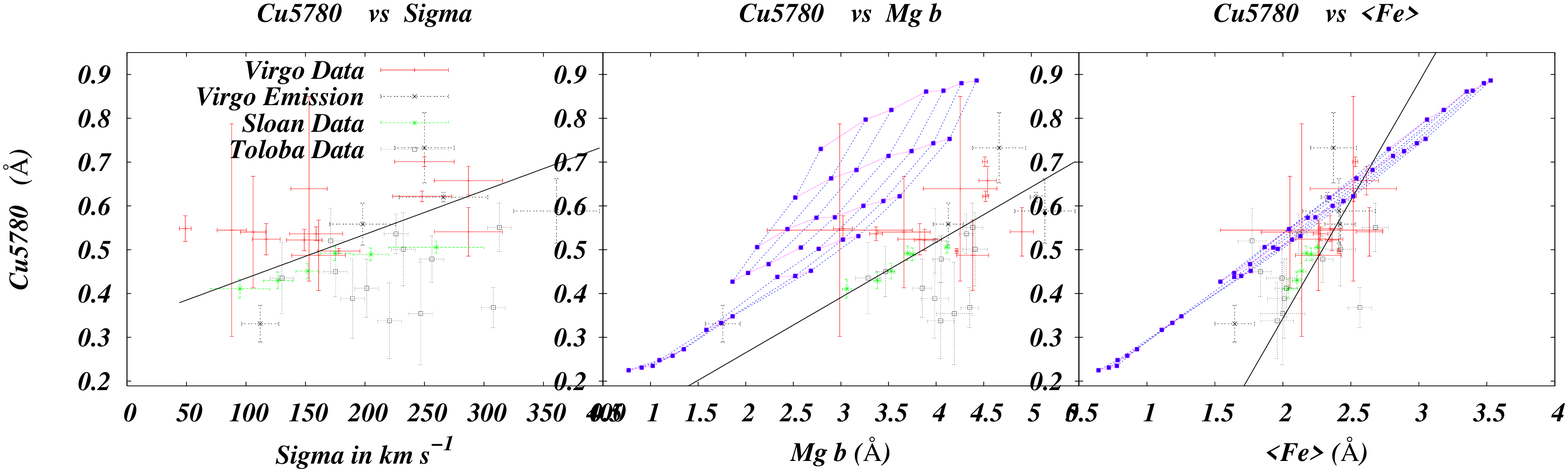}
\caption{}
\end{figure}

\begin{figure}[H]
\includegraphics[width=7in,height=2in]{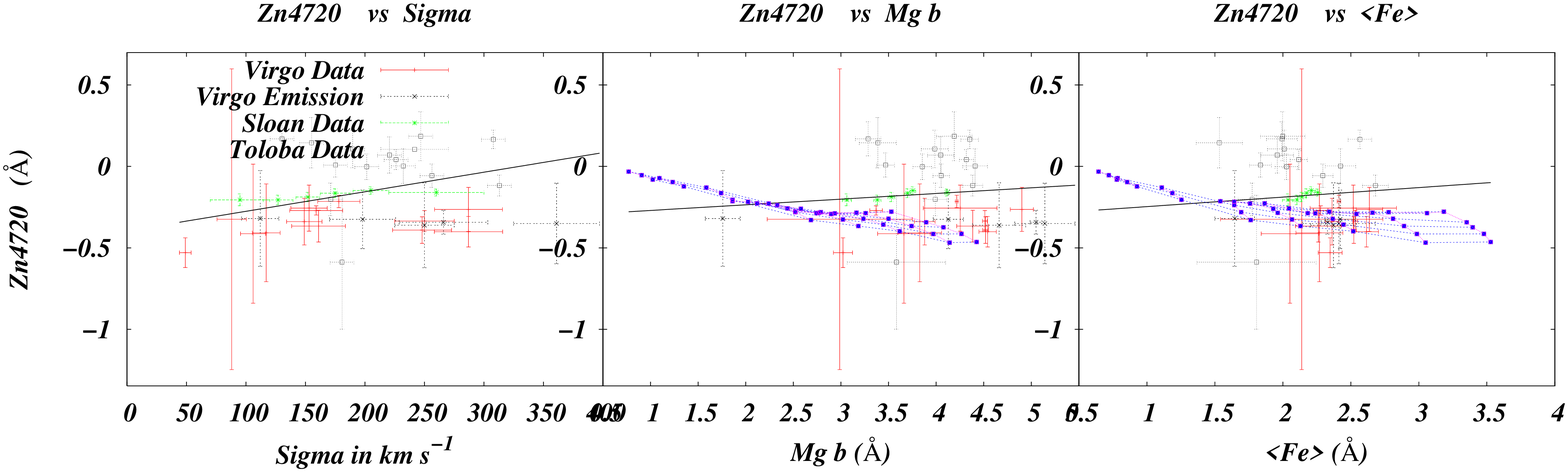}
\caption{}
\end{figure}

\begin{figure}[H]
\includegraphics[width=7in,height=2in]{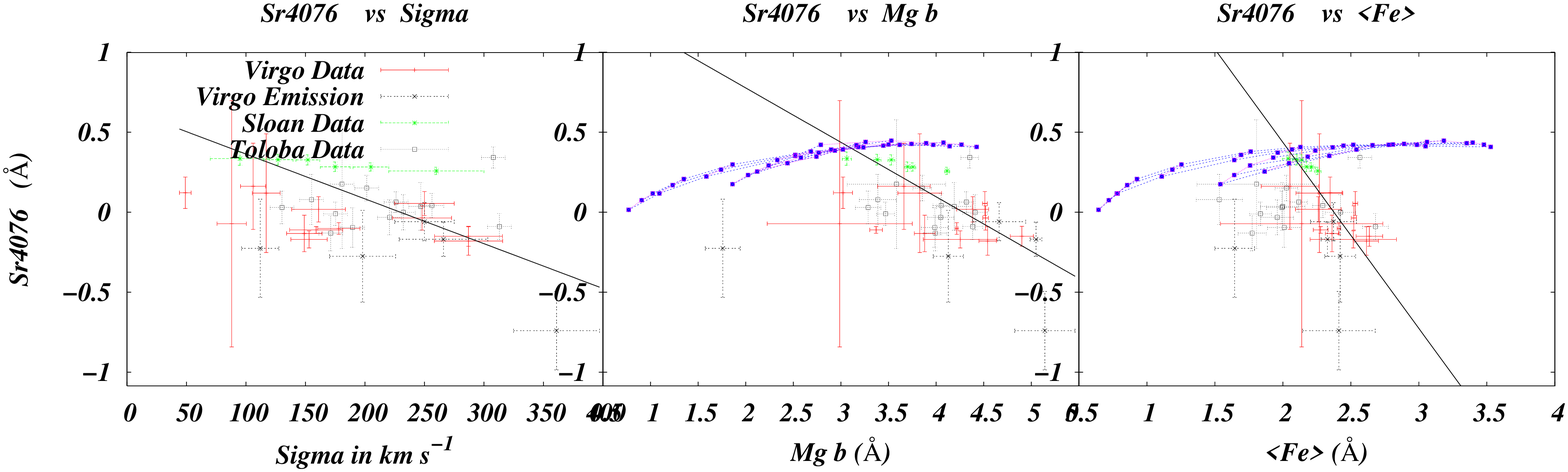}
\caption{}
\end{figure}

\begin{figure}[H]
\includegraphics[width=7in,height=2in]{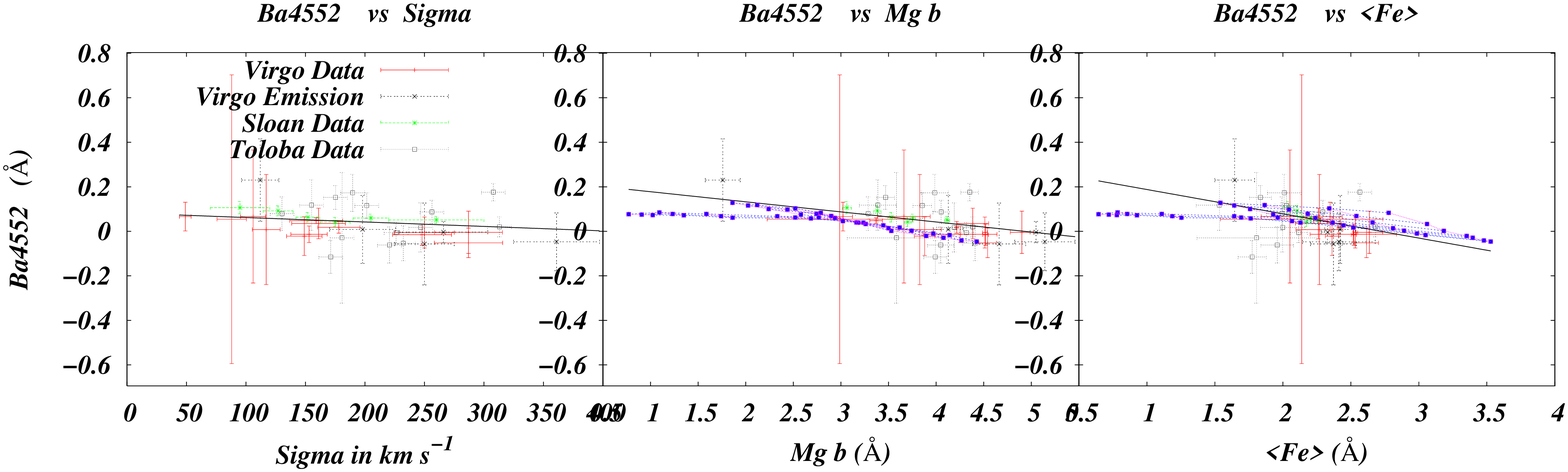}
\caption{}
\end{figure}

\begin{figure}[H]
\includegraphics[width=7in,height=2in]{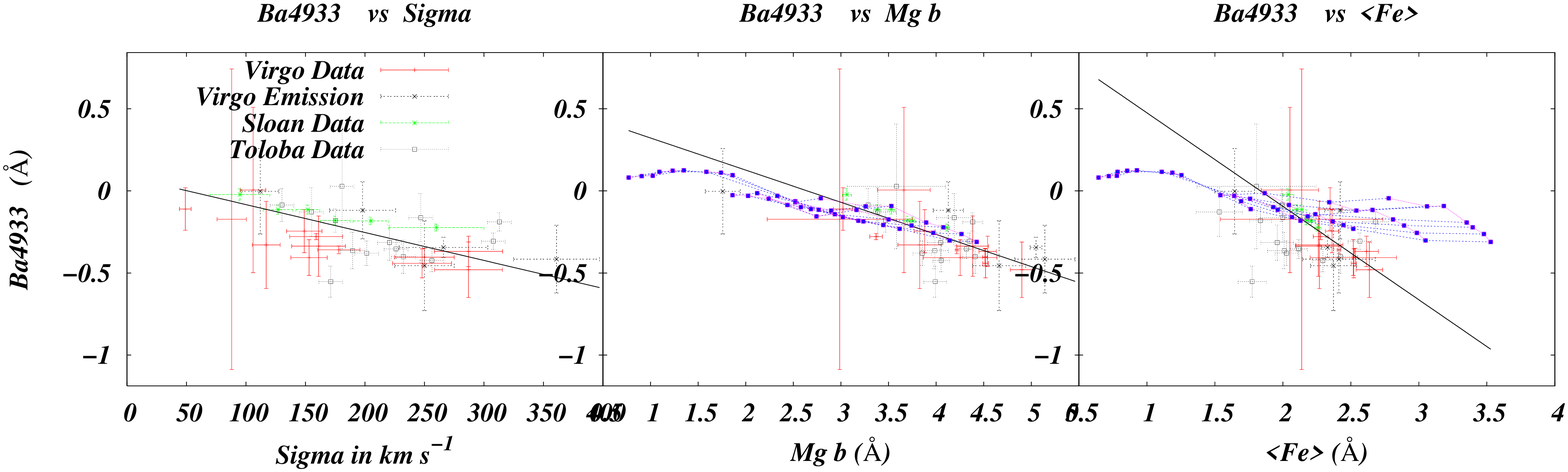}
\caption{}
\end{figure}

\begin{figure}[H]
\includegraphics[width=7in,height=2in]{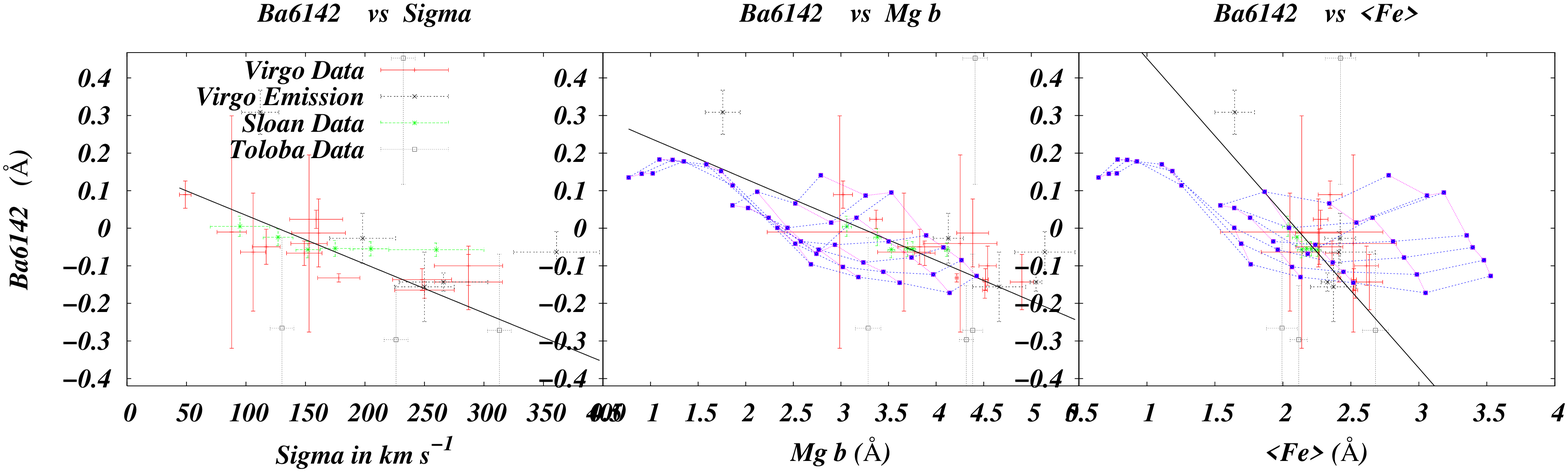}
\caption{}
\end{figure}

\begin{figure}[H]
\includegraphics[width=7in,height=2in]{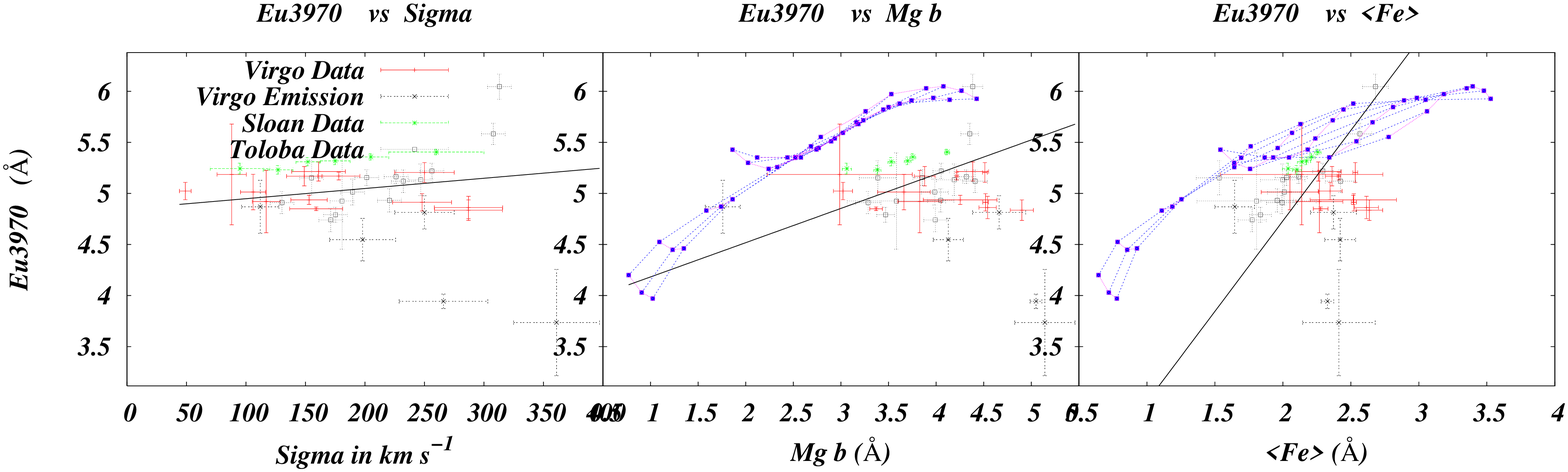}
\caption{}
\end{figure}

\begin{figure}[H]
\includegraphics[width=7in,height=2in]{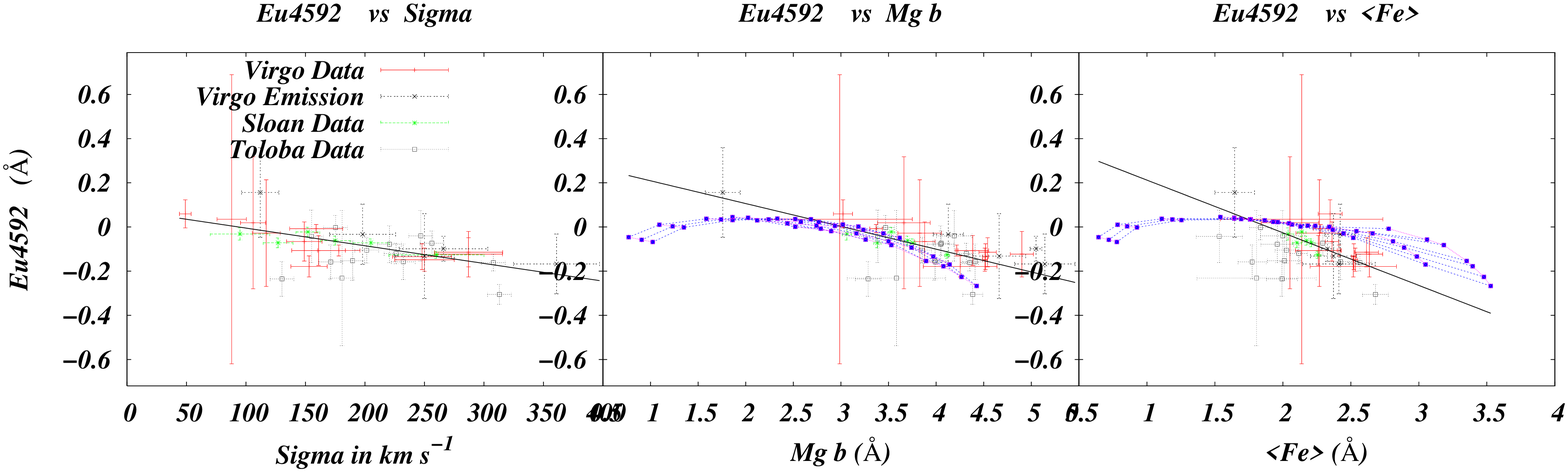}
\caption{}
\end{figure}

\newpage
\pagestyle{plain}

{\centering APPENDIX C\\
VIRGO CLUSTER GALAXY INDEX GRADIENTS\\}

The following 74 figures consist of index measurements of 18 Virgo cluster galaxies plotted as a function of $log(r/R_e)$ where r is the radius and $R_e$ is the effective radius of each individual galaxy. In each figure there are 18 plots one for each galaxy and in each plot is a single black best fit trend line. 

The figures are arranged by order of atomic number starting with hydrogen, where there are more than one index for a given element or elements the indices are further arranged so that the bluest index is first.

The results of this arrangement puts hydrogen first followed by C, N and O then the alpha elements followed by the iron peak elements.

The five galaxies which show hydrogen emission in H$\alpha$ are NGC 1400, NGC 2768, NGC 3156, NGC 4278 and NGC 4486. 

\setcounter{figure}{0}
\begin{figure}[H]
\includegraphics[width=6in,height=7in]{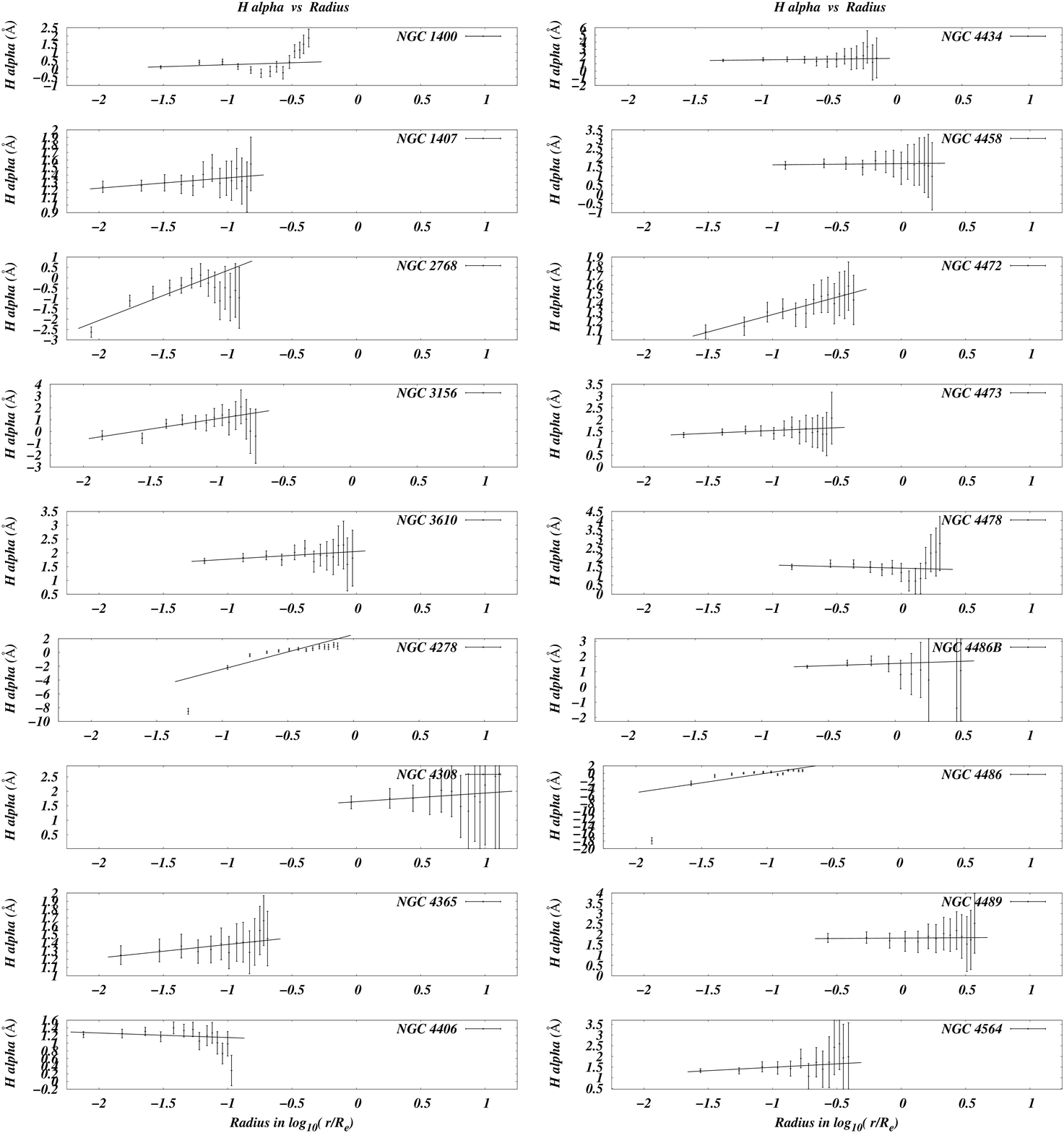}
\caption{
\label{Fig1}}
\end{figure}

\begin{figure}[H]
\includegraphics[width=6in,height=7in]{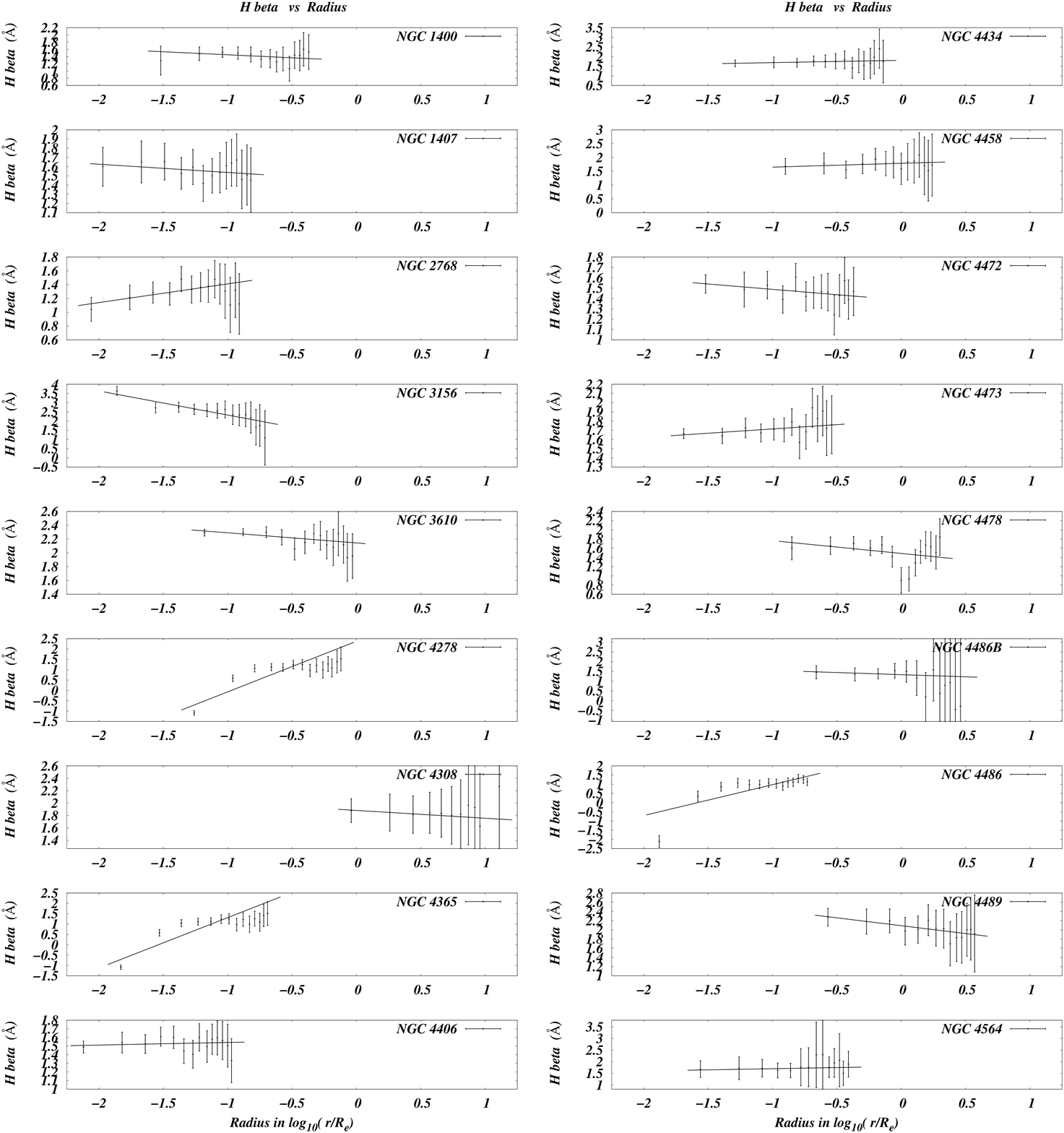}
\caption{}
\end{figure}

\begin{figure}[H]
\includegraphics[width=6in,height=7in]{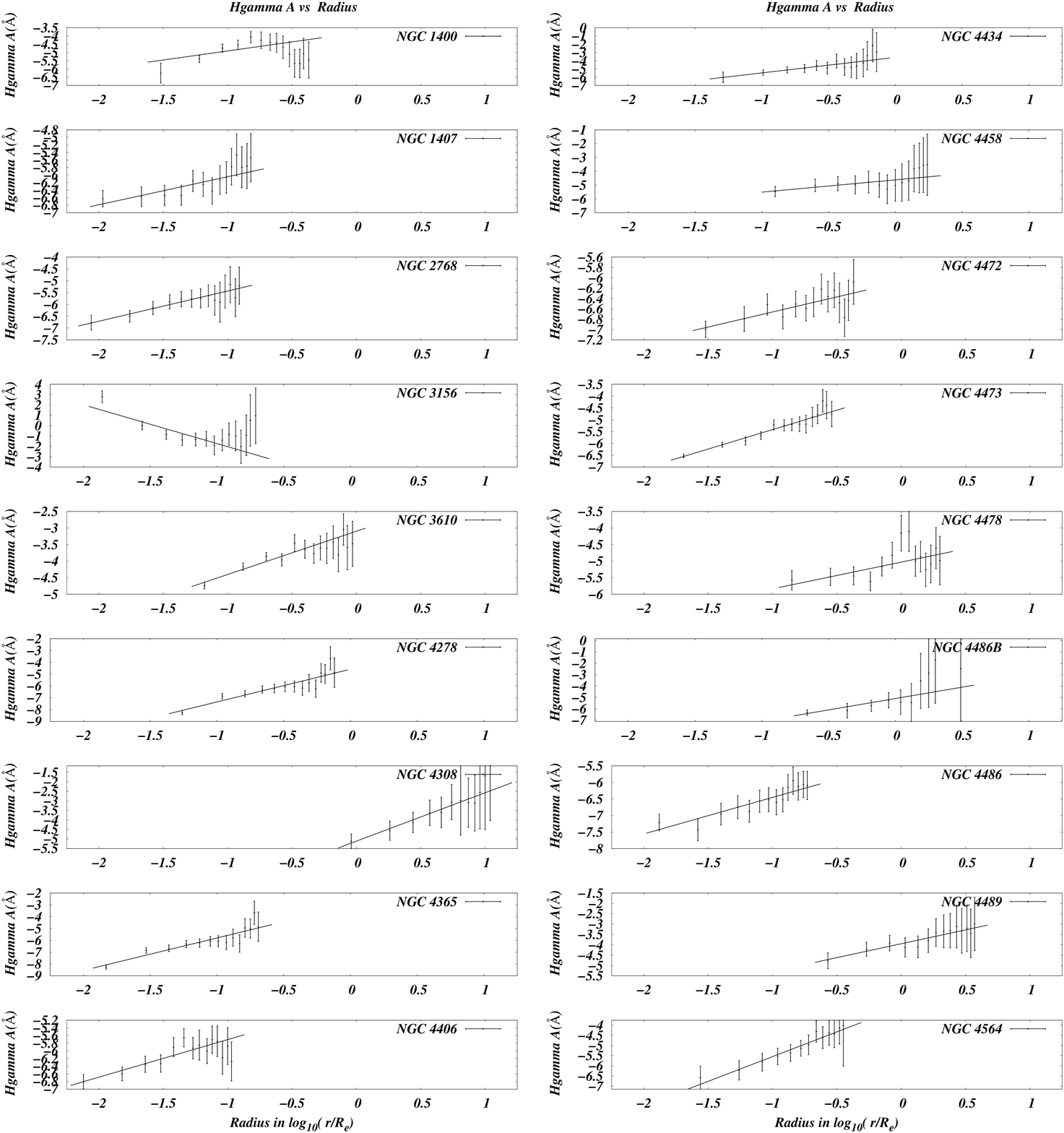}
\caption{
\label{Fig2} }
\end{figure}

\begin{figure}[H]
\includegraphics[width=6in,height=7in]{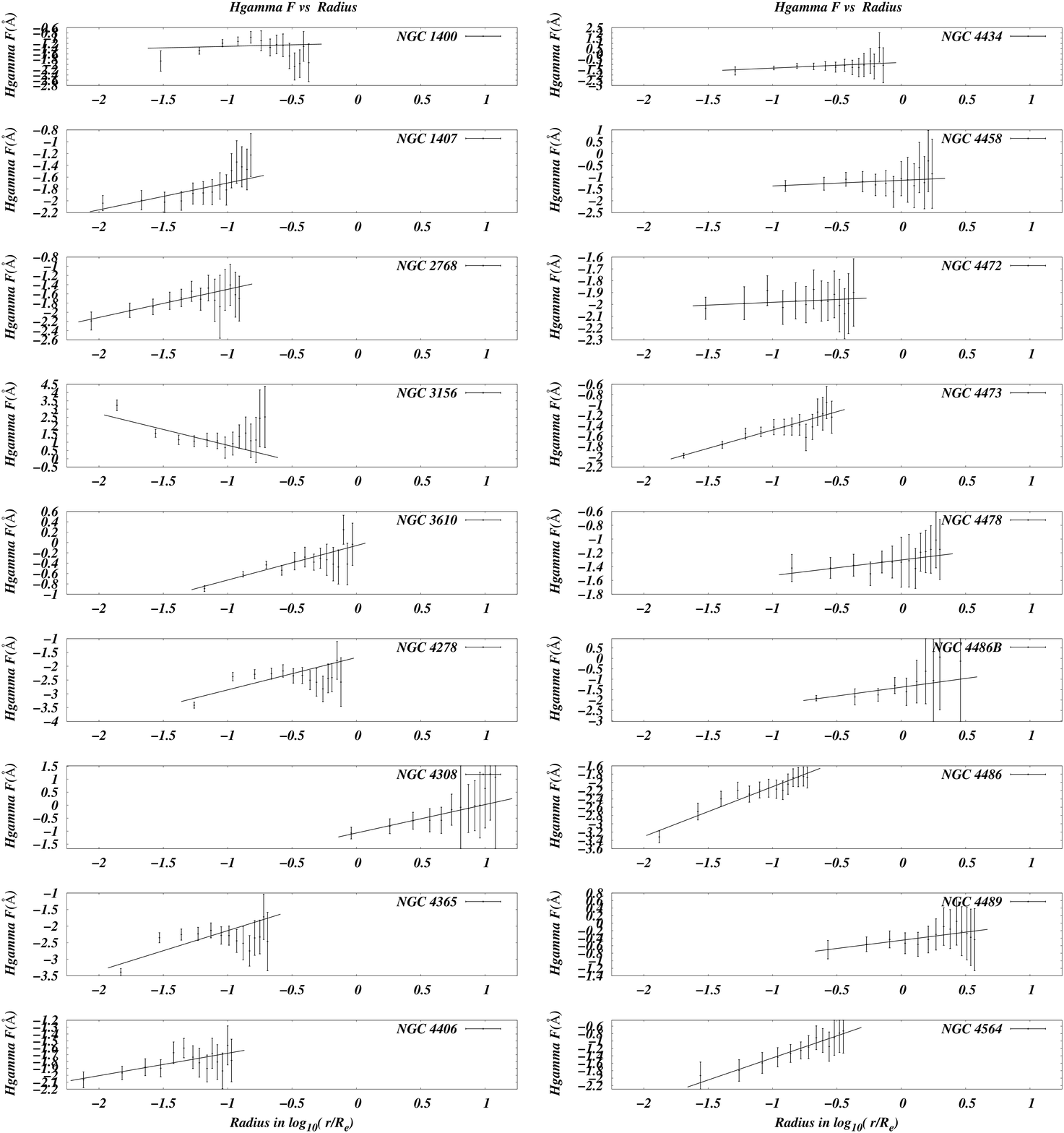}
\caption{
\label{Fig3} }
\end{figure}

\begin{figure}[H]
\includegraphics[width=6in,height=7in]{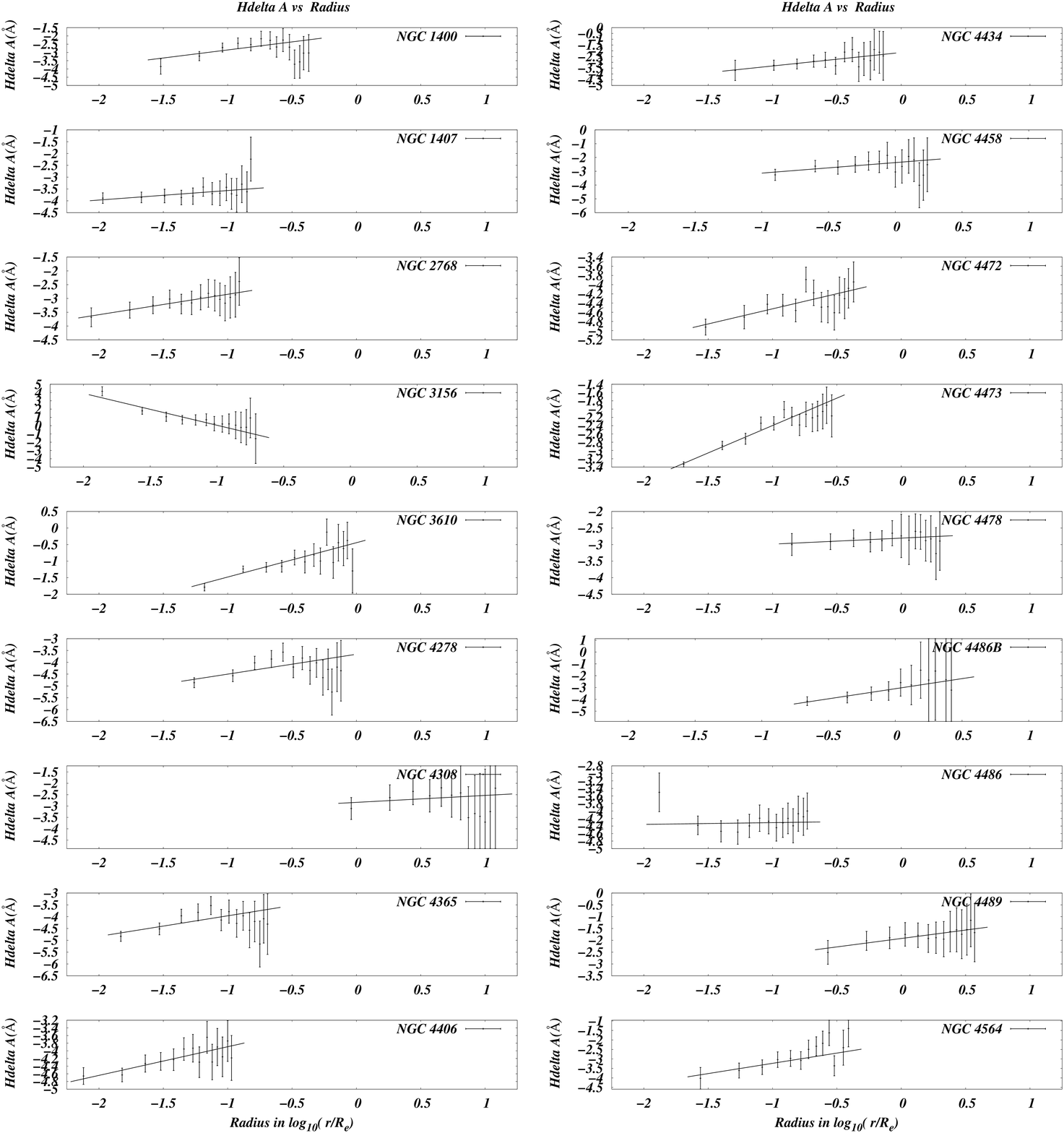}
\caption{
\label{Fig4} }
\end{figure}

\begin{figure}[H]
\includegraphics[width=6in,height=7in]{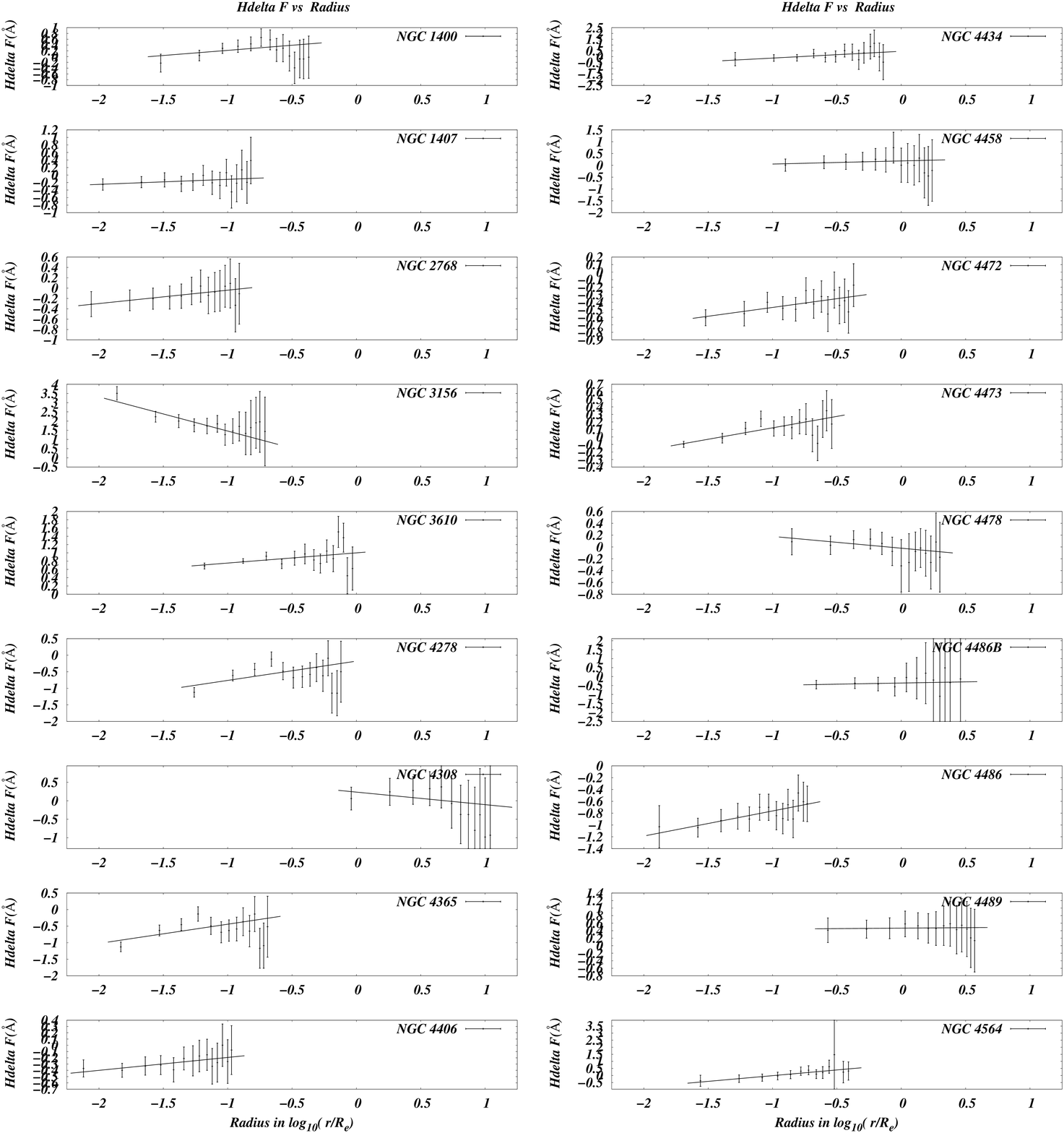}
\caption{
\label{Fig5} }
\end{figure}

\begin{figure}[H]
\includegraphics[width=6in,height=7in]{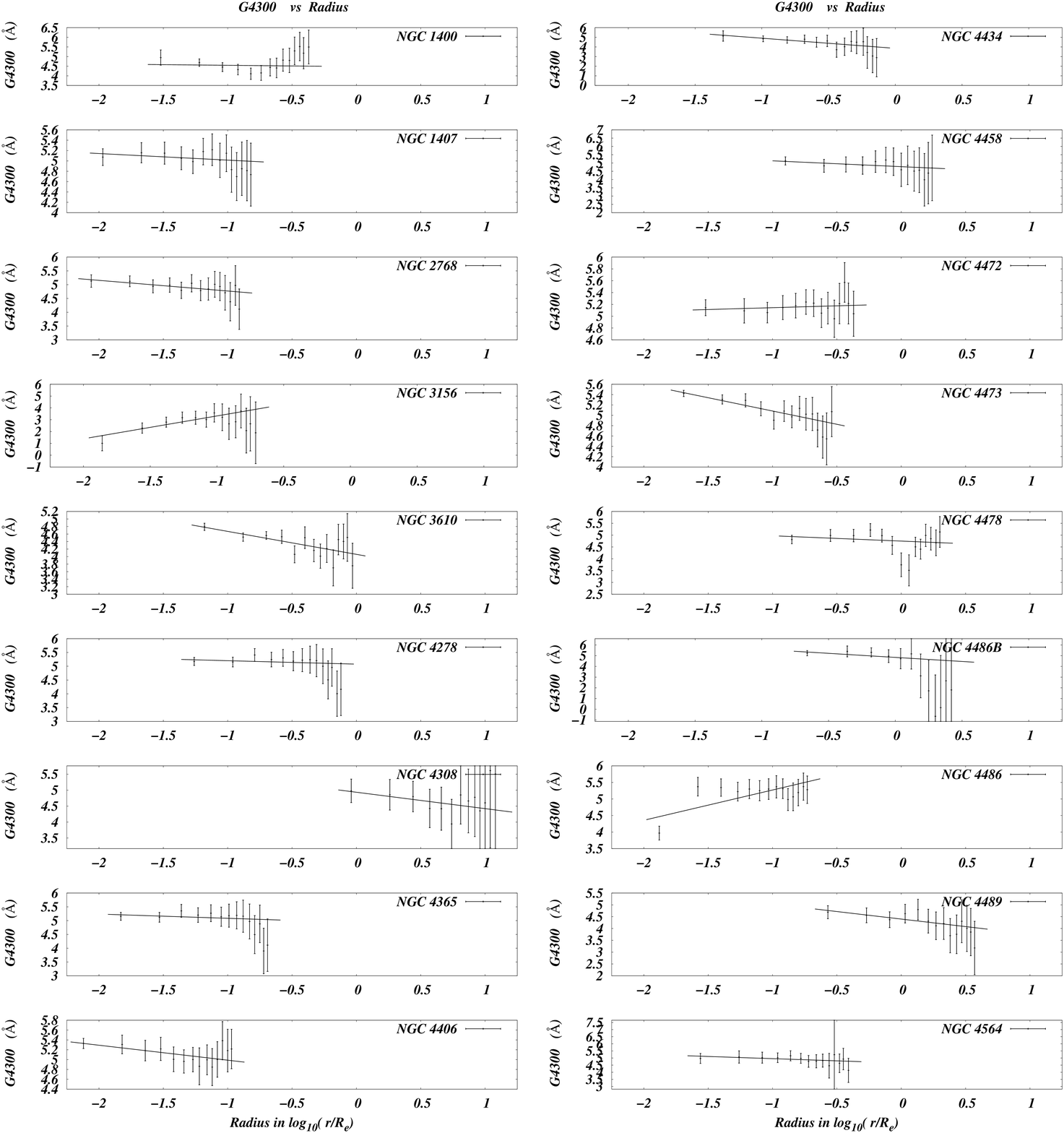}
\caption{}
\end{figure}

\begin{figure}[H]
\includegraphics[width=6in,height=7in]{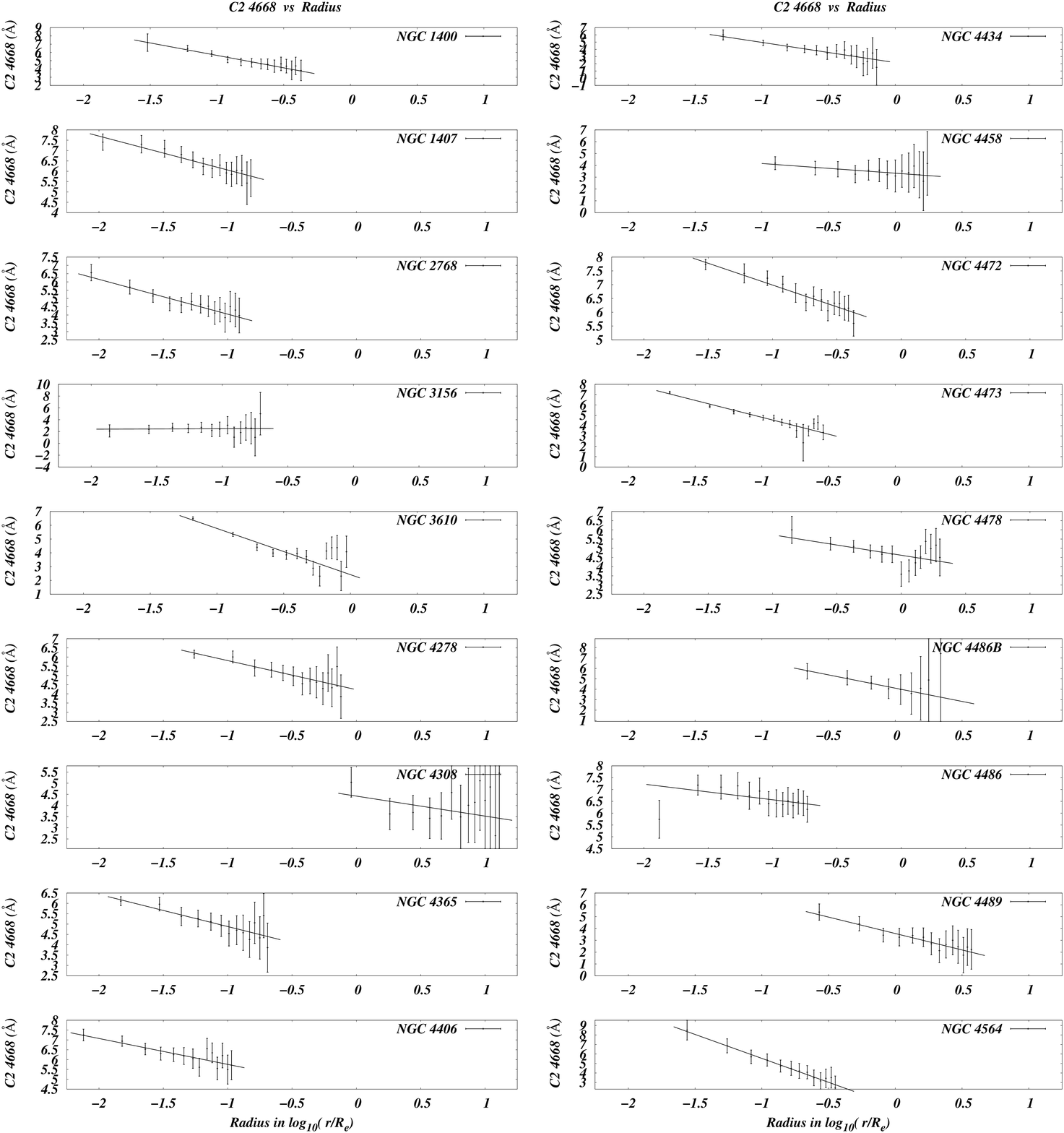}
\caption{}
\end{figure}

\begin{figure}[H]
\includegraphics[width=6in,height=7in]{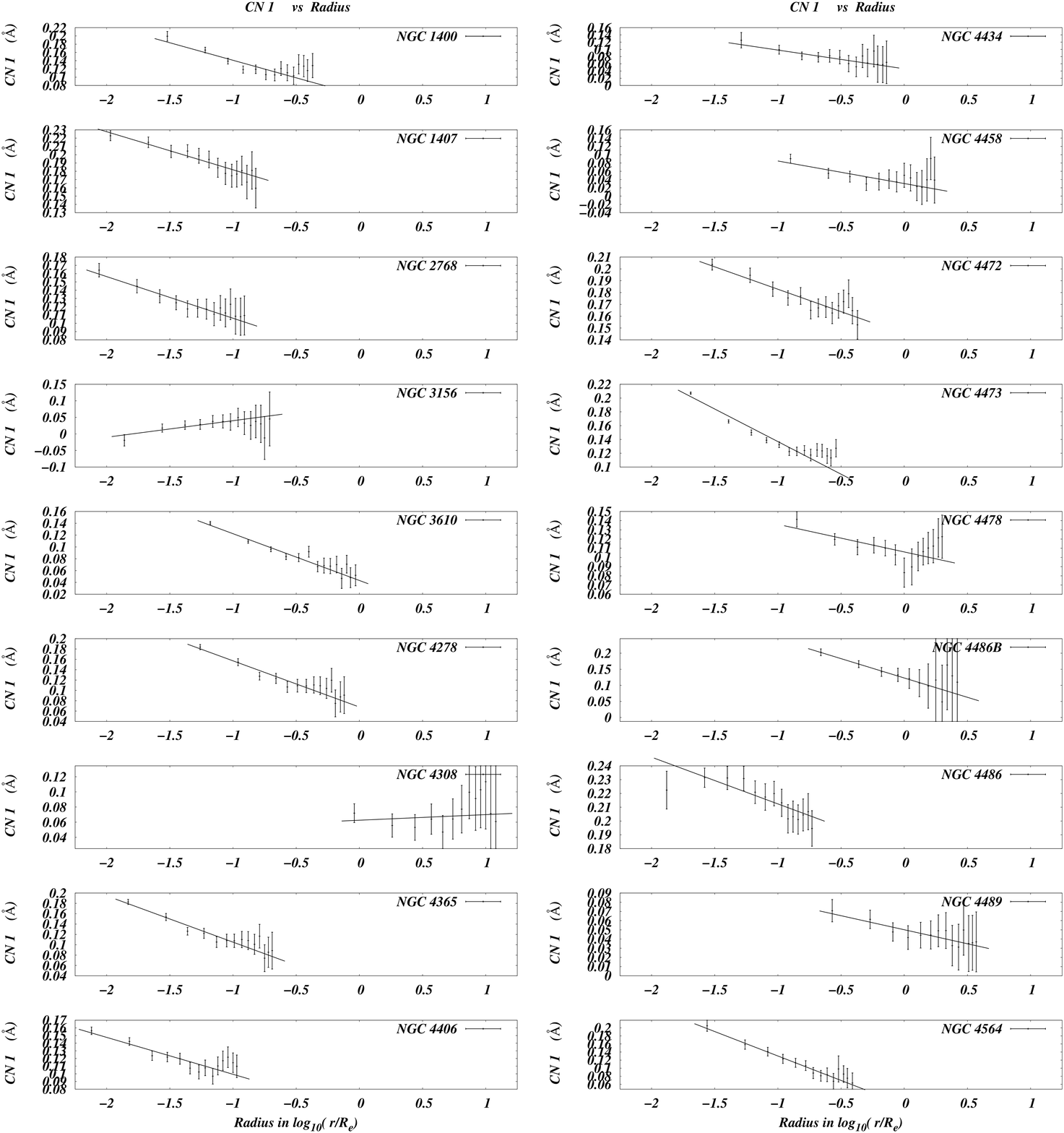}
\caption{}
\end{figure}

\begin{figure}[H]
\includegraphics[width=6in,height=7in]{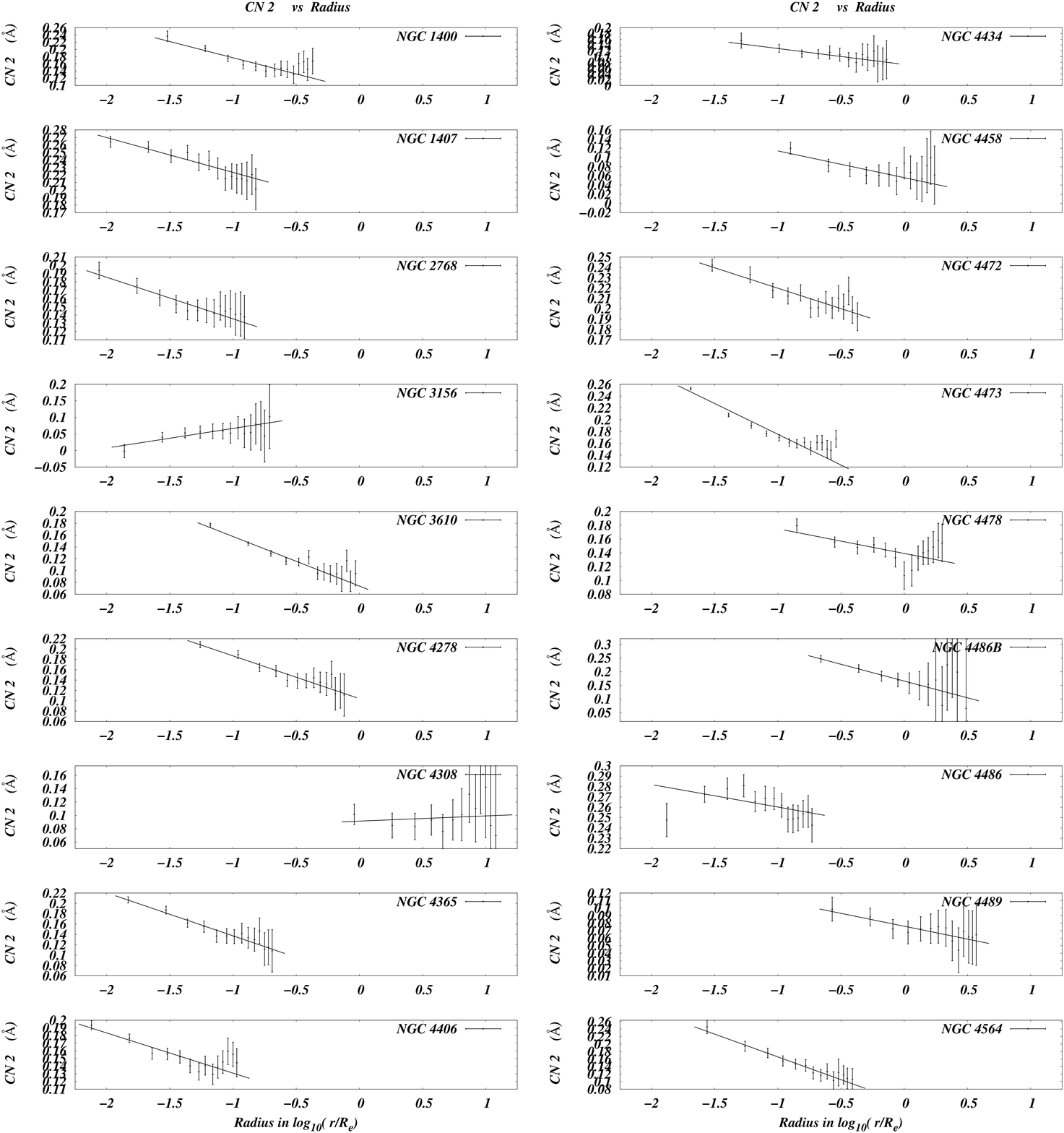}
\caption{}
\end{figure}

\begin{figure}[H]
\includegraphics[width=6in,height=7in]{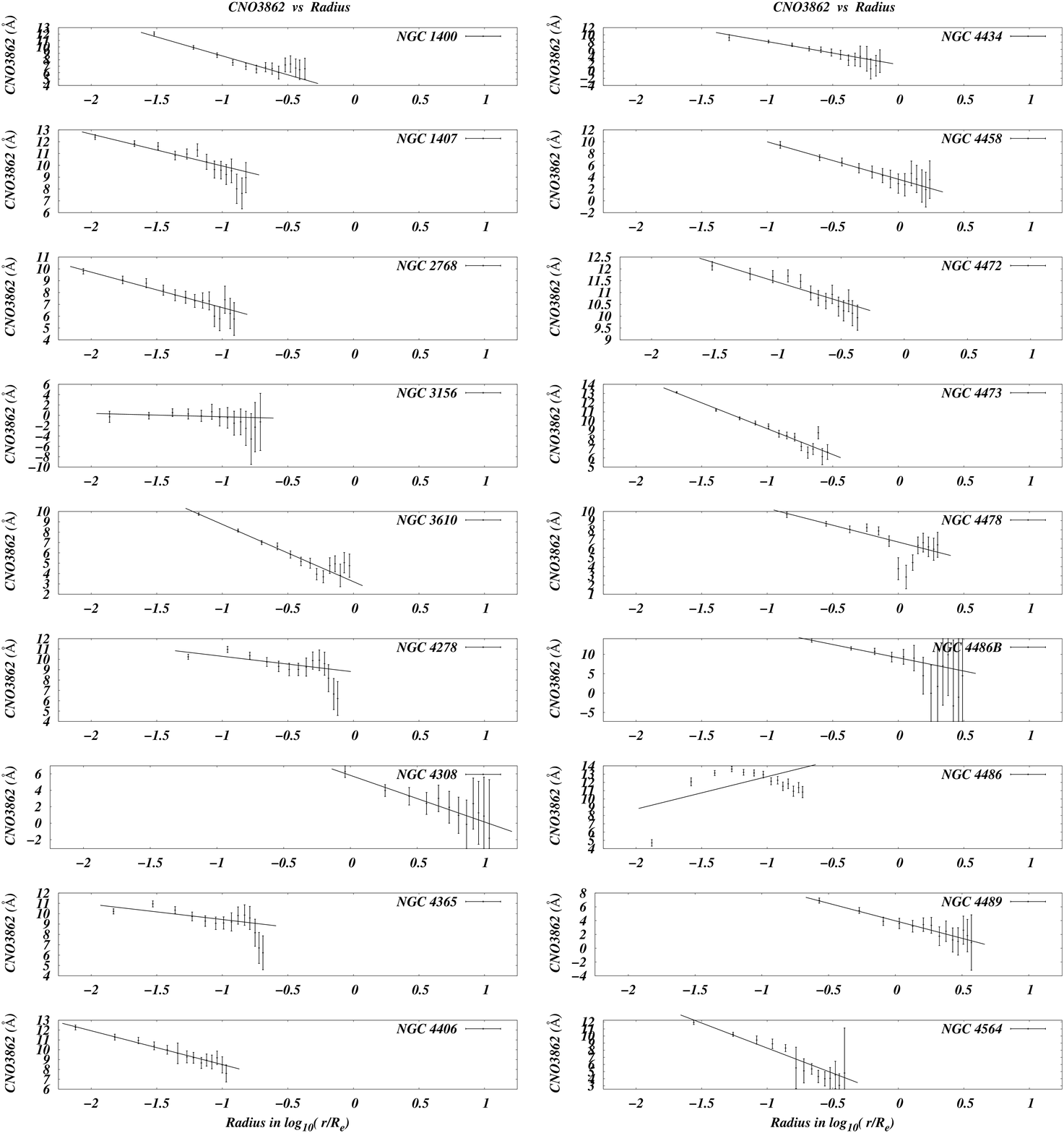}
\caption{}
\end{figure}

\begin{figure}[H]
\includegraphics[width=6in,height=7in]{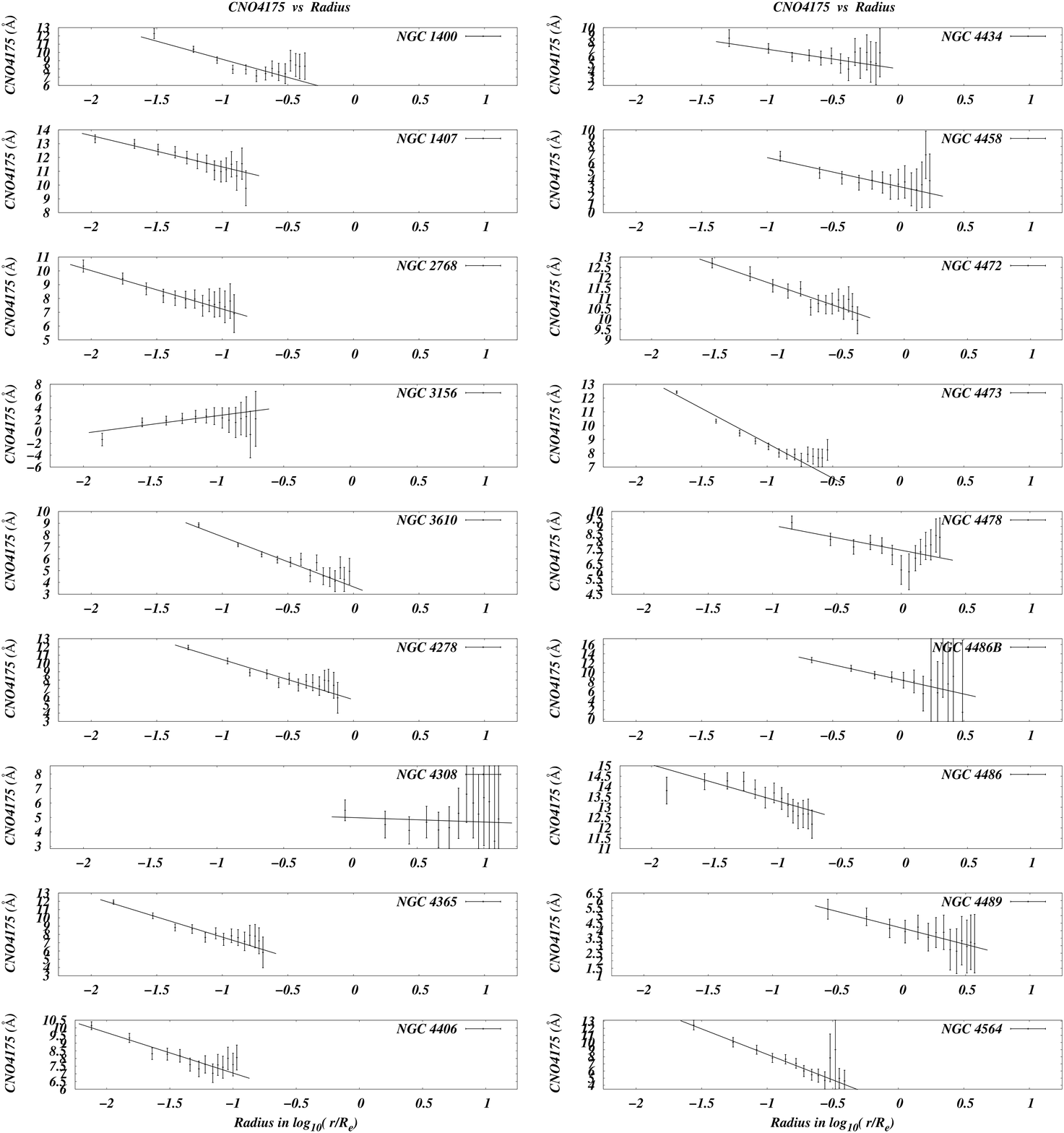}
\caption{}
\end{figure}

\begin{figure}[H]
\includegraphics[width=6in,height=7in]{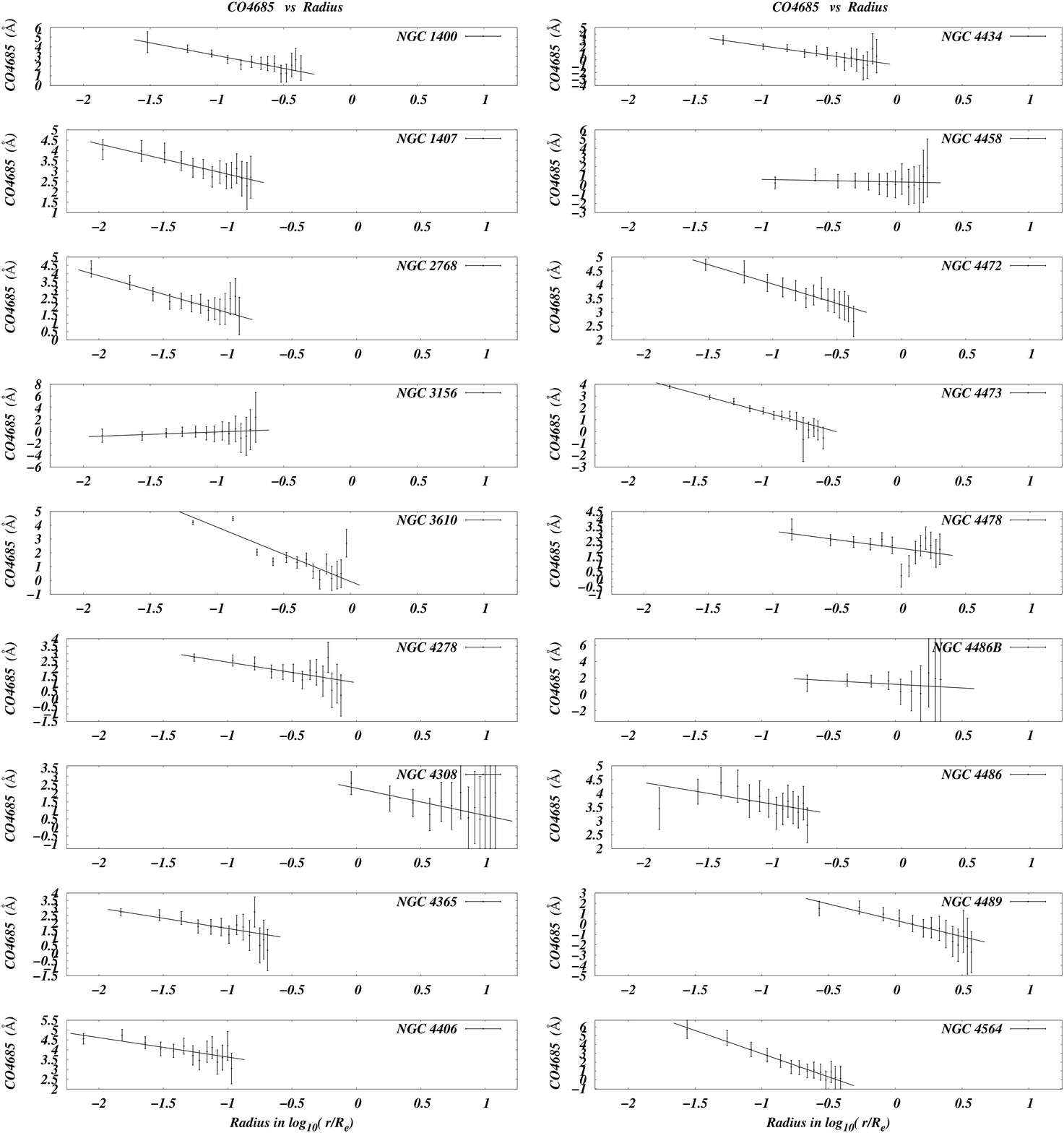}
\caption{}
\end{figure}

\begin{figure}[H]
\includegraphics[width=6in,height=7in]{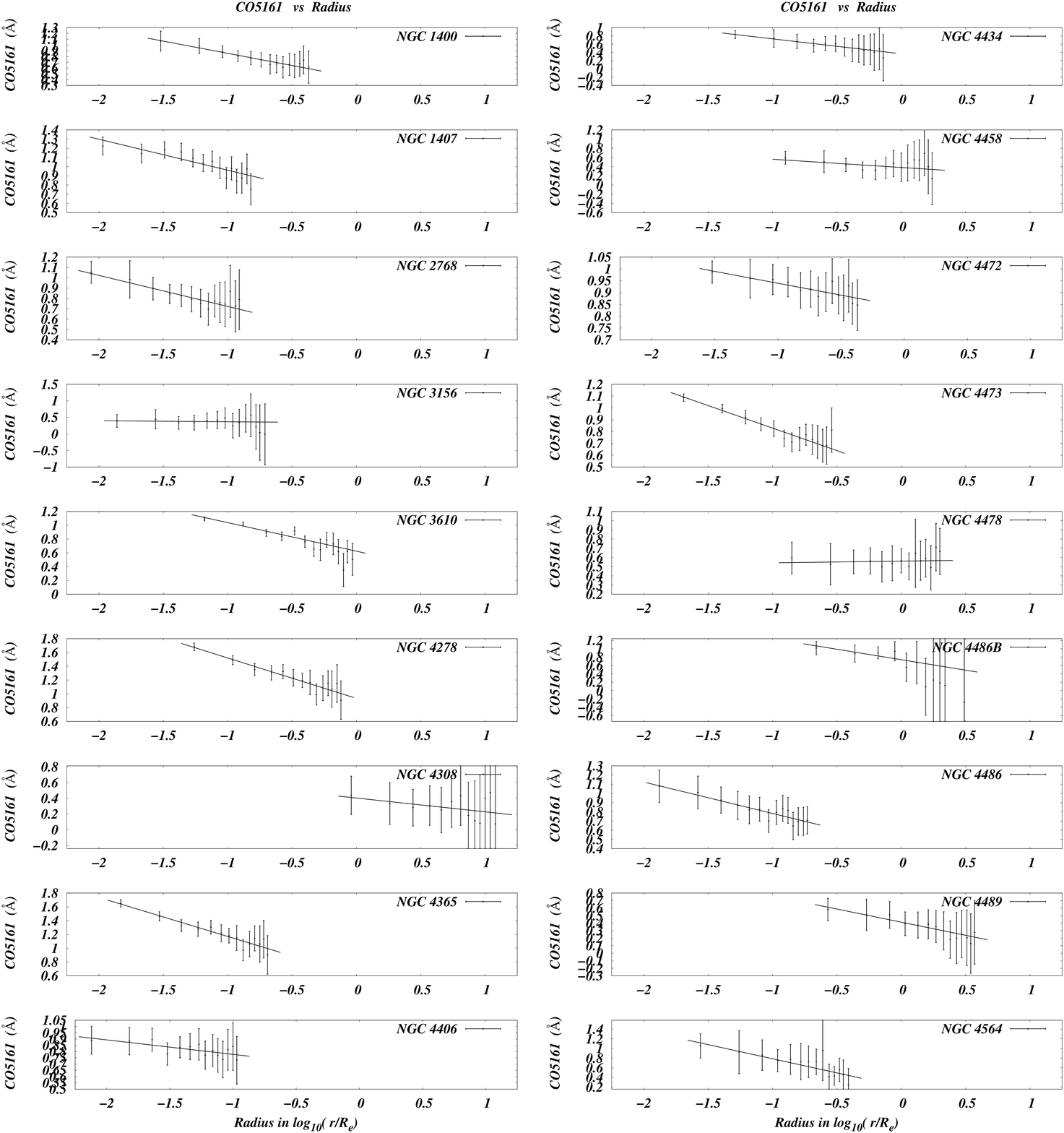}
\caption{}
\end{figure}

\begin{figure}[H]
\includegraphics[width=6in,height=7in]{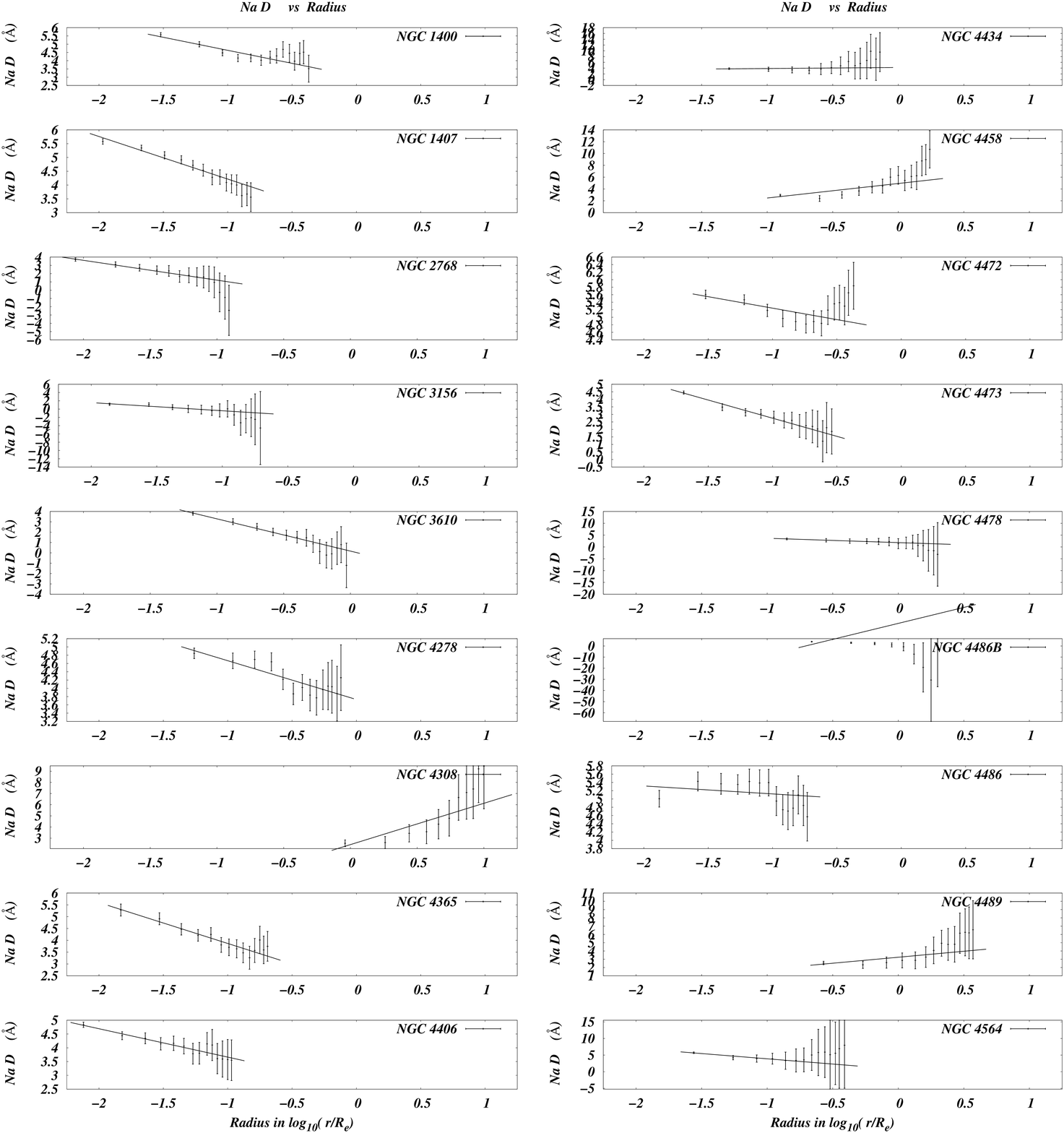}
\caption{}
\end{figure}

\begin{figure}[H]
\includegraphics[width=6in,height=7in]{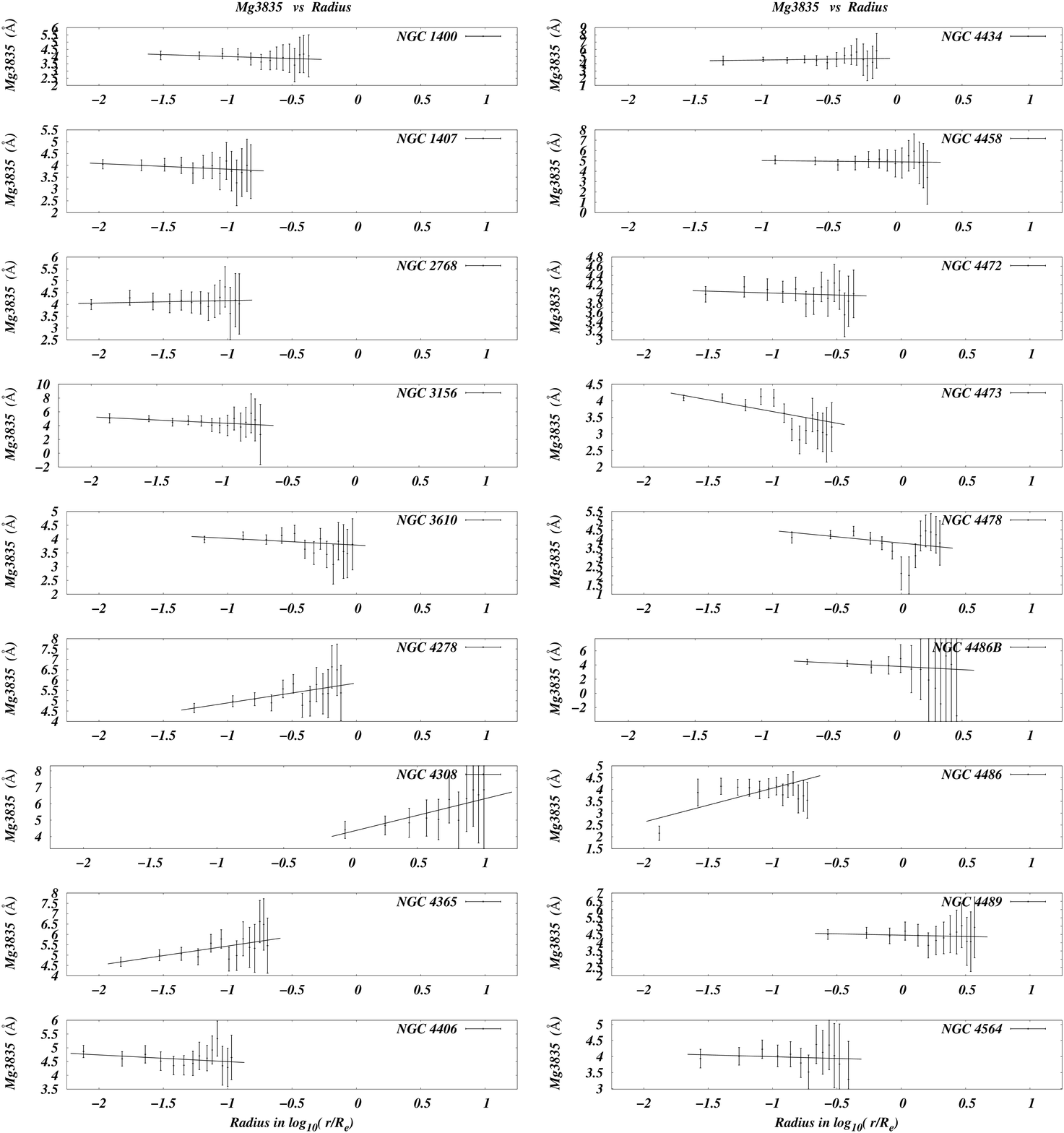}
\caption{}
\end{figure}

\begin{figure}[H]
\includegraphics[width=6in,height=7in]{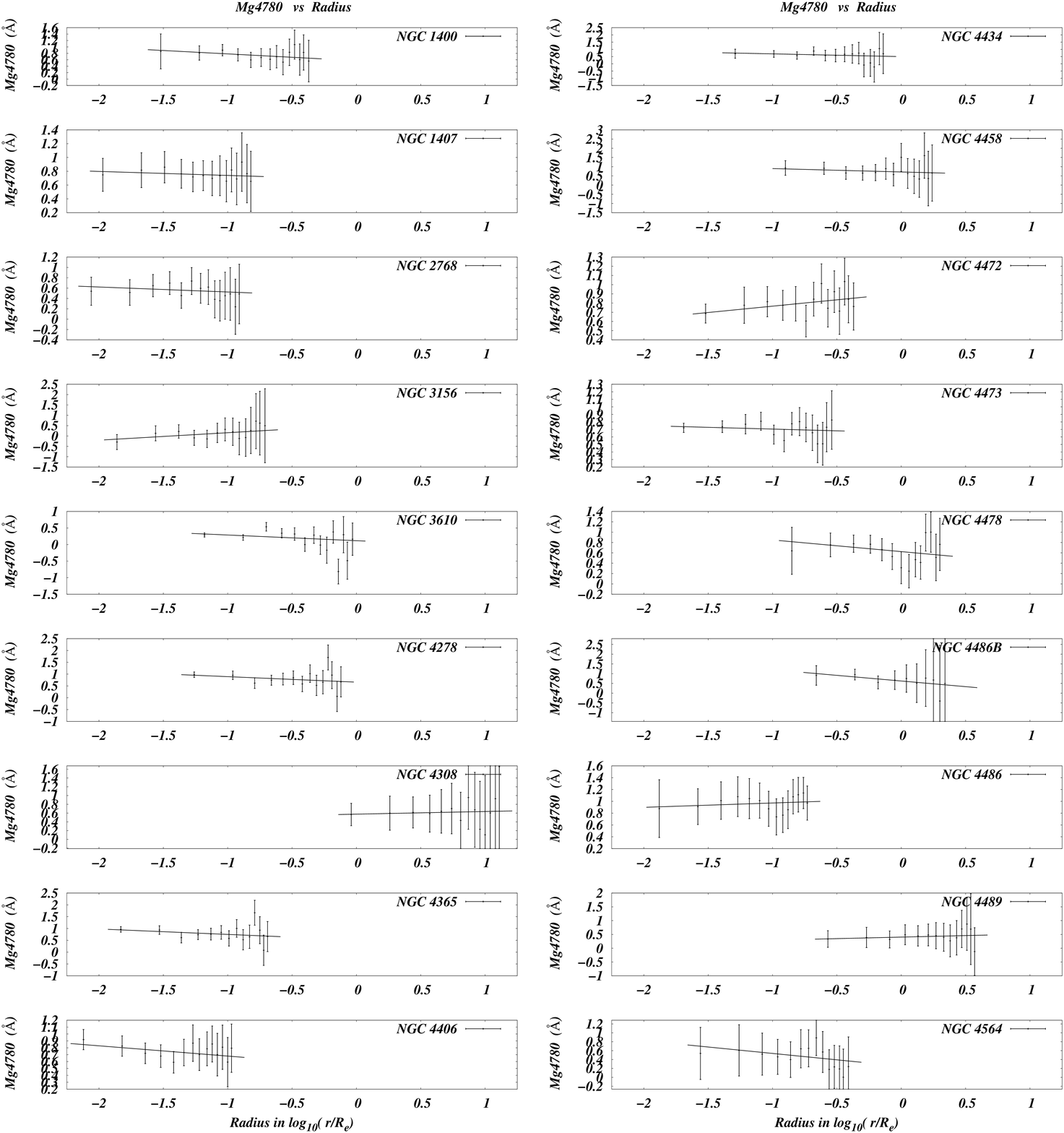}
\caption{}
\end{figure}

\begin{figure}[H]
\includegraphics[width=6in,height=7in]{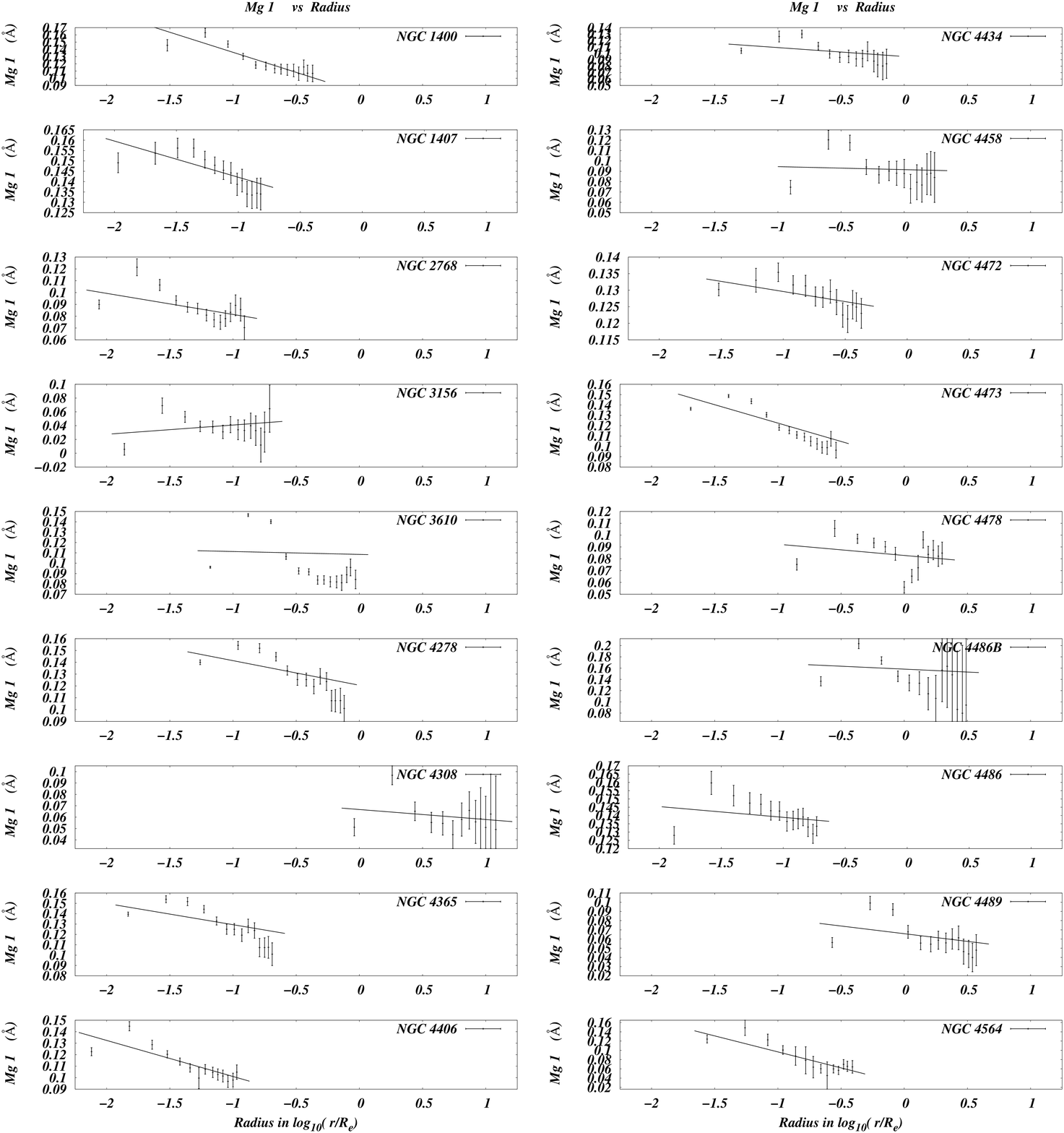}
\caption{}
\end{figure}

\begin{figure}[H]
\includegraphics[width=6in,height=7in]{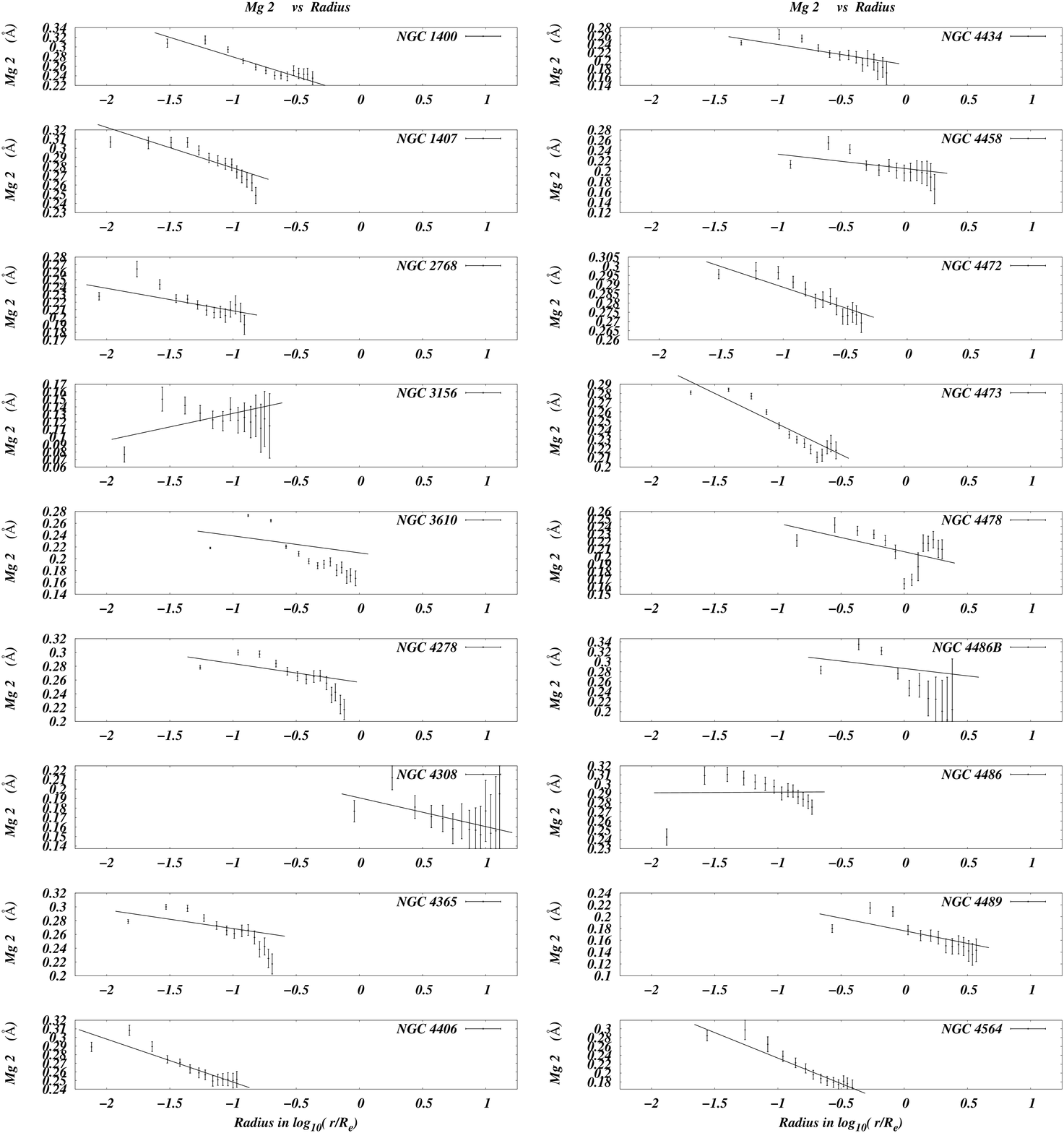}
\caption{}
\end{figure}

\begin{figure}[H]
\includegraphics[width=6in,height=7in]{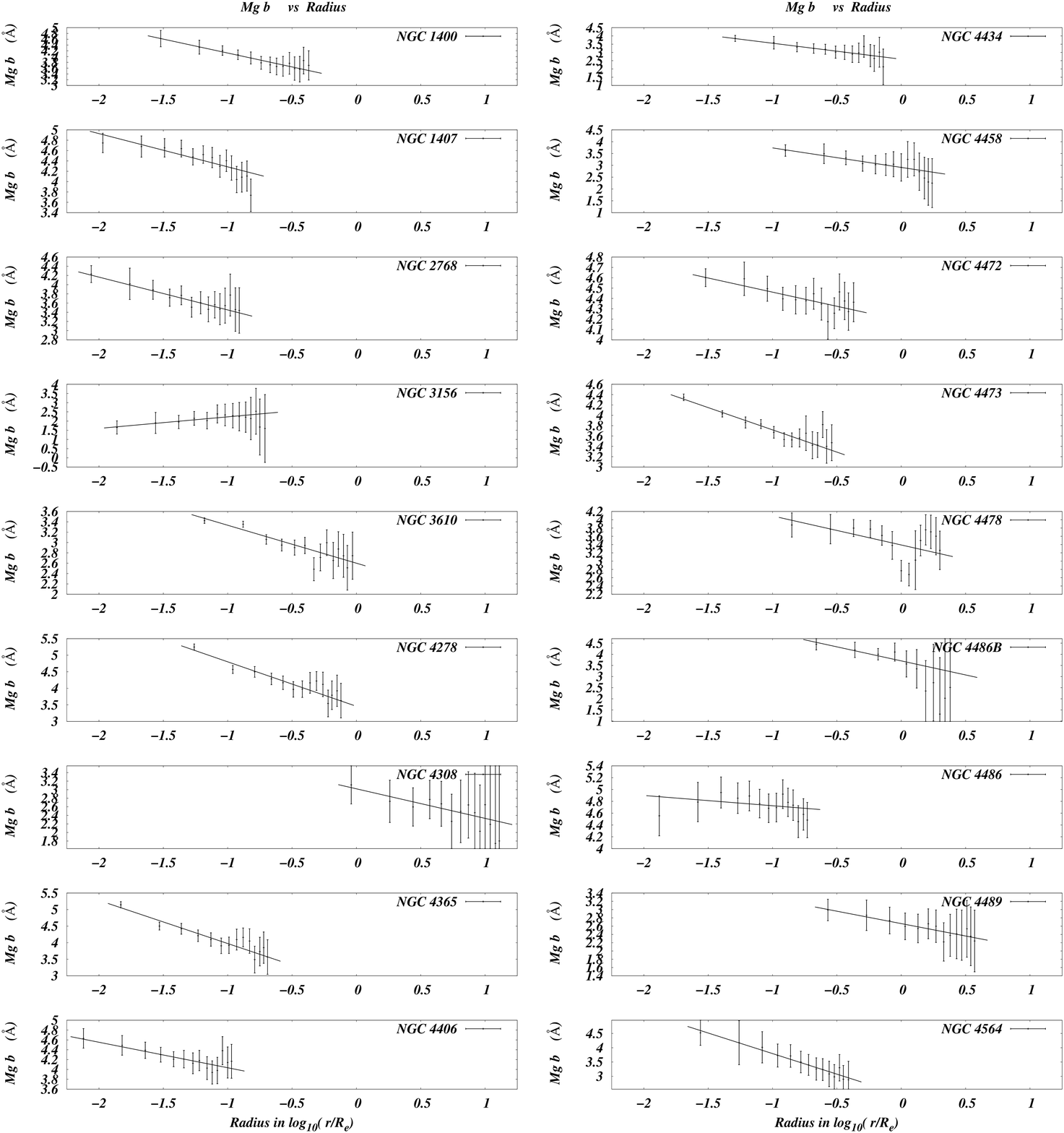}
\caption{}
\end{figure}

\begin{figure}[H]
\includegraphics[width=6in,height=7in]{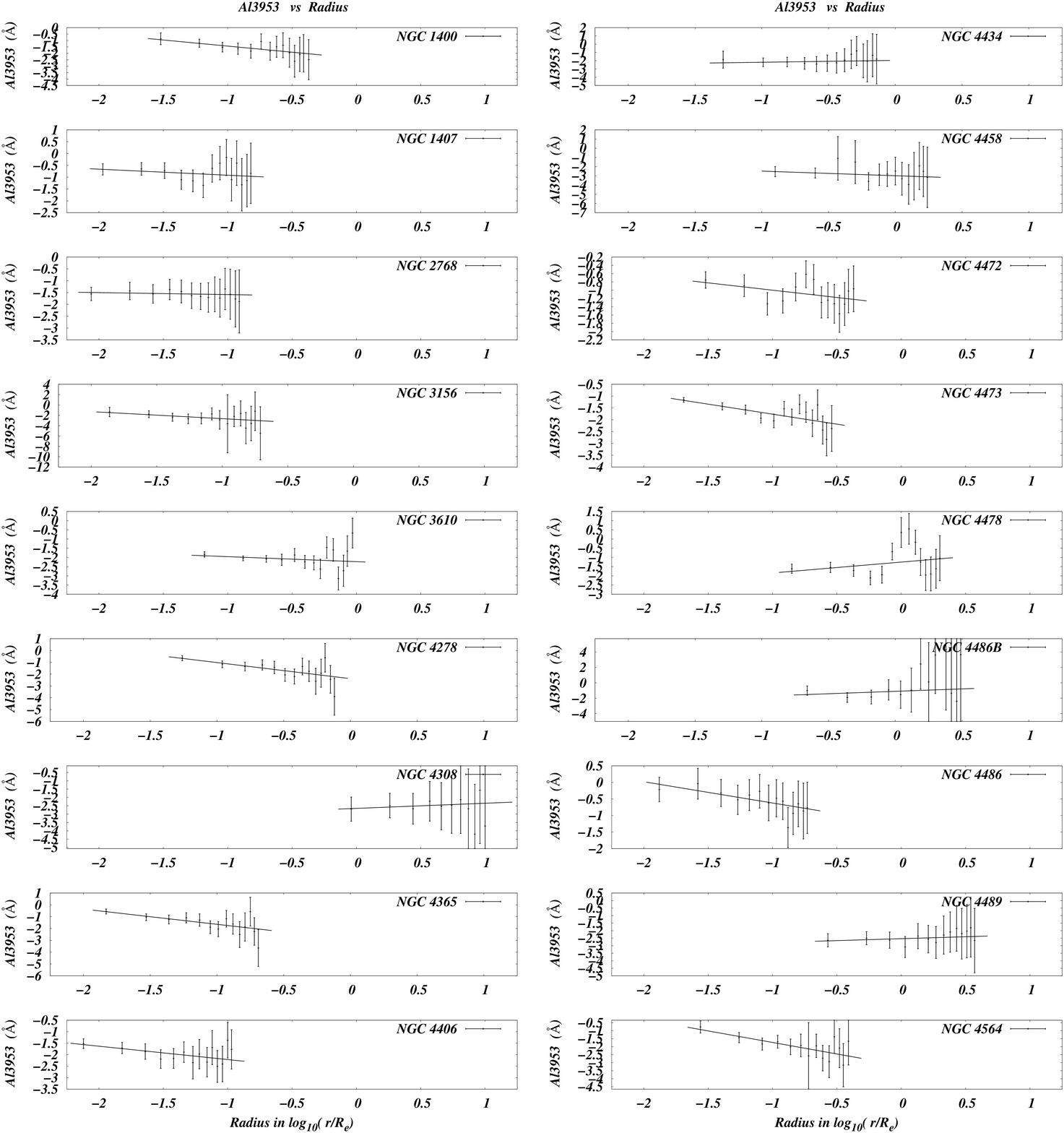}
\caption{}
\end{figure}

\begin{figure}[H]
\includegraphics[width=6in,height=7in]{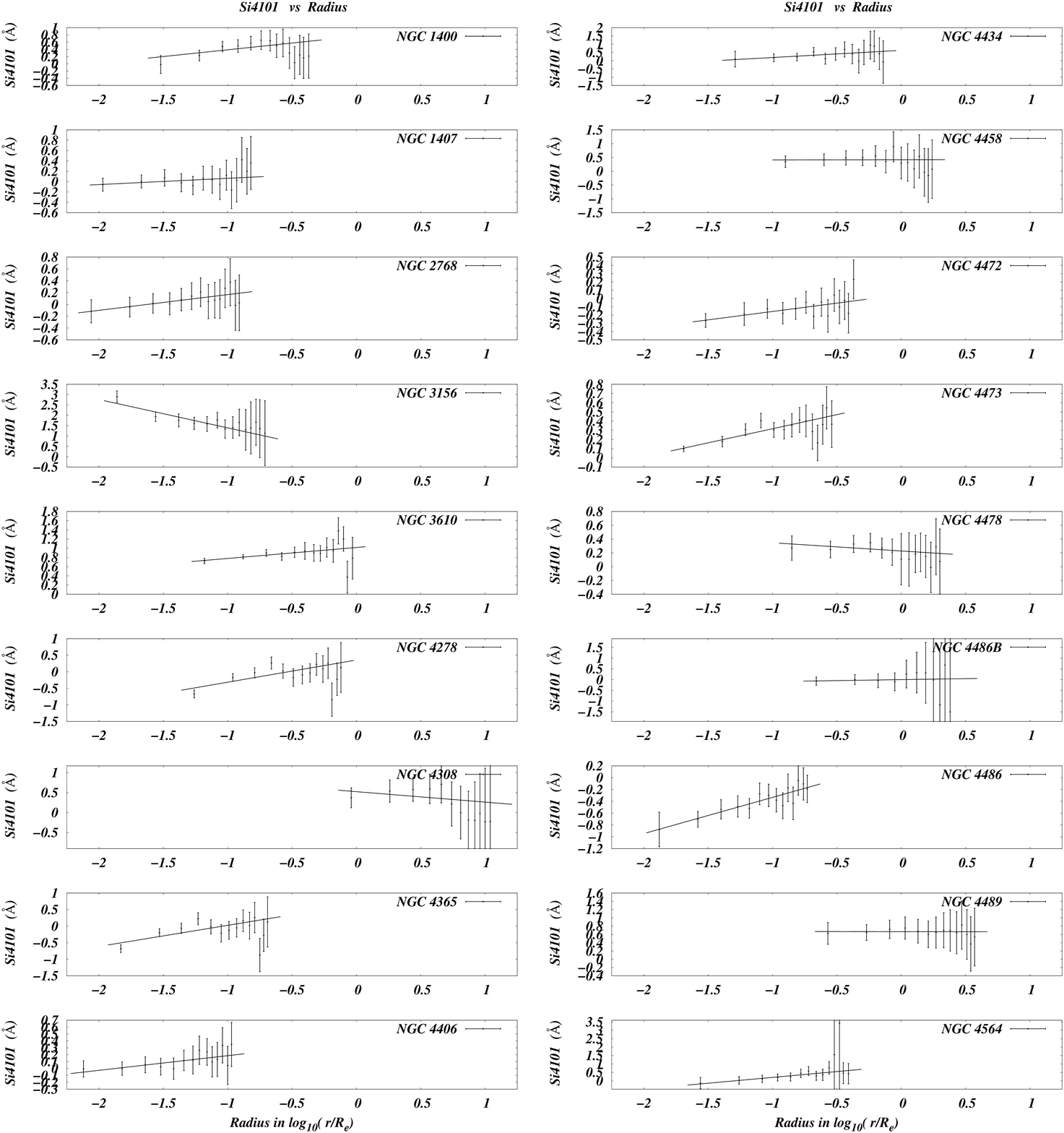}
\caption{}
\end{figure}

\begin{figure}[H]
\includegraphics[width=6in,height=7in]{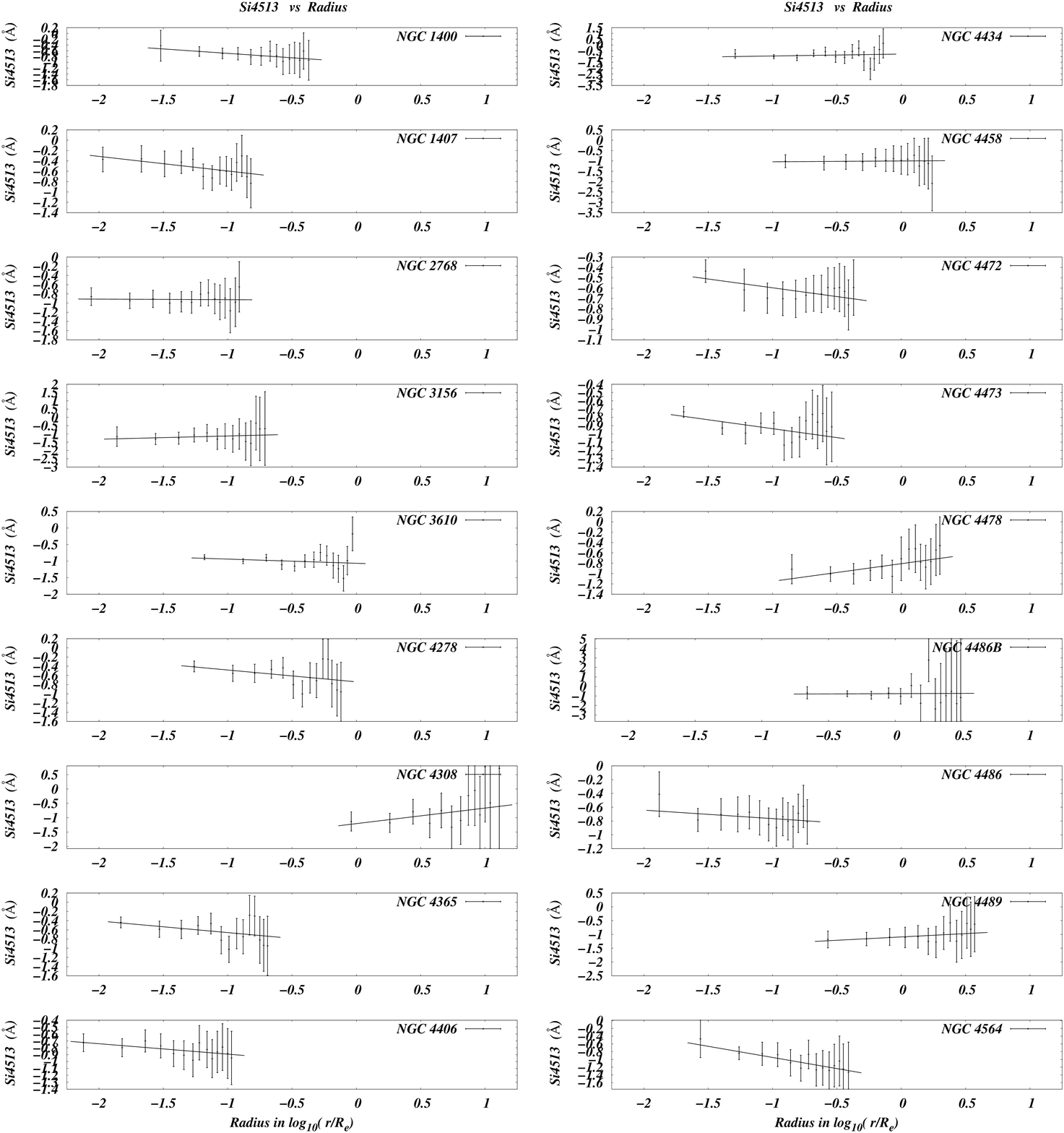}
\caption{}
\end{figure}

\begin{figure}[H]
\includegraphics[width=6in,height=7in]{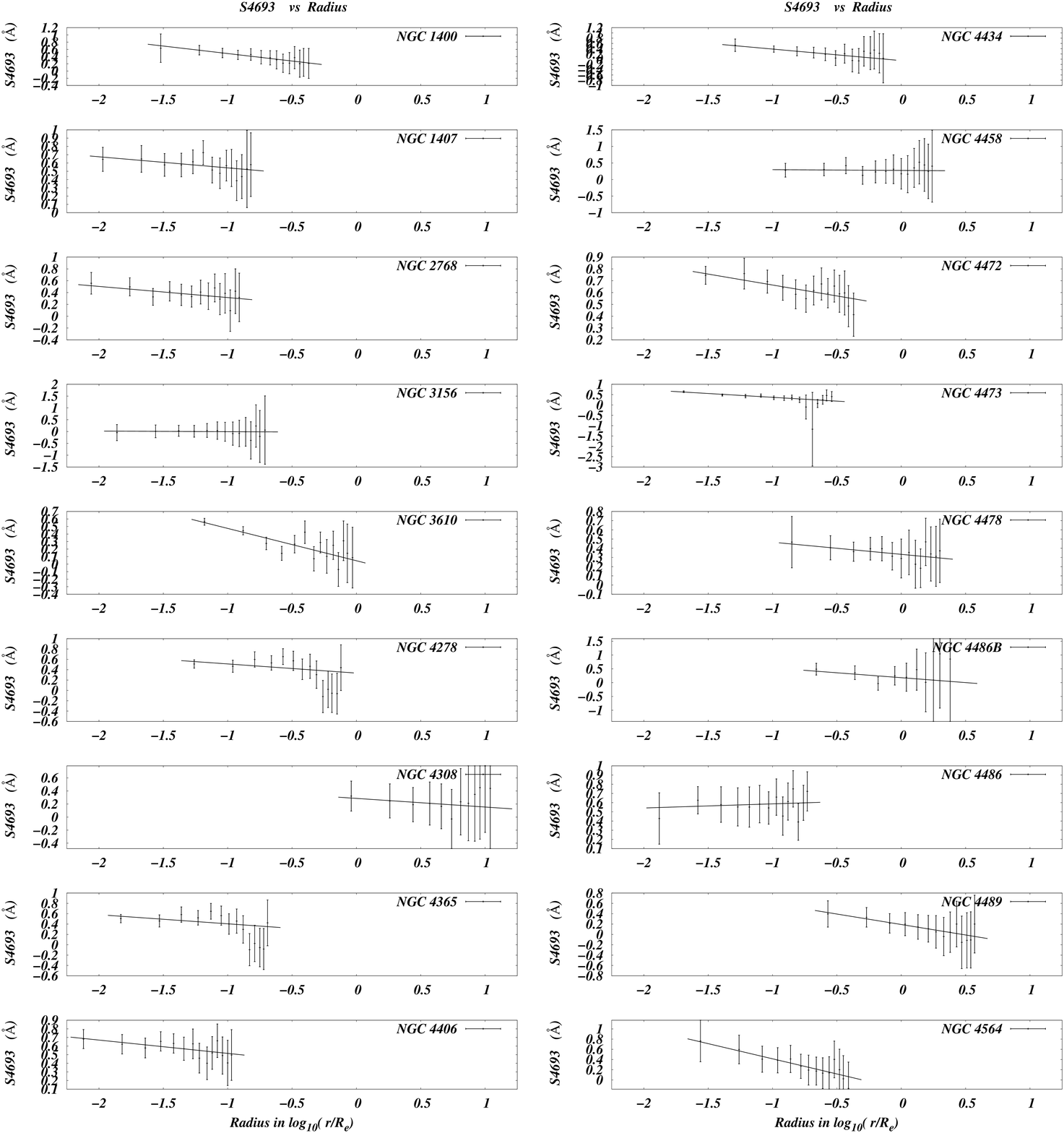}
\caption{}
\end{figure}

\begin{figure}[H]
\includegraphics[width=6in,height=7in]{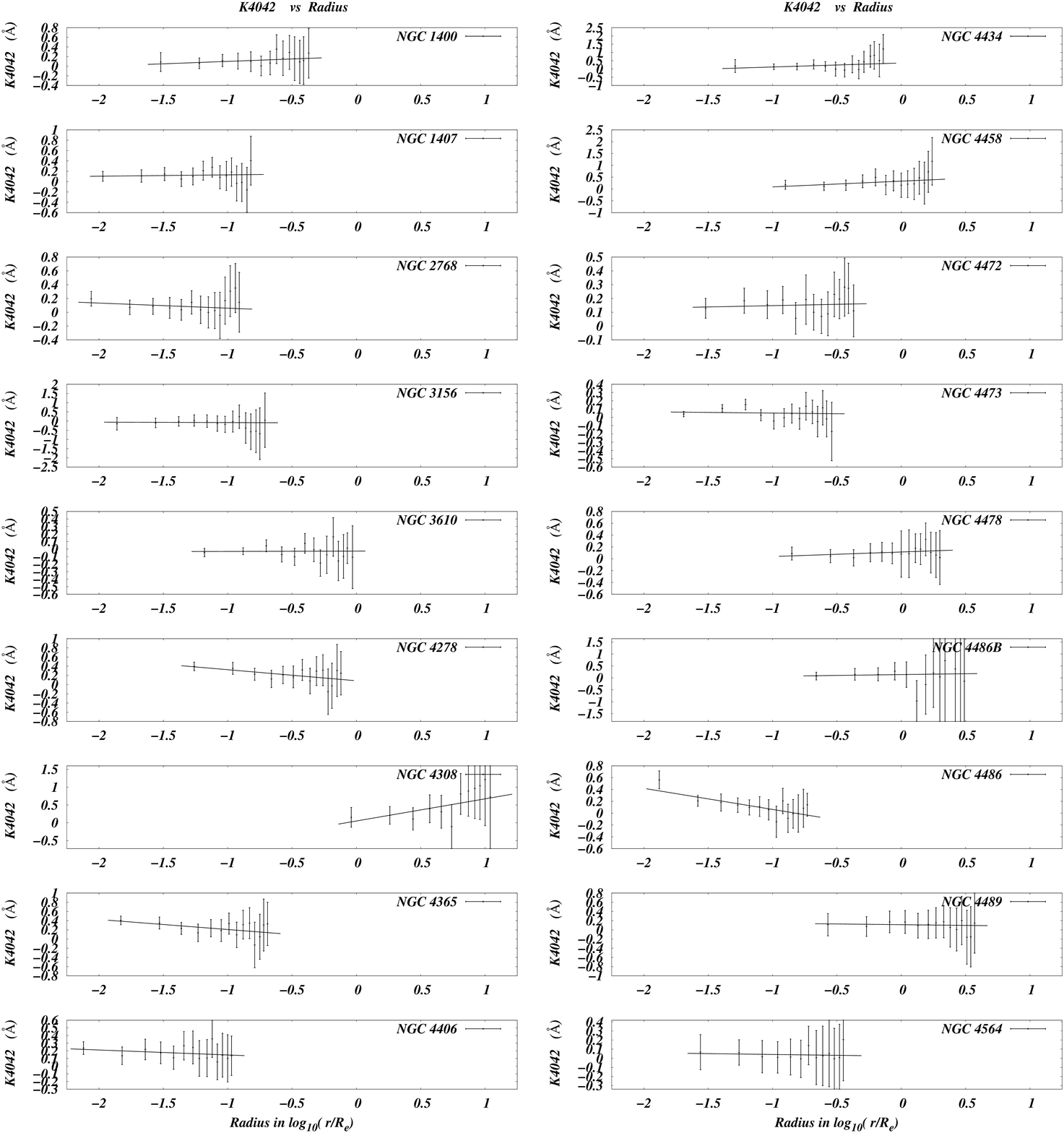}
\caption{}
\end{figure}

\begin{figure}[H]
\includegraphics[width=6in,height=7in]{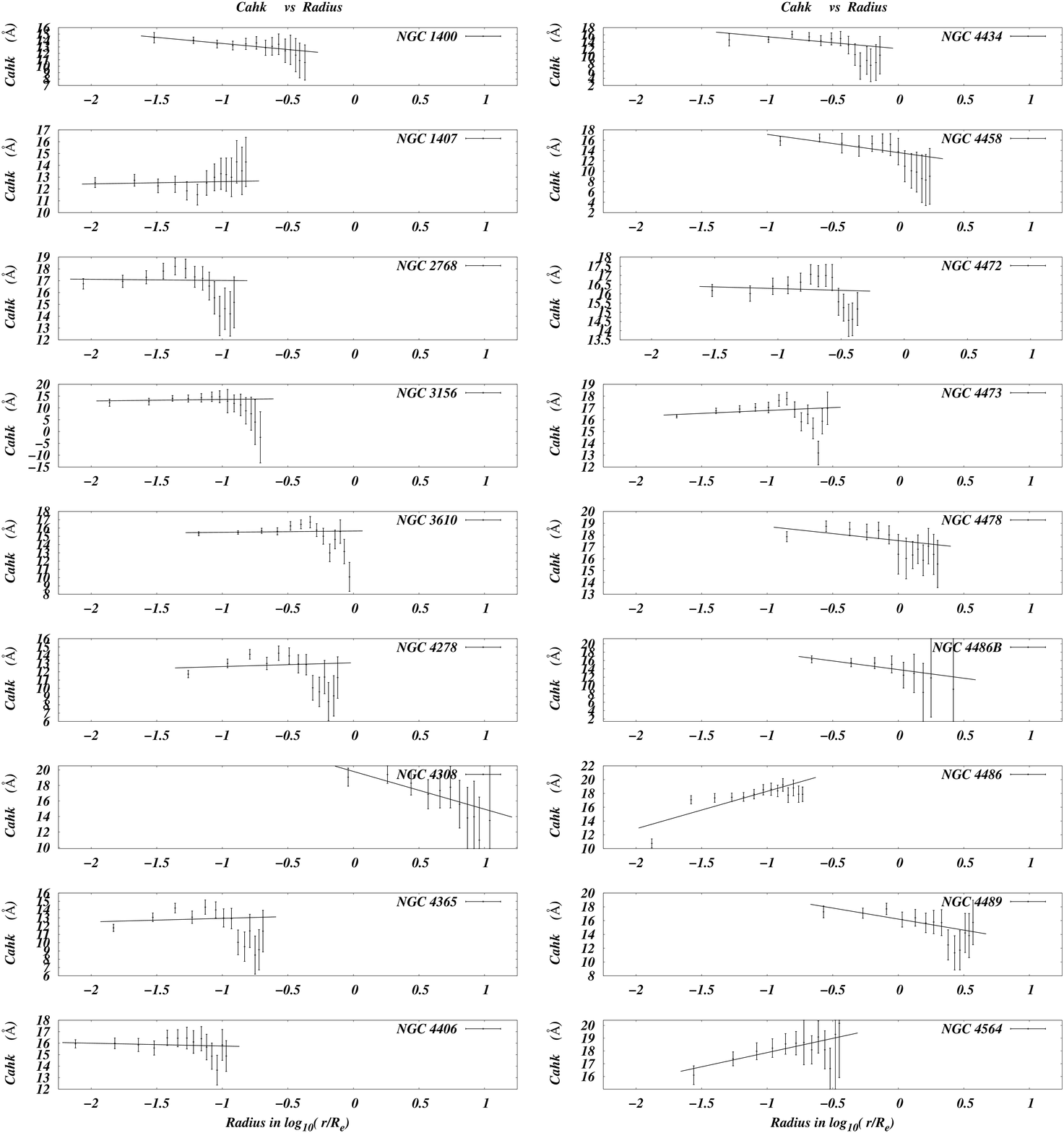}
\caption{}
\end{figure}

\begin{figure}[H]
\includegraphics[width=6in,height=7in]{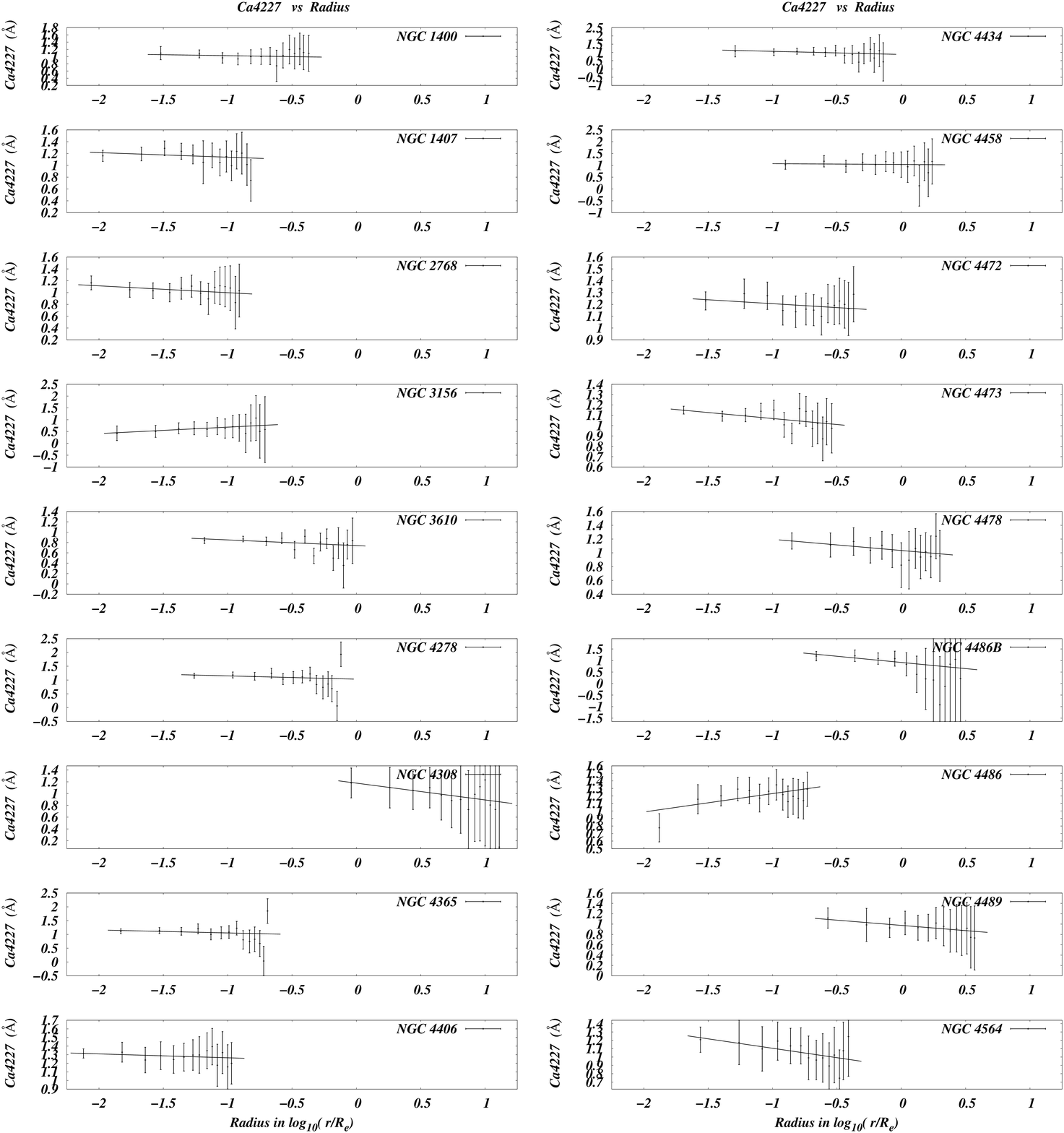}
\caption{}
\end{figure}

\begin{figure}[H]
\includegraphics[width=6in,height=7in]{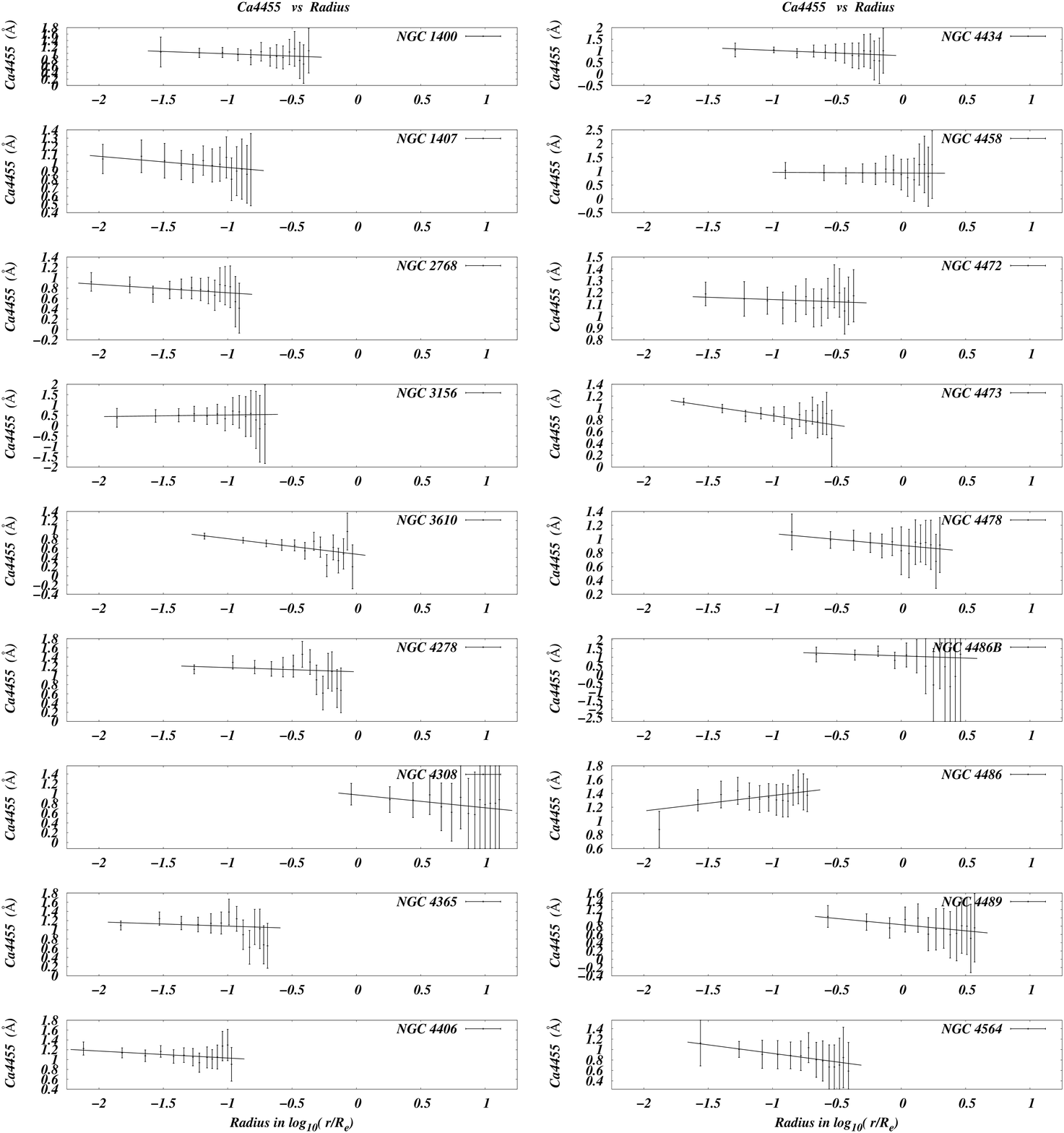}
\caption{}
\end{figure}

\begin{figure}[H]
\includegraphics[width=6in,height=7in]{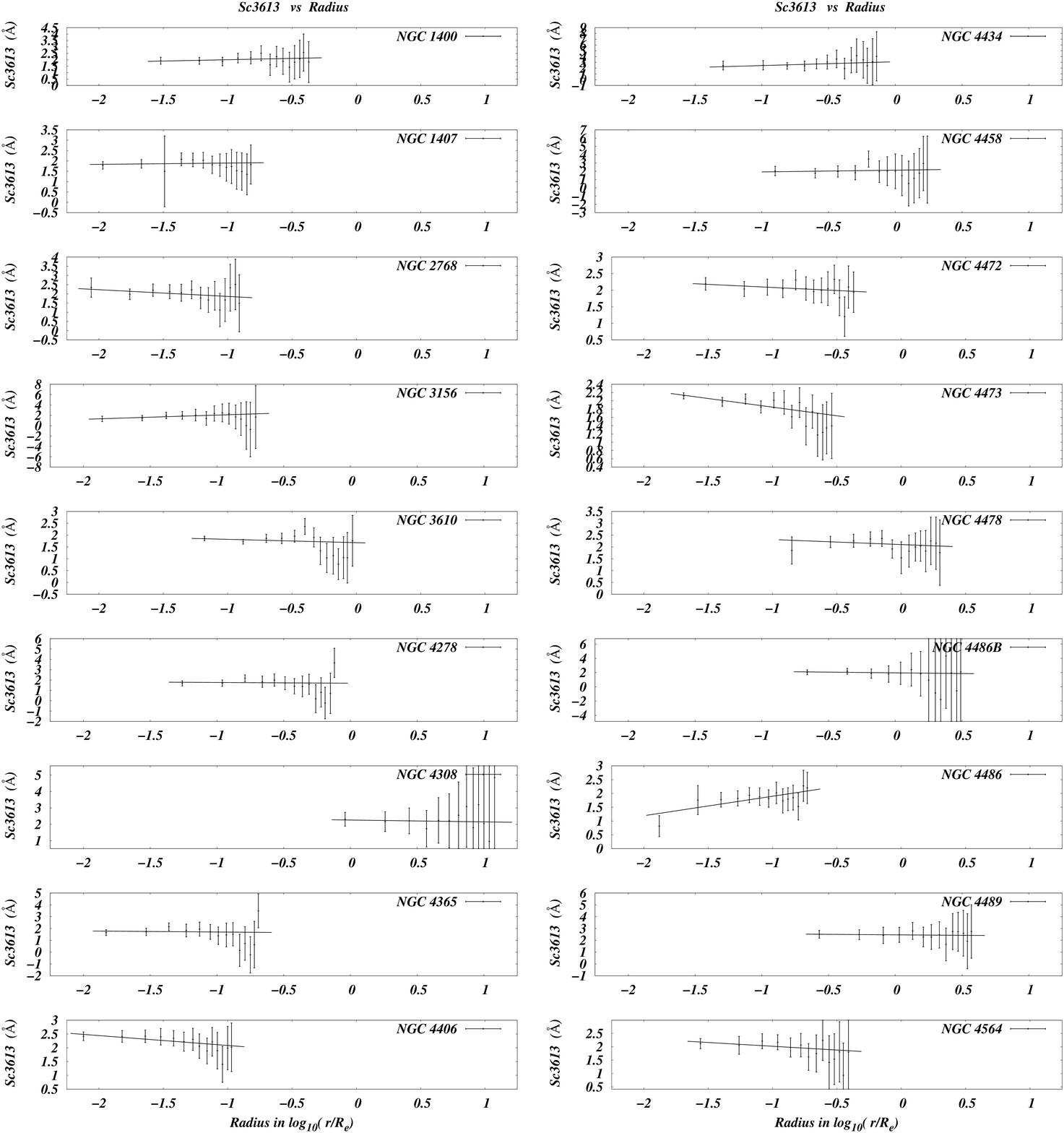}
\caption{}
\end{figure}

\begin{figure}[H]
\includegraphics[width=6in,height=7in]{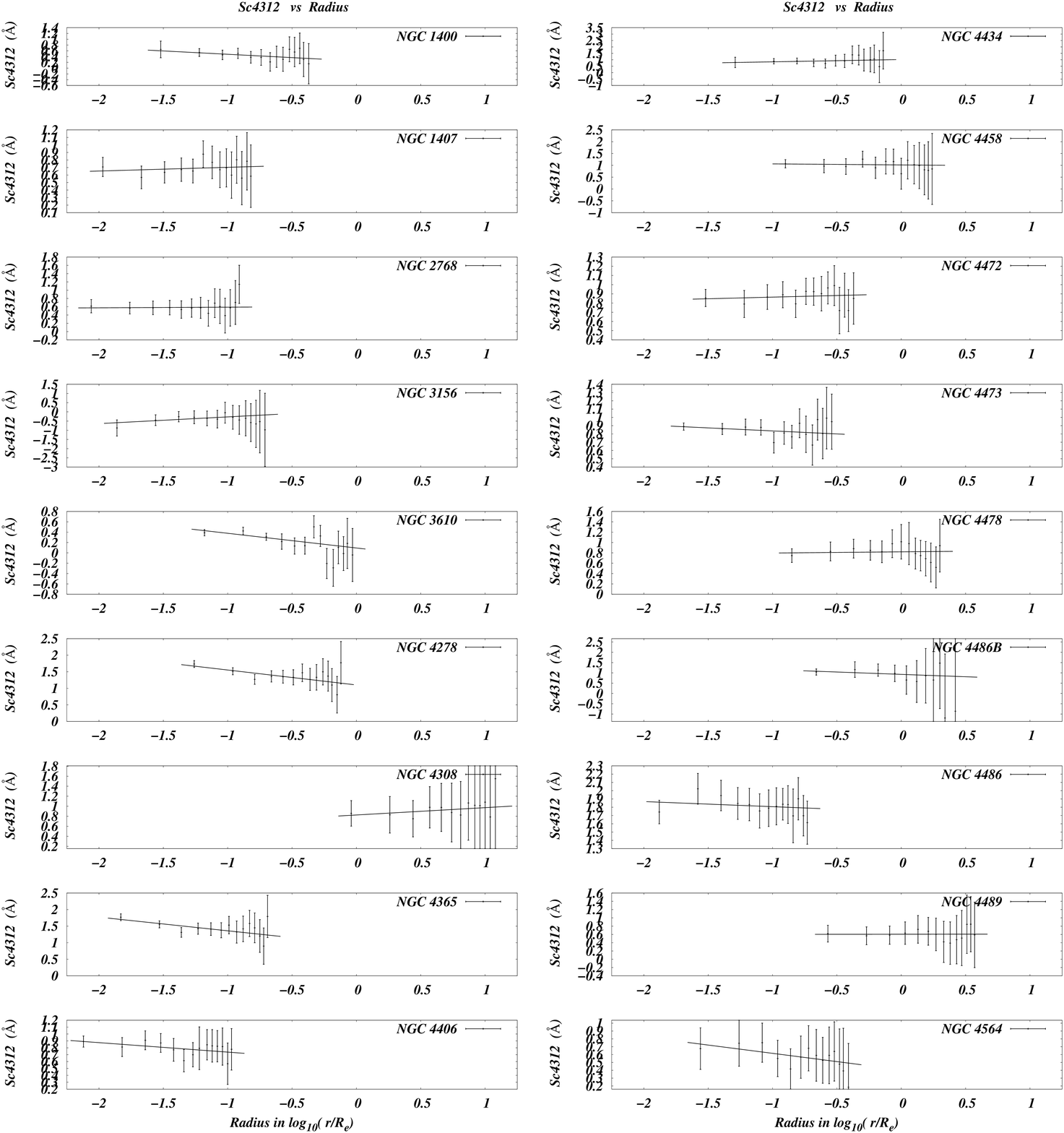}
\caption{}
\end{figure}

\begin{figure}[H]
\includegraphics[width=6in,height=7in]{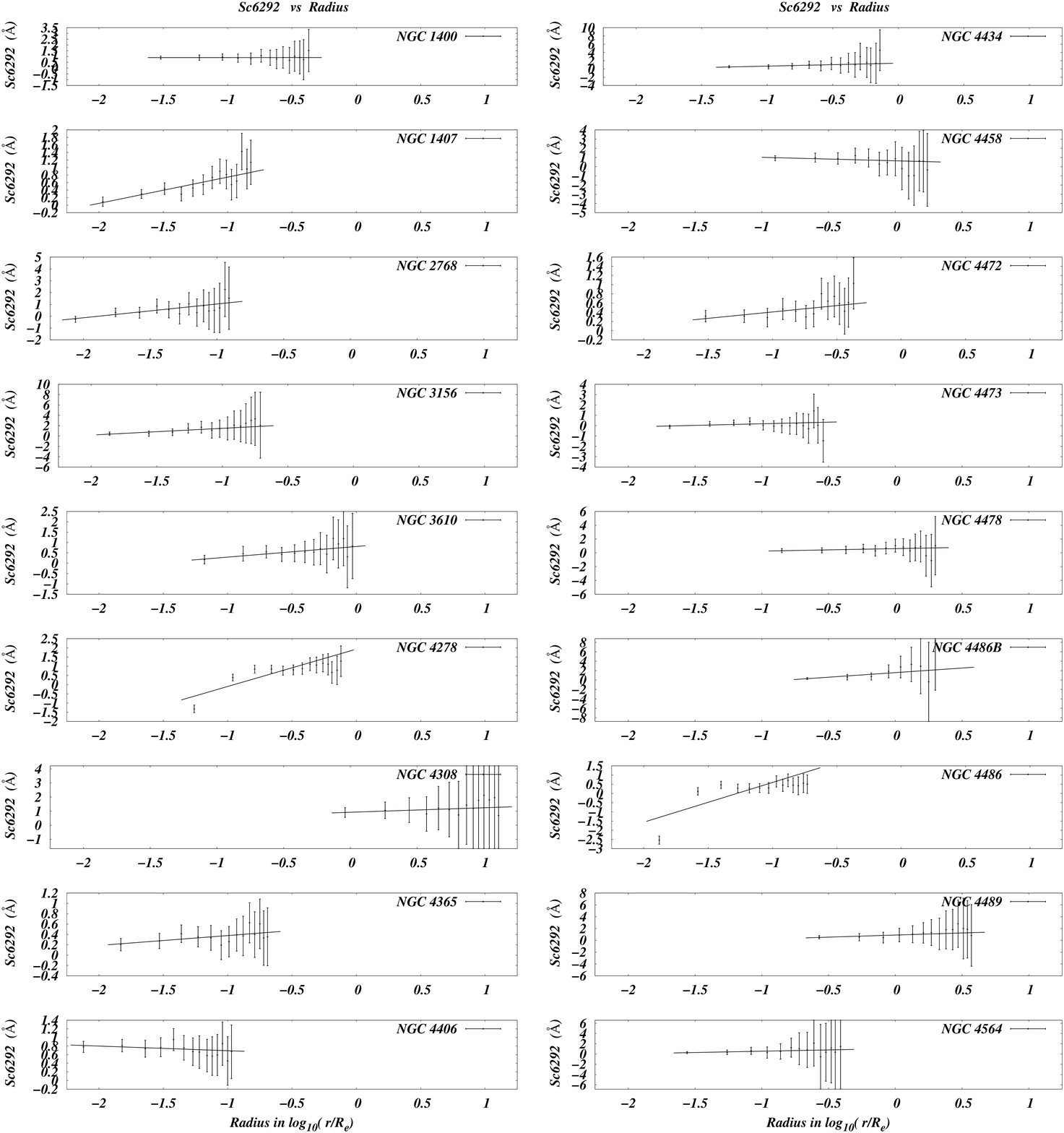}
\caption{}
\end{figure}

\begin{figure}[H]
\includegraphics[width=6in,height=7in]{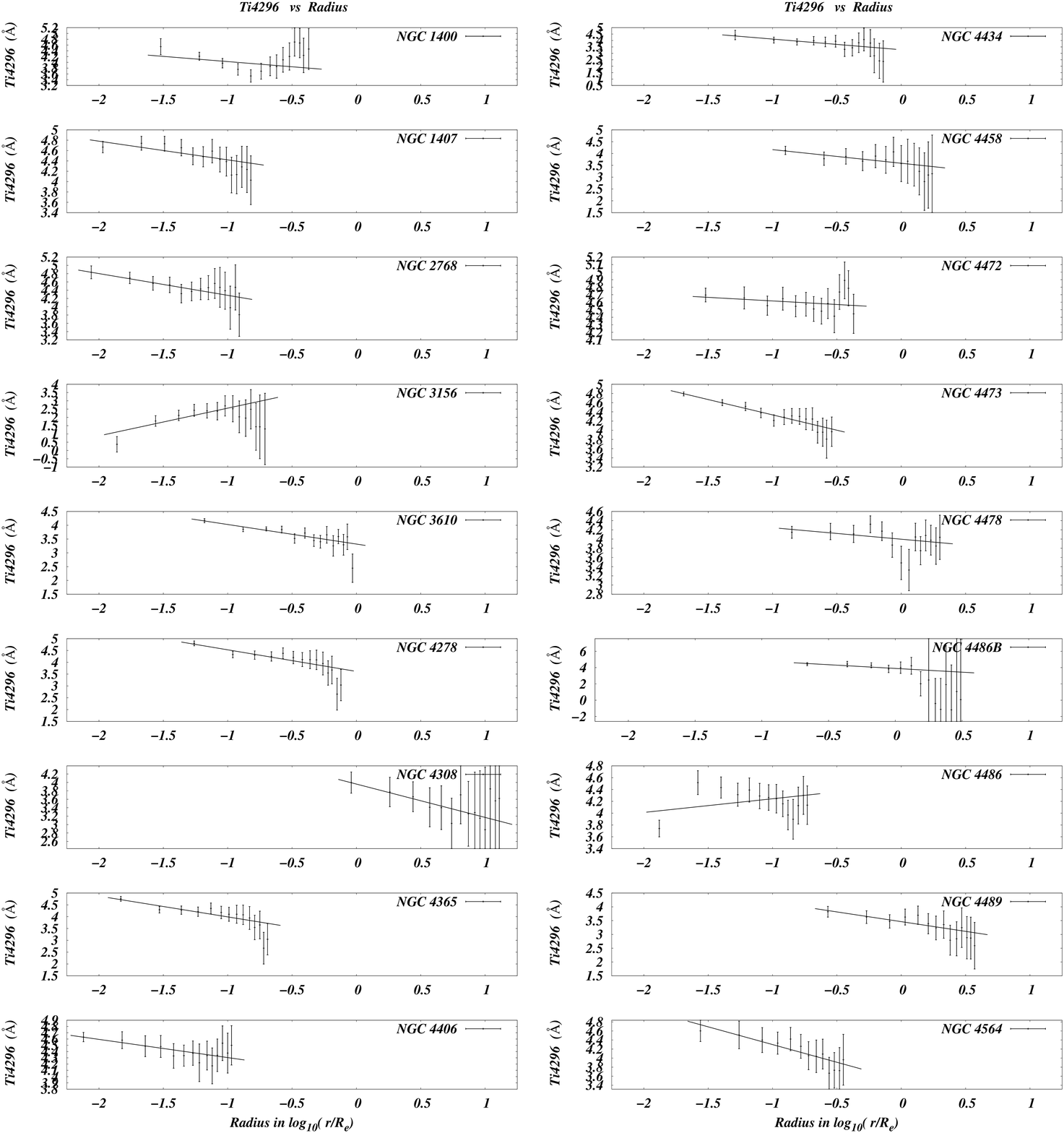}
\caption{}
\end{figure}

\begin{figure}[H]
\includegraphics[width=6in,height=7in]{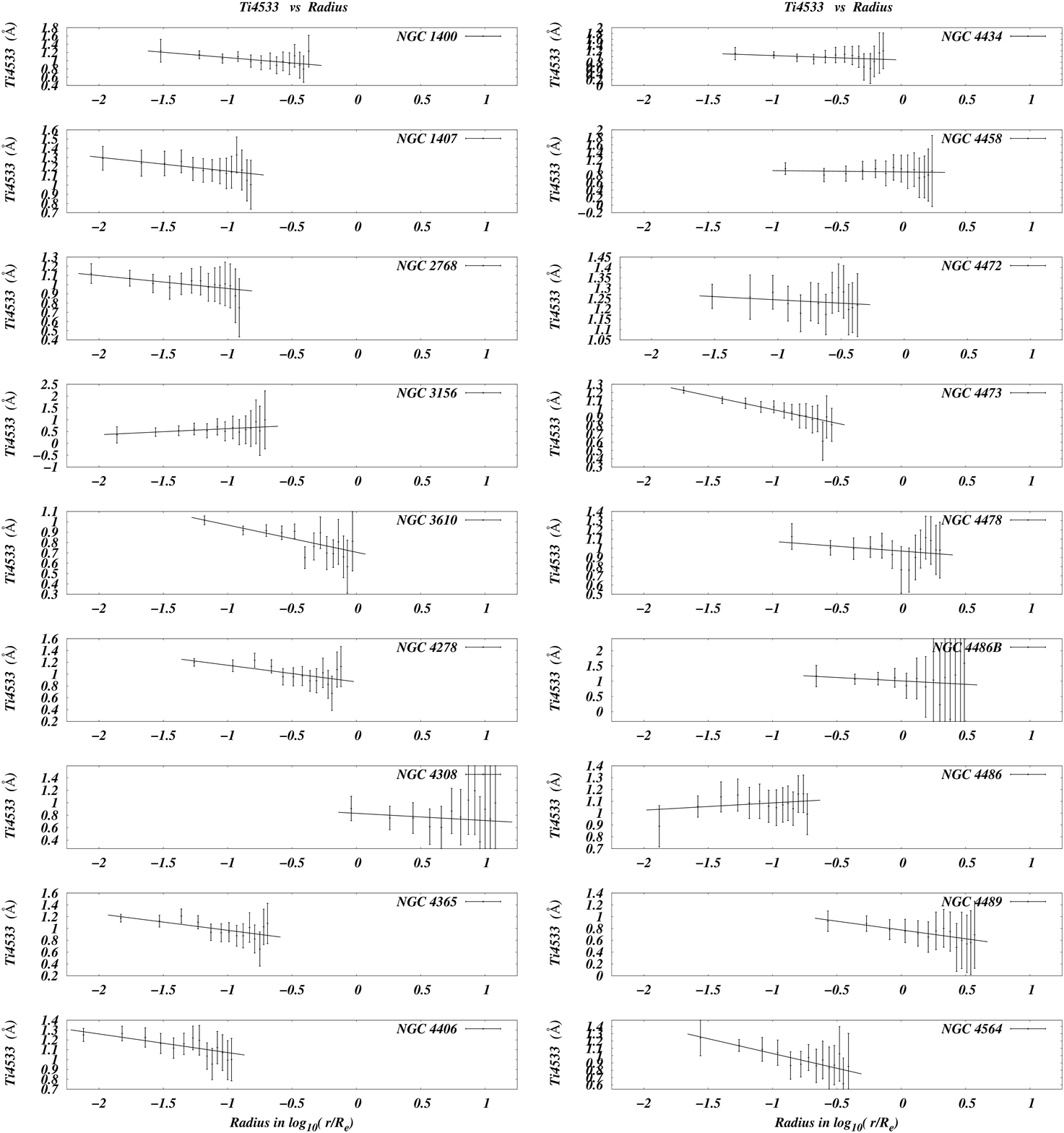}
\caption{}
\end{figure}

\begin{figure}[H]
\includegraphics[width=6in,height=7in]{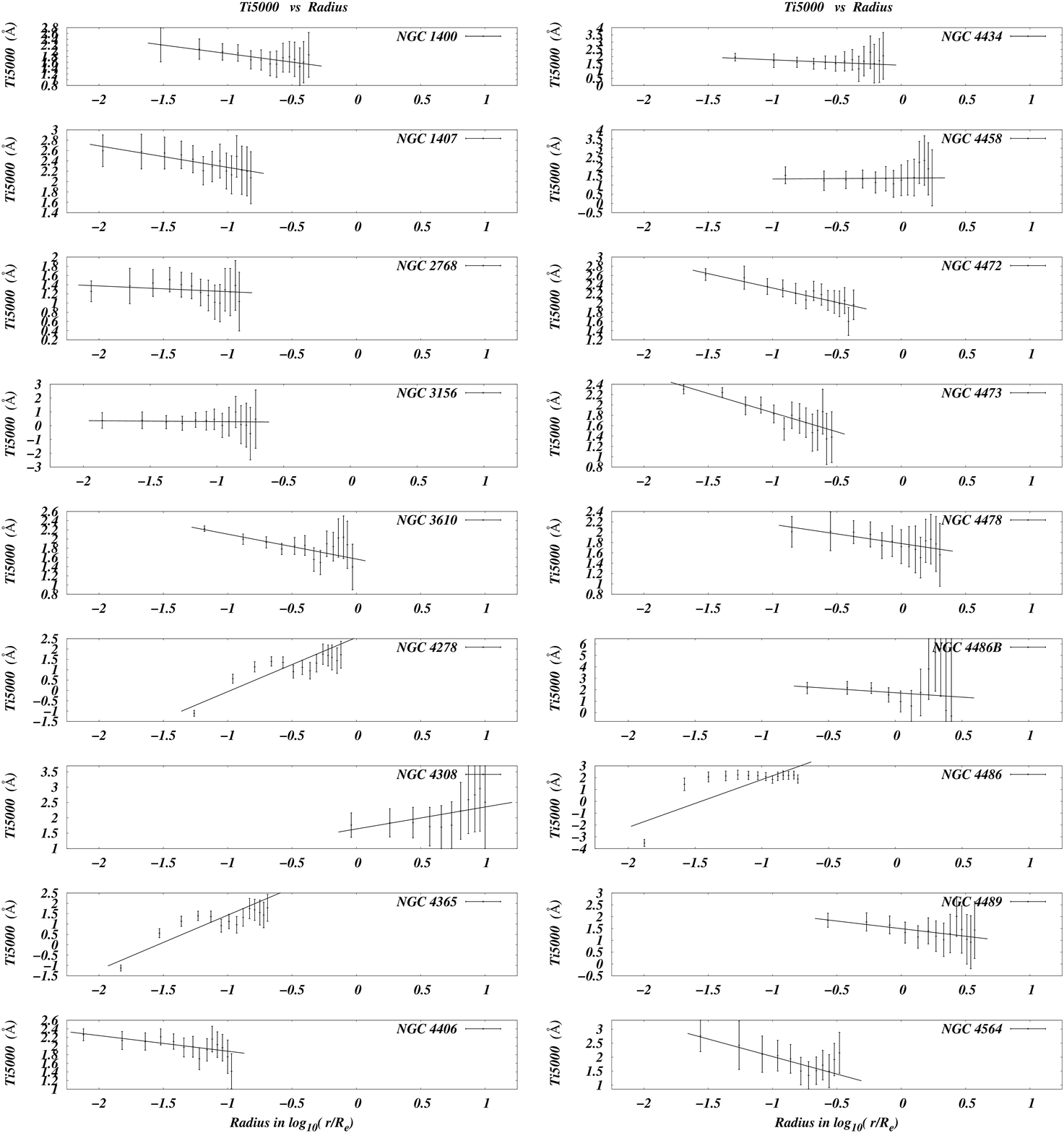}
\caption{}
\end{figure}

\begin{figure}[H]
\includegraphics[width=6in,height=7in]{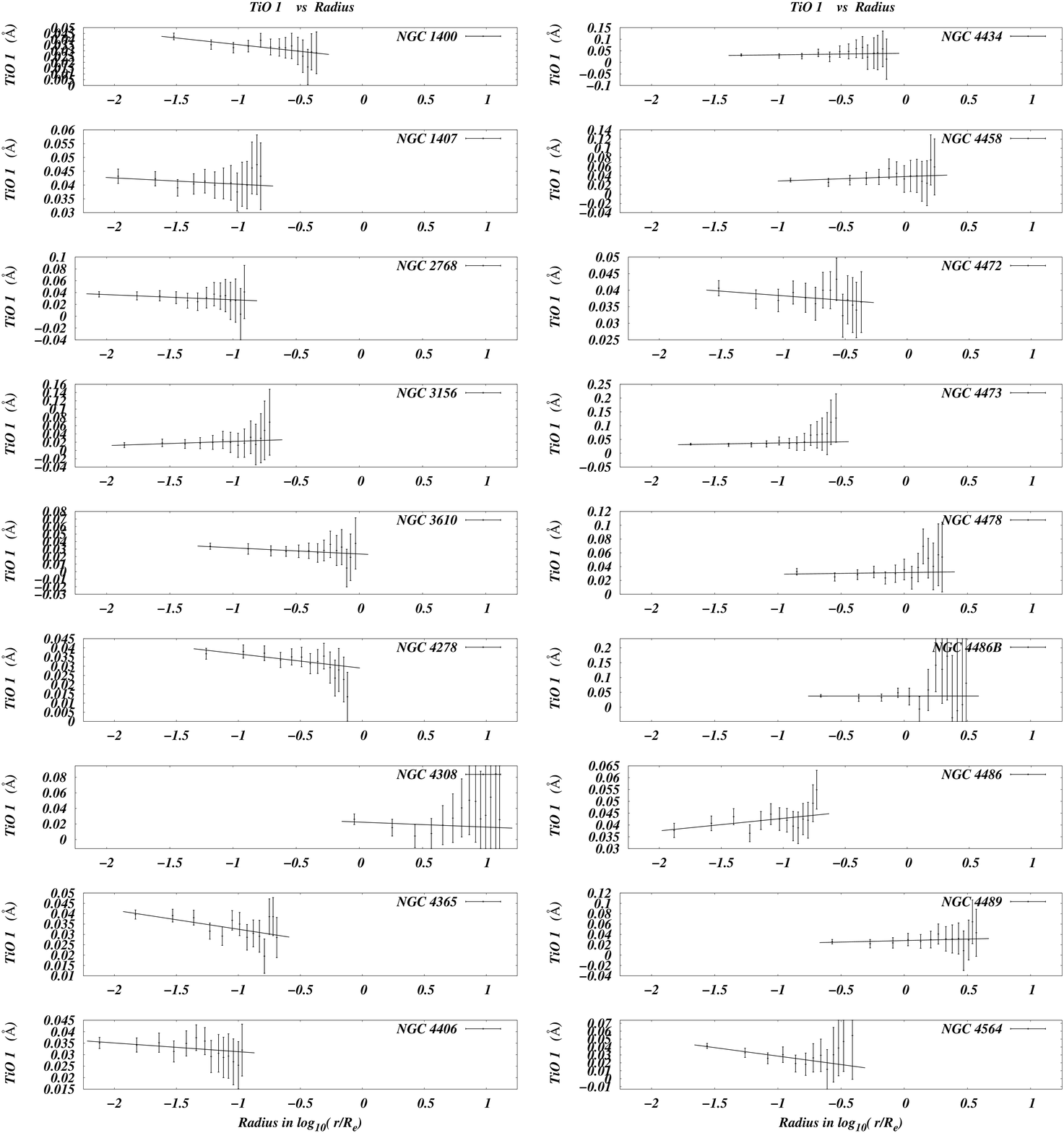}
\caption{}
\end{figure}

\begin{figure}[H]
\includegraphics[width=6in,height=7in]{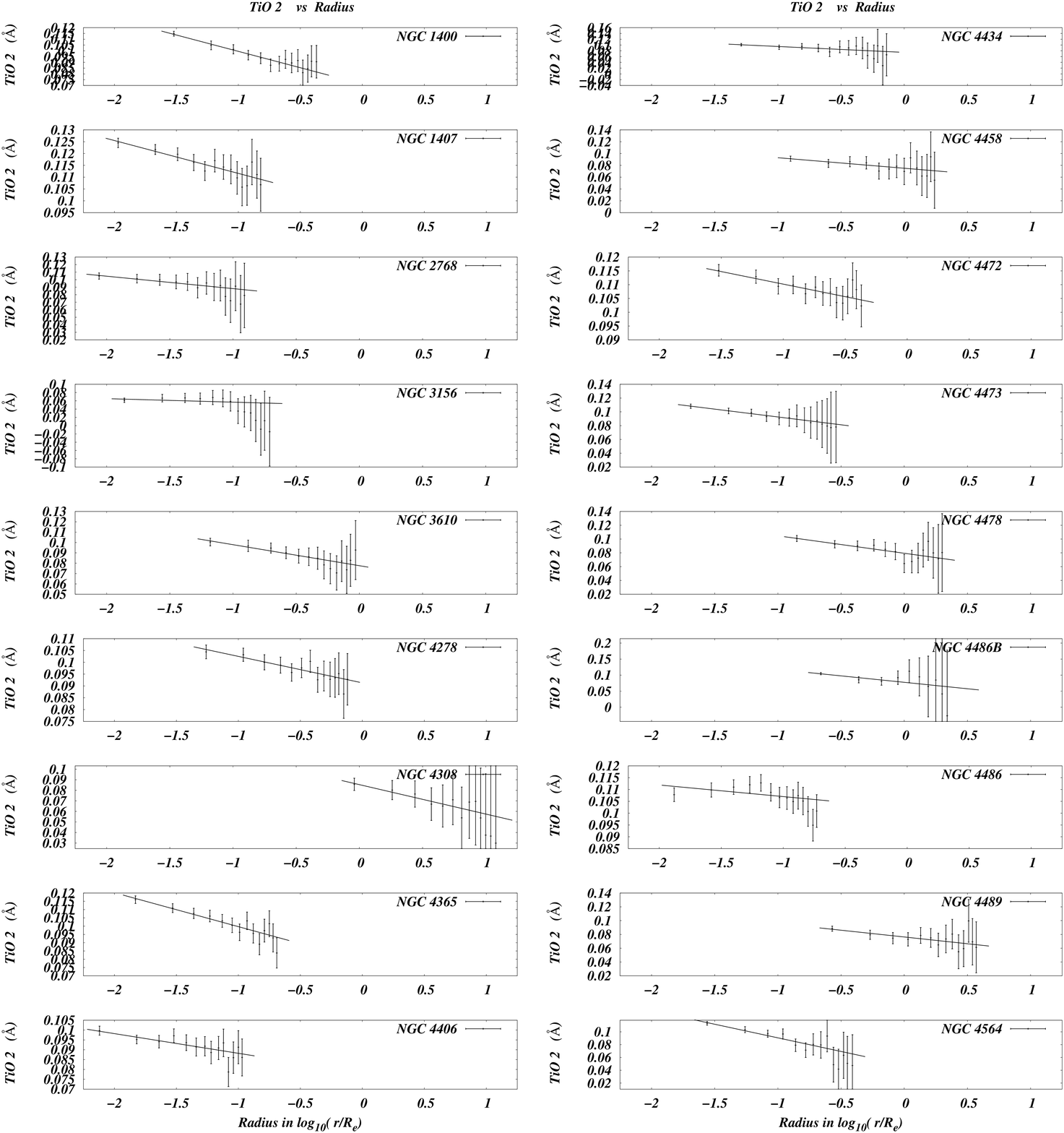}
\caption{}
\end{figure}

\begin{figure}[H]
\includegraphics[width=6in,height=7in]{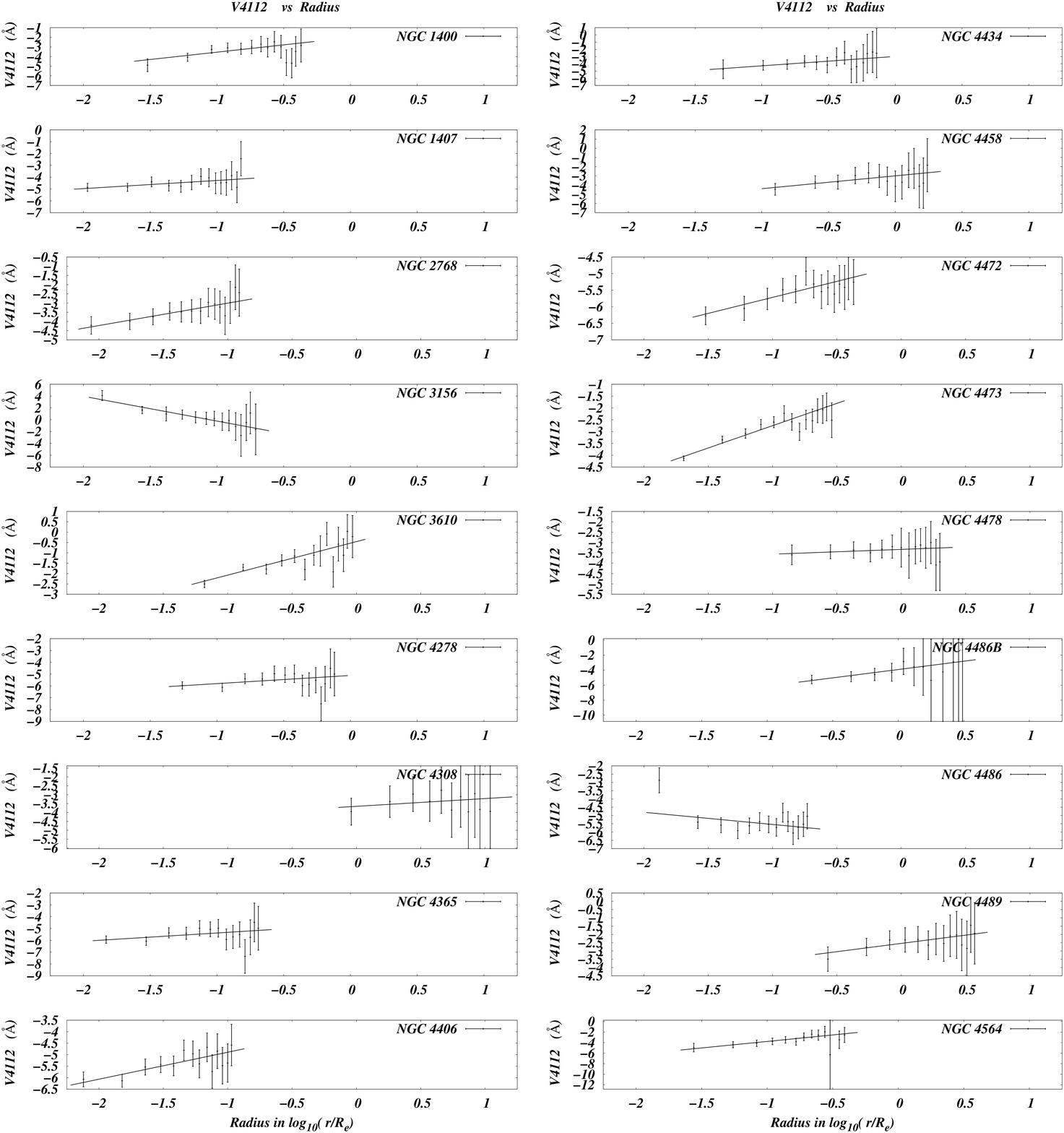}
\caption{}
\end{figure}

\begin{figure}[H]
\includegraphics[width=6in,height=7in]{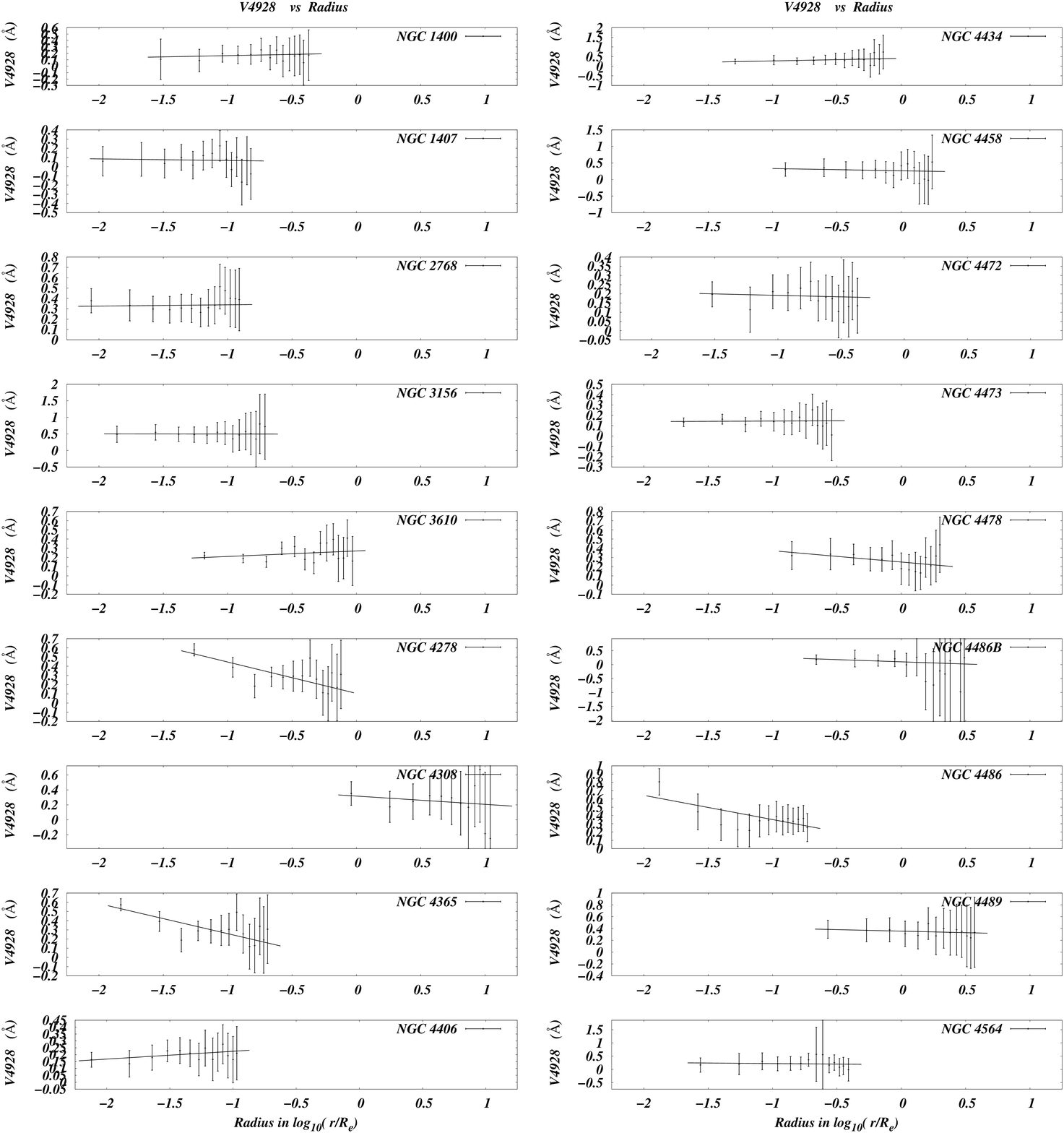}
\caption{}
\end{figure}

\begin{figure}[H]
\includegraphics[width=6in,height=7in]{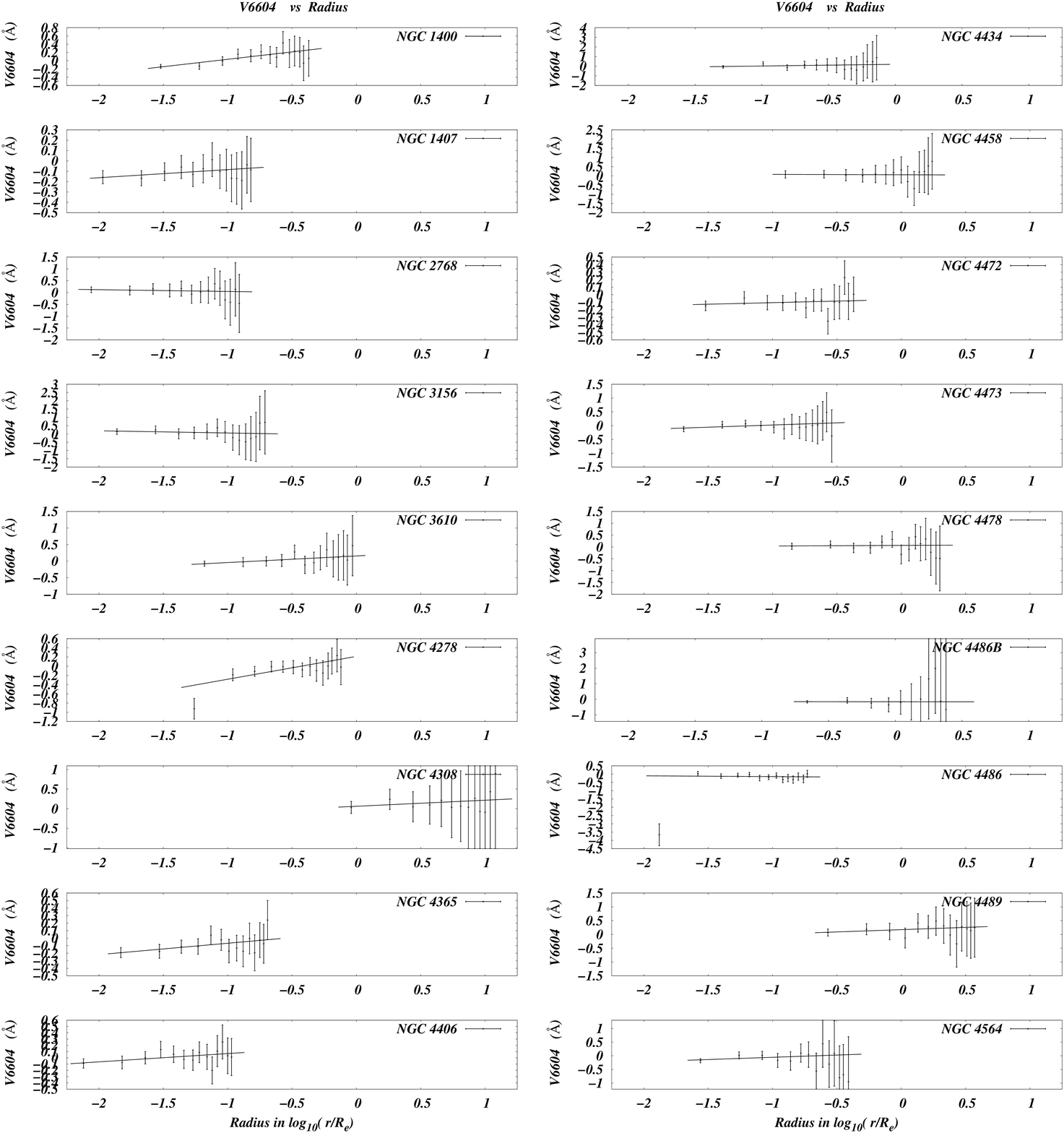}
\caption{}
\end{figure}

\begin{figure}[H]
\includegraphics[width=6in,height=7in]{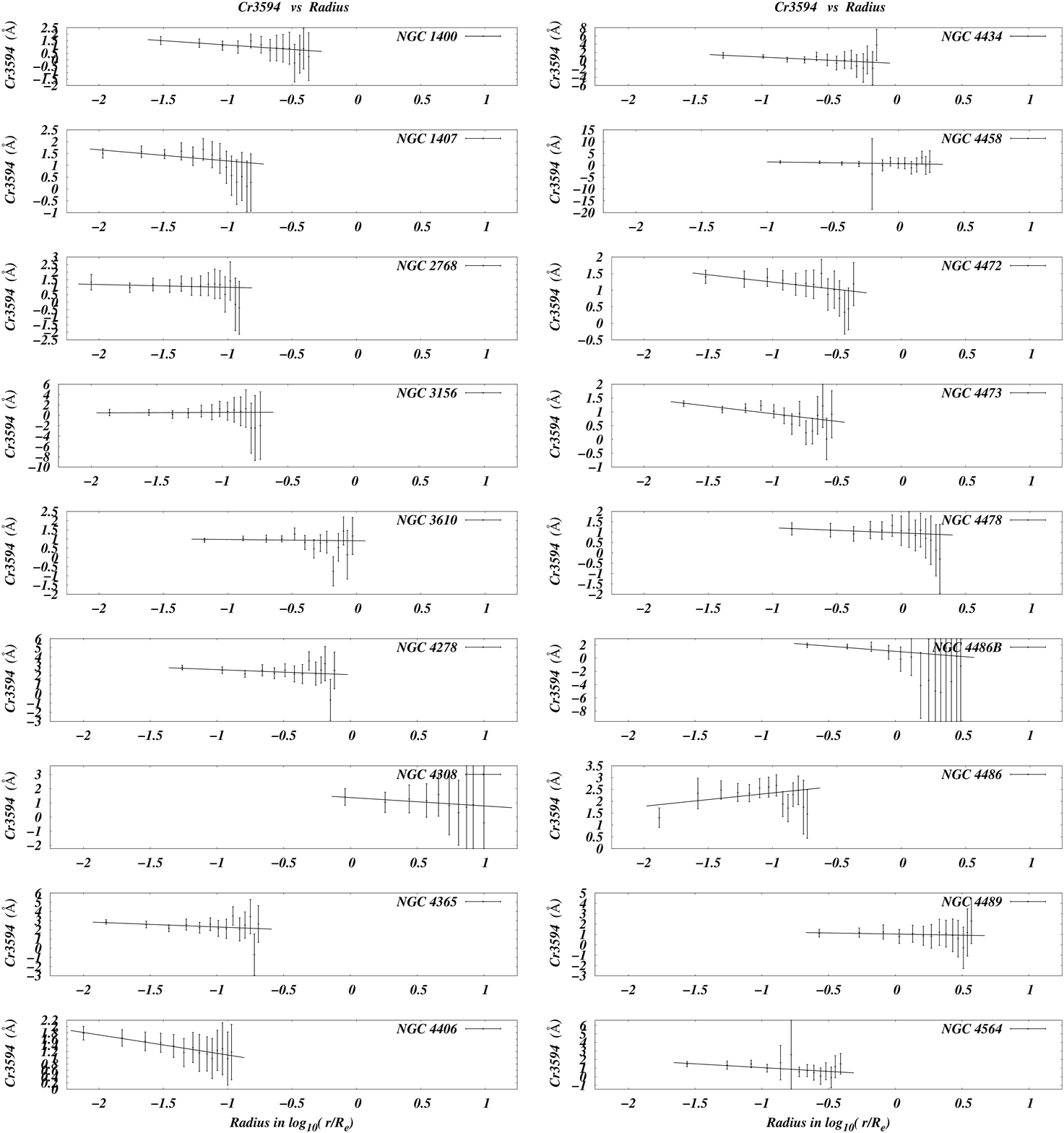}
\caption{}
\end{figure}

\begin{figure}[H]
\includegraphics[width=6in,height=7in]{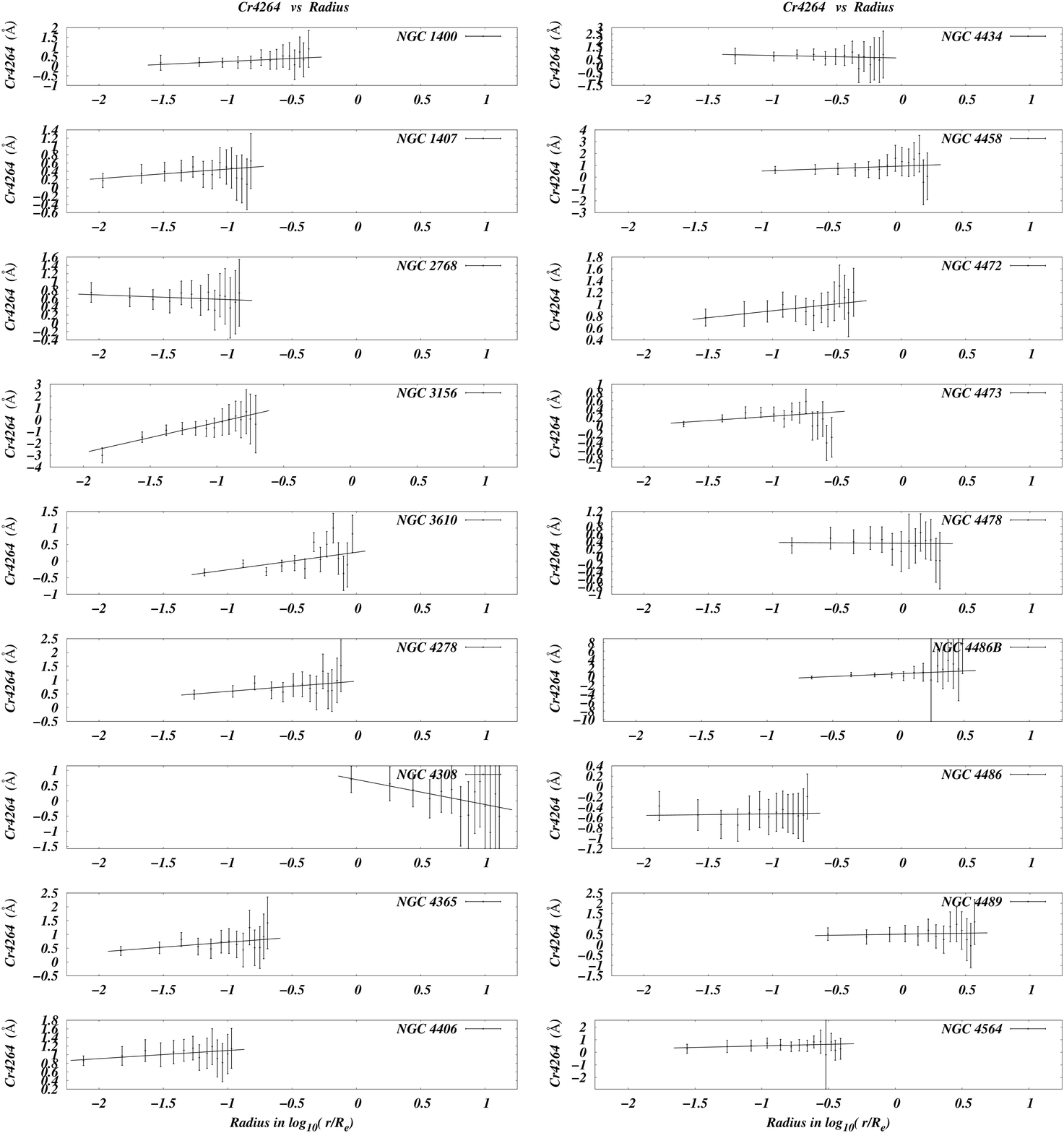}
\caption{}
\end{figure}

\begin{figure}[H]
\includegraphics[width=6in,height=7in]{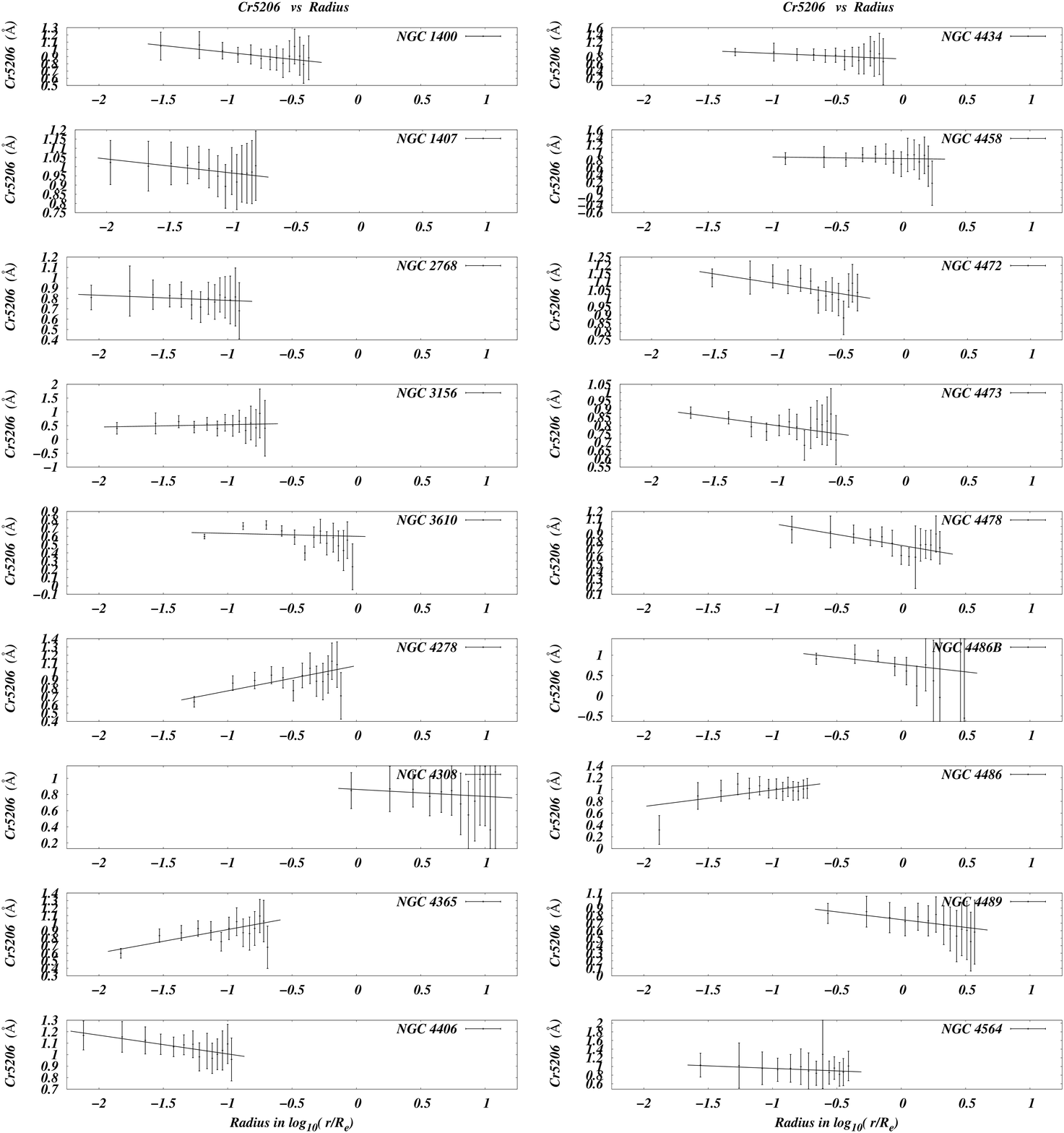}
\caption{}
\end{figure}

\begin{figure}[H]
\includegraphics[width=6in,height=7in]{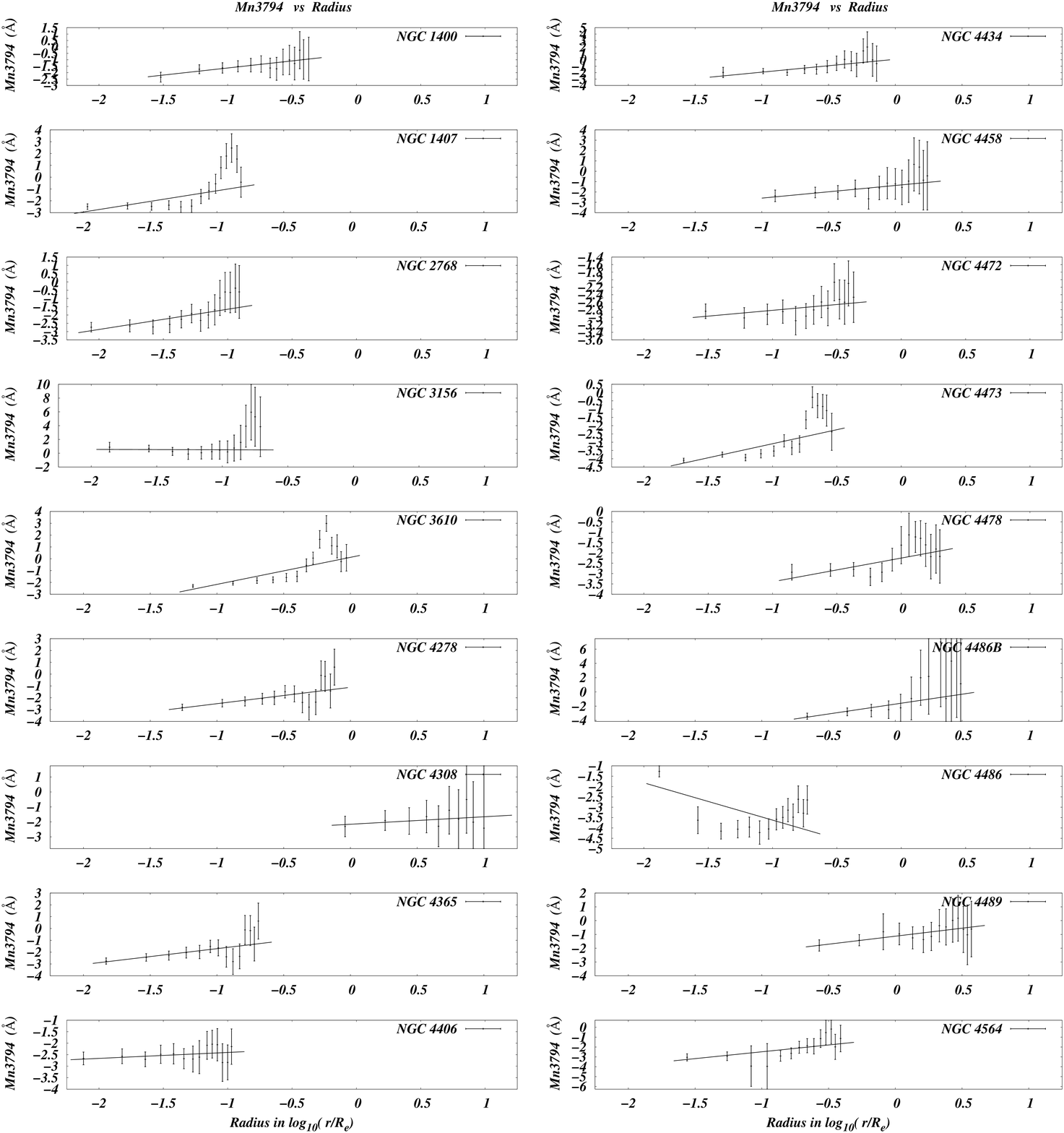}
\caption{}
\end{figure}

\begin{figure}[H]
\includegraphics[width=6in,height=7in]{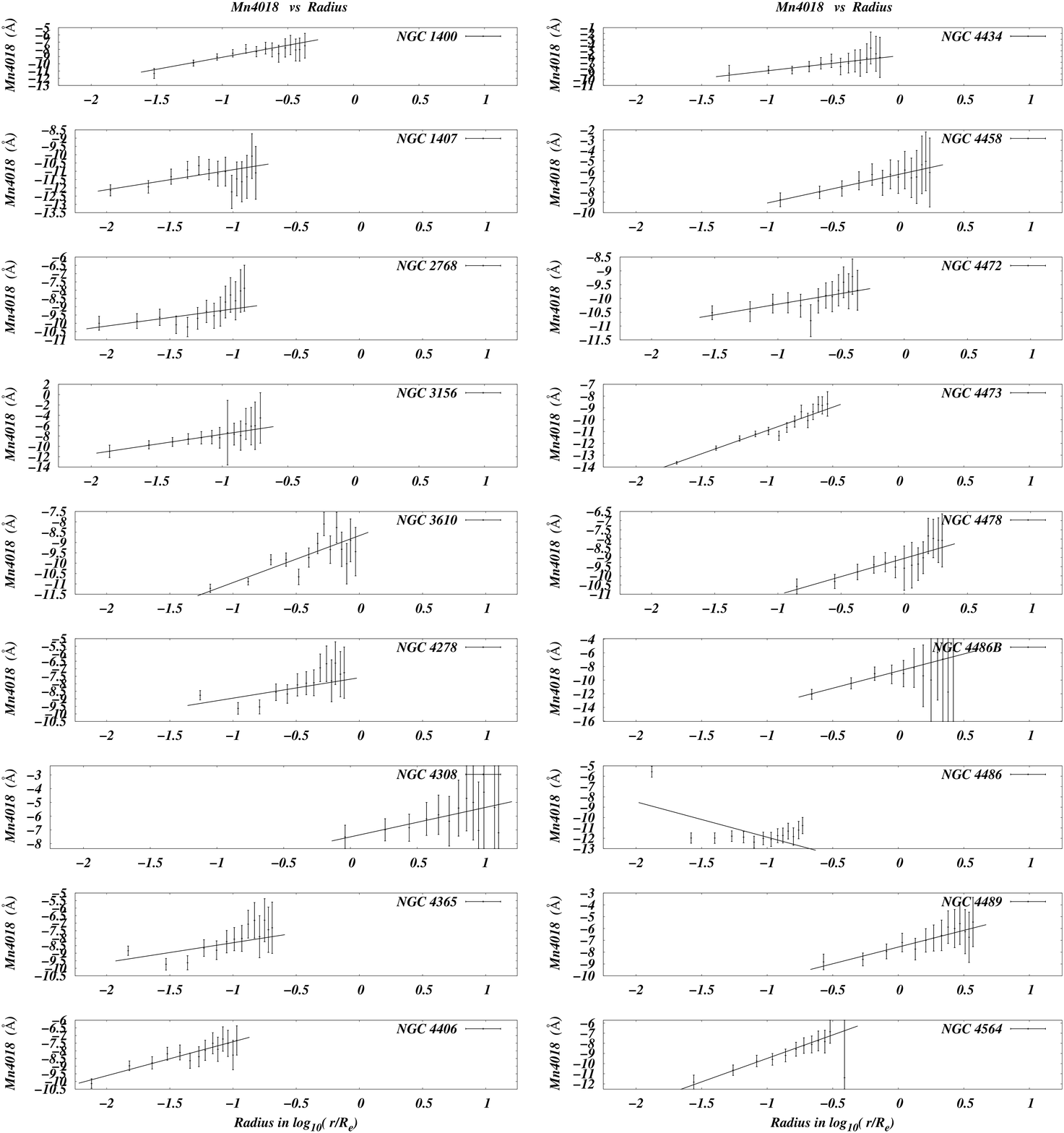}
\caption{}
\end{figure}

\begin{figure}[H]
\includegraphics[width=6in,height=7in]{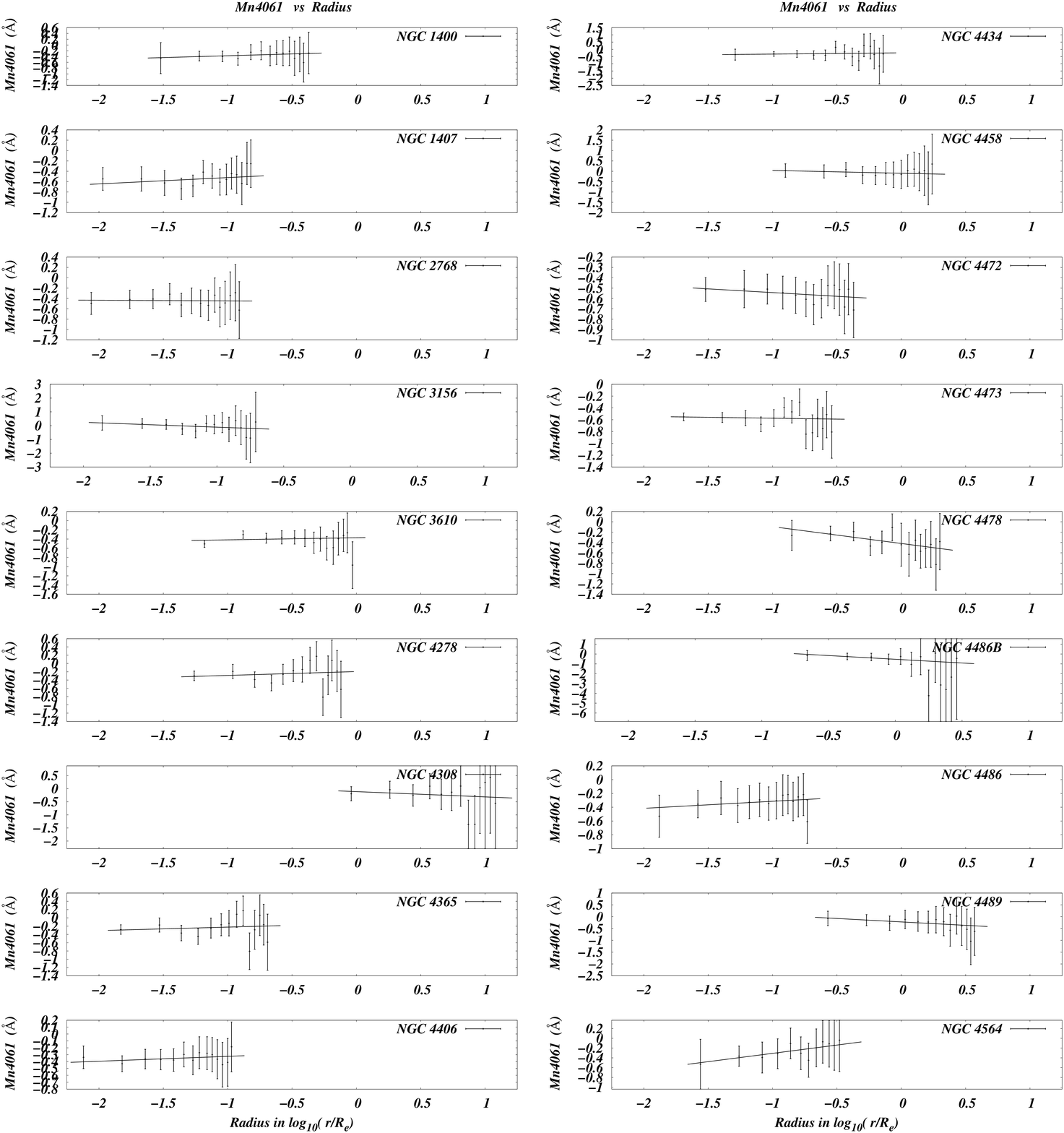}
\caption{}
\end{figure}

\begin{figure}[H]
\includegraphics[width=6in,height=7in]{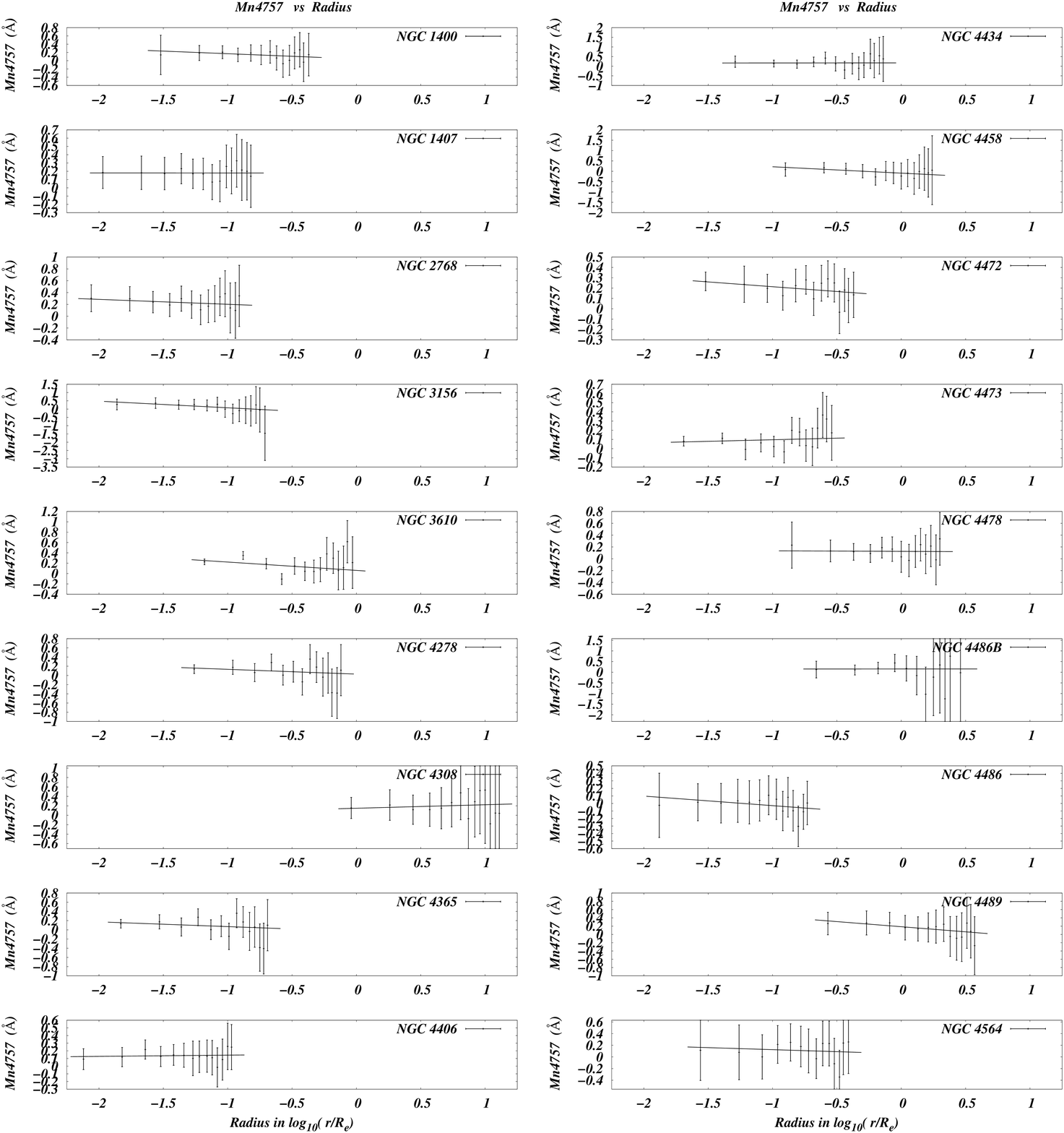}
\caption{}
\end{figure}

\begin{figure}[H]
\includegraphics[width=6in,height=7in]{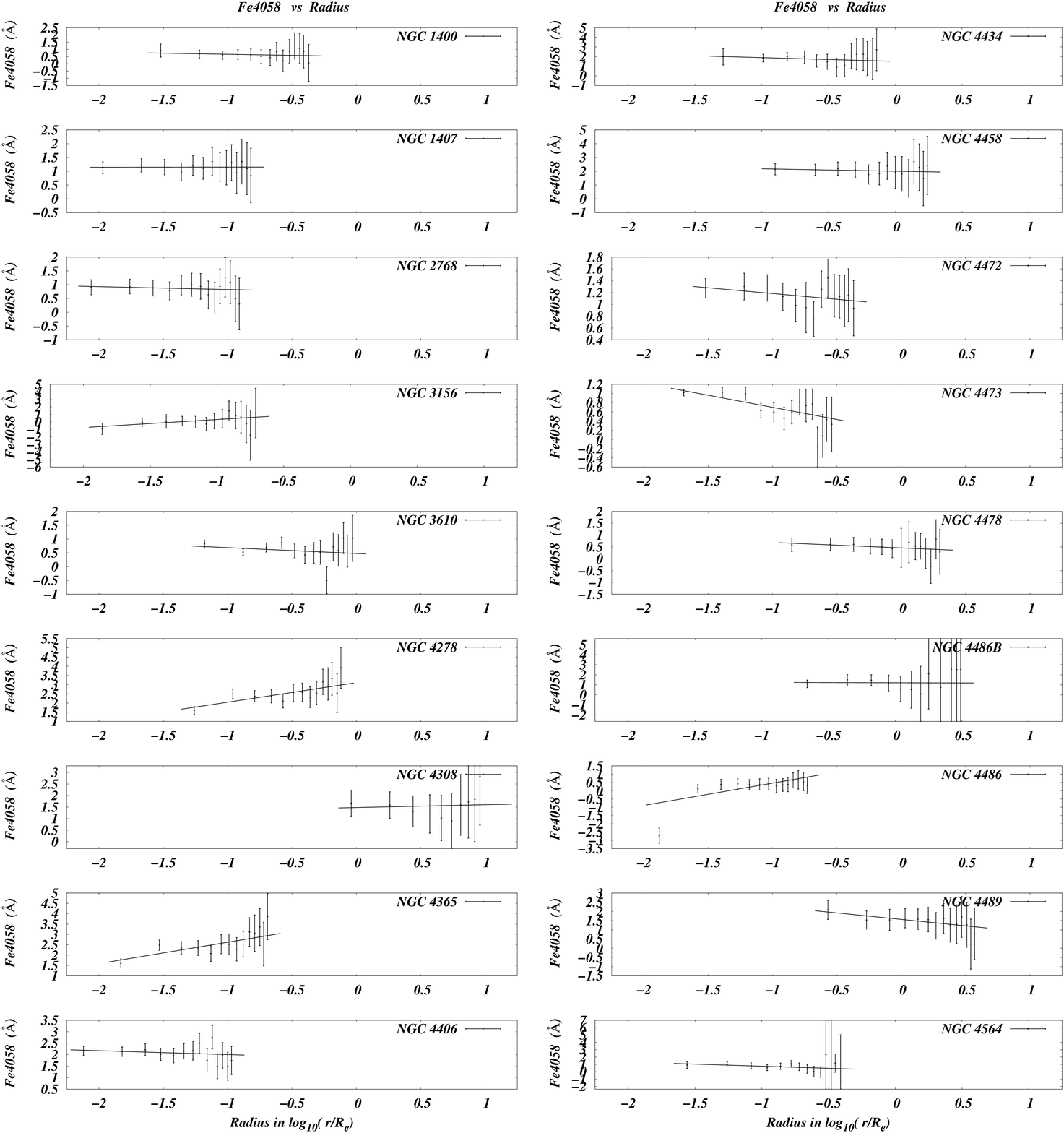}
\caption{}
\end{figure}

\begin{figure}[H]
\includegraphics[width=6in,height=7in]{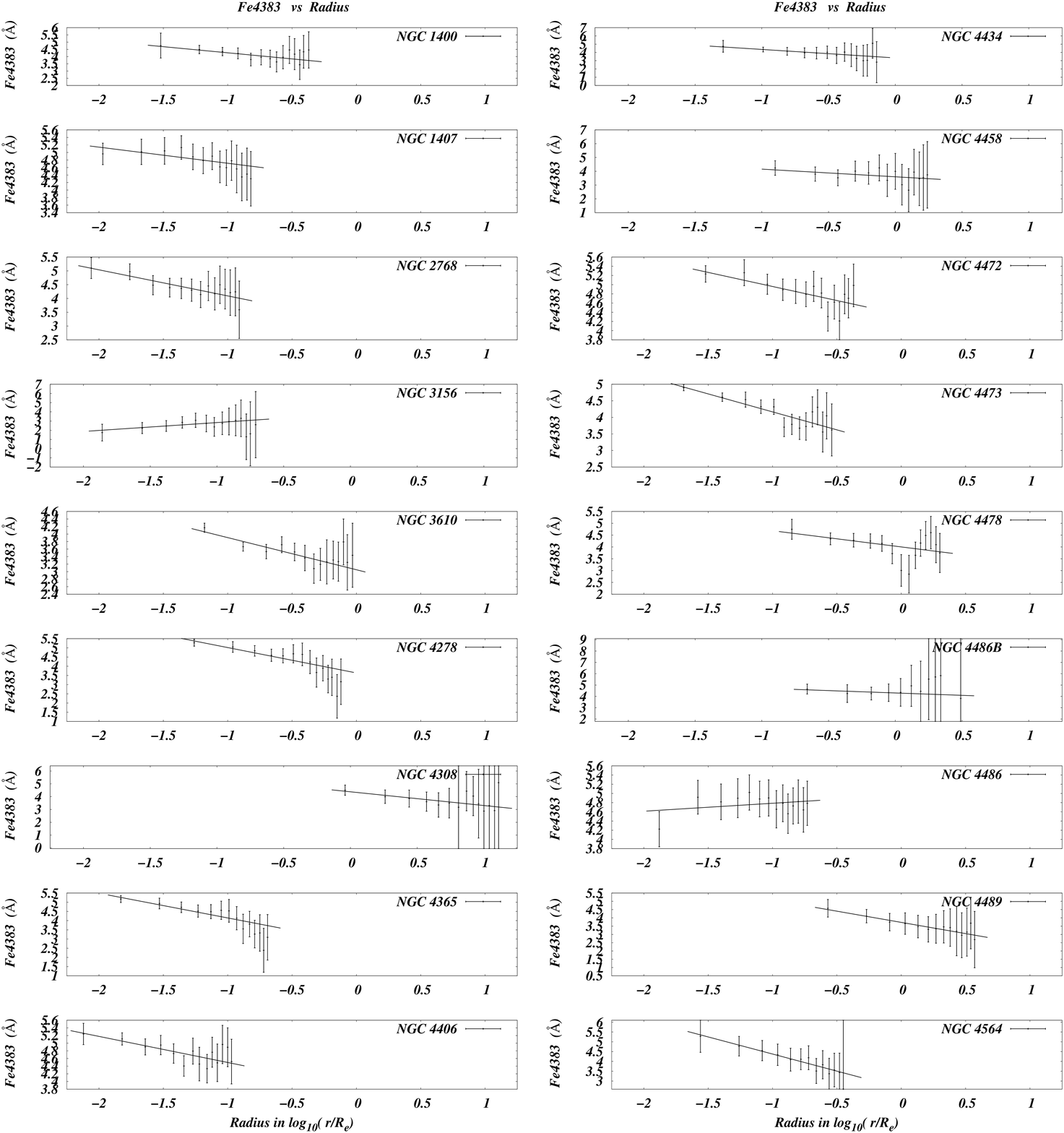}
\caption{}
\end{figure}

\begin{figure}[H]
\includegraphics[width=6in,height=7in]{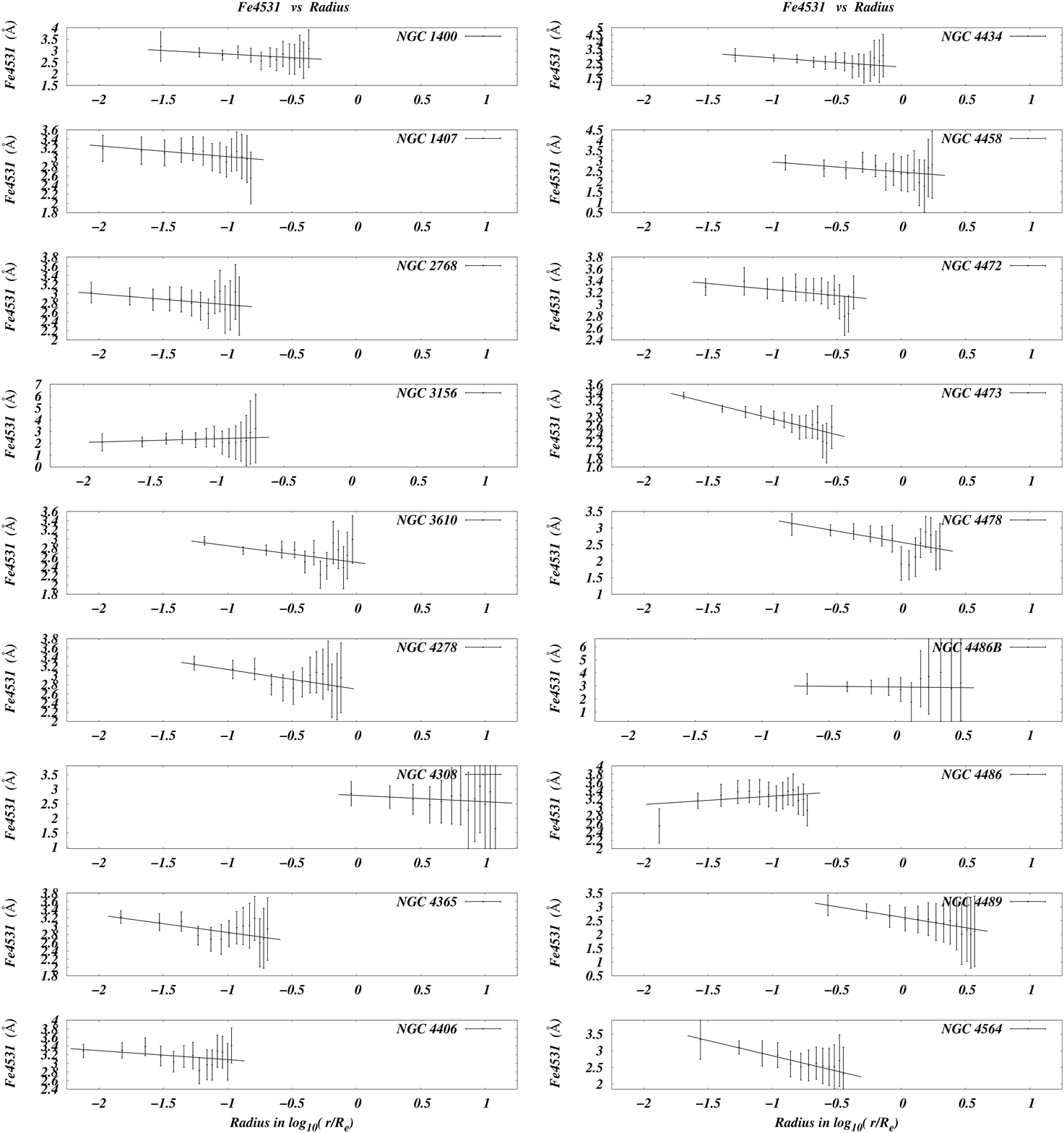}
\caption{}
\end{figure}

\begin{figure}[H]
\includegraphics[width=6in,height=7in]{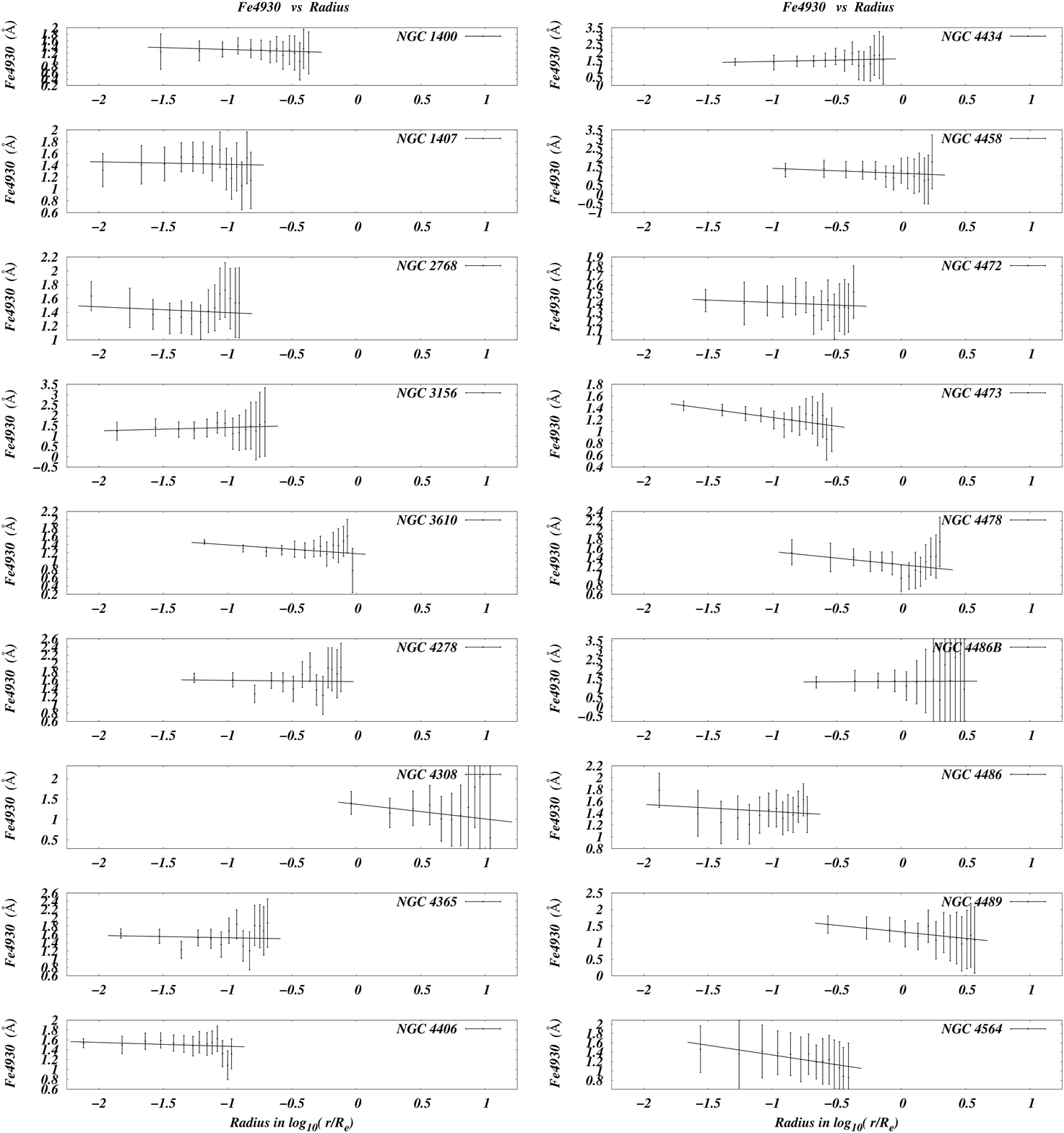}
\caption{}
\end{figure}

\begin{figure}[H]
\includegraphics[width=6in,height=7in]{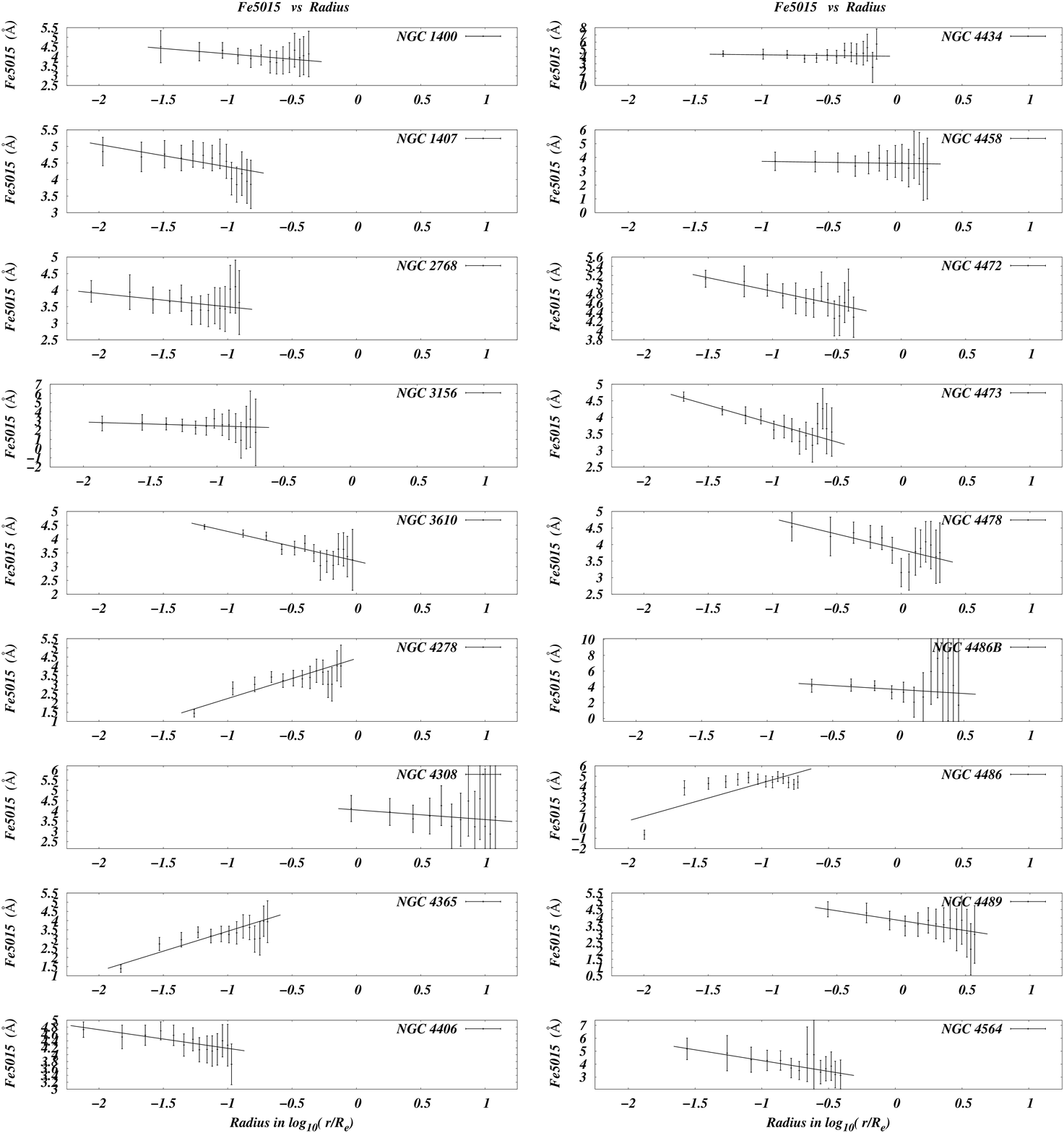}
\caption{}
\end{figure}

\begin{figure}[H]
\includegraphics[width=6in,height=7in]{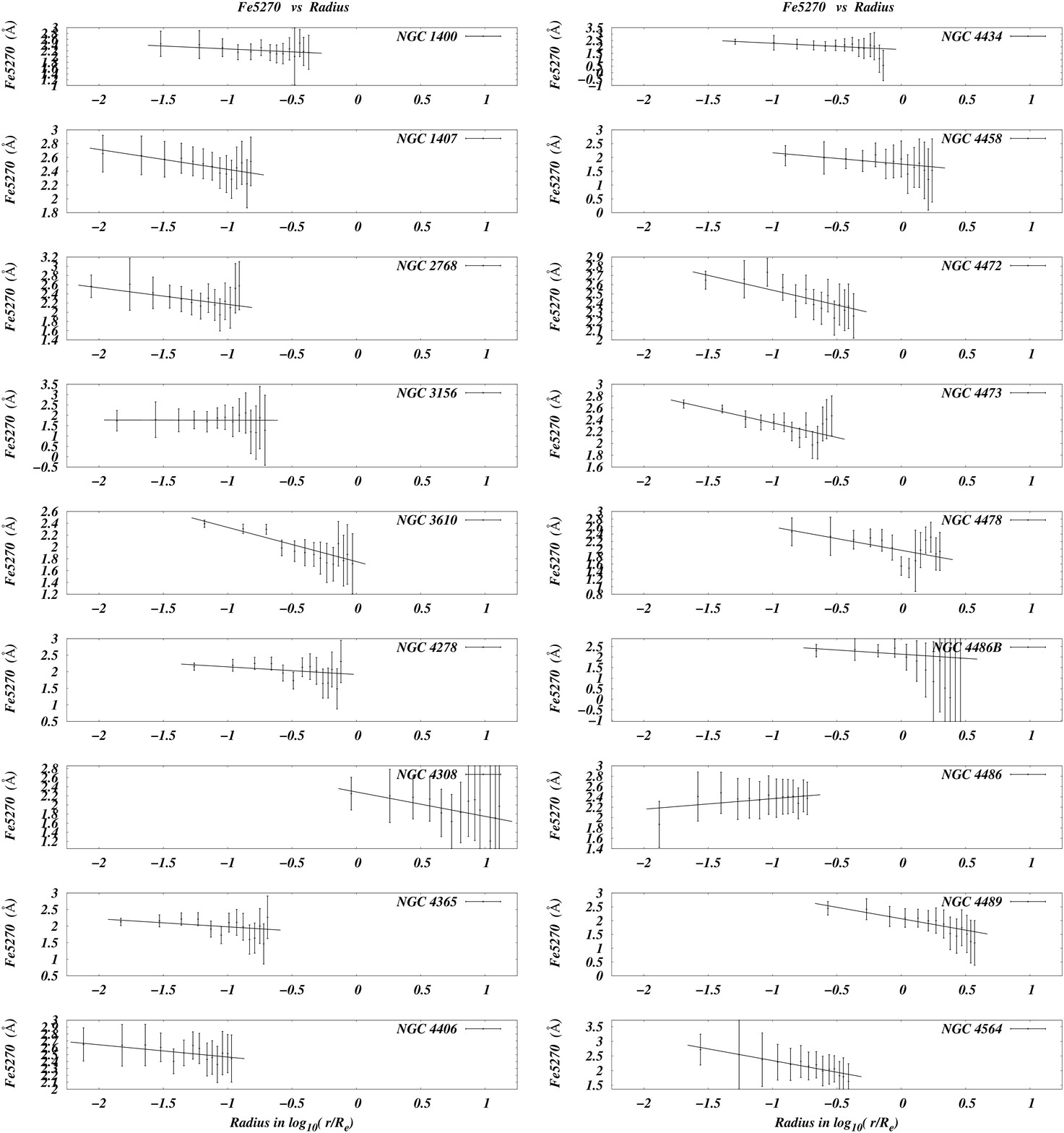}
\caption{}
\end{figure}

\begin{figure}[H]
\includegraphics[width=6in,height=7in]{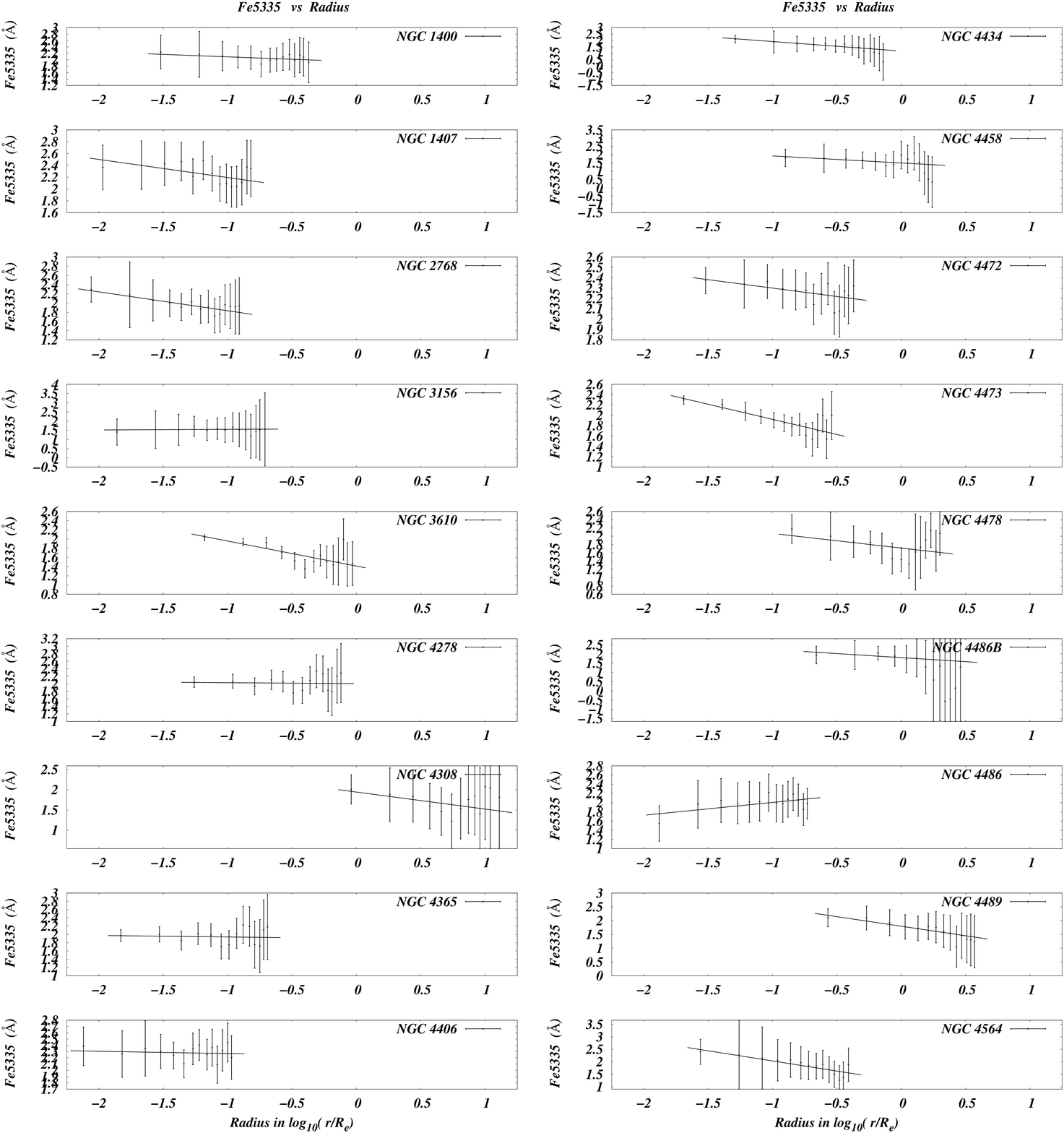}
\caption{}
\end{figure}

\begin{figure}[H]
\includegraphics[width=6in,height=7in]{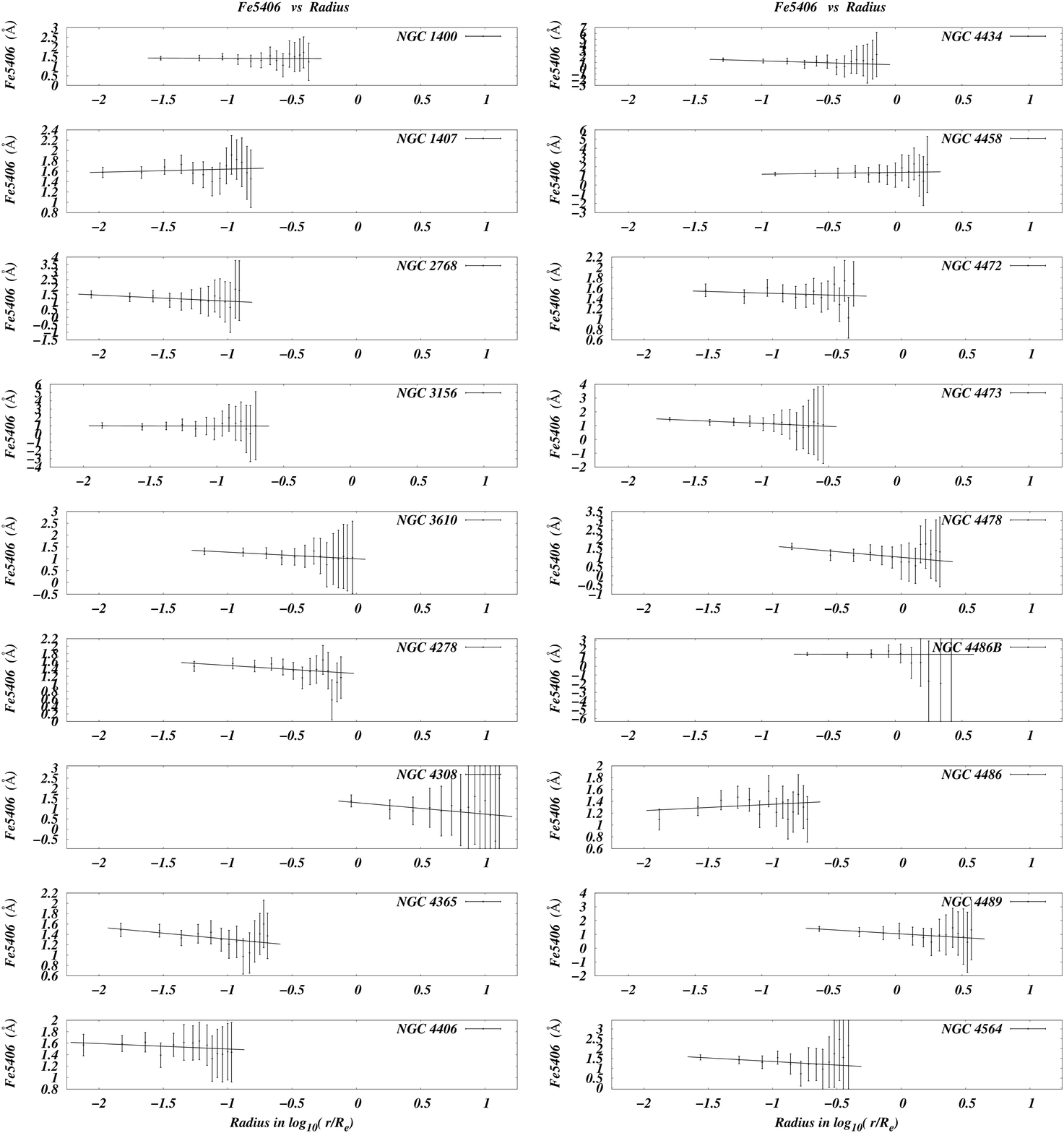}
\caption{}
\end{figure}

\begin{figure}[H]
\includegraphics[width=6in,height=7in]{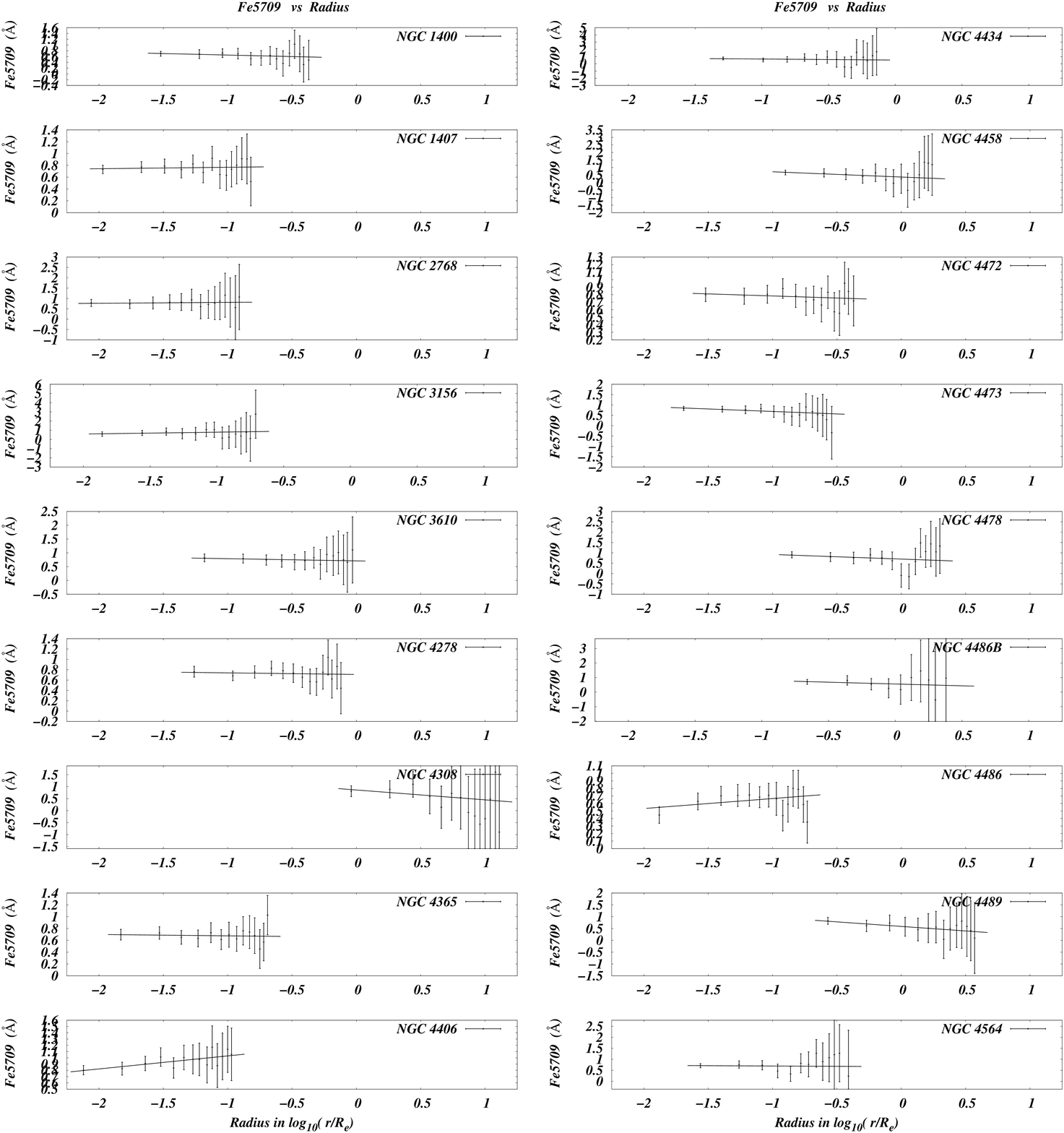}
\caption{}
\end{figure}

\begin{figure}[H]
\includegraphics[width=6in,height=7in]{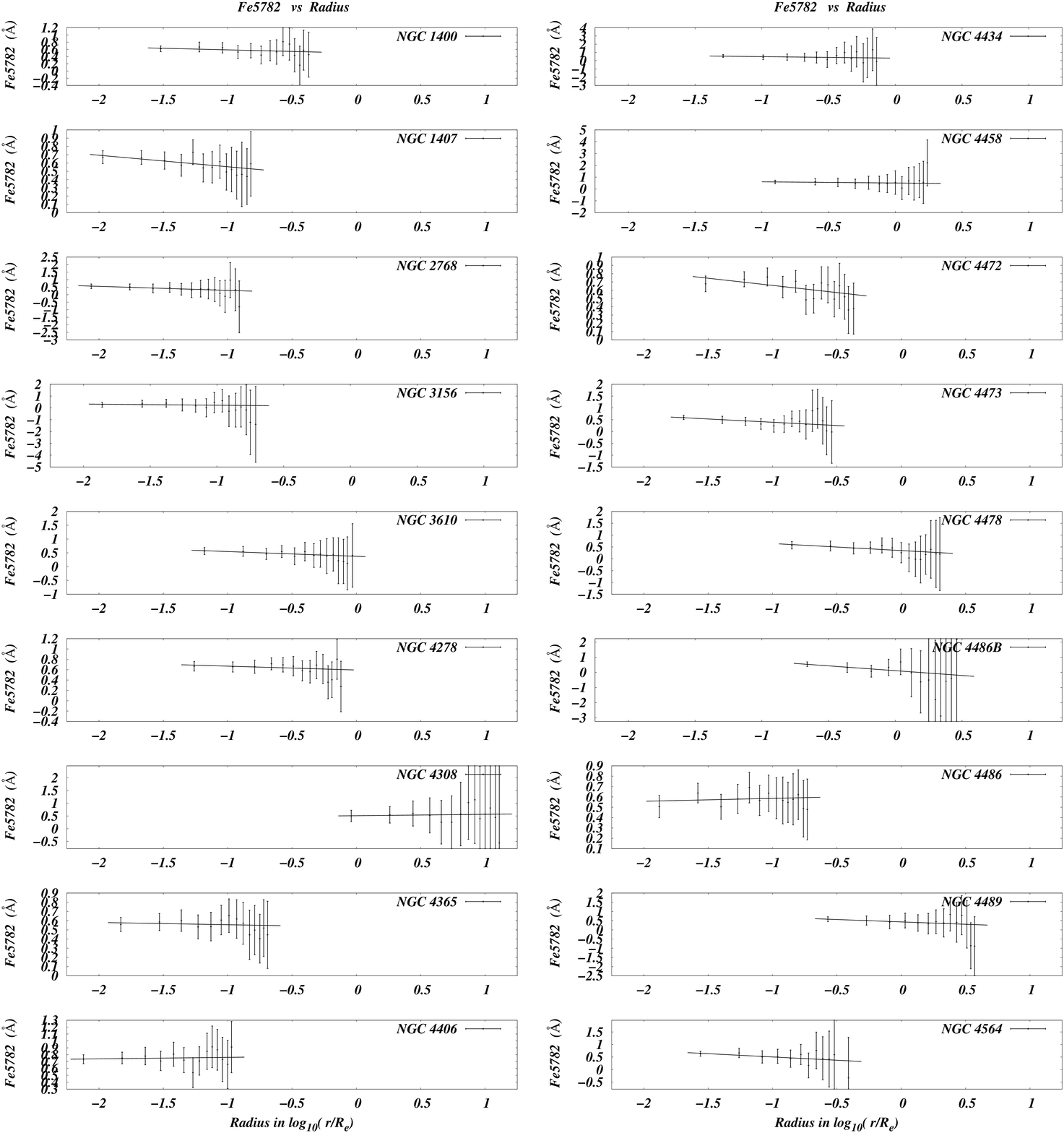}
\caption{}
\end{figure}

\begin{figure}[H]
\includegraphics[width=6in,height=7in]{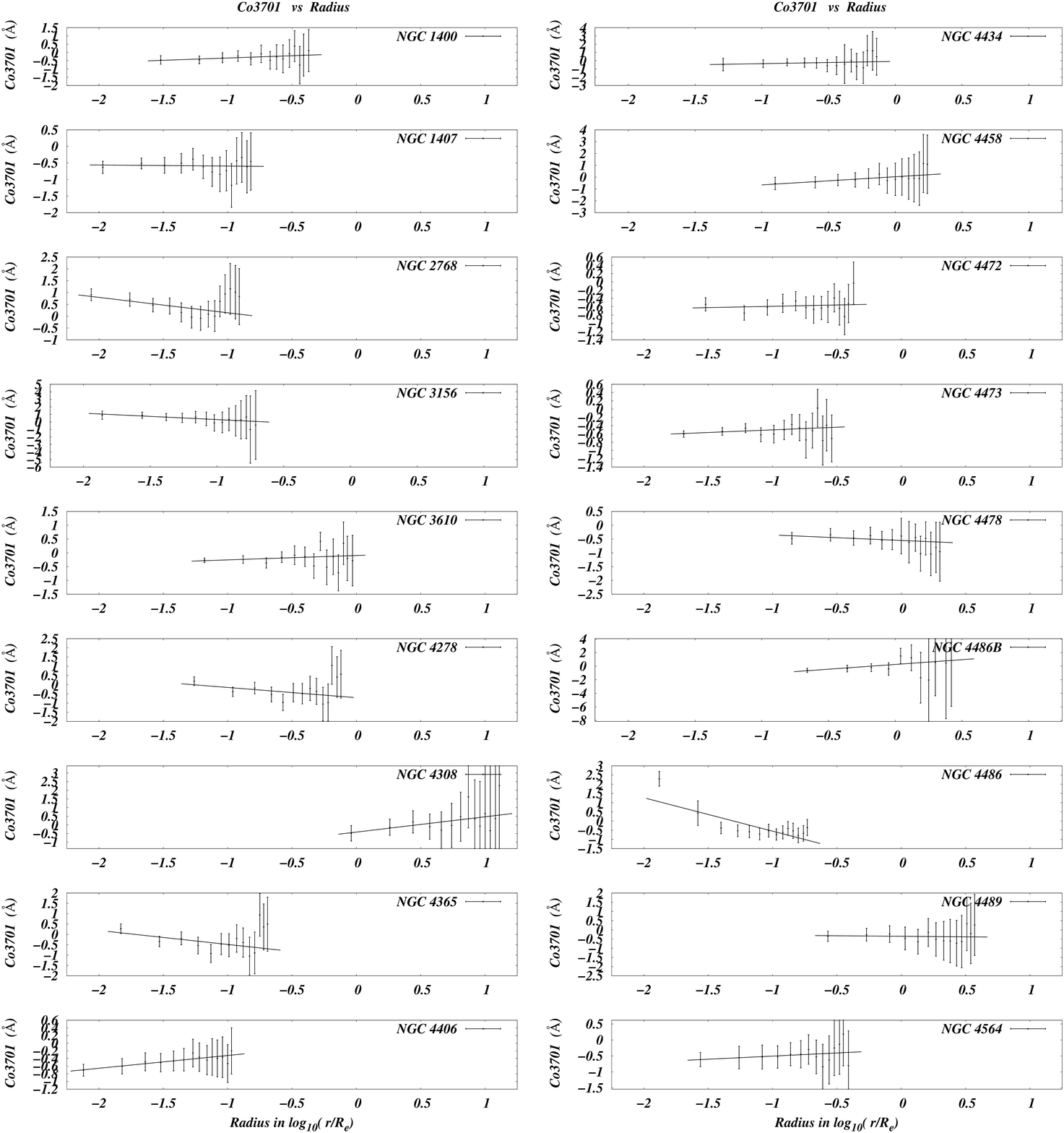}
\caption{}
\end{figure}

\begin{figure}[H]
\includegraphics[width=6in,height=7in]{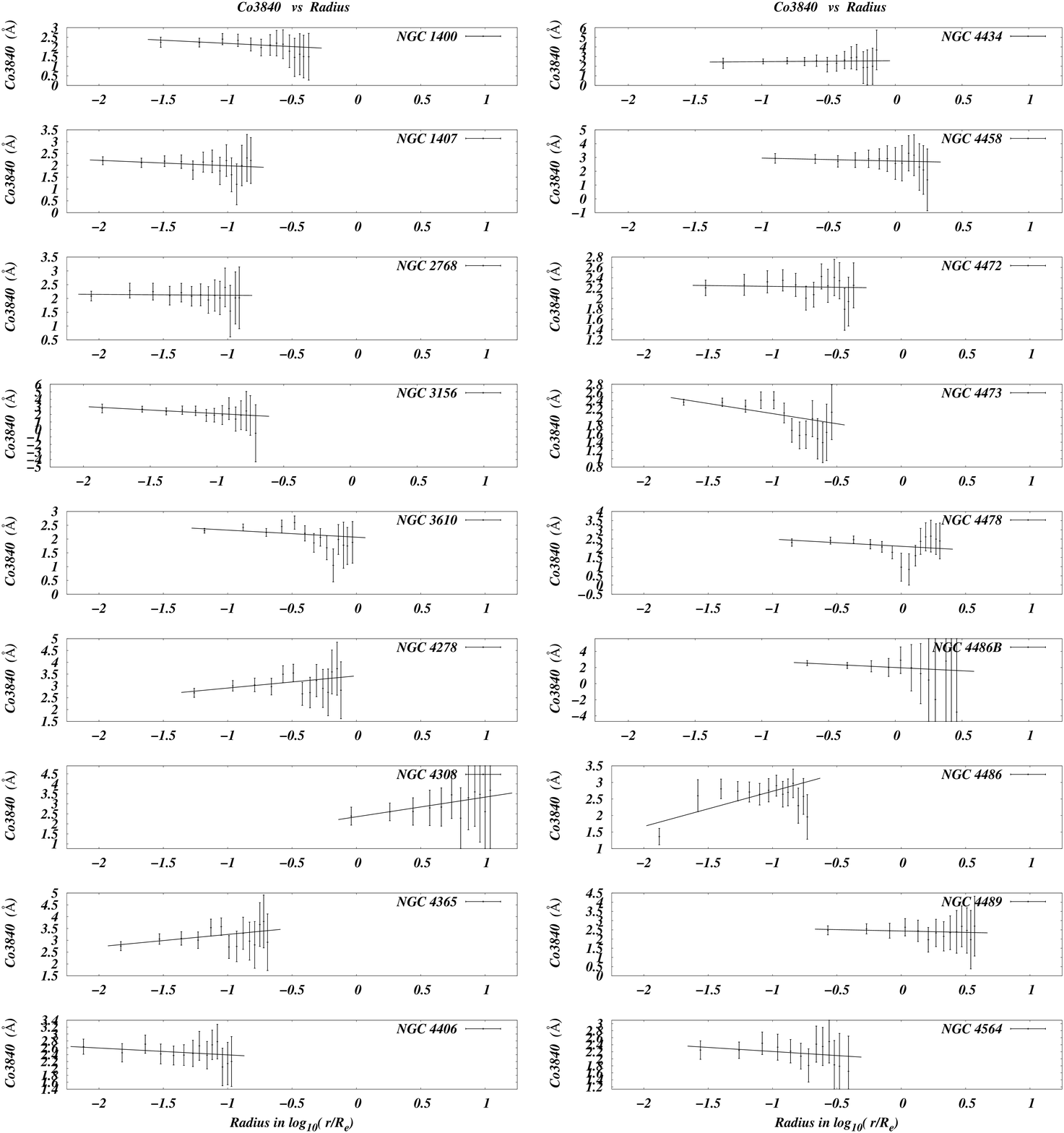}
\caption{}
\end{figure}

\begin{figure}[H]
\includegraphics[width=6in,height=7in]{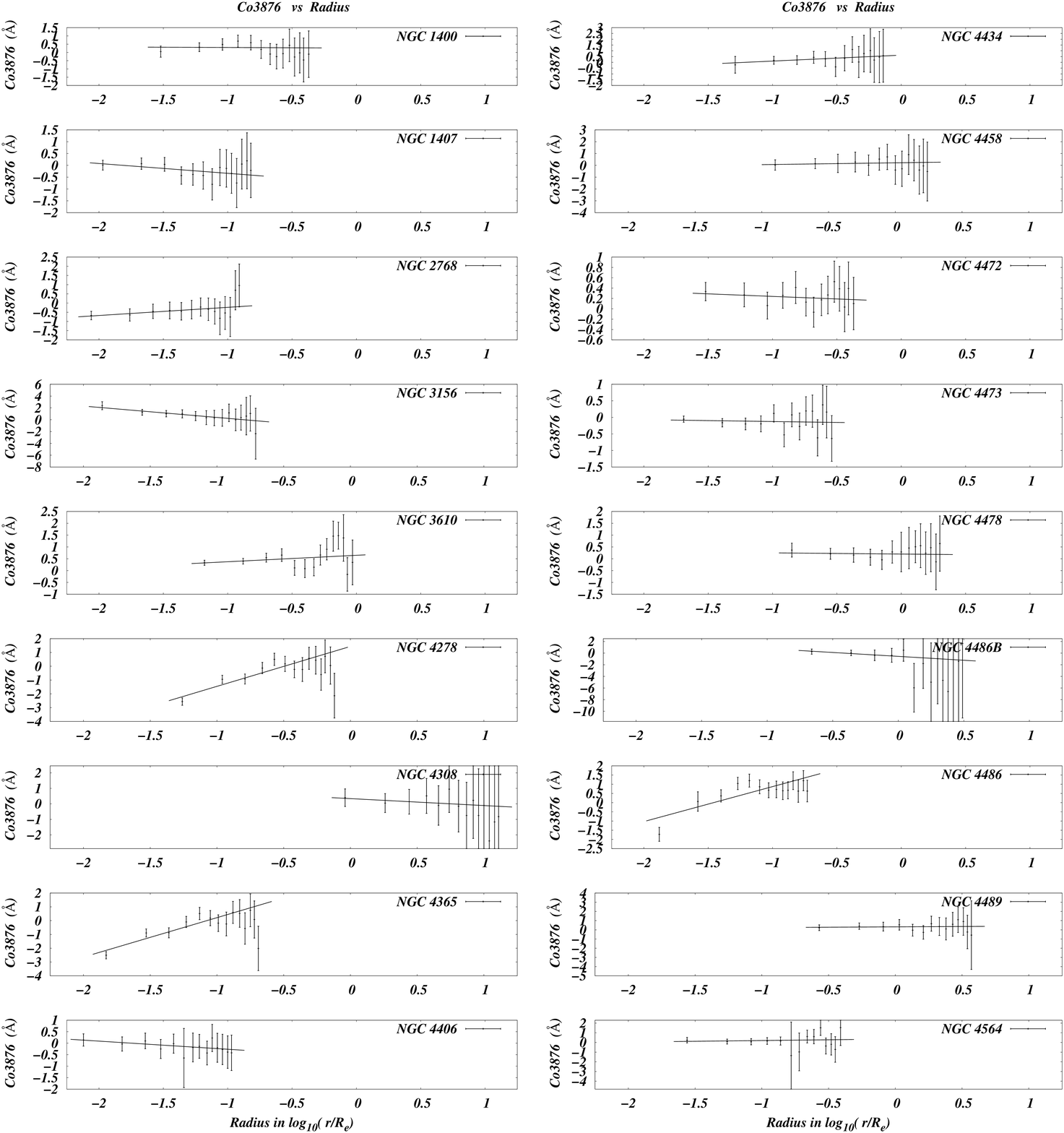}
\caption{}
\end{figure}

\begin{figure}[H]
\includegraphics[width=6in,height=7in]{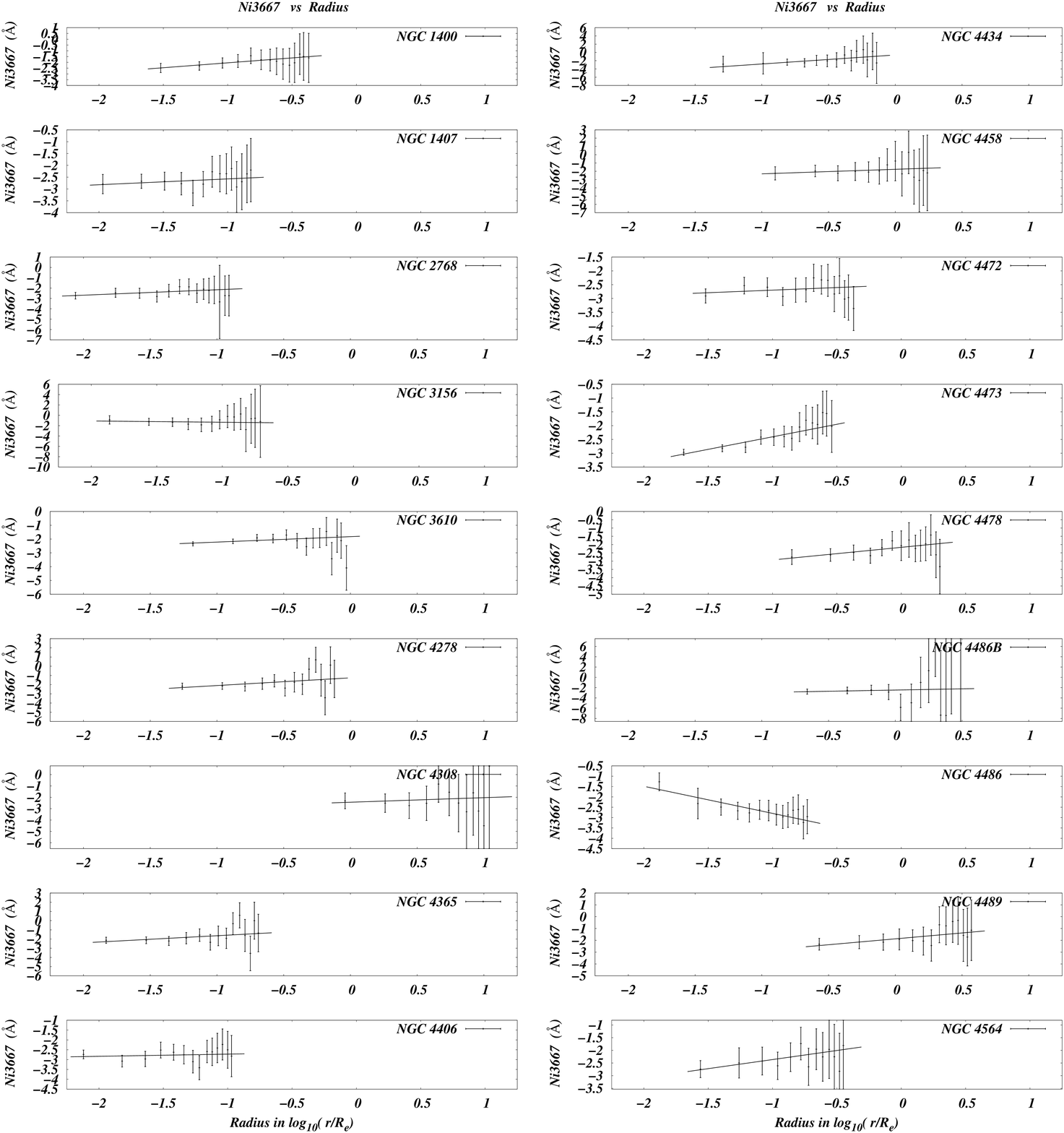}
\caption{}
\end{figure}

\begin{figure}[H]
\includegraphics[width=6in,height=7in]{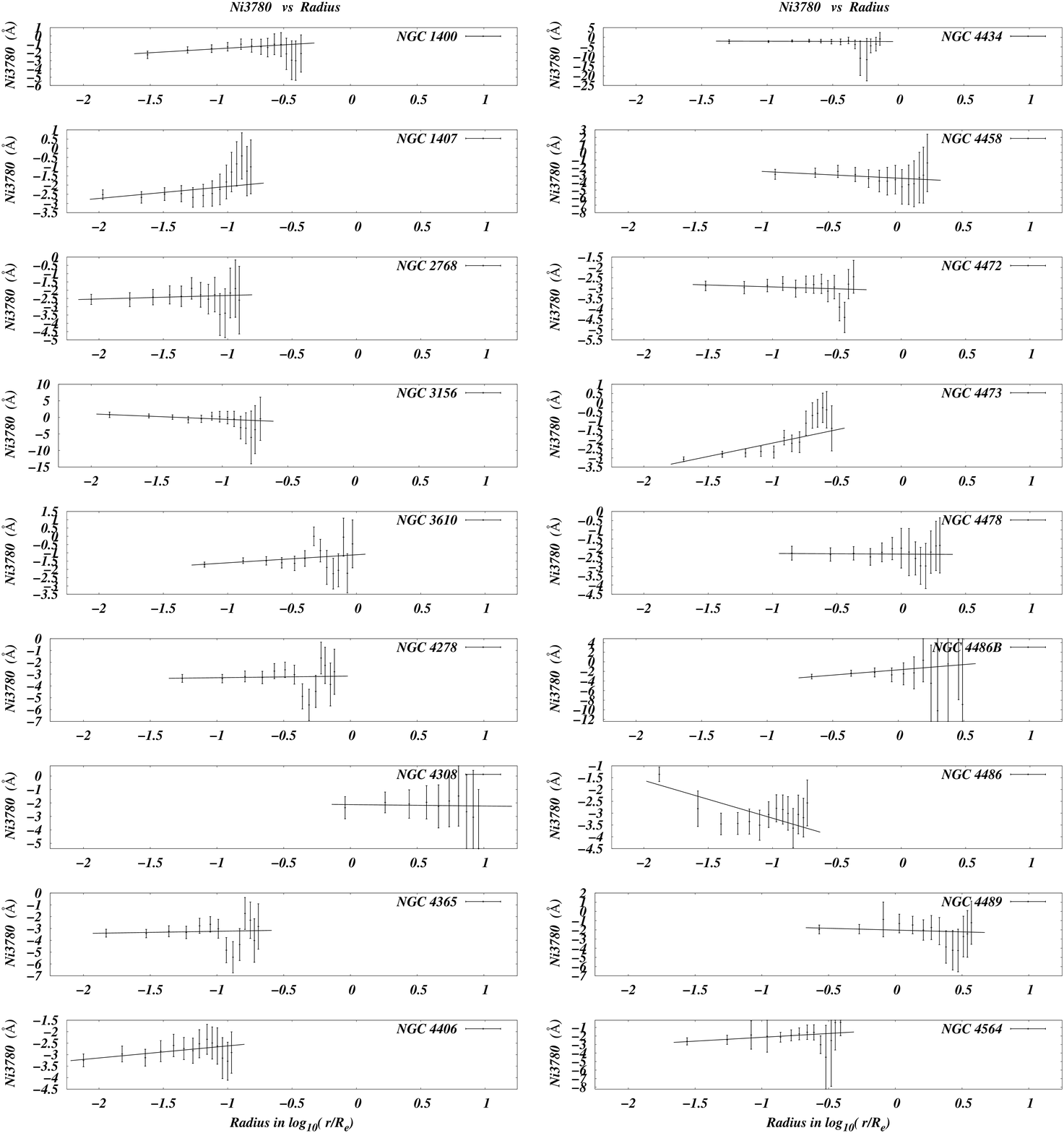}
\caption{}
\end{figure}

\begin{figure}[H]
\includegraphics[width=6in,height=7in]{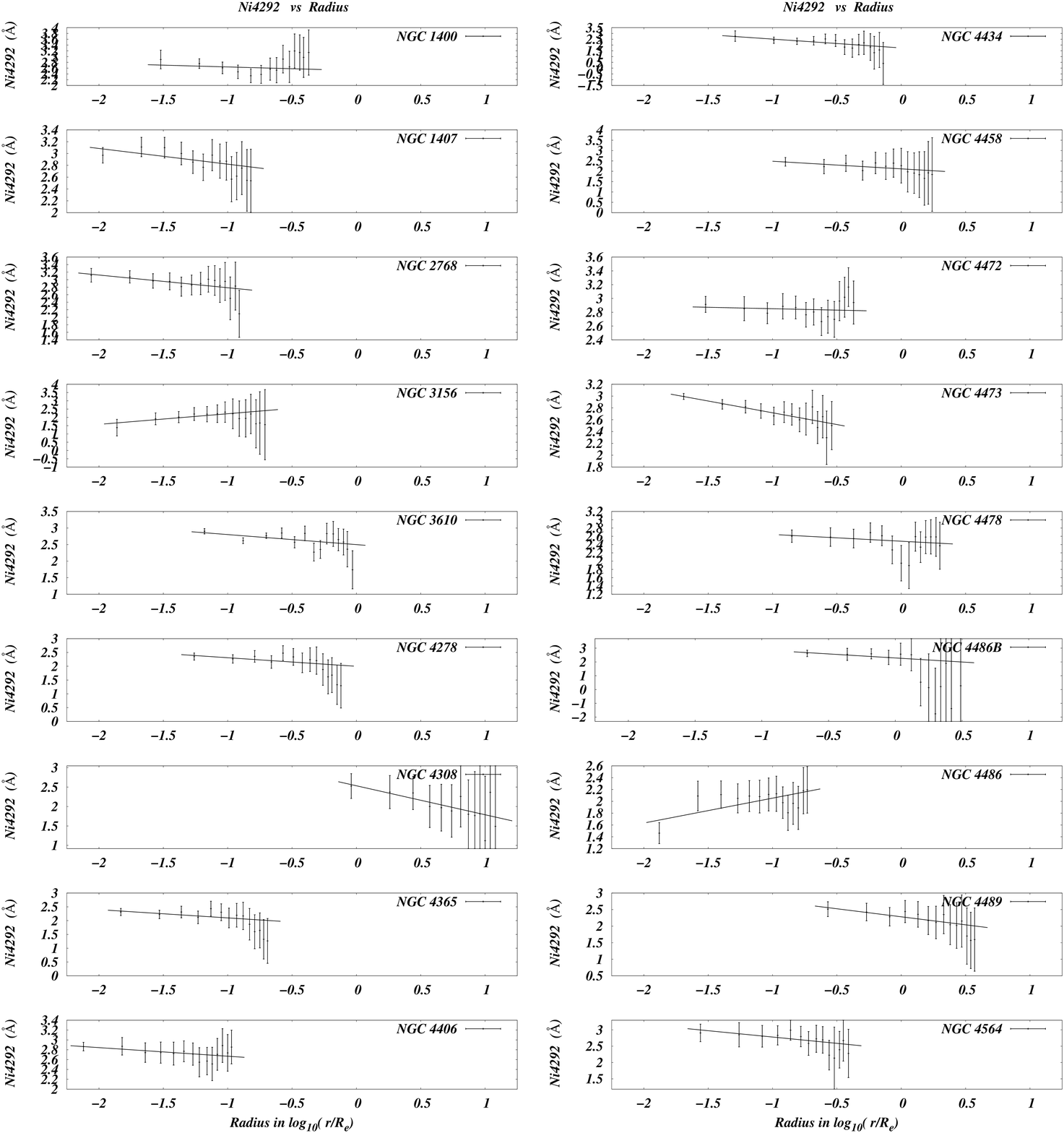}
\caption{}
\end{figure}

\begin{figure}[H]
\includegraphics[width=6in,height=7in]{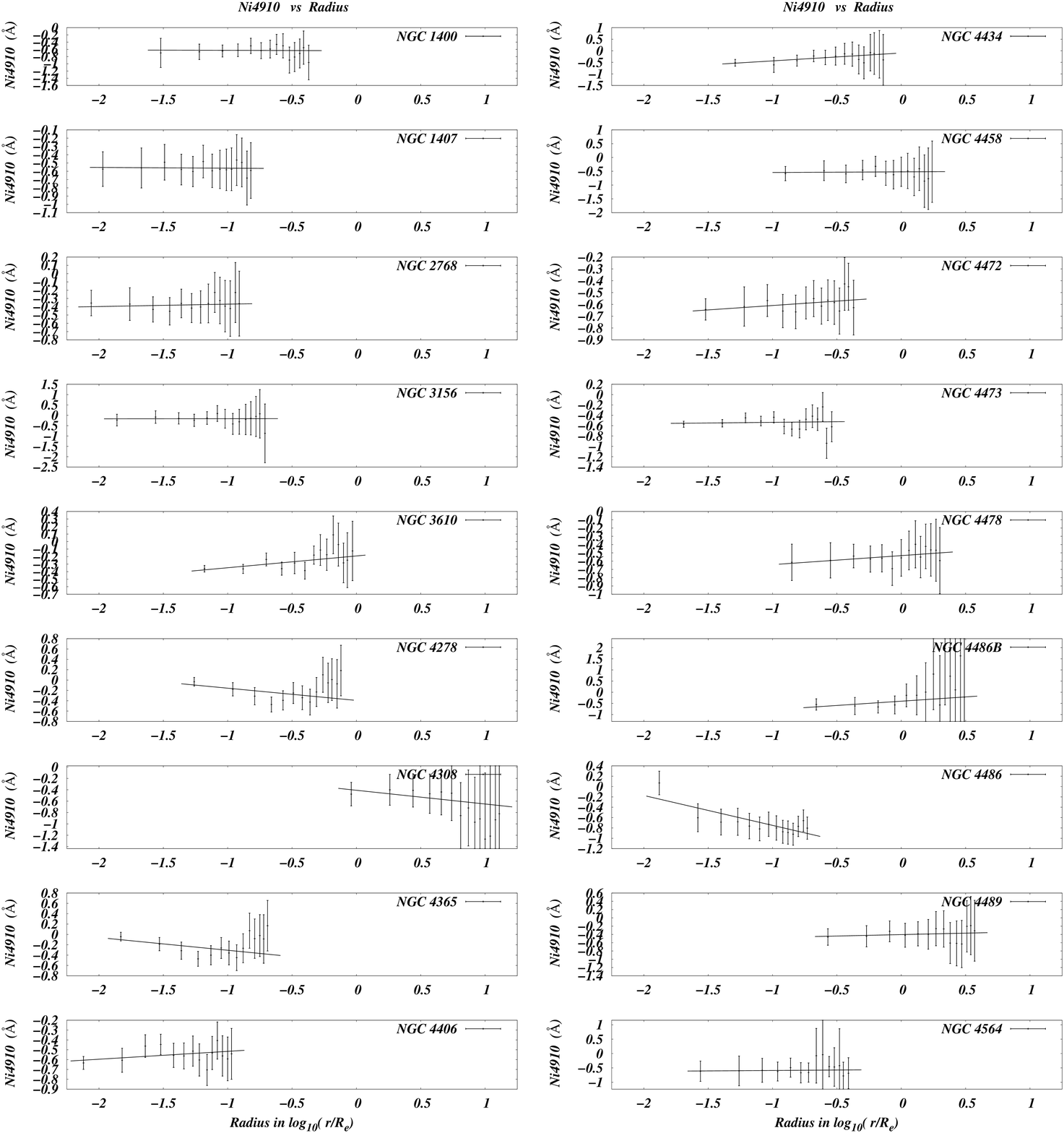}
\caption{}
\end{figure}

\begin{figure}[H]
\includegraphics[width=6in,height=7in]{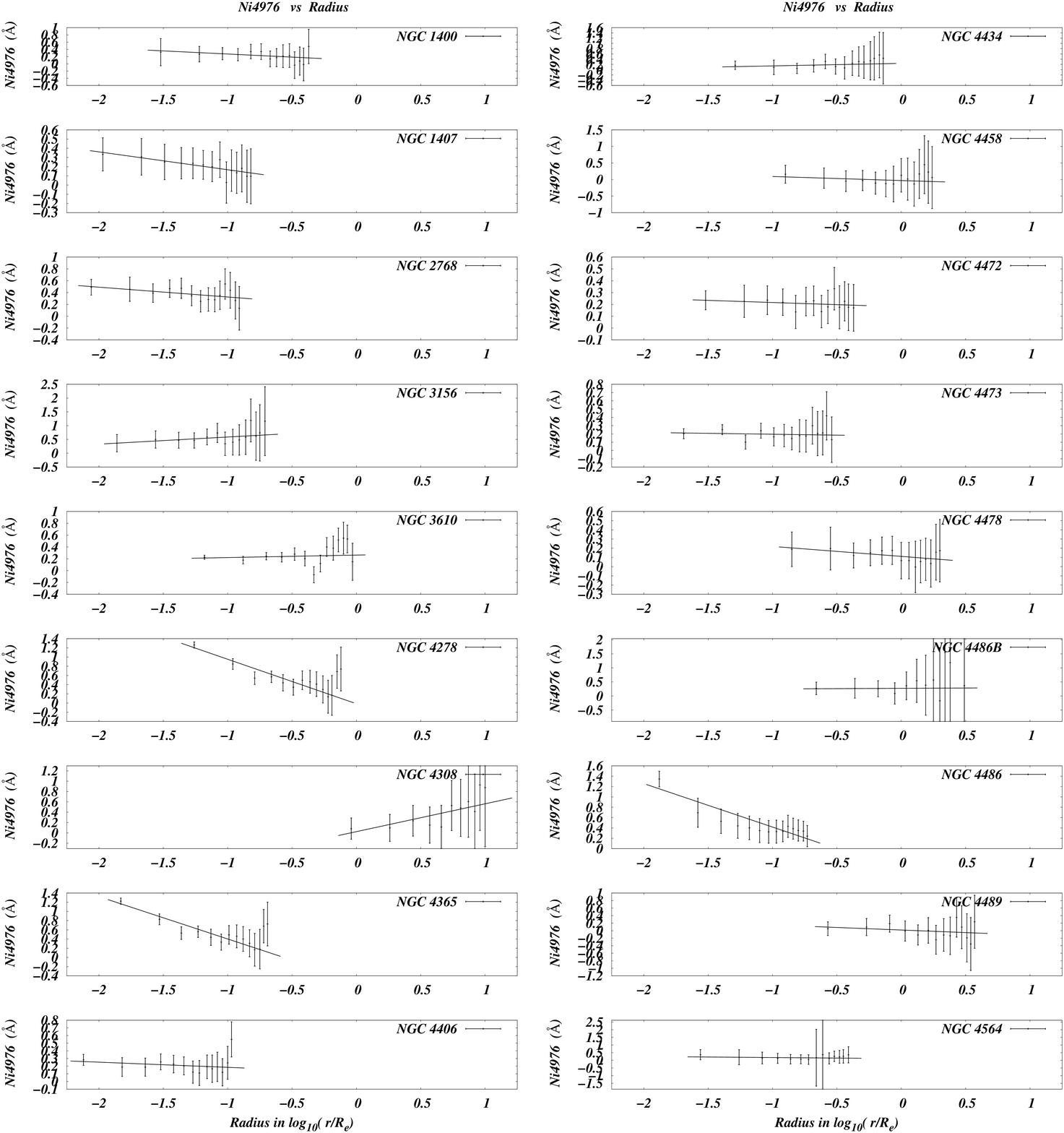}
\caption{}
\end{figure}

\begin{figure}[H]
\includegraphics[width=6in,height=7in]{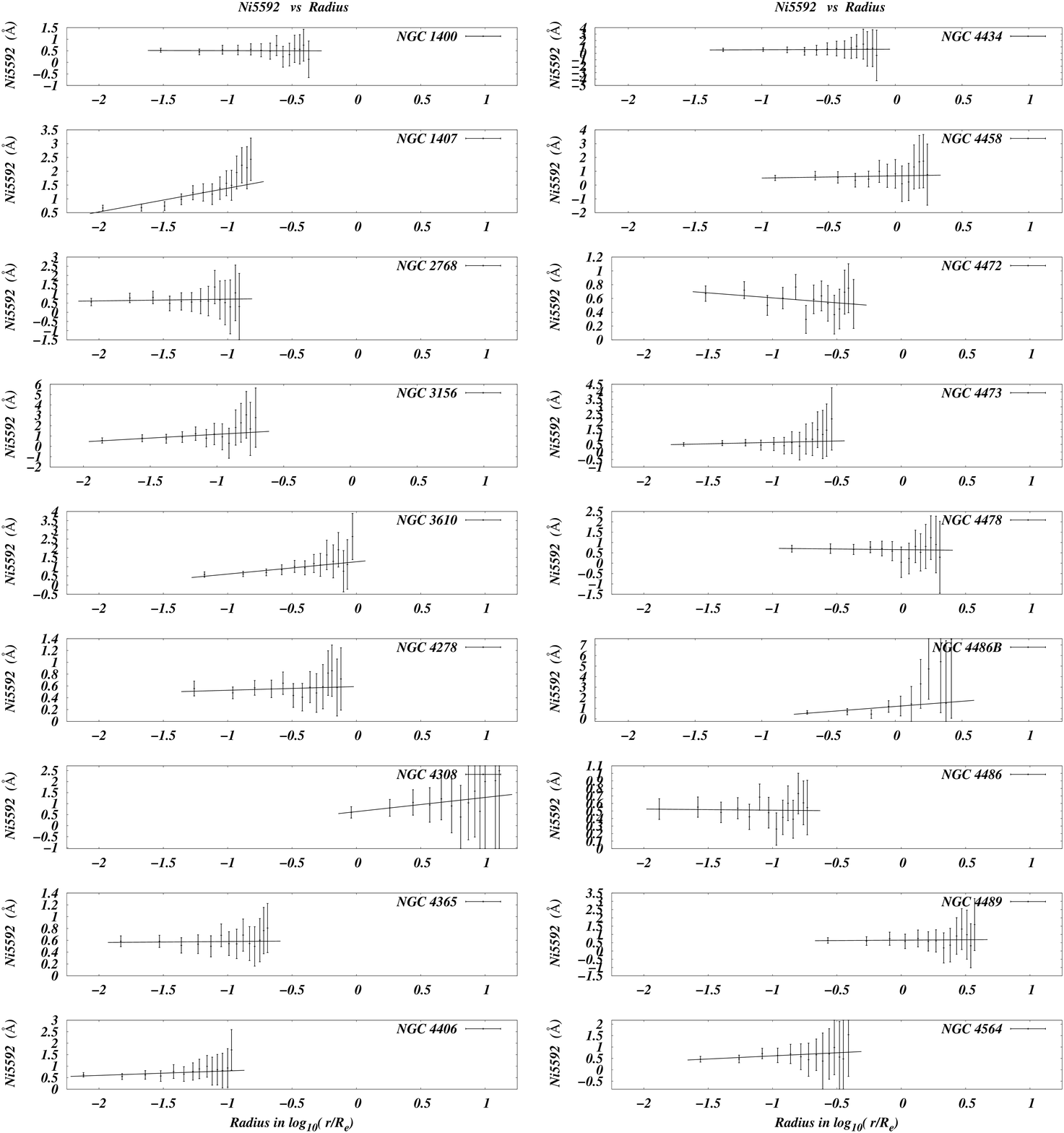}
\caption{}
\end{figure}

\begin{figure}[H]
\includegraphics[width=6in,height=7in]{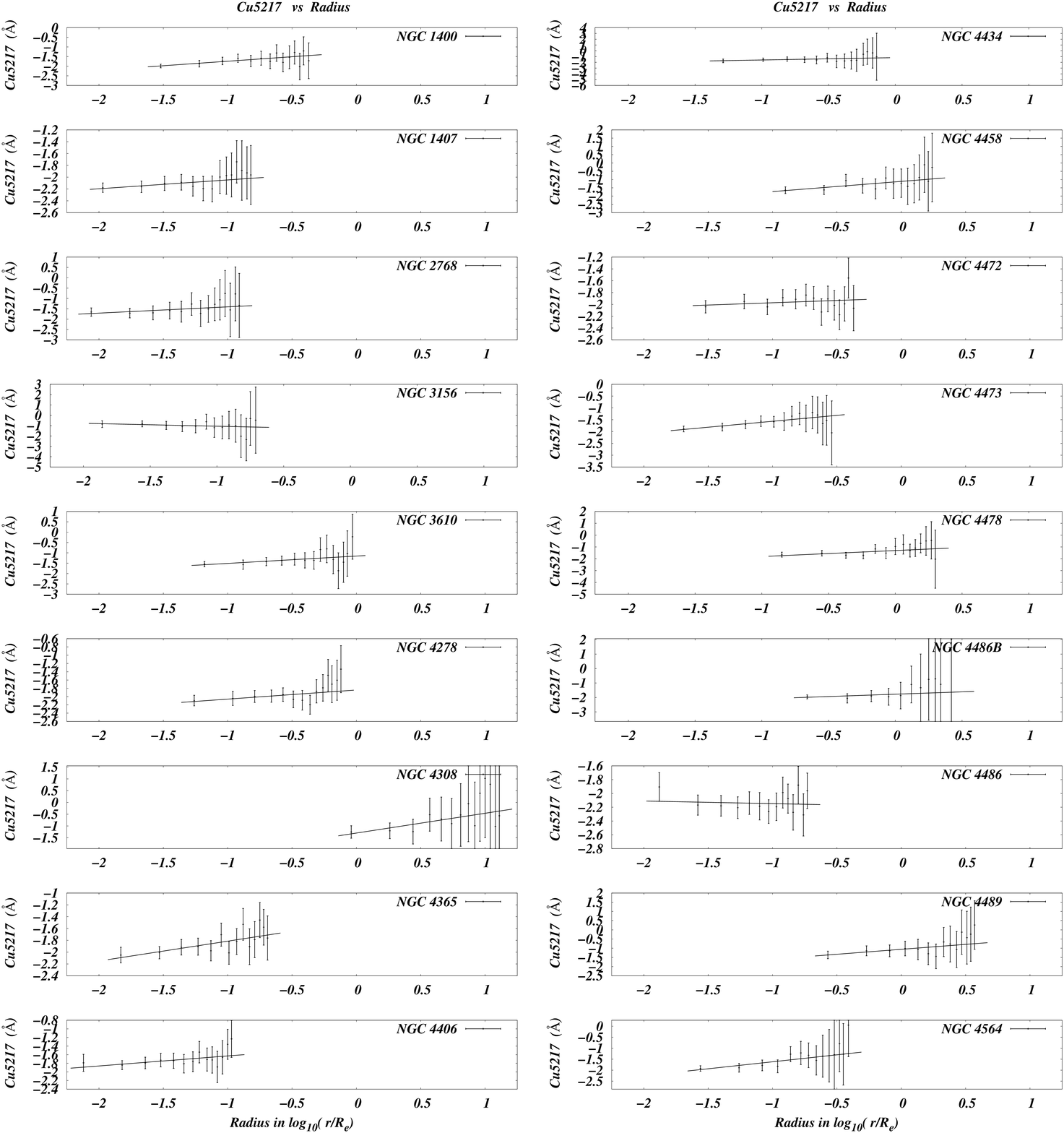}
\caption{}
\end{figure}

\begin{figure}[H]
\includegraphics[width=6in,height=7in]{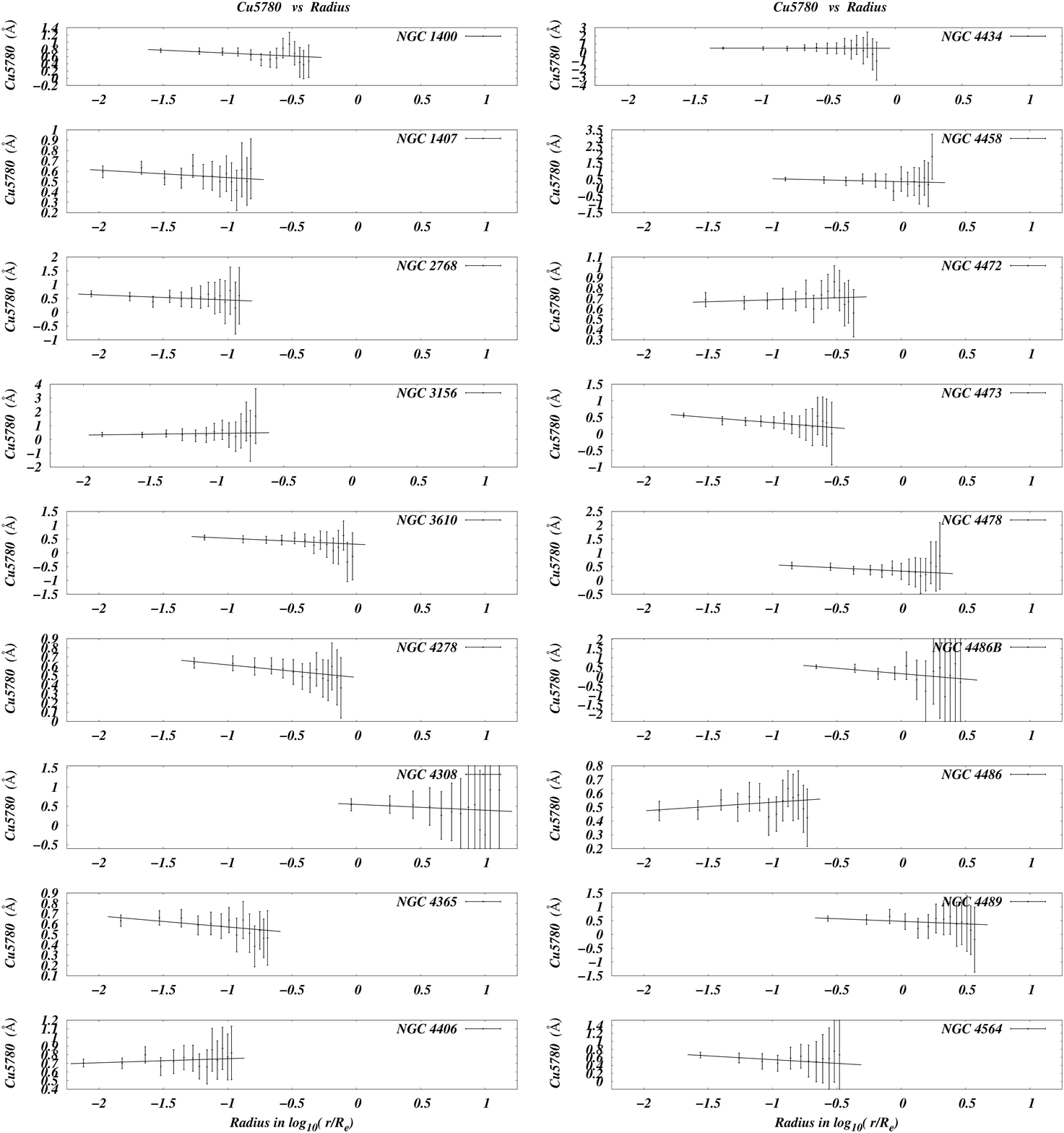}
\caption{}
\end{figure}

\begin{figure}[H]
\includegraphics[width=6in,height=7in]{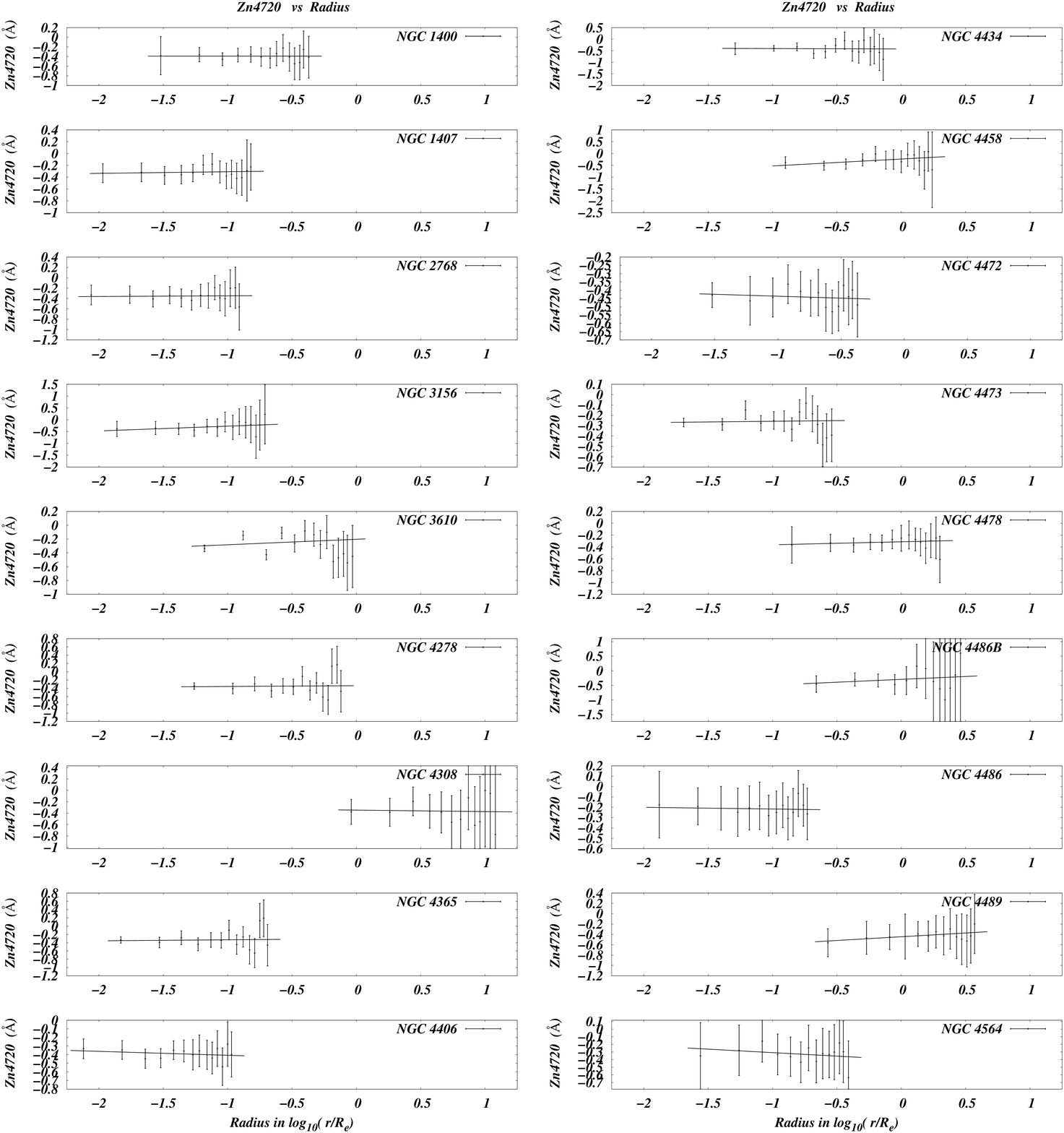}
\caption{}
\end{figure}

\begin{figure}[H]
\includegraphics[width=6in,height=7in]{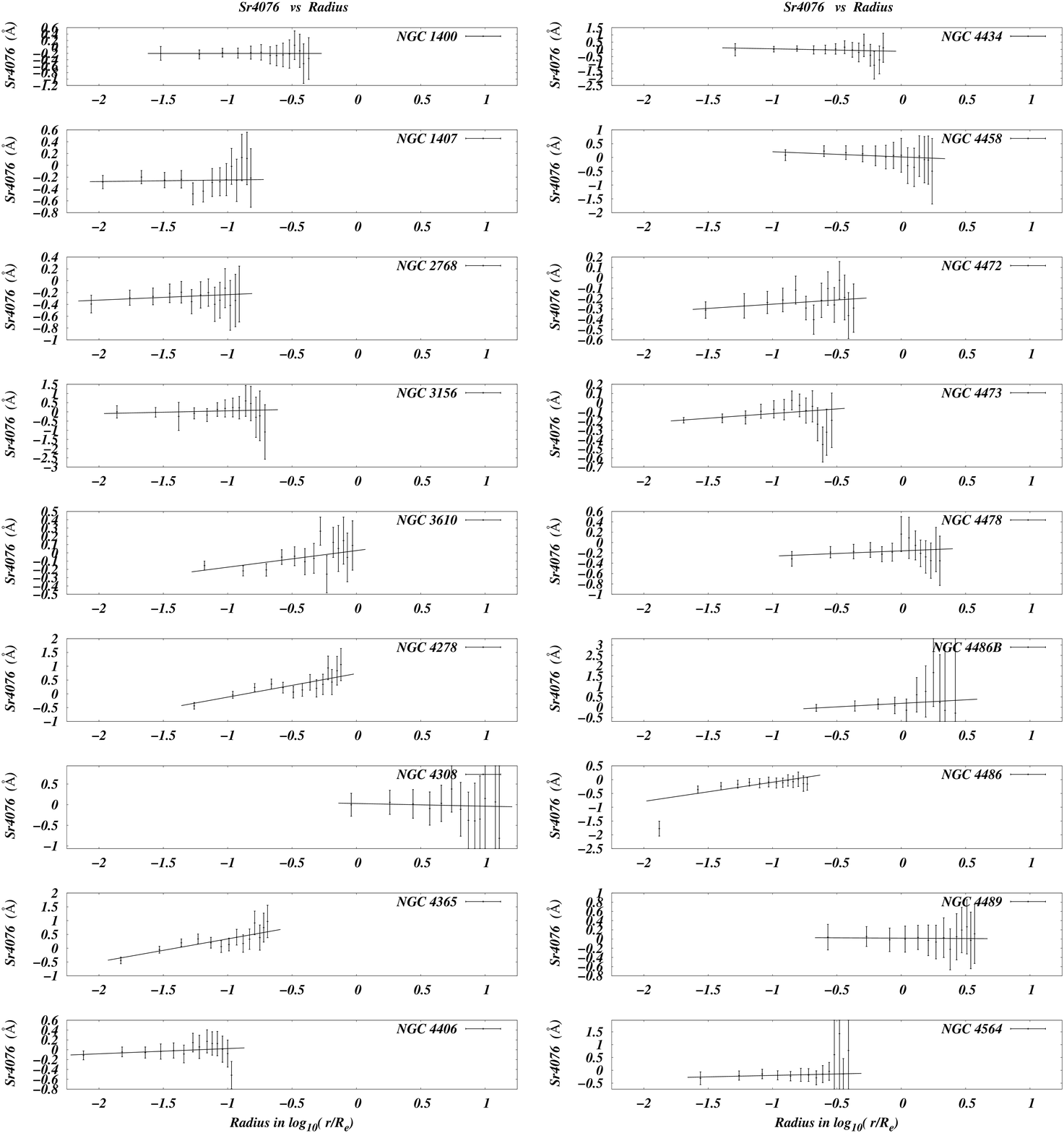}
\caption{}
\end{figure}

\begin{figure}[H]
\includegraphics[width=6in,height=7in]{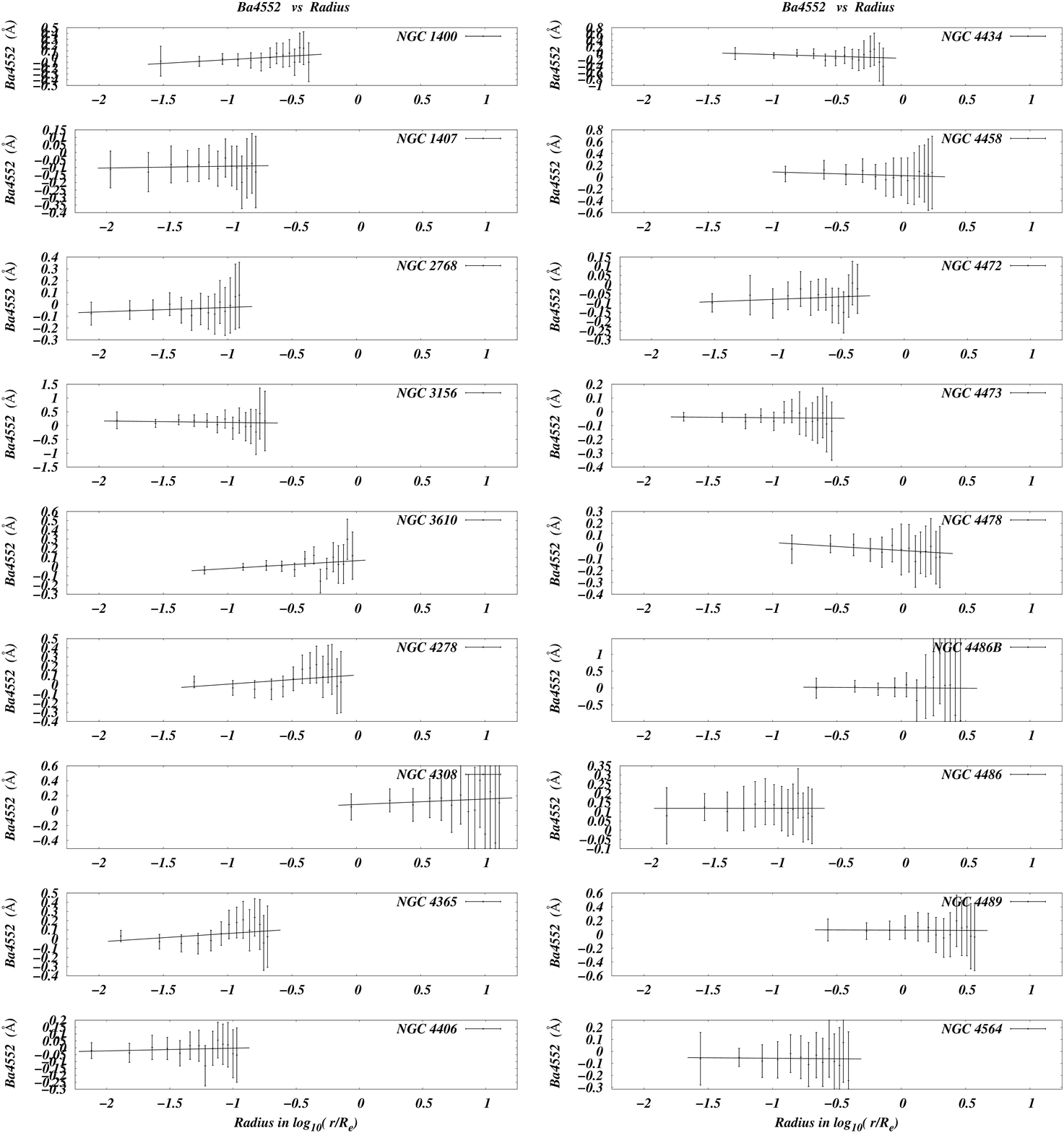}
\caption{}
\end{figure}

\begin{figure}[H]
\includegraphics[width=6in,height=7in]{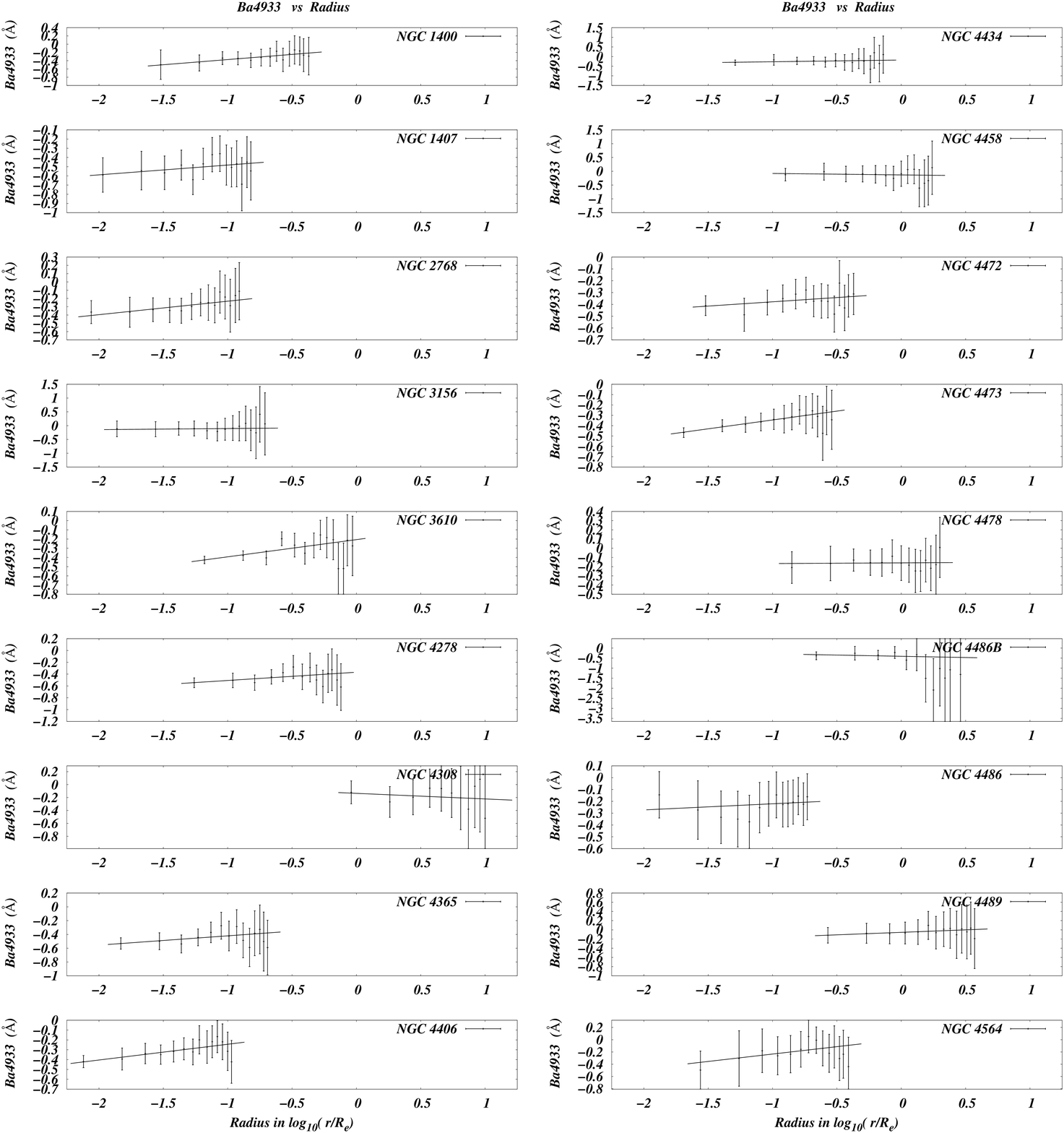}
\caption{}
\end{figure}

\begin{figure}[H]
\includegraphics[width=6in,height=7in]{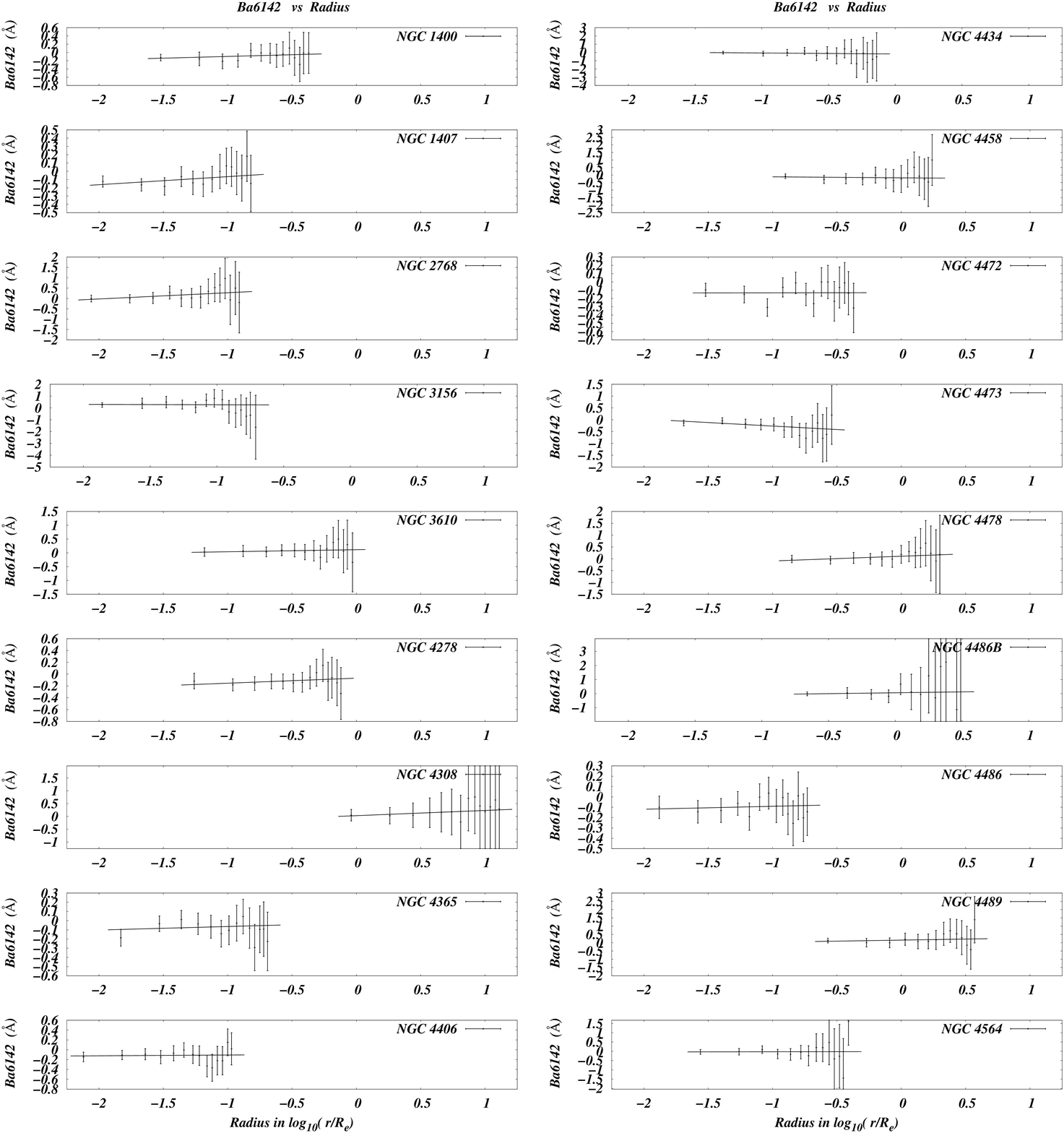}
\caption{}
\end{figure}

\begin{figure}[H]
\includegraphics[width=6in,height=7in]{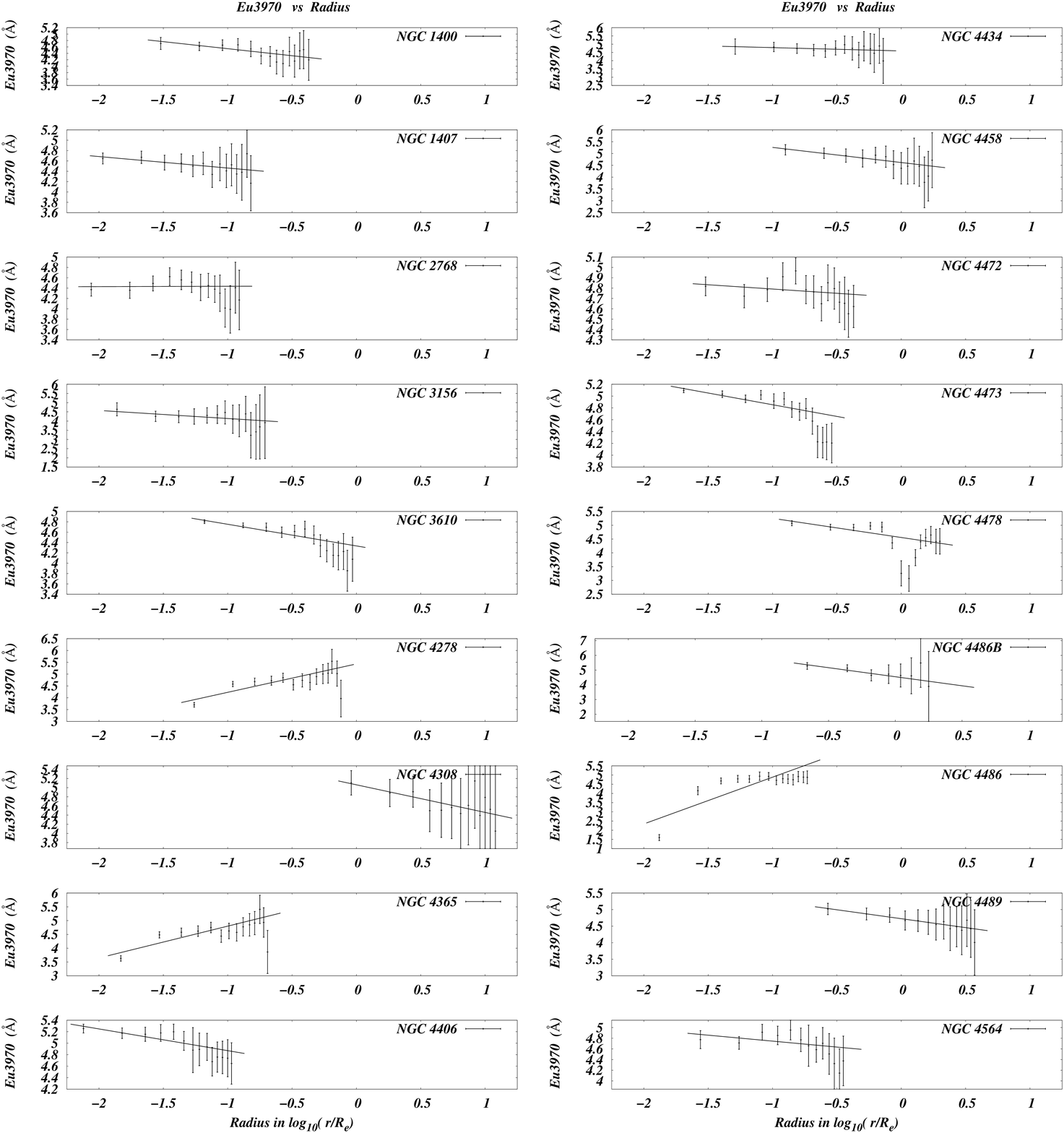}
\caption{}
\end{figure}

\begin{figure}[H]
\includegraphics[width=6in,height=7in]{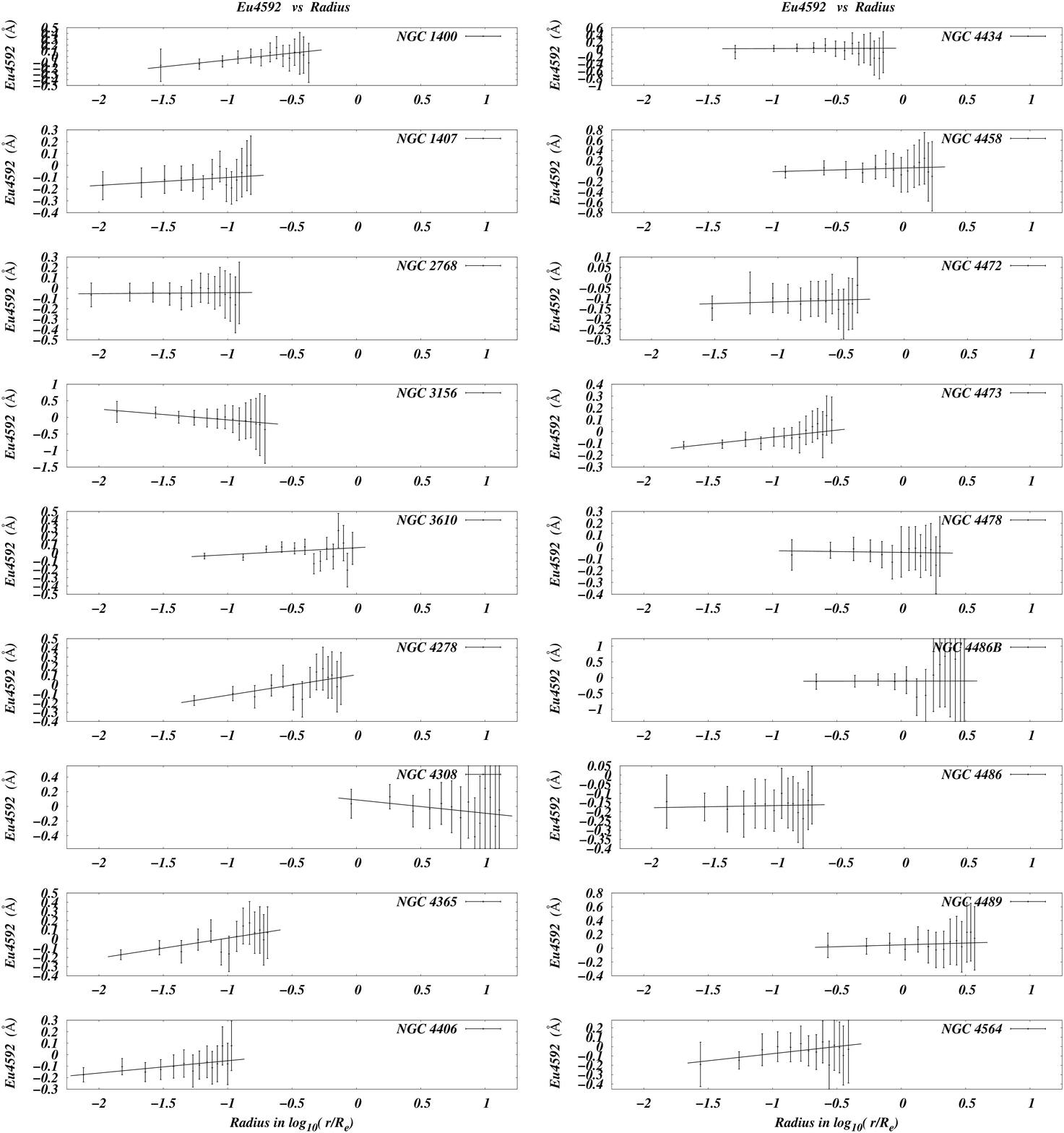}
\caption{}
\end{figure}

\end{document}